\documentclass[prd,superscriptaddress,a4paper,10.5pt,nofootinbib,eqsecnum]{revtex4}

\usepackage{amsfonts,amssymb}
\usepackage{graphicx}
\usepackage{graphics}
\usepackage{epsfig}
\usepackage{dcolumn}
\usepackage{bm}
\usepackage{multirow}
\usepackage{xspace}
\usepackage[export]{adjustbox}

\usepackage{tabularx}
\usepackage{xfrac}
\usepackage{braket}
\usepackage{commath}
\usepackage{epstopdf}
\usepackage[T1]{fontenc}
\usepackage{geometry}
\usepackage{color}
\usepackage{threeparttable}
\usepackage{booktabs}

\usepackage[colorlinks = true,
  linkcolor = blue,
  urlcolor = blue,
  citecolor = blue,
  anchorcolor = blue]{hyperref}

\providecommand{\QQbar}{Q\overline{Q}}
\newcommand{\Vcs}{|V_\mathrm{cs}|}

\newcommand{\Vcd}{|V_\mathrm{cd}|}
\providecommand{\ccbar}{c\overline{c}}
\providecommand{\bbbar}{b\overline{b}}
\providecommand{\ttbar}{t\overline{t}}

\newcommand{\meff}{m{_\mathrm{eff}}}

\newcommand{\alphas}{\alpha_{S}}
\newcommand{\as}{\ensuremath{\alphas}} 
\newcommand{\alphasmZ}{\alphas(m^{2}_\mathrm{Z})}
\newcommand{\asmz}{\alphasmZ}
\newcommand{\lqcd}{\Lambda_{\mathrm{QCD}}}
\newcommand{\MSbar}{\overline{\mathrm{MS}}}
\newcommand{\msbar}{\overline{\mathrm{MS}}}

\newcommand{\pT}{\ensuremath{p_\mathrm{T}}}
\newcommand{\HT}{\ensuremath{H_\mathrm{T}}}
\newcommand{\ptmax}{\ensuremath{p_\mathrm{T}^\mathrm{max}}}
\newcommand{\pTmax}{\ensuremath{p_\mathrm{T}^\mathrm{max}}}
\newcommand{\pTmin}{\ensuremath{p_\mathrm{T}^\mathrm{min}}}

\newcommand{\sqrts}{\sqrt{s}}

\newcommand{\epem}{{\rm e^+e^-}}

\newcommand{\MW}{m_\mathrm{W}}
\newcommand{\MZ}{m_\mathrm{Z}}

\newcommand{\gfitter}{\textsc{gfitter}}

\newcommand{\tr}{\ensuremath{\mathop\mathrm{tr}}}
\renewcommand{\Re}{\ensuremath{\mathop\mathrm{Re}}}

\newcommand{\Rlz}{\mathrm{R}_\mathrm{Z}}
\newcommand{\RZexp}{\mathrm{R}_\mathrm{Z}^\mathrm{exp}}
\newcommand{\RW}{\mathrm{R}_\mathrm{W}}
\newcommand{\so}{\sigma_\mathrm{Z}^\mathrm{had}}
\newcommand{\GZ}{\Gamma_\mathrm{Z}^\mathrm{tot}}

\newcommand{\eg}           {e.g.}
\newcommand{\ie}           {i.e.}

\def\ttt#1{\texttt{\small #1}}

\newcommand{\hrefurl}[1]   {\href{#1}{\url{#1}}}

\newcommand{\com}[1]       {}

\def\lQ{\Lambda_\mathrm{QCD}}
 
\def\MS{\overline{\rm MS}}

\def\MeV{{\rm MeV}}
\def\GeV{{\rm GeV}}

\def\fm{{\rm fm}}
\def\Nf{{N_\mathrm{f}}}
\def\Nl{{N_{\ell}}}
\def\Nh{{N_\mathrm{h}}}
\def\Obs{{O}}

\def\calO{{\mathcal O}}

\newcommand{\syst}{\ensuremath{\,\text{(syst)}}\xspace}
\newcommand{\lum}{\ensuremath{\,\text{(lumi)}}\xspace}
\newcommand{\stat}{\ensuremath{\,\text{(stat)}}\xspace}
\newcommand{\statt}{\ensuremath{\,\text{(num)}}\xspace}

\newcommand{\alphaS}{\alphas}

\newcommand{\zcut}{\ensuremath{z_\text{cut}}}

\newcommand{\Wpm}{\ensuremath{{\rm W^{\pm}}}}

\newcommand{\mcfm}{{\sc mcfm}}

\newcommand{\sherpa}{\textsc{Sherpa}}

\begin{document}

\title{The strong coupling constant: State of the art and the decade ahead\\\vspace{0.2cm} 
}

\author{\textit{Editors:} D.~d'Enterria}
\affiliation{CERN, EP Department, CH-1211 Geneva 23, Switzerland}

\author{S.~Kluth}
\affiliation{Max-Planck-Institut f\"ur Physik, 
80805 M\"unchen, Germany}

\author{G.~Zanderighi}
\affiliation{Max-Planck-Institut f\"ur Physik, 
80805 M\"unchen, Germany}
\affiliation{Physik-Department, Technische Universit\"at M\"unchen, 
85748 Garching, Germany}

\author{\\\textit{Authors:} C.~Ayala}
\affiliation{IAI, Univ.~de Tarapac\'a, Arica, Chile}

\author{M.A.~Benitez-Rathgeb}
\affiliation{University of Vienna, Faculty of Physics, 
1090 Wien, Austria}

\author{J.~Bl\"umlein}
\affiliation{Deutsches Elektronen--Synchrotron DESY, 
15738 Zeuthen, Germany}

\author{D.~Boito}

\affiliation{University of Vienna, Faculty of Physics, 
1090 Wien, Austria}
\affiliation{Instituto de F\'isica de S\~ao Carlos, Universidade de S\~ao Paulo,  CP 369, 13560-970, S\~ao Carlos, SP, Brazil}

\author{N.~Brambilla}
\affiliation{Physik-Department, Technische Universit\"at M\"unchen, 
85748 Garching, Germany}

\author{D.~Britzger}
\affiliation{Max-Planck-Institut f\"ur Physik, 
80805 M\"unchen, Germany}


\author{S.~Camarda}
\affiliation{CERN, EP Department, CH-1211 Geneva 23, Switzerland}

\author{A.~M.~Cooper-Sarkar}
\affiliation{Department of Physics, University of Oxford}

\author{T.~Cridge}
\affiliation{Department of Physics and Astronomy, University College London, London, UK}

\author{G.~Cveti\v{c} }
\affiliation{UTFSM, Valpara\'{\i}so, Chile}

\author{D.~d'Enterria}
\affiliation{CERN, EP Department, CH-1211 Geneva 23, Switzerland}

\author{M.~Dalla Brida}
\affiliation{CERN, TH Department, CH-1211 Geneva 23, Switzerland}

\author{A.~Deur}
\affiliation{Jefferson Lab, Newport News, VA 23606, USA}

\author{F.~Giuli}
\affiliation{CERN, EP Department, CH-1211 Geneva 23, Switzerland}

\author{M.~Golterman}
\affiliation{Department of Physics and Astronomy, 
San Francisco State University, San Francisco, CA 94132, USA}
\affiliation{Department of Physics and IFAE-BIST, Universitat Aut\`onoma de Barcelona, 08193 Bellaterra, Barcelona, Spain}

\author{A.H.~Hoang}
\affiliation{University of Vienna, Faculty of Physics, 
1090 Wien, Austria}
\affiliation{Erwin Schr\"odinger International Institute for Mathematical Physics,
University of Vienna, 1090 Wien, Austria}

\author{J.~Huston}
\affiliation{Department of Physics and Astronomy, Michigan State University, East Lansing, MI 48824 USA}

\author{M.~Jamin}
\affiliation{University of Vienna, Faculty of Physics, 
1090 Wien, Austria}
\affiliation{Central Institute of Mental Health, Medical Faculty Mannheim, Heidelberg University, Mannheim, Germany}

\author{S.~Kluth}
\affiliation{Max-Planck-Institut f\"ur Physik, 
80805 M\"unchen, Germany}

\author{A.~V.~Kotikov}
\affiliation{Joint Institute for Nuclear Research, Dubna, Russia}

\author{V.~G.~Krivokhizhin}
\affiliation{Joint Institute for Nuclear Research, Dubna, Russia}

\author{A.S.~Kronfeld}
\affiliation{Theory Division, Fermi National Accelerator Laboratory, Batavia, IL 60510, USA }

\author{V.~Leino}
\affiliation{Physik-Department, Technische Universit\"at M\"unchen, 
85748 Garching, Germany}

\author{K.~Lipka}
\affiliation{Deutsches Elektronen--Synchrotron DESY, 
22607 Hamburg, Germany}

\author{T.~M{\"a}kel{\"a}}
\affiliation{Deutsches Elektronen--Synchrotron DESY, 
22607 Hamburg, Germany}

\author{B.~Malaescu}
\affiliation{LPNHE, Sorbonne Universit\'e, Universit\'e Paris Cit\'e, CNRS/IN2P3, Paris, France}

\author{K.~Maltman}
\affiliation{Department of Mathematics and Statistics,
York University,  Toronto, ON Canada M3J~1P3}
\affiliation{CSSM, University of Adelaide, Adelaide, SA~5005 Australia}

\author{S.~Marzani}
\affiliation{Dipartimento di Fisica, Universit\`a di Genova and INFN, Sezione di Genova, Via Dodecaneso 33, 16146, Genoa, Italy}

\author{V.~Mateu}
\affiliation{Dept.\ de F\'isica Fundamental e IUFFyM, Universidad de Salamanca, 37008 Salamanca, Spain}
\affiliation{Instituto de F\'isica Te\'orica UAM-CSIC, 28049 Madrid, Spain}

\author{S.~Moch}
\affiliation{II.~Institute for Theoretical Physics, Hamburg University, 
D-22761 Hamburg, Germany}

\author{P.~F.~Monni}
\affiliation{CERN, TH Department, CH-1211 Geneva 23, Switzerland}

\author{P.~Nadolsky}
\affiliation{Department of Physics, Southern Methodist University, Dallas, TX 75275-0181}

\author{P.~Nason}
\affiliation{Max-Planck-Institut f\"ur Physik, 
80805 M\"unchen, Germany}
\affiliation{Universit\`a di Milano\,-\,Bicocca and INFN, Sezione di Milano\,-\,Bicocca, 
20126 Milano, Italy}

\author{A.V.~Nesterenko}
\affiliation{Joint Institute for Nuclear Research, Dubna, Russia}

\author{R.~P\'erez-Ramos}
\affiliation{DRII-IPSA, Bis, 63 Boulevard de Brandebourg, 94200 Ivry-sur-Seine, France}
\affiliation{Sorbonne Universit\'es, UPMC Univ Paris 06, UMR 7589, LPTHE, F-75005, Paris, France}

\author{S.~Peris}
\affiliation{Department of Physics and IFAE-BIST, Universitat Aut\`onoma de 
Barcelona, 08193 Bellaterra, Barcelona, Spain}

\author{P.~Petreczky}
\affiliation{BNL, Upton, New-York}

\author{A.~Pich}
\affiliation{IFIC, CSIC-Universitat de Val\`{e}ncia, 46980 Val\`{e}ncia, Spain}

\author{K.~Rabbertz}
\affiliation{KIT, Karlsruhe}

\author{A.~Ramos}
\affiliation{IFIC, CSIC-Universitat de Val\`{e}ncia, 46980 Val\`{e}ncia, Spain}

\author{D.~Reichelt}
\affiliation{Inst. for Particle Phys. Phenomenology, Dept.\ of Physics, Durham Univ., Durham DH1 3LE, UK}

\author{A.~Rodr\'iguez-S\'anchez}
\affiliation{Université Paris-Saclay, CNRS/IN2P3, IJCLab, 91405 Orsay, France}

\author{J.~Rojo}
\affiliation{Department of Physics and Astronomy, VU Amsterdam, 1081HV Amsterdam, The Netherlands}
\affiliation{Nikhef Theory Group, Science Park 105, 1098XG Amsterdam, The Netherlands}

\author{M.~Saragnese}
\affiliation{Deutsches Elektronen--Synchrotron DESY, 
15738 Zeuthen, Germany}

\author{L.~Sawyer}
\affiliation{Louisiana Tech University, Ruston, LA, USA}

\author{M.~Schott}
\affiliation{Johannes Gutenberg-Universitat Mainz (JGU), Saarstr. 21, 55122 Mainz, Germany}

\author{S.~Schumann}
\affiliation{Institut f{\"u}r Theoretische Physik, Georg-August-Universit{\"a}t G\"ottingen, 
37077 G\"ottingen, Germany}

\author{B.~G.~Shaikhatdenov}
\affiliation{Joint Institute for Nuclear Research, Dubna, Russia}

\author{S. Sint}
\affiliation{School of Mathematics and Hamilton Mathematics Institute, Trinity College Dublin, Dublin 2, Ireland}

\author{G.~Soyez}
\affiliation{Institut de Physique Th\'eorique, Paris Saclay University, CEA, CNRS, F-91191 Gif-sur-Yvette, France}

\author{D.~Teca}
\affiliation{UTFSM, Valpara\'{\i}so, Chile}

\author{A.~Vairo}
\affiliation{Physik-Department, Technische Universit\"at M\"unchen, 
85748 Garching, Germany}

\author{M.~Vos}
\affiliation{IFIC, CSIC-Universitat de Val\`{e}ncia, 46980 Val\`{e}ncia, Spain}

\author{C.~Waits}
\affiliation{Louisiana Tech University, Ruston, LA, USA}

\author{J.~H.~Weber}
\affiliation{HU, Berlin, Germany}

\author{M.~Wobisch}
\affiliation{Louisiana Tech University, Ruston, LA, USA}

\author{K.~Xie}
\affiliation{Department of Physics and Astronomy, University of Pittsburgh, Pittsburgh, PA 15260, USA\looseness=-1}

\author{G.~Zanderighi}
\affiliation{Max-Planck-Institut f\"ur Physik, 
80805 M\"unchen, Germany}
\affiliation{Physik-Department, Technische Universit\"at M\"unchen, 
85748 Garching, Germany}

\begin{abstract}
\vspace{2.2cm}
Theoretical predictions for particle production cross sections and decays at colliders rely heavily on perturbative Quantum Chromodynamics (QCD) calculations, expressed as an expansion in powers of the strong coupling constant $\alphas$. The current $\mathcal{O}(1\%)$ uncertainty of the QCD coupling evaluated at the reference Z boson mass, $\alphasmZ = 0.1179 \pm 0.0009$, is one of the limiting factors to more precisely describe multiple processes at current and future colliders. A reduction of this uncertainty is thus a prerequisite to perform precision tests of the Standard Model as well as searches for new physics. This report provides a comprehensive summary of the state-of-the-art, challenges, and prospects in the experimental and theoretical study of the strong coupling. 
The current $\alphasmZ$ world average is derived from a combination of seven categories of observables: (i)~lattice QCD, (ii)~hadronic $\tau$ decays, (iii)~deep-inelastic scattering and parton distribution functions fits, (iv)~electroweak boson decays, hadronic final-states in (v)~$\epem$,  (vi)~e-p, and (vii)~p-p collisions, and (viii)~quarkonia decays and masses.
We review the current status of each of these seven $\alphasmZ$ extraction methods,
discuss novel $\alphas$ determinations, and examine the averaging method used to obtain the world-average value. Each of the methods discussed provides a ``wish list'' of experimental and theoretical developments required in order to achieve the goal of a per-mille precision on $\alphasmZ$ within the next decade. 
\end{abstract}

\maketitle

\renewcommand{\baselinestretch}{1.05}\normalsize
\tableofcontents
\renewcommand{\baselinestretch}{1.0}\normalsize

\section{Introduction}
\label{sec:intro}

The strong coupling $\alphas$ sets the scale of the strength of the strong interaction, theoretically described by Quantum Chromodynamics (QCD), and is one of the fundamental parameters of the Standard Model (SM) of particle physics. In the chiral limit of zero quark masses and for fixed number of colours $N_c=3$, the $\alphas$ coupling is the only free parameter of QCD. Starting at an energy scale of order $\lqcd\approx0.2$~GeV in the vicinity of the infrared Landau pole of the theory, $\alphas(Q)$ approximately decreases as $1/\log(Q^2/\lqcd^2)$, where $Q$ is the energy scale of the underlying QCD process. 
Its value at the reference Z pole mass amounts today to $\alphasmZ=0.1179 \pm 0.0009$~\cite{ParticleDataGroup:2022pth}, with a $\delta\alphas/\alphas \approx 0.8\%$ uncertainty that is orders of magnitude larger than that of the other three interaction (QED, weak, and gravitational) couplings.\\

\begin{figure}[htbp!]
\centering
\includegraphics[width=0.8\textwidth]{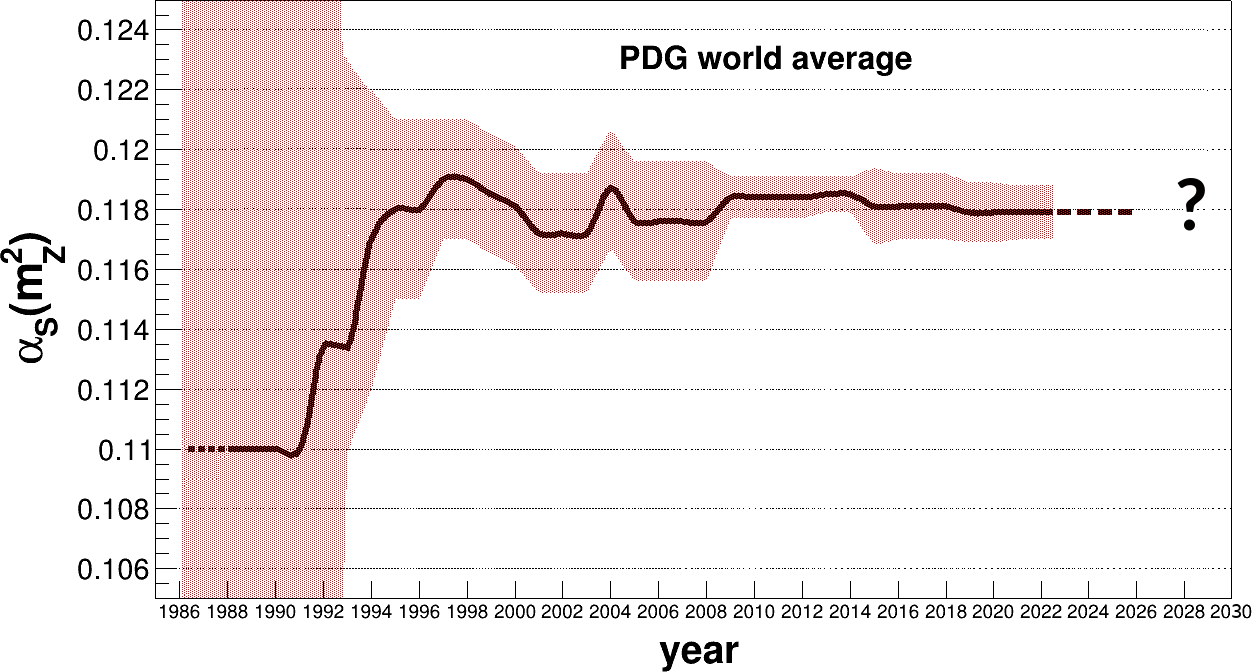}
\caption{Evolution of the average value of the QCD coupling at the Z boson mass scale (with the red bands indicating its associated uncertainty) in the $\MSbar$ renormalization scheme, quoted in the Particle Data Group (PDG) review covering the last four and the current decades~\cite{ParticleDataGroup:2022pth}.
\label{fig:alphas_historic}}
\end{figure}

Our knowledge of the QCD coupling has improved throughout the years (Fig.~\ref{fig:alphas_historic}), from a $\mathcal{O}(100\%)$ uncertainty when it was first constrained from data-versus-theory comparisons at next-to-leading order accuracy in the mid 1980s -- exploiting, already then, a variety of observables (deep-inelastic scattering cross sections, total hadronic cross section in $\epem$ annihilation, the distribution of 3-jet events in $\epem$ collisions, the energy flow of energy in $\epem$ annihilation, $\Upsilon$ branching fractions, the behavior of baryon form factors at large momentum transfer, measurements of the photon structure functions, and the hyperfine splittings in the J/$\psi$ state)~\cite{ParticleDataGroup:1986kuw}-- to the present $\mathcal{O}(1\%)$ precision~\cite{ParticleDataGroup:2022pth}. 
Further improving our knowledge of $\alphas$ is fundamental, among other things, to reduce the theoretical ``parametric'' uncertainties in the calculations of all perturbative QCD (pQCD) processes whose cross sections or decay rates depend on powers of $\alphas$, as is the case for virtually all those measured in proton and nuclear collisions at the Large Hadron Collider (LHC), as well as in $\epem$ annihilation at future high-precision colliders. In the Higgs sector, our imperfect knowledge of $\alphas$ propagates today into total final uncertainties for key processes such as the Higgs $gg$-fusion and $\ttbar$-associated production cross sections of 
$\sim$2-3\%~\cite{Anastasiou:2016cez,Cooper-Sarkar:2020twv}, or of $\sim$4\% 
for the H\,$\to gg$ partial decay width~\cite{Proceedings:2019vxr,Heinemeyer:2021rgq}. In the electroweak (EW) sector of the SM, the input $\alphasmZ$ value is the leading source of uncertainty in the computation of crucial precision pseudo-observables such as the total and partial hadronic Z boson widths~\cite{Blondel:2018mad,Heinemeyer:2021rgq}. The QCD coupling plays also a fundamental role in the calculation of key quantities in top-quark physics, such as the top mass, width, and its Yukawa coupling~\cite{Hoang:2020iah}. Last but not least, the value of $\alphasmZ$ and its energy evolution have also far-reaching implications including the stability of the electroweak vacuum~\cite{Chetyrkin:2016ruf}, the existence of new coloured sectors at high energies~\cite{Llorente:2018wup}, and in our understanding of physics approaching the Planck scale, such as \eg\ in the precise energy at which the interaction couplings may unify.\\

This report, co-authored by experimental and theoretical experts from all relevant subfields who participated at the $\alphas(2022)$ workshop in Feb.~2022 (ECT$^*$, Trento, \ttt{https://indico.cern.ch/e/alphas2022}), explores in depth the latest developments in the determination of $\alphas$ from the key categories where high-precision measurements and calculations are currently available. 
%
The following main questions are addressed in detail for each of the $\alphas$ extraction methods: What is the current state-of-the-art? What is the expected theoretical and experimental precision in about ten years from now (indicated by the interrogation symbol in Fig.~\ref{fig:alphas_historic}), and what needs to be achieved in order to reach a $\mathcal{O}(0.1\%)$ precision? 
%
In particular, this report examines, for the different calculations of $\alphas$-sensitive observables involved, what the current state-of-the-art is with regards to higher-order (pQCD, mixed QCD-EW) corrections, and what the impact of nonperturbative corrections/uncertainties is. Whenever there are new ideas or techniques to reduce them, these are illustrated. %
From an experimental point of view, the report discusses what the current leading systematic and statistical uncertainties of the $\alphas$-sensitive observables are, and what future reductions of them are expected with current (pp) and future ($\epem$, e-p) machines. New observables are also suggested.\\

The review is organized as follows. Sections~\ref{sec:latt}--\ref{sec:QQbar} discuss $\alphas$ determinations based on, consecutively, lattice-QCD methods; hadronic tau-lepton decays; deep-inelastic scattering and parton densities fits; electroweak fits; hadronic final states in $\epem$, and in e-p and p-p collisions; and quarkonium bound states. Section~\ref{sec:average} discusses the averaging method used to currently obtain the world-average $\alphasmZ$ value. The last section~\ref{sec:summ} ends with a summary of the discussions of the $\alphas(2022)$ workshop, and with a ``wish-list'' assessment in data/theory developments needed to reach a precision of $\alphas$ at the per-mille level in the upcoming years.




\clearpage
\section{\texorpdfstring{\boldmath$\alphasmZ$ from lattice QCD}{alphasmZ from lattice QCD}}
\label{sec:latt}

\subsection{Prospects of lattice determinations of \texorpdfstring{$\alphas$}{alphas} from the FLAG perspective \protect\footnote{A\lowercase{uthors:} S. S\lowercase{int} (T\lowercase{rinity} C\lowercase{ollege} D\lowercase{ublin})}}

\subsubsection{Overview of the current situation}

FLAG stands for ``Flavour Lattice Averaging Group'' and constitutes an effort by the lattice QCD community to supply qualified information on lattice results for selected physical quantities to the wider particle physics community. These include the QCD parameters, \ie\ $\alphas$ and the quark masses. 
An extensive report aimed at a general particle physics audience is
published every 2--3 years~\cite{Aoki:2021kgd,FlavourLatticeAveragingGroup:2019iem} and in the meantime occasional updates are made to the online version maintained at the University of Bern\footnote{\url{http://flag.unibe.ch}}. 
We strongly encourage readers to explore this report, in particular, 
the $\alphas$ chapter. For a complementary pedagogical introduction 
cf.\ the review~\cite{DelDebbio:2021ryq}.

In the FLAG $\alphas$ working group we have recently provided the updated lattice QCD average
\begin{equation}
  \alphasmZ = 0.1184(8) \,, 
  \label{eq:FLAG21}
\end{equation}
based on results published\footnote{It is FLAG policy to require that the original works entering any FLAG 
averages to be always cited alongside the FLAG report!} 
in~\cite{Maltman:2008bx,Aoki:2009tf,McNeile:2010ji,Chakraborty:2014aca,Bruno:2017gxd,Bazavov:2019qoo,Ayala:2020odx,Cali:2020hrj}.
This represents a minimal change from FLAG 2019 \cite{FlavourLatticeAveragingGroup:2019iem}, 
and no reduction in the error. The FLAG criteria for $\alphas$ have remained unchanged since FLAG 2019 and there are now some indications that the criteria may need to be revised in the future~\cite{Aoki:2021kgd}.

Lattice determinations of $\alphas$  use up, down, and strange quarks (and 
sometimes the charm quark) in the sea, and perturbatively evolve across the charm and bottom thresholds 
to obtain $\alphas$ in the 5-flavour theory.
The perturbative matching across quark thresholds has been put to a nonperturbative test in~\cite{Athenodorou:2018wpk} 
which demonstrates that the perturbative description of decoupling 
(known to 4-loop order~\cite{Bernreuther:1981sg,Grozin:2011nk,Chetyrkin:2005ia,Schroder:2005hy,Kniehl:2006bg,Gerlach:2018hen}) 
provides an excellent quantitative description even for the charm quark. 
Hence, one may avoid the potentially large cutoff effects
associated with the charm quark mass, which is usually not so small compared 
to the lattice cutoff scale $1/a$ ($a$ denotes the lattice spacing).

A lattice determination of $\alphas$  starts with the choice of an observable $O(\mu)$ 
depending on a single scale $\mu$ with a perturbative expansion of the form
\begin{equation}
   O(\mu) =  c_0 + c_1 \alpha_{\msbar}(\mu) + c_2 \alpha_{\msbar}^2(\mu)\, + \ldots
\end{equation}
It is convenient to normalize the observable as an effective coupling 
\begin{equation}
 \alpha_\text{eff} =  (O(\mu) -c_0)/c_1 = \alpha_{\msbar} + d_1 \alpha^2_{\msbar}\,+ \ldots
\end{equation}
We then refer to $d_1 = c_2/c_1$  as the 1-loop matching coefficient, even though 
some choices of $O(\mu)$ may require a 1-loop-diagram to obtain $c_0$ and thus 2 more loops to obtain $d_1$.  
It should be clear that each choice of $O(\mu)$ leads to a different lattice
determination of $\alphas$, in close analogy to phenomenological determinations.
The main difference therefore is whether the original data are produced by
a lattice simulation or taken from experiment. A bonus of the lattice setup is the 
access to observables which are not experimentally measurable, for instance,
one may study QCD in a finite Euclidean space-time volume $L^4$ and define
the observable $O$ through a finite volume effect~\cite{Luscher:1992an}. 
However, many renormalization schemes for the coupling assume large volume, 
\ie\ in practice one needs to show that the necessarily finite $L$ is causing 
negligible effects on the chosen observable. Note also that quark masses can be varied, 
and one may naturally define mass-independent couplings
(such as the $\msbar$ coupling) by imposing renormalization conditions in the chiral limit~\cite{Weinberg:1973xwm}.

A common problem of most lattice determinations of $\alphas$ has been dubbed
the ``window problem'': In order to match to hadronic physics, the spatial volume $L^3$ must be large enough to avoid significant finite volume effects due to pion polarization ``around the world''. On the other hand, the matching to the coupling requires perturbative expansions to be reliable, so one needs to reach as high a scale $\mu$ as possible, but still significantly below the cutoff scale $1/a$ 
such as to avoid large lattice distortions. If taken together this means
\begin{equation}
   \frac{1}{L} \ll m_\pi \ll \mu \ll \frac{1}{a} \quad \Rightarrow \quad \frac{L}{a} = \mathcal{O}(10^3),  
\end{equation}
(where the hadronic scale $m_{\pi}$ is the pion mass) which is just a reflection of the fact that very different energy scales cannot be resolved simultaneously on a single lattice of reasonable size\footnote{$L/a=100$ would be considered a large lattice by today's standards.}. In addition, the continuum limit $a\to 0$ requires a range of lattice sizes satisfying the above constraint, and one would like to have a range of scales $\mu$ such as to verify that the perturbative regime has been reached. This window problem enforces various compromises; in most cases the energy scales reached for perturbative matching to the $\msbar$-coupling is therefore rather low. As a consequence, even in the best cases (with 3-loop matching to the $\msbar$-coupling) systematic errors due to truncation of the perturbative series and/or contributions from nonperturbative effects 
are dominant~\cite{Aoki:2021kgd,DelDebbio:2021ryq}.

A solution to the window problem is however known since the 1990's, in the form
of the step-scaling method~\cite{Luscher:1991wu}. The method is based on a finite volume
renormalization scheme with $\mu=1/L$. It is then possible to recursively step up the energy scale by 
a fixed scale factor $s=2$, and a scale difference of $\mathcal{O}(10^2)$ is thus covered in 5 to 6 steps. 
The window problem is by-passed, as the approach uses multiple (pairs of) lattices with size $L/a$ and $2L/a$, 
covering a wide range of physical scales $\mu=1/L$ without the need 
to represent simultaneously any hadronic scale, except at the lowest scale reached, 
$\mu_\text{had} = 1/L_\text{max}$. Once the scale $\mu_{\text{had}}$ is matched 
to a hadronic quantity such as the proton mass, all the higher scales are known too, 
as the scale ratios are powers of two. At the high energy end, now orders of
magnitude above the hadronic scales, perturbation theory can be safely applied
to match to the $\msbar$-coupling, or, equivalently to extract the 
3-flavour $\Lambda$-parameter.  The method has been applied in Ref.~\cite{Bruno:2017gxd},
with the result $\Lambda_{\msbar}^{(3)}=341(12)$~MeV, which
translates to $\alphasmZ=0.1185(8)$. It is important to realize that this
is the only lattice determination of $\alphas$ where the error is still statistics dominated. 
For this reason, we quote this error for the FLAG average~(\ref{eq:FLAG21}) 
as a conservative estimate of the uncertainty, 
instead of combining the (mostly systematic) errors in quadrature.

\subsubsection{Future prospects and conclusions}

Most lattice determinations of $\alphas$ are now limited by systematic errors, due
to the relatively low energy scales where perturbation theory is applied, and
the limited range of available energy scales. 
The one exception is the step-scaling method which enables the nonperturbative
scale evolution up to very high energies. Perturbation theory can be tested 
and then safely applied.

Is it possible to incorporate at least some elements of the step-scaling method into some of the 
other lattice determinations? In its original form, the step-scaling method uses a
finite volume renormalization scheme, \ie\ $\mu=1/L$, which means that
finite volume effects are part of the scheme definition, rather than systematic
effects to worry about. In fact, all lattice determinations of $\alphas$ could incorporate the step
scaling approach by simply working at fixed $\mu L$. Unfortunately, this means that perturbation theory
needs to be set up in a finite volume too, which can be rather complicated, 
depending on the choice of boundary conditions. 
In particular, the perturbative results computed for infinite volume
could not be used anymore, and adapting these to finite volume is no minor change.
As a side remark, it would be highly desirable that experts in perturbation theory
cooperate with lattice QCD practitioners to adapt and develop perturbative techniques 
for some selected finite volume schemes.

In the short to medium term future some progress may still be possible
by taking a few steps with the  step-scaling methods in large volume, perhaps with smaller scale factors, \eg\ $s=3/2$.
This would require that finite volume effects in the chosen observable are controlled
and eliminated at each step. In fact, it is quite plausible that finite volume effects are smaller
in the chosen observables for the coupling than in some of the other hadronic observables. 
An example of this strategy was presented in Ref.~\cite{Husung:2020pxg} for the force between static quarks 
(there for $\Nf=0$ quark flavours). It is worth emphasizing that such progress requires a dedicated effort, 
including computational resources for the production of additional
gauge configurations, which may have spatial volumes too small to serve other goals
of a lattice collaboration. In the past this need for dedicated simulations has been a practical
obstacle in some cases, but it seems now evident that it cannot be avoided if real progress
is to be achieved.

The required computational resources for the full step-scaling method are
substantial, too. In particular the bulk of the statistical error 
in the step-scaling result of~\cite{Bruno:2017gxd}
is accumulated by the scale evolution at {\em high} energy scales.
While a further reduction of the error would be feasible by brute force, the ALPHA collaboration
has instead proposed the decoupling strategy~(\cite{DallaBrida:2019mqg}, 
for an introduction cf.~\cite{DallaBrida:2020pag}). The scale evolution is traced in the $\Nf=0$ 
pure gauge theory,
with less resources and to better precision than in $\Nf=3$ QCD~\cite{DallaBrida:2019wur,Nada:2020jay}. 
When combined with a nonperturbative computation for the simultaneous decoupling of $\Nf=3$ dynamical quarks, 
the resulting error is currently comparable to the direct $\Nf=3$ result of 
Ref.~\cite{Bruno:2017gxd} and still statistics dominated~\cite{DallaBrida:2022eua}. I refer to Sec.~\ref{sec:latt_DallaBrida_Ramos}
for details. 
One aspect of this strategy is the importance acquired by results in the $\Nf=0$ theory. 
Rather than being a mere test bed for the development of new methods, 
the pure gauge theory now indirectly contributes to $\alphas$. 
The FLAG $\alphas$ working group will therefore keep monitoring $\Nf=0$ results for the $\Lambda$-parameter.

In conclusion, significant progress in lattice determinations of $\alphasmZ$ will require at least some elements
of the step-scaling method in order to reach larger energy scales and at least partially evade the window problem.
A total error clearly below half a percent for $\alphasmZ$ seems achievable within the next few years
by pushing the step-scaling method further, possibly in combination with the decoupling strategy. In order to corroborate such results it is very desirable to apply the step-scaling method to further observables in a finite space-time volume. Developing the necessary perturbative techniques then constitutes a challenge where cooperation with experts in perturbation theory might have a significant impact.\\

\noindent{\it Acknowledgments---}
I thank my colleagues in the FLAG $\alphas$-WG, Peter Petreczky and Roger Horsley for the pleasant collaboration and feedback on a draft of this contribution. Partial support by the EU unter grant agreement H2020-MSCA-ITN-2018-813942 (EuroPLEx) is gratefully acknowledged.

\subsection{A precise determination of \texorpdfstring{$\alphas$}{alphas} from lattice QCD using decoupling of heavy quarks
\protect\footnote{A\lowercase{uthors:} M. D\lowercase{alla} B\lowercase{rida} (CERN), A. R\lowercase{amos} (IFIC, V\lowercase{al\`encia})}}
\label{sec:latt_DallaBrida_Ramos}

The determination of the strong coupling using lattice QCD uses a
nonperturbatively defined quantity $\Obs(q)$ that depends on a single
short distance scale $1/q$. Using the perturbative expression for this
quantity\footnote{It is convenient to normalize the observable $\Obs(q)$
  so that the perturbative expansion starts with $\alphas(q)$.}
\begin{equation}
  \label{eq:Oq}
          \Obs(q) \overset{q\to\infty}{\sim} \alphas(q) + \sum_{n=2}^Nc_n\alphas^n(q) +
          \calO\left(  \alphas^{N+1}(q)\right)  + {\calO}\left( \frac{\Lambda}{q} \right)^p + \dots\,,
\end{equation}
where $N$ is the number of known coefficients of the perturbative series, 
one can estimate the value of $\alphas(q)$. 
On the lattice, apart from the value of $\Obs(q)$, one needs to determine
the value of the scale $q$ in units of some well-measured hadronic
quantity (\eg~the ratio $q/m_p$ with $m_p$ being the proton mass).

There are two types of corrections in Eq.~(\ref{eq:Oq}). 
First, we have the nonperturbative (``power'') corrections. 
They are of the form $(\Lambda/q)^p$. Second, we have the perturbative
corrections. Their origin is the truncation of the perturbative
series to a finite order $N$, and parametrically these corrections are of the form
$\alphas^{N+1}(q)$. 
\begin{figure}[htpb!]
  \centering
  \includegraphics[width=\textwidth]{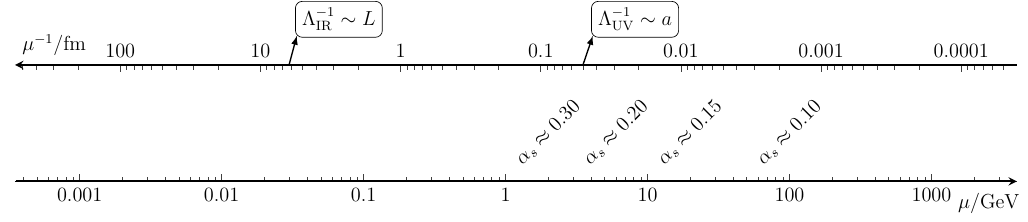}
  \caption{Lattice QCD simulations come with an IR cutoff $\Lambda_{\rm IR}=L^{-1}$ (where $L$
   is the size of the finite space-time volume of the simulation), and an ultraviolet cutoff $\Lambda_{\rm UV}=a^{-1}$
   (where $a$ is the lattice spacing). All relevant physical scales must lie far away 
   from both these cutoffs in order to be free from finite-volume and discretization effects. 
   As a result, given the current computational and algorithmic capabilities, if one wants to compute 
   some hadronic quantities, the range of high-energy scales that can simultaneously be probed
   is limited.}
  \label{fig:lqcdscales}
\end{figure}

In principle, both kind of uncertainties decrease by taking $q\to\infty$. 
Unfortunately, though, a single lattice simulation can only cover a
limited range of energy scales (Fig.~\ref{fig:lqcdscales}). Thus,
if one insists on determining the hadronic scale (\eg~$m_p$) and the value of
the observable $\Obs(q)$ using the \emph{same} lattice simulation, the
volume has to be large, $L \gtrsim 1/m_\pi$, and therefore the energy
scales $q$ that can be reached are at most a few GeV. Power corrections 
decrease quickly with the energy scale $q$, but due to the logarithmic 
dependence of the strong coupling with the energy scale,
\begin{equation}
  \alphas(q) \overset{q\to\infty}{\sim}  \frac{1}{\log (q/\Lambda_{\rm QCD})}\,,
\end{equation}
perturbative uncertainties decrease very slowly. 
In fact, most lattice QCD extractions of the strong coupling are
dominated by the truncation uncertainties of the perturbative series. 

What can be said about such perturbative uncertainties? First, it has
to be noted that due to the asymptotic nature of perturbative
expansions, it is in general very difficult to estimate the difference
between the truncated series for $\Obs(q)$ and its nonperturbative value
(see the original works~\cite{DallaBrida:2018rfy, Brida:2016flw} and
the review~\cite{DelDebbio:2021ryq}). 
Second, an idea of the size of such uncertainties can be obtained 
by using the scale-variation method. 
If one assumes that at scales $q$ power corrections are negligible,
one can use an arbitrary renormalization scale in the perturbative
expansion Eq.~(\ref{eq:Oq}), 
\begin{equation}
  \label{eq:Oqs}
          \Obs(q) \overset{q\to\infty}{\sim} \alphas(\mu) + \sum_{n=2}^Nc_n(\mu/q)\alphas^n(\mu) +
          \calO\left(  \alphas^{N+1}(\mu)\right)  + \dots \,.
\end{equation}
The dependence on $\mu$ on the r.h.s.~is spurious, and due to the
truncation of the perturbative series. 
This dependence can be exploited to estimate the truncation effects:
the value of $\alphas$ can be extracted for different choices of
$\mu$, and the differences among these extractions will give us an estimate 
of the effects due to missing higher-order pQCD corrections. Moreover, following Ref.~\cite{DelDebbio:2021ryq}, 
the estimate of the truncation uncertainties \emph{can be obtained from 
the known perturbative coefficients alone}. 
No nonperturbative data for $\Obs(q)$ is needed to estimate these uncertainties. 

In Ref.~\cite{DelDebbio:2021ryq} a detailed analysis of several lattice
methods to extract the strong coupling is performed along these lines. 
What are their conclusions? Most ``large volume'' approaches (those
that aim at computing the scale $q$ in physical units and the
observable $\Obs(q)$ using the same lattices) have perturbative
uncertainties between about $1\%$ and $3\%$ in $\alphasmZ$.

It is important to emphasize that this generic approach cannot say what 
the errors of a specific determination are. However, given the fact that 
\emph{scale uncertainties can underestimate the true
truncation errors} (see Ref.~\cite{DallaBrida:2018rfy} for a concrete
example), 
this exercise draws a clear picture: A
\emph{substantial reduction in the uncertainty of the strong coupling will
only come from dedicated approaches},
where the multiscale problem discussed above is solved. 
In this contribution, we will summarize the efforts of the ALPHA
collaboration in solving this challenging problem (see Refs.~\cite{Sommer:2015kza,DallaBrida:2020pag} for a review).

\subsubsection{Finite-volume schemes}

A first step towards solving the difficult multiscale problem of extracting $\alphas$ using 
lattice QCD comes from the following simple realization. The computation of the hadronic quantities 
$m_{\rm had}$, needed for fixing the bare parameters of the lattice QCD action, and the 
determination of the nonperturbative coupling $\alpha_\Obs(q)\equiv\Obs(q)$ at large $q$, from 
which we extract $\alphas$, are two distinct problems.%
\footnote{In this and the following sections, we find convenient to use the notation 
$\alpha_\Obs(q)\equiv\Obs(q)$, and interpret the extraction of $\alphas$ as the matching 
between the nonperturbative scheme for the QCD coupling $\alpha_\Obs$ and the $\MSbar$-scheme.}
Hence, in order to best keep all relevant uncertainties under control, we need dedicated lattice 
simulations for the calculation of $\alpha_\Obs(q)$. In fact, as mentioned above, despite being 
convenient in practice, determinations of $\alpha_\Obs(q)$ based on simulations originally intended 
for the computation of low-energy quantities come with severe limitations on the energy 
scales $q$ at which $\alphas$ can be extracted. As a concrete example, consider a typical 
state-of-the-art hadronic lattice simulation, with \eg~$L/a=128$ points in each of the 
four space-time dimensions, and a spatial size $L$ large enough to comfortably fit all the 
relevant low-energy physics, say, $m_\pi L\approx 4$ with $m_\pi\approx135\,\MeV$. This results in 
a lattice spacing $a\approx 0.045\,\fm$, which sets the constraint, $q\ll a^{-1}\approx 4\,\GeV$. 
With such a low upper-limit on $q$, reaching high precision on $\alphas$ is very likely 
impeded by the systematic uncertainties related to perturbative truncation errors and 
nonperturbative corrections.

The most effective way to determine nonperturbatively the coupling $\alpha_\Obs(q)$ at high-energy 
is to consider a \emph{finite-volume} renormalization scheme~\cite{Luscher:1991wu}. These schemes are 
built in terms of observables $\Obs$ defined in a finite space-time volume. The renormalization scale  
of the coupling $q$ is then identified with the inverse spatial size of the finite volume, \ie\ 
$q=L^{-1}$.  In order words, a running coupling is defined through the response of some correlation 
function(s) as the volume of the system is varied. As a result, finite volume effects are 
part of the definition of the coupling, rather than a systematic uncertainty in its determination. 
This is clearly an advantage, since for these schemes lattice systematics are under control once 
a single condition, $L^{-1}=q\ll a^{-1}$ $\Rightarrow$ $L/a\gg 1$, is met. This is a much simpler 
situation than having to simultaneously satisfy: $L^{-1}\ll q\ll a^{-1}$. In principle, there is 
lots of freedom in choosing a finite-volume scheme. However, in practical applications several 
technical aspects need to be taken intro consideration. We refer the reader to Ref.~\cite{Sommer:2015kza} 
for a detailed discussion about these points and for concrete examples of schemes.

\subsubsection{Step-scaling strategy}
\label{subsec:StepScaling}

The way we exploit finite-volume schemes for the determination of $\alphas$ can be summarized 
in a few key steps, which we typically refer to as \emph{step-scaling} strategy~\cite{Luscher:1991wu,Jansen:1995ck,Sommer:2015kza}. 
{\bf 1)} We begin at low-energy by implicitly defining a hadronic scale $\mu_{\rm had}$ through a 
specific value of a chosen finite-volume coupling. Taking $\alpha_\Obs(\mu_{\rm had}=L^{-1}_{\rm had})\sim1$, 
we expect $\mu_{\rm had}\approx\Lambda_{\rm QCD}$. {\bf 2)} Using results from hadronic, large volume 
simulations, we can accurately establish the value of $\mu_{\rm had}$ in physical units. 
This is done by computing the ratio $\mu_{\rm had}/m_{\rm had}\sim\mathcal{O}(1)$, where $m_{\rm had}$ 
is an experimentally measurable low-energy quantity, \eg~a hadronic mass or decay constant. 
{\bf 3)} We simulate pairs of lattices with physical sizes $L$ and $L/2$, and determine the 
nonperturbative running of the finite-volume coupling $\alpha_\Obs(1/L)$ with the energy scale.%
\footnote{In practice, several pairs of lattices with fixed spatial sizes $L$ and $L/2$ but different 
lattice spacing $a$ are simulated in order to extrapolate the lattice results to the continuum limit, $a\to0$.
Within this approach the simulated lattices cover at most a factor of two in energy. This allows for
having control on discretization errors at any energy scale.}
This is encoded in the (inverse) step-scaling function: $\sigma^{-1}_\Obs(u)=\alpha_\Obs(2/L)|_{u=\alpha_\Obs(1/L)}$,
which measures the variation of the coupling as the energy scale is increased by a factor of two.
{\bf 4)} Starting from  $\mu_{\rm had}=L_{\rm had}^{-1}\sim\mathcal{O}(100)\,\MeV$, after $n\sim\mathcal{O}(10)$ steps as in 3), 
we reach nonperturbatively high-energy scales, $\mu_{\rm PT}=2^n/L_{\rm had}\sim\mathcal{O}(100)\,\GeV$.
{\bf 5)} Using the perturbative expansion of $\alpha_\Obs(\mu_{\rm PT})$ in terms of $\alpha_{\MSbar}(\mu_{\rm PT})$
we extract the latter (cf.~Eq.~(\ref{eq:Oq})). {\bf 6)} Given $\alpha_{\MSbar}(\mu_{\rm PT})$, through the perturbative 
running in the $\MSbar$-scheme, we obtain a value for $\Lambda_{\MSbar}/\mu_{\rm PT}$,
from which $\Lambda_{\MSbar}/m_{\rm had}$ can be readily inferred.

\subsubsection{\texorpdfstring{$\alphas$}{alphas} from a nonperturbative determination of $\Lambda_{\MSbar}^{(\Nf=3)}$}
\label{sec:Nf3QCD}

Following a step-scaling strategy, the ALPHA collaboration has obtained a subpercent precision determination of 
$\alphasmZ$ from a nonperturbative determination of $\Lambda_{\MSbar}^{(3)}$~\cite{Brida:2016flw,DallaBrida:2016kgh,Bruno:2017gxd,DallaBrida:2018rfy}. 
We refer the reader to these references for the details on our previous calculation. Here we simply want to recall 
a few points which are relevant for the following discussion. In refs.~\cite{DallaBrida:2016kgh,DallaBrida:2018rfy} 
the nonperturbative running of some convenient finite-volume couplings in $\Nf=3$ QCD was obtained, 
from $\mu_{\rm had}=197(3)\,\MeV$, up to $\sim$\,140\,\GeV.%
\footnote{The physical units of $\mu_{\rm had}$ were accurately established from a combination of pion and kaon decay constants~\cite{Bruno:2017gxd}.} 
Using NNLO perturbation theory, $\alpha^{(3)}_{\MSbar}(\mu_{\rm PT})$ was then extracted at $\mu_{\rm PT}\approx70\,\GeV$, and from it the result: $\Lambda_{\MSbar}^{(3)}=341(12)\,\MeV$. Thanks to the fact that we covered nonperturbatively a wide range of high-energy scales, a careful assessment of the accuracy of perturbation theory in matching the finite-volume and $\MSbar$ schemes was possible. The result is that the error on $\Lambda_{\MSbar}^{(3)}$ is entirely dominated by the statistical uncertainties associated with the determination of the nonperturbative running from low to high energy~\cite{Bruno:2017gxd}. In particular, perturbative truncation errors and nonperturbative corrections are completely negligible within the statistical uncertainties~\cite{DallaBrida:2018rfy}. 
Finally, from the result for $\Lambda_{\MSbar}^{(3)}$, using perturbative decoupling relations, we included the effect of the charm and bottom quarks in the running to arrive at: $\alpha_{\MSbar}^{(5)}(m_\mathrm{Z})=0.1185(8)$~\cite{Bruno:2017gxd}. 
From these observations, it is clear that an improved determination of $\alphas$ may be obtained by reducing the statistical uncertainties on $\Lambda_{\MSbar}^{(3)}$ due to the nonperturbative running of the finite-volume couplings. 
On the other hand, to which extent this is possible very much depends on how accurate it is to rely on perturbative decoupling for including charm effects. 
 
\subsubsection{How accurate is \texorpdfstring{$\Nf=3$}{Nf=3} QCD?} 
\label{subsec:Nf3QCD}
 
Including the charm is particularly challenging in hadronic, large volume simulations. The charm has a mass 
$m_c\approx 1.3\,\GeV$. In units of the typical cutoffs set in hadronic simulations this means $am_c\gtrsim 0.3$. 
It is thus difficult to comfortably resolve the characteristic energy scales associated with the charm in 
current large volume simulations: it requires very fine lattice spacings, which are computationally very demanding.%
\footnote{We recall that the computational cost of lattice simulations grows roughly $\propto a^{-7}$ as $a\to0$.}
In addition, simulations become more expensive as the number of quarks increases, and the tuning of the 
parameters in the lattice QCD action is more involved. It is therefore important to assess whether the 
computational effort required to include the charm is actually needed to significantly improve 
our current precision on $\alphas$. If that is not the case, the resources are better invested in improving the results from $\Nf=3$ QCD.

In order to answer this question, we must study the systematics that enter in the $\Lambda_{\MSbar}^{(3)}\to\Lambda_{\MSbar}^{(4)}$ step. 
The first class is related to the use of perturbative decoupling relations for estimating the ratio 
$\Lambda_{\MSbar}^{(3)}/\Lambda_{\MSbar}^{(4)}$. As we shall recall below, the ratio of $\Lambda$-parameters
with different flavour content is given by a function $P_{\ell,\rm f}(M/\Lambda_{\MSbar}^{(\Nf)})=\Lambda_{\MSbar}^{(\Nl)}/\Lambda_{\MSbar}^{(\Nf)}$, 
which depends on the renormalization group invariant (RGI) mass $M$ of the decoupling quark(s) and the theories considered. 
The function $P_{\ell, \rm f}$ can in principle be nonperturbatively defined (Sec.~\ref{subsec:DecNP}). 
In phenomenological applications, however, we approximate it with its asymptotic, perturbative expansion at some 
finite loop-order, \ie\ $P_{\ell,\rm f}(M/\Lambda_{\MSbar}^{(\Nf)})\sim P_{\ell,\rm f}^{(n\text{-loop})}(M/\Lambda_{\MSbar}^{(\Nf)})+
\mathcal{O}(\alpha^{n-1}(M))+\mathcal{O}(M^{-2})$. Whether this is appropriate, it all depends on the size of the perturbative 
and nonperturbative corrections to this approximation for values of the quark masses $M\sim M_c$, with $M_c$ 
the RGI charm mass. 

The second class of charm effects concerns the hadronic renormalization of the lattice theory. 
Normally, the  bare parameters entering the lattice QCD action are tuned by requiring that a number of ratios of 
hadronic quantities $R_H$ (as many as parameters to tune), reproduce their experimental counter-parts. Typical 
examples are, for instance, $R_H=m_\pi/f_\pi, m_K/f_\pi, \ldots$%
\footnote{The experimental numbers for the hadronic quantities of interest are usually ``corrected'' for QED and 
strong  isospin breaking effects, if these are not included in the lattice simulations (cf.~ref.~\cite{Aoki:2021kgd}).}
On the other hand, $\Nf=3$ QCD simulations do not include charm effects, while these are obviously present 
in the experimental determinations. Whether this mismatch is relevant in practice, it all depends on the actual 
size of charm effects in the ratios $R_H$.

\subsubsection{Effective theory of decoupling and perturbative matching}
\label{subsec:EFT}

For the ease of presentation, we define our \emph{fundamental theory} as $\Nf$-flavour QCD (${\rm QCD}_{\Nf}$) with 
$\Nl$ massless quarks, and $\Nh=\Nf-\Nl$ mass-degenerate heavy quarks of mass $M$. In the limit where $M$ is larger than 
any other scale in the problem, this theory can be approximated by an \emph{effective theory} (EFT) with Lagrangian~\cite{Weinberg:1980wa}
\begin{equation}
	\mathcal{L}_{\rm dec}=\mathcal{L}_{\rm QCD_\Nl} + 
	{1\over M^2}\sum_{i} \omega_i \Phi_i + \mathcal{O}(M^{-4})\,.
\end{equation}
The leading order term in the $1/M$ expansion is the Lagrangian $\mathcal{L}_{\rm QCD_\Nl}$ of massless ${\rm QCD}_{\Nl}$, 
while the effect of the heavy quarks is represented by nonrenormalizable interactions $\Phi_i$ suppressed by higher powers 
of $1/M$. Massless ${\rm QCD}_{\Nl}$ has a single parameter, the gauge coupling $\bar{g}^{(\Nl)}$. The EFT is hence 
predictive once its coupling is given as a proper function of the coupling of the fundamental theory, $\bar{g}^{(\Nf)}$, 
and the quark masses $M$. In this situation, we say that the couplings are matched,
\begin{equation}
	\label{eq:MatchingCouplings}
	\bar{g}^{(\Nl)}(\mu)=F_\Obs(\bar{g}^{(\Nf)}(\mu),M/\mu)\,.
\end{equation}	
The function $F_\Obs$ depends in principle on the observable $\Obs$ that is used to establish the matching.
At leading order in the EFT, however, it is consistent to drop any correction of $\mathcal{O}(M^{-2})$ in the relation 
(\ref{eq:MatchingCouplings}), which thus becomes universal, \ie\ it only depends on the renormalization 
scheme chosen for the couplings. In the $\MSbar$-scheme, the so-called decoupling relation (\ref{eq:MatchingCouplings}) is 
known at 4-loop order~\cite{Bernreuther:1981sg,Schroder:2005hy,Chetyrkin:2005ia,Kniehl:2006bg,Grozin:2011nk,Gerlach:2018hen}, 
and it is usually evaluated at the scale $\mu=m_\star$, where $m_\star=\overline{m}_{\MSbar}(m_\star)$ 
with $\overline{m}_{\MSbar}(\mu)$ the mass of the heavy quarks in the $\MSbar$-scheme. In formulas,
\begin{equation}
	\bar{g}_{\overline{\text{\rm MS}}}^{(\Nl)}(m_\star) 
	=g_\star\,{\xi}\big(g_\star)\,,
	\qquad
	g_\star\equiv\bar{g}_{\overline{\text{\rm MS}}}^{(\Nf)}(m_\star)\,,
	\qquad
	\xi(g)=1+c_2 g^4 + c_3 g^6 + c_4 g^8+\mathcal{O}(g^{10})\,.
\end{equation}
The relation between the couplings can be reexpressed in terms of the corresponding $\Lambda$-parameters of the 
two theories,
\begin{equation}
	\label{eq:MatchingLambdasPT}
	P_{\ell,\rm f}(M/\Lambda_{\MSbar}^{(\Nf)})={\Lambda_{\MSbar}^{(\Nl)}/\Lambda_{\MSbar}^{(\Nf)}}=
	{\varphi^{(\Nl)}_{\overline{\rm MS}}\big(g_\star\,{\xi}(g_\star)\big)/\varphi^{(\Nf)}_{\overline{\rm MS}}(g_\star)}\,,
\end{equation}
where the RGI-parameters are given by ${\Lambda^{(\Nf)}_{\rm X}}=\mu\,\varphi^{(\Nf)}_{\rm X}(\bar{g}^{(\Nf)}_{\rm X}(\mu))$ 
and $M=\overline{m}_{\rm X}(\mu)\varepsilon^{(\Nf)}_{\rm X}(\bar{g}^{(\Nf)}_{\rm X}(\mu))$. Explicit 
expressions for the functions $\varphi^{(\Nf)}_{\rm X}$ and $\varepsilon^{(\Nf)}_{\rm X}$ in terms 
of the $\beta$- and $\tau$-functions in a generic scheme ${\rm X}$ can be found in \eg~\cite{Sommer:2015kza}.

\begin{figure}[hptb]
	\includegraphics[scale=1]{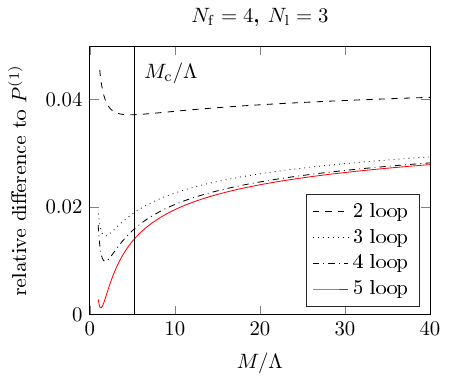}
	\quad
	\includegraphics[scale=1]{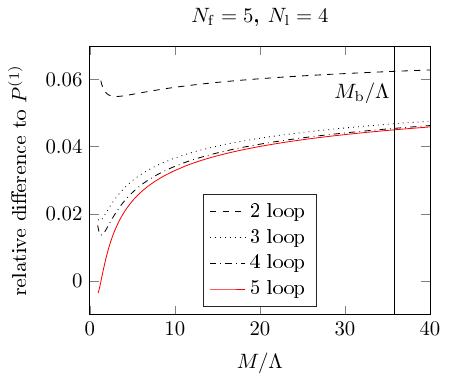}
	\caption{Relative differences from the (unsystematic) 1-loop approximation $P^{(1)}_{\ell,\rm f}(M/\Lambda) = (M/\Lambda)^{\eta_0}$, 
	$\eta_0 = 2\Nh/(33 - 2\Nl)$, for different orders	of the perturbative expansion of $P_{\ell,\rm f}(M/\Lambda)$
	as a function of $M/\Lambda$~\cite{Athenodorou:2018wpk}. The results for $\Nl = 3$, $\Nf = 4$ ($\Nl = 4$, $\Nf = 5$) are given in 
	the left (right) panel. The values for the RGI charm ($M_c$) and bottom ($M_b$) quark masses in units of the proper $\Lambda$-parameters 
	are marked by vertical lines.}
	\label{fig:P}
\end{figure}

In Fig.~\ref{fig:P} we present the perturbative results from Ref.~\cite{Athenodorou:2018wpk} for $P_{\ell, \rm f}(M/\Lambda_{\MSbar}^{(\Nf)})$,
for the phenomenological relevant cases of $\Nl=3$, $\Nf=4$ and $\Nl=4$, $\Nf=5$. More precisely, we show the relative 
deviation with respect to the 1-loop approximation, $P^{(1)}_{\ell,\rm f}$, for different orders of the perturbative 
expansion of $P_{\ell,\rm f}$. Focusing on the case of $P_{3,4}$, we see how the perturbative corrections at 4- and 5-loop 
order are very small already for values of $M$ comparable to that of the charm. Judging from the perturbative behavior alone, 
the series thus appears to be well within its regime of applicability. As a result, any estimate for the perturbative 
truncation errors on $P_{\ell,\rm f}$ based on the last-known contributions to the series leads to very small uncertainties. 
When translated to the coupling we find, for instance, $\alpha^{(5)}_{\overline{\text{\rm MS}}}(m_\mathrm{Z}) = 0.1185(8)(3)_{\rm PT dec}$,
where the second error is estimated as the sum of the full 4- and 5-loop contributions due to the decoupling of 
both charm and bottom quarks~\cite{Bruno:2017gxd}. As anticipated, the perturbative error estimate is well-below the uncertainties 
from other sources.

\subsubsection{How perturbative are heavy quarks?}
\label{subsec:DecNP}

Above we have shown how, within perturbation theory, the perturbative decoupling of the charm appears to be very accurate. 
A natural question to ask at this point is how reliable this picture is at the nonperturbative level. In other words, 
can we quantify the accuracy of perturbative decoupling for the charm by comparing it directly to nonperturbative results, 
thus getting an estimate for the size of nonperturbative corrections? In order to answer this question, we start by 
formulating the matching of $\Lambda$-parameters (\ref{eq:MatchingLambdasPT}) in nonperturbative terms~\cite{Bruno:2014ufa,Athenodorou:2018wpk},
\begin{equation}
	\label{eq:MatchingLambdasNP}
	{\Lambda^{(\Nl)}_{\MSbar} \over m_{\rm had,1}^{(\Nl)}} =
	P^{\rm had,1}_{\ell,\rm f}\big({M/\Lambda^{(\Nf)}_{\MSbar}}\big)
	{\Lambda^{(\Nf)}_{\MSbar} \over m_{\rm had,1}^{(\Nf)}(M)}\,.
\end{equation}
We thus say that the EFT is matched to the fundamental theory if its $\Lambda$-parameter $\Lambda^{(\Nl)}_{\MSbar}\equiv
\Lambda^{(\Nl)}_{\MSbar}(M,\Lambda^{(\Nf)}_{\MSbar})$ in units of a hadronic quantity $m_{\rm had,1}^{(\Nl)}$,
is a proper function of the heavy quark masses $M$, and the $\Lambda$-parameter of the fundamental theory 
$\Lambda^{(\Nf)}_{\MSbar}$ in units of the same hadronic quantity $m_{\rm had,1}^{(\Nf)}(M)$, computed in
${\rm QCD}_{\Nf}$ where $\Nh$ quarks are heavy with mass $M\gg\Lambda_{\rm QCD}$.%
\footnote{Here we use the $\MSbar$-scheme as reference scheme for the $\Lambda$-parameters. As we have seen in Sec.~\ref{subsec:StepScaling}, 
$\Lambda_{\MSbar}$ can be indirectly expressed in terms of any nonperturbative scheme of choice.}
Once the theories are matched, decoupling predicts that for any other hadronic quantity $m_{\rm had,2}^{(\Nl)}$ computed 
in the EFT, we should expect: $m_{\rm had,2}^{(\Nl)}=m_{\rm had,2}^{(\Nf)}(M)+\mathcal{O}(M^{-2})$ (cf.~Sec.~\ref{subsec:EFT}). 
As anticipated by our notation, the function $P^{\rm had}_{\ell,\rm f}$ depends on the hadronic quantity considered for 
the matching. If we were to choose a different quantity, we expect: 
$P^{\rm had,1}_{\ell,\rm f}\big({M/\Lambda^{(\Nf)}_{\MSbar}}\big)\sim P^{\rm had,2}_{\ell,\rm f}\big({M/\Lambda^{(\Nf)}_{\MSbar}}\big)+\mathcal{O}(M^{-2})$.

The relation (\ref{eq:MatchingLambdasNP}) leads to the interesting \emph{factorization formula}~\cite{Bruno:2014ufa,Athenodorou:2018wpk},
\begin{equation}
	\label{eq:Factorization}
	{m_{\rm had}^{(\Nf)}(M)\over m_{\rm had}^{(\Nf)}(0)}=
	{Q}^{\rm had}_{\ell,\rm  f} 
	\times 	
	P^{\rm had}_{\ell, \rm f}\big({M/\Lambda^{(\Nf)}_{\MSbar}}\big)\,,
	\qquad	
	{Q}^{\rm had}_{ \ell, \rm f}
	= {m_{\rm had}^{(\Nl)}/\Lambda^{(\Nl)}_{\MSbar} \over m_{\rm had}^{(\Nf)}(0)/\Lambda^{(\Nf)}_{\MSbar}}\,.
\end{equation}
On the l.h.s.~of this equation we have the hadronic quantity $m_{\rm had}^{(\Nf)}(M)$ computed in ${\rm QCD}_{\Nf}$ 
where $\Nh$ quarks have mass $M$, over the same hadronic quantity $m_{\rm had}^{(\Nf)}(0)$ computed in the chiral limit, 
\ie\ where all $\Nf$ quarks are massless. This ratio can be expressed as the product of a nonperturbative, 
$M$-independent factor $Q^{\rm had}_{\ell, \rm f}$, and the function $P^{\rm had}_{\ell, \rm f}\big({M/\Lambda^{(\Nf)}_{\MSbar}}\big)$, 
which encodes all the $M$-dependence. As we recalled in Sec.~\ref{subsec:EFT}, for $M\to\infty$ the function $P^{\rm had}_{\ell, \rm f}$ 
admits an asymptotic perturbative expansion. Hence, by measuring nonperturbatively on the lattice the l.h.s.~of Eq.~(\ref{eq:Factorization}), we can compare the $M$-dependence of several such ratios with what is predicted by a perturbative 
approximation for $P^{\rm had}_{\ell, \rm f}$. In this way, we can assess the typical size of nonperturbative effects
in $P^{\rm had}_{\ell, \rm f}$ as a function of $M$. In Ref.~\cite{Athenodorou:2018wpk} a careful study was carried out considering 
the case of the decoupling of two heavy quarks with mass $M\sim M_c$, for the case of $\Nl=0$, $\Nf=2$.%
\footnote{The reason why the authors of Ref.~\cite{Athenodorou:2018wpk} considered $\Nl=0$, $\Nf=2$ instead of $\Nl=3$, $\Nf=4$, 
has to do with the technical difficulties of simultaneously simulating heavy and light quarks on the lattice (cf.~Sec.~\ref{subsec:Nf3QCD}). 
The effect of decoupling two heavy quarks rather than just one is however expected to more than compensate the effects on decoupling
induced by the presence of the light quarks (cf.~refs.~\cite{Athenodorou:2018wpk} for a detailed discussion about this point).}
 Under very reasonable assumptions, it is possible to extract from these results quantitative information 
on the nonperturbative corrections in the phenomenological interesting case of $P_{3,4}(M_c/\Lambda^{(4)}_{\MSbar})$.
The conclusions of Ref.~\cite{Athenodorou:2018wpk} are that these effects are very much likely below the per-cent level. 
This means that it is safe to use a perturbative estimate for $P_{3,4}(M_c/\Lambda^{(4)}_{\MSbar})$ in transitioning  from
$\Lambda_{\MSbar}^{(3)}\to\Lambda_{\MSbar}^{(4)}$, as long as, say, $\delta\Lambda_{\MSbar}^{(3)}\gtrsim 1.5\%$ or so.

If we now consider ratios of hadronic quantities, where the dependence on the $\Lambda$-parameters drops out, 
we immediately conclude that 
\begin{equation}
	{m_{\rm had,1}^{(\Nl)}/ m_{\rm had,2}^{(\Nl)}}=
	{m_{\rm had,1}^{(\Nf)}(M)/ m_{\rm had,2}^{(\Nf)}(M) }
	+ 
	{\mathcal O}(M^{-2})\,.
\end{equation}
In this case, one can obtain estimates for the typically size of $\mathcal{O}(M^{-2})$ effects in these ratios by comparing 
both sides of the above equation computed through lattice simulations. In refs.~\cite{Knechtli:2017xgy,Hollwieser:2020qri} 
careful studies have been conducted considering both the case of the decoupling of two heavy quarks of mass $M\sim M_c$ with $\Nl=0$, 
$\Nf=2$, and more recently for the decoupling of a single charm quark with $\Nl=3$, $\Nf=3+1$, \ie\ in the presence of three 
mass-degenerate lighter quarks. From these calculations, the authors conclude that the typical effects due to the charm on 
dimensionless ratios of low-energy quantities are well-below the per-cent level. This means that these effects are not relevant 
for a per-cent precision determinations of $\Lambda_{\MSbar}^{(3)}$.

In summary, thanks to recent dedicated studies, we are able to conclude that it is safe to rely on perturbative decoupling for the charm quark in connecting $\Lambda_{\MSbar}^{(3)}$ and $\Lambda_{\MSbar}^{(4)}$, as long as  $\delta\Lambda_{\MSbar}^{(3)}\gtrsim 1.5\%$. The 0.7\% precision extraction of $\alphas$ from Ref.~\cite{Bruno:2017gxd} is based on a determination of $\Lambda_{\MSbar}^{(3)}$ with an uncertainty of $\delta\Lambda_{\MSbar}^{(3)}\approx 3.5\%$ (cf.~Sec.~\ref{sec:Nf3QCD}). Hence, there is still about a factor of two of possible improvement within the $\Nf=3$ flavour theory.

\subsubsection{The strong coupling from the decoupling of heavy quarks}

The previous section suggests a method to relate the $\Lambda$-parameters 
in theories with a different number of quark flavours (cf.~Eq.~(\ref{eq:MatchingLambdasNP})). 
Taking this relation to the extreme, one should be able to determine 
$\Lambda_{\overline{\rm MS} }^{(\Nf) }$ from the pure-gauge one, $\Lambda_{\overline{\rm MS} }^{(0)}$. 
Of course this requires to decouple $\Nf $ heavy quarks with
$M\gg \Lambda^{(\Nf )}_{\overline{\rm MS} }$. 
The physical world is very different from this limit, but lattice QCD
can simulate such \emph{unphysical} situation. 

A possible strategy for the determination of the strong coupling based on this idea is the
following:
\begin{enumerate}
\item Starting from a scale $\mu_{\rm dec}$ in ${\rm QCD}_{\Nf}$, one
  determines the value of a nonperturbatively defined coupling at
  such scale in a massless renormalization scheme, $\bar g^{(\Nf)}_{\Obs}(\mu_{\rm dec})$.
\item By performing lattice simulations at
  increasing values of the quark masses, one is able to determine the
  value of the coupling in a massive scheme, $\bar g^{(\Nf )}_{\Obs}(\mu_{\rm dec}, M)$. 
\item For $M$ larger than any other scale in the problem (\ie\ $M\gg \Lambda_{\rm QCD}, \mu_{
    \rm dec}$), the massive coupling is, up to heavy-mass corrections,
  the same as the corresponding coupling in the  pure-gauge theory, \ie\ 
  \begin{equation}
    \label{eq:deceq}
    \bar g^{(\Nf)}_{\Obs}(\mu_{ \rm dec}, M) = \bar g^{(0)}_{\Obs}(\mu_{ \rm dec})\,,
  \end{equation}
  where corrections of $\mathcal{O}(1/M^2)$ have been dropped. 
  
\item From the running of the coupling $\bar g^{(0)}_{\Obs}(\mu_{ \rm dec})$
  in the pure-gauge theory we can determine the pure-gauge $\Lambda$-parameter in units of
  $\mu_{\rm dec}$,
  \begin{equation}
    \frac{\Lambda^{(0)}_{\overline{\rm MS} }}{\mu_{\rm dec}} = 
    \frac{\Lambda^{(0)}_{\overline{\rm MS} }}{\Lambda^{(0)}_{\Obs}} 
    \times \varphi^{(0)}_{\Obs}(\bar g^{(0)}_{\Obs}(\mu_{ \rm dec}))\,.
  \end{equation}
  (Note that the ratio of $\Lambda$-parameters in different schemes
  can be exactly determined via a perturbative, 1-loop computation (cf.~\eg~\cite{Sommer:2015kza}).) 

  \item Since all $\Nf $ quarks are heavy, one can employ the
  decoupling relations to estimate the $\Nf\geq 3$ flavour
  $\Lambda$-parameter as (cf.~Eq.~(\ref{eq:MatchingLambdasNP})),
  \begin{equation}
    \label{eq:masterdec}
    \Lambda^{(\Nf )}_{\overline{\rm MS} } = \mu_{\rm dec} \times
    \frac{\Lambda^{(0)}_{\overline{\rm MS} }}{\Lambda^{(0)}_{\Obs}} \times 
    \varphi^{(0)}_{\Obs}(\bar g^{(0)}_{\Obs}(\mu_{ \rm dec})) \times
    \frac{1}{P^{(5\text{-loop})}_{0,\Nf }(M/\Lambda^{(\Nf )}_{\overline{\rm MS} })} 
    + \mathcal{O}(\alphas^4(m_{\star})) + \mathcal{O}(M^{-2}) 
    \,.
  \end{equation}
\end{enumerate}

Equation~(\ref{eq:masterdec}) is our master formula to determine $\Lambda^{(\Nf=3)}_{\overline{\rm MS} }$ 
from precise results for the nonperturbative running of the gauge coupling in the pure-gauge theory 
(\ie\ for the function $\varphi^{(0)}_\Obs$). Several comments are in order at this point:
\begin{itemize}
\item There are two types of corrections in Eq.~(\ref{eq:masterdec}). 
  First, ``perturbative'' ones of $\mathcal{O}(\alphas^4(m_\star))$. 
  They come from using perturbation theory for the function $P_{0,N_{\rm
      f} }$. Second,  we have nonperturbative ``power'' corrections. 
  Their origin is the decoupling condition, Eq.~(\ref{eq:deceq}),
  as well as the use of a perturbative approximation for the function $P_{0,\Nf }$.
  
\item Both perturbative and power corrections vanish for $M\to\infty$. 
  In fact, the following relation is exact
  \begin{equation}
    \Lambda^{(\Nf )}_{\overline{\rm MS} } = \lim_{M\to\infty}\mu_{\rm dec} \times
    \frac{\Lambda^{(0)}_{\overline{\rm MS} }}{\Lambda^{(0)}_{\Obs}} \times \varphi^{(0)}_{\Obs}(\bar g^{(0)}_{\Obs}(\mu_{ \rm dec})) 
    \times \frac{1}{P^{(n\text{-loop})}_{0,\Nf }(M/\Lambda^{(\Nf )}_{\overline{\rm MS} })}\,,
  \end{equation}
  with $\bar g^{(0)}_{\Obs}(\mu_{ \rm dec})=\bar g^{(\Nf)}_{\Obs}(\mu_{ \rm dec}, M)$.

\item The situation and challenges in this approach might look similar to those
  present in more ``conventional'' extractions of the strong coupling (cf.~Eq.~(\ref{eq:Oq})). 
  The subtle difference however is that, in the present case, perturbative corrections
  are very small even for scales $\sim M_c$ (cf.~Sec.~\ref{subsec:EFT}). 
  In particular, if one is considering quarks with masses of a few GeV, they are completely 
  negligible in practice, and one has only to deal with the power corrections, which decrease
  much faster with the relevant energy scale.
  
\end{itemize}

This method to extract the strong coupling was proposed
in Ref.~\cite{DallaBrida:2019mqg} (for a recent review see Ref.~\cite{DallaBrida:2020pag}). 
Here we present first results using this strategy~\cite{DallaBrida:2022eua}.%
\footnote{At the time of the workshop only a preliminary analysis of these results was available \cite{Brida:2021xwa}.}
We follow closely the strategy described above, skipping the technical details. 
The reader interested in the details is encouraged to look at the
original references~\cite{DallaBrida:2019mqg,DallaBrida:2020pag,DallaBrida:2022eua}.

\begin{enumerate}
\item A scale $\mu_{\rm dec} = 789(15)$ MeV is determined in a finite-volume
renormalization scheme in three-flavour QCD using results from Ref.~\cite{DallaBrida:2016kgh}. 
This corresponds to a value of the renormalized nonperturbative coupling
$\bar g^{(3)}(\mu_{\rm dec}) \approx 1.9872$. 

\item For technical reasons the massive
coupling is determined in a slightly different renormalization scheme than the massless one. 
The value of the massive coupling is then determined by keeping the value of the bare
coupling (and therefore the lattice spacing) fixed and increasing the
value of the quark masses. This determination is performed for several
values of the quark masses, $z = M/\mu_{\rm dec} = 1.972, 4, 6, 8, 10, 12$,
and several values of the lattice spacing with $1/(a\mu_{\rm dec}) = 12,\dots,48$. 
The results are extrapolated to the continuum limit for each value of the
quark masses labeled by $z$ (see Fig.~\ref{fig:aextr}). We refer the reader to 
\cite{DallaBrida:2022eua} for a detailed discussion.

\begin{figure}[htbp!]
\centering
  \includegraphics[width=0.48\textwidth]{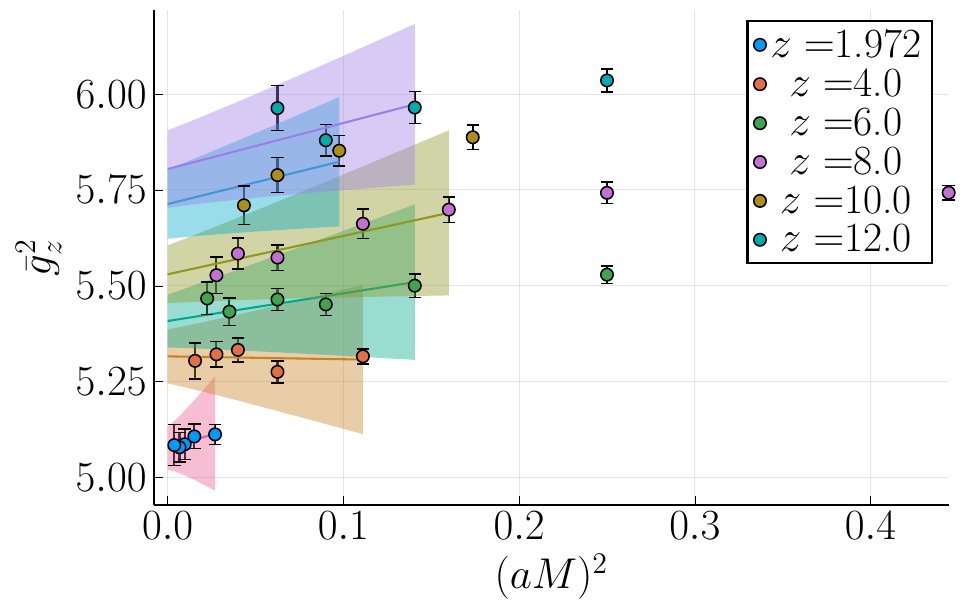}    
  \includegraphics[width=0.48\textwidth]{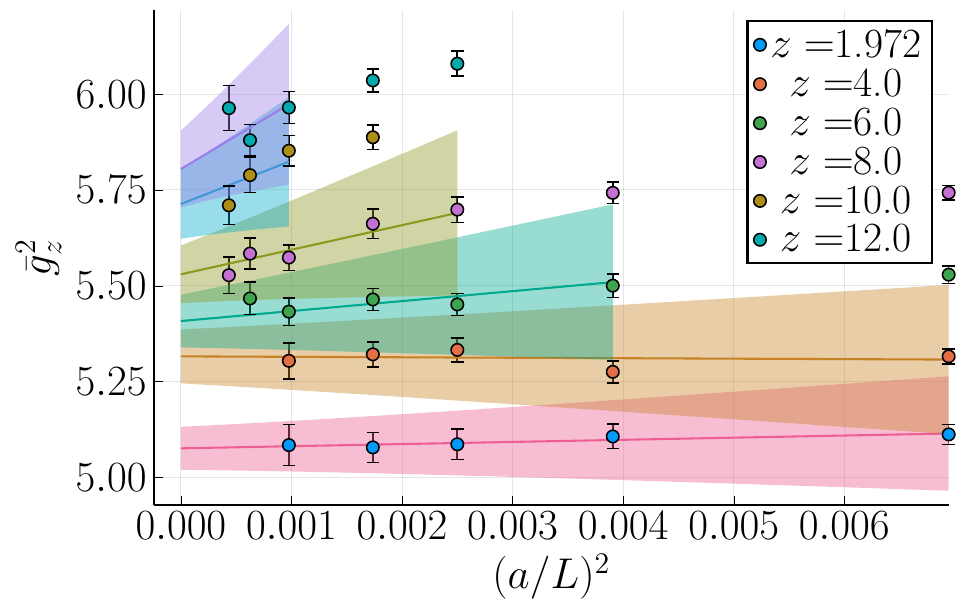}    
\caption{Continuum extrapolations of the massive coupling $\bar g^2_z = [\bar g^{(3)}(\mu_{\rm
    dec}, M)]^2$ for different values of the quark masses labeled by
  $z=M/\mu_{\rm dec}$. Only data at fine enough lattice spacings (\ie\ for which ($aM)^2
  < 0.16$) are included in the fit. Note that the fit error bands include an estimate for the remaining $\mathcal O(a)$ effects (see~\cite{DallaBrida:2022eua} for a full discussion).} 
  \label{fig:aextr}
\end{figure}

\item The values of the massive coupling are used as
  matching condition between ${\rm QCD}_{\Nf}$ and the pure-gauge theory
  (cf.~Eq.~(\ref{eq:deceq})). 
  The running in the pure-gauge theory is known very
  precisely from the literature~\cite{DallaBrida:2019wur}.
  The ratio ${\Lambda^{(0)}_{\overline{\rm MS} }}/{\mu_{\rm dec}}$
  can thus be accurately determined.

\begin{table}
\centering
\caption{Values of the massive three-flavour coupling $\bar g^{(3)}(\mu_{\rm dec},M)$ and the corresponding values of $\Lambda^{(3)}_{\overline{\rm MS},{\rm eff} }$ obtained by applying our  master formula Eq.~(\ref{eq:masterdec}) and ignoring perturbative and  power corrections.\vspace{0.2cm}}
\label{tab:lambda_eff}
\tabcolsep=3.mm
\begin{tabular}{cll}
  \toprule
  $z$ & $[\bar g^{(3)}(\mu_{\rm dec},M)]^2$ & $\Lambda^{(3)}_{\overline{\rm MS}, {\rm eff}}$ [MeV]\\
  \midrule
  1.972&   5.076(56)  &             426(14) \\
  4  &   5.316(70)  &             388(13) \\
  6  &   5.408(69)  &             363(12) \\
  8  &   5.530(76)  &             351(12) \\
  10 &   5.713(90)  &             349(12) \\
  12 &    5.80(10)  &             343(12) \\
\bottomrule
\end{tabular}
\end{table}

\item Given this result, by applying the master formula
  Eq.~(\ref{eq:masterdec}), one obtains the estimates, $\Lambda^{(3)}_{\overline{\rm MS},{\rm eff} }$, for
  $\Lambda^{(3)}_{\overline{\rm MS} }$ given in Table~\ref{tab:lambda_eff}.
  These estimates should approach the correct three-flavour
  $\Lambda$-parameter in the limit $M\to\infty$. 
  Figure~\ref{fig:mtoinfty} shows that this is indeed the case. 
  Our data are consistent with a linear extrapolation in $\mu_{\rm
    dec}^2/M^2$, which results in
  \begin{equation}
    \Lambda^{(3)}_{\overline{\rm MS}} = 336(10)(6)(3)\, {\rm MeV}\,,
  \end{equation}
  where the first uncertainty is statistical, the second is due to an estimate of the $\mathcal O(a)$ effects present in our setup, and the last is due to
  different parameterizations of the $M\to\infty$ limit. 
  
  \begin{figure}
    \centering
    \includegraphics[width=0.7\textwidth]{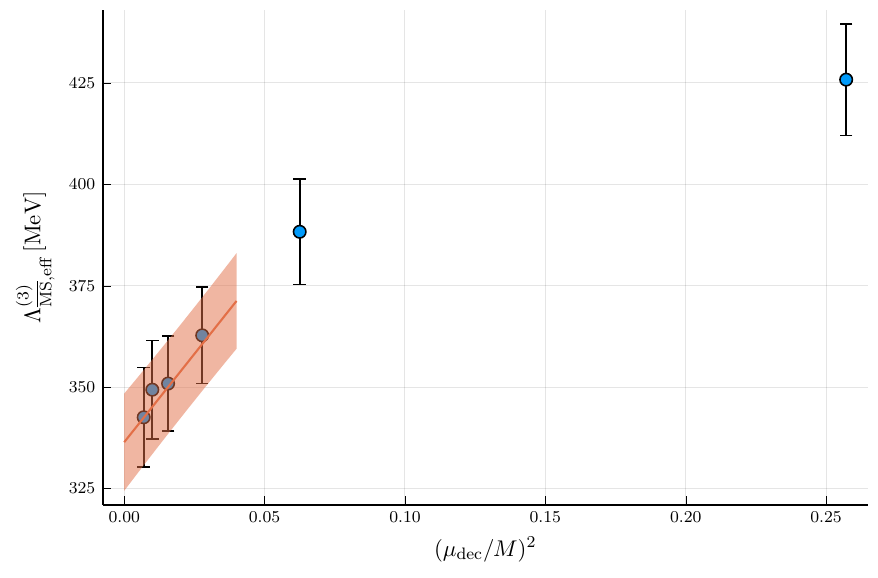}
    \caption{Estimates of the three-flavour $\Lambda$-parameter (Table~\ref{tab:lambda_eff}) and its  $M\to\infty$ extrapolation yielding our final result for $\Lambda^{(3)}_{\overline{\rm MS} }$.}
    \label{fig:mtoinfty}
  \end{figure}

\end{enumerate}

This result for the three-flavour $\Lambda$-parameter translates,
after crossing the charm and bottom thresholds, into
\begin{equation}
  \alphasmZ = 0.11823(84)\,.
\end{equation}
The result has a 0.7\% uncertainty, which is in fact dominated by the statistical uncertainties in the
scale $\mu_{\rm dec} = 789(15)$ MeV and in the running of the pure-gauge theory. 
These statistical uncertainties can be
reduced with a modest computational effort.  
The effects of the truncation of the perturbative series are
subdominant in our approach. 
A non-perturbative determination of the relevant improvement parameter would eliminate entirely the systematics related to $\mathcal{O}(a)$ effects, 
and make possible a further reduction of the uncertainty by a factor of two with this approach
(see \cite{DallaBrida:2022eua} for a full discussion). 
Going beyond this precision would require including charm effects nonperturbatively 
and also some serious thinking on how to include electromagnetic effects in the
determination of both the scale and the running.

\subsection{Strong coupling constant \texorpdfstring{$\alphas$}{alphas} from moments of quarkonium correlators 
\protect\footnote{A\lowercase{uthors:} P. P\lowercase{etreczky} (BNL), J. H. W\lowercase{eber} (HU, B\lowercase{erlin})}}
\label{sec:latt:moments}

A lattice method conceptually similar to the determination of $\alphasmZ$ or 
heavy-quark masses from the $R$-ratio via quarkonium sum rules~\cite{Dehnadi:2015fra, Boito:2020lyp} 
uses heavy-quark two-point correlators; for a recent review see Ref.~\cite{Komijani:2020kst}. 
Renormalization group invariant pseudoscalar correlators $G(t)$ and their $n$th time moments $G_n$ are defined as 
\begin{align}
G_n = \sum_t t^n (G(t)+G(N_t-t)),
\quad
G(t)=a^6 \sum_{\bm{x}} (a m_{h0})^2 \braket{ j_5(\bm{x},t) j_5(0,0) },
\quad
j_5(\bm{x},t) = \bar{\psi}(\bm{x},t) \gamma_5 \psi(\bm{x},t).
\end{align}
The symmetrization accounts for (anti)periodic boundary conditions in time. 
Here, $am_{h0}$ is the corresponding bare heavy-quark mass in the lattice scheme; 
$am_{h0}$, resp.\ $m_h$, can be varied quite liberally for valence quarks in the 
charm- and bottom-quark regions and in between. Since sea quark mass variation 
is expensive, most results have partially quenched heavy quarks, \ie\ heavy-quark 
masses in sea and valence sectors can be different in (2+1+1)-flavour QCD, or heavy 
quarks are left out from the sea in (2+1)-flavour QCD.\\

The weak-coupling series of $G_n$, which are finite for $n\ge 4$, is known 
up to $\mathrm{N^3LO}$, resp.\ $\mathcal{O}(\alphas^3(\nu_h))$, for $\Nf$ 
massless and one massive quark flavour~\cite{Sturm:2008eb, Kiyo:2009gb, Maier:2009fz},
\begin{align}
G_n=\tfrac{g_n(\alphas(\nu),\nu/m_h)}{a m_h^{n-4}(\nu_m)},
\end{align}
where $\nu= x m_h$, proportional to the heavy-quark mass $m_h$, 
is the $\mathrm{\MSbar}$ renormalization scale of the coupling; and
$\nu_m=x_m m_h$, the scale at which the $\mathrm{\MSbar}$ heavy-quark mass 
$m_h(\nu_m)$ is defined, could be different from $\nu$~\cite{Dehnadi:2015fra}. 
In published weak-coupling results, heavy-quark masses on internal (sea) and external 
(valence) quark lines must agree~\cite{Sturm:2008eb, Kiyo:2009gb, Maier:2009fz}. 
This restriction has profound implications for $\alphas$ extractions in (2+1+1)-flavour QCD. 

As the bare heavy-quark mass in lattice units, $am_{h0}$, is a large parameter, 
improved quark actions are needed; so far most calculations have employed the highly 
improved staggered quarks (HISQ)~\cite{HPQCD:2008kxl, McNeile:2010ji, Chakraborty:2014aca, 
Maezawa:2016vgv, Petreczky:2019ozv, Petreczky:2020tky}, while domain-wall fermions 
have been used as well~\cite{Nakayama:2016atf}. 
Some data sets involved values of $am_{h0}$ corresponding to different heavy-quark 
masses $m_h$~\cite{McNeile:2010ji, Chakraborty:2014aca, Petreczky:2019ozv, 
Petreczky:2020tky}; for this reason even (2+1+1)-flavour QCD lattice results still 
involve partially quenched heavy quarks~\cite{Chakraborty:2014aca}. 
Enforcing an upper bound $am_{h0} \lesssim 1$ to limit the size of lattice artifacts 
implies that fewer independent ensembles can constrain the data at larger $m_h$ and entail 
larger errors of the respective continuum limit. 

The so-called (tree-level) reduced moments, known perturbatively at $\mathcal{O}(\alphas^3(\nu_h))$, 
\begin{align}
&
R_4 = \left(\tfrac{G_4^{\mathrm{QCD}}}{G_4^{\mathrm{0}}}\right)
\hskip2em,&&
R_4(\alphas(\nu),\tfrac{\nu}{m_h}) = 
1 + \sum_{j=1}^3 r_{nj}(\tfrac{\nu}{m_h}) \left(\tfrac{\alphas(\nu)}{\pi}\right)^j
\\
&
R_n = \left(\tfrac{G_n^{\mathrm{QCD}}}{G_n^{\mathrm{0}}}\right)^{\frac{1}{(n-4)}}
,&&
R_n(\alphas(\nu),\tfrac{\nu}{m_h}) =
\left(\tfrac{m_{h0}}{m_h(\nu_m)}\right) 
\left( 1 + \sum_{j=1}^3 r_{nj}(\tfrac{\nu}{m_h}) \left(\tfrac{\alphas(\nu)}{\pi}\right)^j \right),~n>4,
\end{align} 
eliminate the tree-level contribution, thus cancelling the leading lattice artifacts~\cite{HPQCD:2008kxl}. 
The coefficients $r_{nj}$ are numbers of order $1$ without any obvious pattern, see, \eg\ Table~1 of 
Ref.~\cite{Komijani:2020kst}. 
On the one hand, the lowest reduced moment $R_4$ and ratios of higher reduced moments, \ie\  
$\sfrac{R_{6}}{R_{8}}$ or $\sfrac{R_{8}}{R_{10}}$, are dimensionless; their respective 
continuum extrapolations turned out to be challenging, in particular due to 
$\log(a\Lambda_{\mathrm{QCD}})$ and $\log(am_{h0})$ dependence~\cite{Petreczky:2020tky}. 
Such $\log(a)$ dependence is manifest in slopes that decrease for larger $am_{h0}$. 
On the other hand, higher moments $\sfrac{R_{n}}{m_{h0}},~n \ge 6$, are dimensionful 
and scale with $\sfrac{1}{m_{h0}}=\sfrac{a}{(am_{h0})}$; thus, because the scale uncertainty 
and the uncertainty of the tuned bare heavy-quark mass ($am_{h0}$) strongly impact the results, 
these are insensitive to any $\log(a)$ effects, and continuum extrapolations are straightforward. 
For the improved actions used in published results, lattice spacing dependence could be 
parameterized for functions of the reduced moments, \ie\ 
$R(m_h) = \{ R_4(m_h), \sfrac{R_n(m_h)}{m_{h0}}, \sfrac{R_n(m_h)}{R_{n+2}(m_h)}, \},~n \ge 6$, as 
\begin{align}
R(m_h)
&=R^\mathrm{cont}(m_h)+\sum_{i=1}^N \sum_{j=1}^{M_i} b_{ij}^{(R)} (\alphas^b)^i 
\Bigl[1+\sum_{k=1}^i d_{ijk}^{(R)} \ln^k(a m_{h0})\Bigr](a m_{h0})^{2j}, 
\label{Rn_adep}
\end{align}
where $\alphas^b = \sfrac{g_0^2}{(4\pi u_0^4)}$ is the boosted lattice 
coupling~\cite{Lepage:1992xa}; the tadpole factor $u_0$ is defined in terms of the plaquette, $u_0^4=\braket{\sfrac{\mathrm{Tr}~{U_{\mu\nu}}}{N_c}}$, 
and partially accounts for the expected $\log(a \Lambda_{\mathrm{QCD}})$ dependence. 

For $R_4$ and $\sfrac{R_{6}}{R_{8}}$, separate continuum extrapolations 
for each heavy-quark mass proved feasible only for $m_h \le 1.5m_c$~\cite{Petreczky:2019ozv}. 
Continuum extrapolation for $m_h \ge 2m_c$ required joint fits including $m_h < 2m_c$~\cite{Petreczky:2020tky}; 
similar joint fits with Bayesian priors were used in Refs.~\cite{McNeile:2010ji, Chakraborty:2014aca}. 
The published continuum results for $R_4$ and $\sfrac{R_{6}}{R_{8}}$ at $m_h=m_c$  or $1.5m_c$ 
are consistent among each other~\cite{Komijani:2020kst, Petreczky:2020tky}; 
any significant deviations can be traced back to known deficiencies in the respective 
analyses~\cite{Maezawa:2016vgv, Nakayama:2016atf}, see \eg\ Table~65 of~Ref.~\cite{Aoki:2021kgd}. 
For $\sfrac{R_{n}}{m_{h0}},~n \ge 6$ fits with $N=1,~M_1=2$ are sufficient for any $m_h$, and published results for 
$R_n(m_c)/m_{c0},~n\ge 6$ are consistent. 
Severe finite volume effects affect $\sfrac{R_{8}}{R_{10}}$; $\sfrac{R_{8}}{R_{10}}$ at $m_c$ is systematically low (and 
inconsistent with $R_4$), while continuum extrapolation with joint fits proved reliable for $m_h \ge 1.5m_c$~\cite{Petreczky:2020tky}. 
With the aforementioned exceptions the continuum results in (2+1)-flavour QCD span the region $m_h=m_c,\ldots,4m_c$ 
for valence quarks, see Tables~1 and 4 of Ref.~\cite{Petreczky:2020tky}; 
with increasing heavy-quark mass $m_h$, $R_4$ and the ratios decrease towards $1$, while the errors of the 
lattice calculations increase. 
Continuum results in (2+1+1)-flavour QCD have not been published. 

Comparing $R_4(m_h)$ to $R_4(\alphas(\nu),\sfrac{\nu}{m_h})$ permits extraction of $\alphas(\nu)$; truncation errors, 
estimated via inclusion of terms $\pm 5 r_{n3}\,\alphas^4(\nu)$, turn out to be too large for the ratios to provide 
more than a cross-check~\cite{Petreczky:2020tky}. 
Once $\alphas(\nu)$ is given, $R_{n}/m_{h0}$ permits obtaining the $\MSbar$ heavy-quark mass $m_h(\nu_m)$; 
combining both yields $\Lambda_{\MSbar}^{(\Nf)}$. 
Whether or not the two scales $\nu_h = x m_h$ and $\nu_m =x_m m_h$, should be varied jointly ($x = x_m$) or independently 
($x \neq x_m$) is under investigation; the latter has yielded in a reanalysis of published lattice results at $m_h=m_c$ 
about 50\% larger estimates of truncation errors~\cite{Boito:2020lyp}. 
$\alphas(\nu)$ or $m_h(\nu_m)$ for different $am_{h0}=x_\mathrm{lat} am_{c0}$ are consistent with perturbative 
running~\cite{Petreczky:2020tky}. 
Nonperturbative contributions ---parametrized in terms of QCD condensates--- are negligible for $m_h \ge 1.5m_c$ due to 
suppression by powers of at least $1/m_h^4$; similarly, truncation errors diminish dramatically for $m_h \ge 1.5m_c$~\cite{Petreczky:2020tky}. 

\begin{table}
\centering
\caption{$\Lambda_{\MSbar}^{(\Nf=3)}$ obtained for different values of $m_h$ and different
choices of the renormalization scale $\nu$ (using $\nu=\nu_m$). The first error comes from the 
lattice calculations, the second error is the perturbative error, and the last error is due to 
the gluon condensate. From Ref.~\cite{Petreczky:2020tky}.\label{tab:Lambda from quarkonium moments}\vspace{0.2cm}}
\tabcolsep=4.mm
\begin{tabular}{c|c|c|c|c|c}
\hline
$\sfrac{m_h}{m_c}$  &  $\sfrac{\nu}{m_h}=\sfrac{2}{3}$  & $\sfrac{\nu}{m_h}=1$   &  $\sfrac{\nu}{m_h}=\sfrac{3}{2}$ &  $\sfrac{\nu}{m_h}=2$    &  $\sfrac{\nu}{m_h}=3$   \\
\hline
1.0        &        --       & 323(4)(6)(3)  &  323(4)(7)(3)  &  327(4)(13)(3)  &  340(4)(21)(3)  \\
1.5        & 314(8)(23)(1)   & 326(9)(4)(1)  &  326(8)(5)(1)  &  329(8)(10)(1)  &  341(9)(18)(1)  \\
2.0        &        --       & 327(13)(3)(0) &  327(13)(4)(0) &  330(13)(9)(0)  &  341(14)(16)(0) \\
3.0        & 325(20)(20)(0)  & 332(21)(2)(0) &  332(21)(4)(0) &  335(22)(22)(0) &  344(22)(14)(0) \\
4.0        &        --       & 336(26)(2)(0) &  336(26)(3)(0) &  339(27)(7)(0)  &  347(28)(17)(0) \\
\hline
\end{tabular}
\end{table}

A recent analysis has exposed that the choice of the lattice scale ratio, \ie\ $x_\mathrm{lat}=\sfrac{m_{h}}{m_{c}}=\sfrac{(am_{h0})}{(am_{c0})}$, 
and the perturbative scale ratio, $x_\mathrm{pert}=\sfrac{\nu}{m_h}$, both have a significant and systematic impact on the extracted $\Lambda_{\MSbar}^{(\Nf)}$, and consequently on $\alphasmZ$~\cite{Petreczky:2020tky}, while the composition of the error budget is very different (Table~\ref{tab:Lambda from quarkonium moments}).
Neglecting the spread due to varying either of these two scale ratios led in most past determinations to significantly underestimated errors~\cite{HPQCD:2008kxl, McNeile:2010ji, Chakraborty:2014aca, Maezawa:2016vgv, Petreczky:2019ozv}. Nonetheless, the central value of Ref.~\cite{McNeile:2010ji} is in good agreement with the corresponding entry of Table~\ref{tab:Lambda from quarkonium moments}. 

The current FLAG sub-average ---taking the latest results~\cite{Petreczky:2020tky} partially into account, and using error estimates from independent scale variation~\cite{Boito:2020lyp}--- reports 
\begin{align}
 &\alphasmZ = 0.11826(200)&&\text{(FLAG sub-average for heavy-quark two-point correlators)~\cite{Aoki:2021kgd}}.
\end{align}

Table~\ref{tab:Lambda from quarkonium moments} suggests that a viable approach on the lattice side to reducing the errors in the next decade may be by performing more accurate lattice calculations using masses $m_h \ge 2m_c$, where the truncation errors are subleading in current results. 
The corresponding continuum extrapolations could be made more robust in two ways. 
First, by including more intermediate heavy-quark mass values (\eg\ $\sfrac{m_h}{m_c}=1.75,~2.25,~2.5,~2.75,~3.25$, etc.) 
in the joint fits, Eq.~\eqref{Rn_adep}, one may hope to significantly reduce the lattice errors of the continuum results for $m_h \ge 2m_c$.  
Second, by relying on one-loop instead of tree-level reduced moments at finite lattice spacing as suggested in Ref.~\cite{Weber:2020gfh}, one may simplify the approach to the continuum limit; 
while cumbersome, these calculations in lattice perturbation theory are in principle straightforward and are expected to eliminate all $i=1$ terms from the series corresponding to Eq.~\eqref{Rn_adep}. 
The availability of lattice results in (2+1+1)-flavour QCD  with partially quenched heavy quarks suggests that a viable approach on the perturbative side to reducing the errors may be to permit partially quenched heavy quarks, \ie\ different heavy-quark masses on internal (sea) and external (valence) lines. 
In particular, application of this method in (2+1+1)-flavour QCD permits no $\alphas$ extraction for valence quarks at $m_h > m_c$, the only readily available strategy to alleviate the truncation error, on the basis of the currently available weak-coupling calculations that require $m_\text{h,sea} = m_\text{h,val}$. 
If this deficiency were remedied, the constraining power could be improved in a joint analysis of partially quenched lattice calculations with different $m_h$ in (2+1+1)-flavour QCD similar to the recent analysis in (2+1)-flavour QCD~\cite{Petreczky:2020tky}, or by even combining (2+1)- and (2+1+1)-flavour QCD continuum results in a joint analysis that assumes perturbative decoupling of the heavy quark. 
Last but not least, the expected accuracy would obviously benefit from $\mathrm{N^4LO}$, resp.\ $\mathcal{O}(\alphas^4(\nu_h))$, calculations.\\

\noindent{\it Acknowledgments---} PP was supported by U.S. Department of Energy under Contract No.\ DE-SC0012704. JHW's reserch was funded by the Deutsche Forschungsgemeinschaft (DFG, German Research Foundation) - Projektnummer 417533893/GRK2575 ``Rethinking Quantum Field Theory''.

\subsection{Strong coupling constant \texorpdfstring{$\alphas$}{alphas} from the static energy, the free energy and the force
\protect\footnote{A\lowercase{uthors:} N. B\lowercase{rambilla} (TUM), V. L\lowercase{eino} (TUM), 
P. P\lowercase{etreczky} (BNL), A.  V\lowercase{airo} (TUM)   J. H. W\lowercase{eber} (HU, B\lowercase{erlin})
}}

QCD observables that have been computed with high precision in perturbative- and lattice-QCD with 2+1 or 2+1+1 flavours are well suited to provide determinations of $\alphas$ in the kinematic regions where pQCD applies. The advantage of looking at observables is that continuum analytical and lattice results may be compared without having to deal with renormalization issues and change of scheme.
Moreover, if in the kinematic regions where pQCD is used the perturbative series 
converges well, and nonperturbative corrections turn out to be small, and if 
lattice data can be extrapolated to continuum, then a very precise extraction of  $\alphas$ is possible even by 
comparing few lattice data with the perturbative expression. 
The comparison provides $\Lambda_{\MSbar}$ times the lattice scale. By supplying a precise determination of the 
lattice scale, one obtains $\Lambda_{\MSbar}$. Finally, $\Lambda_{\MSbar}$ may be traded with $\alphas$ conventionally expressed at the mass of the Z, $\alphasmZ$.

\subsubsection{The QCD static energy}

The QCD static energy $E_0(r)$, \ie\ the energy
between a static quark and a static antiquark separated by a distance $r$, is one of these golden observables for 
the extraction of $\alphas$ and it is also a basic object to study the strong interactions~\cite{Wilson:1974sk}.
The QCD static energy, $E_0(r)$, is defined (in Minkowski spacetime) as
\begin{equation}
E_0(r) = \lim_{ T\to\infty}\frac{i}{T} \ln \, \left\langle {\rm Tr} \,{\rm P} 
\exp\left\{i g \oint_{r\times T} dz^\mu \, A_\mu\right\} \right\rangle
\equiv\lim_{ T\to\infty}\frac{i}{T} \ln \, \left\langle    W_{r \times T} \right\rangle,
\end{equation}
where the integral is over a rectangle of spatial length $r$, the distance between the static quark and the static 
antiquark, and time length $T$; $\langle \dots \rangle$ stands for the path integral over the gauge fields $A_\mu$ and 
the light quark fields, P is the path-ordering operator of the colour matrices and $g$ is the SU(3) gauge coupling 
($\alphas = g^2/(4\pi)$); $W_{r\times T}$ is the static Wilson loop.
The above definition of $E_0(r)$ is valid at any distance $r$.

On the lattice side, the Wilson loop is one of the most accurately known quantities that has been computed since 
the establishment of lattice QCD.
In the short distance range, $r\lQ \ll 1$ for which $\alphas(1/r) \ll 1$, $E_0(r)$ may be computed in pQCD  
and expressed as a series in $\alphas$ (computed at a typical scale of order $1/r$):
\begin{equation}
E_0(r) = \Lambda_s - \frac{4\alphas}{3r}(1 + \dots),
\end{equation}
where $\Lambda_s$ is a constant that accounts for the normalization of the static energy and the dots stand for higher-order terms.
The expansion of $E_0(r)$ in powers of $\alphas$ has been computed, in the continuum in the $\MSbar$ scheme, using perturbative 
and effective field theory techniques, in particular potential Nonrelativistic QCD \cite{Brambilla:1999xf}.
It is currently known at next-to-next-to-next-to-leading-logarithmic (N$^3$LL) accuracy, 
\ie\ including terms up to order $\alphas^{4+n}\ln^n\alphas$ with $n\ge0$.  
At three loops, a contribution proportional to $\ln\alphas$ appears for the first time. 
This three-loop logarithm has been computed in~\cite{Brambilla:1999qa}.
The complete three-loop contribution has been computed in~\cite{Anzai:2009tm,Smirnov:2009fh}.
The leading logarithms have been resummed to all orders in~\cite{Pineda:2000gza}, 
providing, among others, also the four-loop contribution proportional to $\alphas^5\ln^2 \alphas$.
The four-loop contribution proportional to $\alphas^5 \ln \alphas$ has been computed in~\cite{Brambilla:2006wp}.
Next-to-leading logarithms have been resummed to all orders in~\cite{Brambilla:2009bi}.
However the constant coefficient of the $\alphas^5$ term is not yet known.
$E_0(r)$ is, therefore, one of the best known quantities in pQCD
lending to a perfect playground for the extraction of $\alphas$.

The $\ln\alphas$ terms in $E_0(r)$ signal the cancellation of contributions coming from the soft energy scale $1/r$ and the ultrasoft (US) 
energy scale $\alphas/r$. The ultrasoft scale is generated in loop diagrams by  
the emission of virtual ultrasoft gluons changing the colour state of the quark-antiquark pair~\cite{Brambilla:1999qa,Appelquist:1977tw}. 
The resummation of these logarithms accounts for the running from the soft to the ultrasoft scale.

The soft and ultrasoft energy scales may be effectively factorized in potential Nonrelativistic QCD~\cite{Pineda:1997bj,Brambilla:1999xf}.
The factorization of scales leads to the formula~\cite{Brambilla:1999qa,Brambilla:1999xf}:
\begin{eqnarray}
E_0(r) &=& \Lambda_s ~~~ + ~~~V_{s}(r, \mu) - i \frac{V_A^2}{3} \int_0^\infty dt \, e^{-i t (V_o-V_s)}
\,\langle {\rm Tr} \,\{ {\bf r} \cdot g{\bf E}(t,{\bf 0}) \, {\bf r} \cdot g{\bf E}(0,{\bf 0})\}  \rangle(\mu) + \dots \;,
\label{E0}
\end{eqnarray}
where $\mu$ is the US renormalization scale, $V_s(r,\mu) = -4\alphas/(3r) + \dots$ is the colour-singlet static potential, 
$V_o(r,\mu) = \alphas/(6r) + \dots$ is the colour-octet static potential, $V_A = 1 + \mathcal{O}(\alphas^2)$ is a Wilson coefficient giving the 
chromoelectric dipole coupling, and ${\bf E}$ is the chromoelectric field. 
The colour-singlet and colour-octet static potentials encode the contributions from the soft scale $1/r$, whereas the low-energy 
contributions are in the term proportional to the two chromoelectric dipoles.  
While at short distances, $r\lQ \ll 1$, the potentials $V_s$ and $V_o$ may be computed in perturbation theory, 
low-energy contributions include nonperturbative contributions.

The perturbative expansion of $V_s$ is affected by a renormalon ambiguity of order $\lQ$. 
The ambiguity reflects in the poor convergence of the perturbative series. 
A first method to cure the poor convergence of the perturbative series of $V_s$ consists in subtracting a (constant) 
series in $\alphas$ from $V_s$ and reabsorb it into a redefinition of the normalization constant $\Lambda_s$.
This is the strategy we followed, for instance, in~\cite{Bazavov:2012ka}. 
A second possibility consists in considering the force 
\begin{equation}
F(r,\nu) = \frac{d}{d r}E_0(r).
\label{force}
\end{equation}
It does not depend on $\Lambda_s$ and is free from the renormalon of order $\lQ$~\cite{Necco:2001xg,Pineda:2002se}.
Once integrated upon the distance, the force gives back the static energy
\begin{equation}
E_0(r)=\int_{r_*}^{r}dr'\, F(r',1/r'),
\label{E0force}
\end{equation}
up to an irrelevant constant determined by the arbitrary distance $r_*$, 
which can be reabsorbed in the overall normalization when comparing with lattice data.
This is the strategy followed, for instance, in~\cite{Bazavov:2019qoo}.
We note that Eq.~\eqref{E0} provides also the explicit form of the nonperturbative 
contributions encoded in the chromoelectric correlator. They are proportional to $r^3$ at very short distances and to $r^2$ at 
somewhat larger distances.

In summary, $E_0(r)$ is one of the best known quantities in pQCD lending an ideal observable for the extraction of $\alphas$ by comparing lattice data and perturbative calculation in the appropriate short distance window.
This way of extracting of $\alphas$ has been developed in a 
series of papers \cite{Brambilla:2010pp,Bazavov:2012ka,Bazavov:2014soa}.
Here we report about our best determination from \cite{Bazavov:2019qoo}.
The method provides one of the most precise low energy determinations of $\alphas$.
The strong coupling constant extracted in this way relies typically on low energy data because the lattice cannot explore too small distances. 
It therefore provides a precise check of the running of the coupling constant and a determination of it that is complementary to high-energy determinations.

Concerning the power counting of the perturbative series a remark is in order.
Upon inspection of the numerical size of the contributions coming from the soft and the US scale at each order, in the analysis of \cite{Bazavov:2019qoo} 
it was decided to count the leading US resummed terms along with the three loop terms,  
since the $\alphas^4\log\alphas$ terms appear to be of the same size as the $\alphas^4$ terms, and, moreover, 
to partially cancel each other.
It was also decided not to include subleading US logarithms in the analysis, as the finite four loop contribution
is unknown and a cancellation similar to the one happening at three loops may also happen at four loops.
Nevertheless, it may be also legitimate 
to count leading US resummed terms as if they were parametrically  of order $\alphas^3$ and count subleading US logarithms 
of order $\alphas^5 \log \alphas$ as if they were parametrically of order $\alphas^4$, including them in the analysis.
This is the procedure adopted in~\cite{Ayala:2020odx}.
Most of the difference between the central value of $\alphas$ obtained in the analysis of \cite{Bazavov:2019qoo} 
and in the one of~\cite{Ayala:2020odx} is due to this different counting of the perturbative series.
The two analyses are consistent once errors, in particular those due to the truncation of the perturbative series, are accounted for.

The static energy can be computed on the lattice as the ground state of Wilson loops or 
temporal Wilson line correlators in a suitable gauge, typically in Coulomb gauge. 
Polyakov loops at sufficiently low temperatures could be employed as well. 
All energy levels from any of these correlators are affected by a constant, lattice spacing dependent self-energy contribution 
that diverges in the continuum limit. 
It can be removed by matching the static energy at each finite lattice spacing to a finite value at some distance. 
In calculations with an improved action, all these correlators, which are obtained from spatially 
extended operators, are affected by nonpositive contributions at very small distance and time, which 
cannot be resolved on coarse lattices or with insufficient suppression of the lowest excited states.
Although Wilson line correlators retain an advantage in terms of the excited state suppression, the relative disadvantage of 
Wilson loops could be alleviated to some extent with smeared spatial links. 

The ground state energy $E_0(\bm{r},a)$ can be extracted from such correlators, \eg\ via multi-exponential fits, in the large 
Euclidean time region for each lattice three-vector $\sfrac{\bm{r}}{a}=(n_1,n_2,n_3)$, where $~0\le n_i \le \sfrac{L_i}{2a}$.  
Obviously, small lattice spacing $a$ is indispensable in order to access small distances $|\bm{r}| \lesssim 0.15\,\mathrm{fm}$~\cite{Bazavov:2014soa}. 
Yet simulations with periodic boundary conditions fail to sample the different topological sectors of the QCD vacuum properly at small $a$. 
This topological freezing, which is known to be a quantitatively small but significant problem in low-energy hadron physics, does 
not lead in high-energy quantities (such as $E_0(\bm{r},a)$ at small distances) to statistically significant effects due to small 
changes of the topological charge~\cite{Weber:2018bam}. 
Although significant effects in $E_0(\bm{r},a)$ due to large changes of the topological charge cannot be completely ruled out, 
they seem very unlikely, given that the topology does not contribute at all in the weak-coupling calculations used in the comparison. 
Furthermore, sea quark mass effects due to light or strange quarks do not play a role in this range~\cite{Bazavov:2017dsy, Weber:2018bam}, 
while dynamical charm effects are significant~\cite{Steinbeisser:2021jgc,Brambilla:2022het}. 
At distance larger than $|\bm{r}| \gtrsim 0.15\,\mathrm{fm}$ light sea quark mass effects become nonnegligible~\cite{Bazavov:2017dsy, Weber:2018bam}.

Continuum extrapolation is only possible for somewhat larger distances probed by multiple lattice spacings, 
or if the functional form of $E_0(\bm{r},a)$ were known to sufficient 
accuracy to predict the shape at small $\sfrac{\bm{r}}{a}$ on the fine lattices. 
Due to the breaking of the continuous $O(3)$-symmetry group to the discrete $W_3$-symmetry group 
the lattice gluon propagator, and hence $E_0(\bm{r},a)$, is a non-smooth function of $\sfrac{|\bm{r}|}{a}$;
geometrically inequivalent combinations of the $n_i$, \ie\ belonging to different representations of $W_3$ but 
corresponding to the same geometric distance $\sfrac{|\bm{r}|}{a}=\sqrt{n_1^2+n_2^2+n_3^2}$, \eg\ $(3,0,0)$ or $(2,2,1)$, yield 
inconsistent $E_0(|\bm{r}|,a)$ due to lattice artifacts. 
Moreover, the same $|\bm{r}|$ accessed through different lattice spacings $a$ is generally affected by different types of non-smooth 
lattice artifacts corresponding to the different underlying $\sfrac{\bm{r}}{a}$. 
For this reason, a continuum extrapolation of $E_0(|\bm{r}|,a)$ at fixed $|\bm{r}|$ utilizing a parametrization of lattice 
artifacts in terms of a smooth function in $|\bm{r}|$ is incapable of describing the small $\sfrac{|\bm{r}|}{a}$ region, 
see \eg\ Ref.~\cite{Komijani:2020kst}. 
These inconsistencies are much larger than the statistical errors at small $\sfrac{|\bm{r}|}{a}$, but covered by the statistical 
errors at large $\sfrac{|\bm{r}|}{a}$. 
A tree-level correction (TLC) procedure defines the (tree-level) improved distance $\sfrac{r_I}{a} = f(n_1,n_2,n_3)$ 
and alleviates these inconsistencies somewhat: at $\sfrac{|\bm{r}|}{a} \gtrsim 3$ this is sufficient, while further effort is 
needed at smaller $\sfrac{|\bm{r}|}{a}$. 
A nonperturbative correction (NPC) procedure heuristically estimates the lattice artifacts remaining in $E_0(r_I,a)$ by 
comparing to a suitable smooth function, either obtained at a finer lattice spacing, or in a continuum calculation, which, 
however, potentially introduces systematic errors. 
Both approaches have been used yielding consistent results; for details see Ref.~\cite{Bazavov:2019qoo}.

\begin{figure}[htpb!]
\centering
\includegraphics[width=0.49\textwidth]{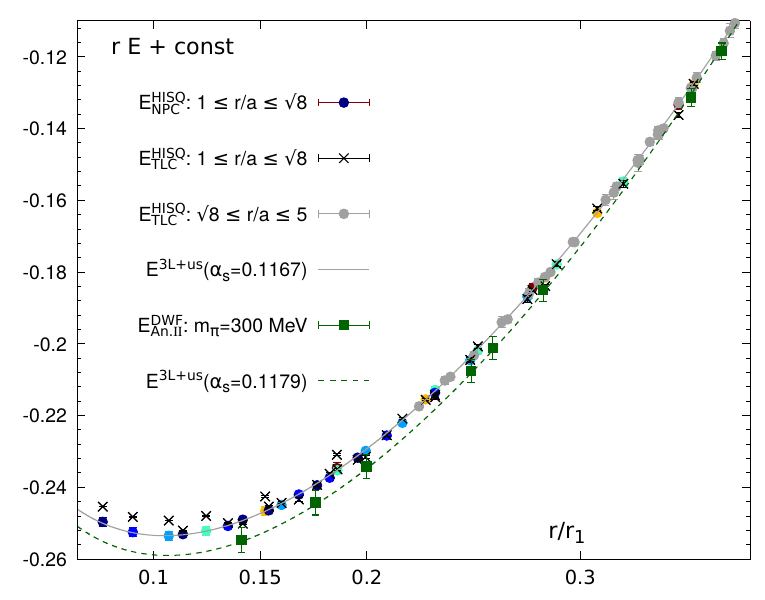}
\caption{
The nonperturbative lattice and the perturbative continuum results for the static energy multiplied 
by the distance, $r E(r)$ ($E\equiv E_0$). 
The HISQ data~\cite{Bazavov:2019qoo} are {nonperturbatively corrected} (NPC, coloured bullets) or 
{tree-level corrected} (TLC, black crosses and gray bullets). 
The colour indicates the lattice spacing in units of the scale $r_1$, $\sfrac{a}{r_1}$. 
The DWF data~\cite{Takaura:2018vcy,Takaura:2018lpw} are from a one-step analysis {\rm II} that mixes the 
continuum extrapolation with the fit to the OPE result at $\mathrm{N^3LO}$, using a parametrization 
of discretization artifacts (green squares). 
The lines represent the {three-loop result with resummed leading ultrasoft logarithms}, corresponding 
to $\alphasmZ=0.1167$ (gray, solid) or $\alphasmZ=0.1179$ (green, dashed). 
The former uses the central value $\alphasmZ=0.1167$ of the analysis of the (TLC or NPC) HISQ data with 
$\sfrac{|\bm{r}|}{a} \ge \sqrt{8}$ (gray bullets), the latter uses the central value $\alphasmZ=0.1179$ 
of the OPE-based one-step analysis {\rm II} of the DWF data~\cite{Takaura:2018vcy,Takaura:2018lpw}. 
The NPC HISQ data with $\sfrac{|\bm{r}|}{a} < \sqrt{8}$ are well-aligned with the fit excluding these data, 
while the TLC HISQ data with $\sfrac{|\bm{r}|}{a} < \sqrt{8}$ cannot be consistently described by a continuum result for any value of $\alphasmZ$. 
The figure is taken from Ref.~\cite{Komijani:2020kst}.
}
\label{fig:static energy}
\end{figure}

For Wilson line correlators, all combinations of $0\le n_1,n_2,n_3 \le \sfrac{L}{2a}$ 
are accessible, thus permitting access to noninteger distances in units of the lattice spacing; 
this entails all of the aforementioned complications. 
After the nonperturbative correction, $E_0^\mathrm{NPC}(r_I,a)$ from Wilson line correlators computed 
in (2+1)-flavour QCD on lattices with highly improved staggered quarks (HISQ) was shown to exhibit no 
statistically significant lattice artifacts anymore~\cite{Bazavov:2019qoo} (Fig.~\ref{fig:static energy}).
With this data set the range of interest can be probed using a rather large number of $\sfrac{r_I}{a}$ 
values and many lattice spacings (up to six spacings in the range $a \in (0.024,\ldots,0.06)\,\mathrm{fm}$), 
where the underlying high statistics ensembles had been generated for studies of the QCD equation of 
state~\cite{HotQCD:2014kol, Bazavov:2017dsy}. 
The uncertainty due to the lattice scale, lattice spacing dependence, estimates of the uncertainty due to 
treating residual lattice artifacts with the nonperturbative correction, or due to changes of the fit range 
are within the the statistical error and subleading in the error budget. 
Instead it was found that estimates of the continuum perturbative  truncation error dominate the error budget. 
In Ref.~\cite{Bazavov:2019qoo} these have been estimated by a scale variation between $\sfrac{2}{r}$ and $\sfrac{1}{2r}$, 
inclusion of a parametric estimate of a higher order term $\pm \sfrac{4}{3} \sfrac{\alphas^5}{r}$, 
and variation between resummation or no resummation of the leading ultrasoft logarithms $\sfrac{\alphas^{3+n} \ln^n\alphas}{r}$. 
The scale dependence becomes nonmonotonic below $\sfrac{1}{\sqrt{2}r}$ at large $r$, which makes robust error estimates 
challenging unless the range is restricted to $\max{(|\bm{r}|)} \le 0.1\,\mathrm{fm}$. 
As lattice data at larger $|\bm{r}|$ are discarded, the statistical error increases while the truncation error 
decreases. Eventually, for $\max{(|\bm{r}|)} \ll 0.1\,\mathrm{fm}$, the nonperturbative correction to the lattice data 
becomes essential to having enough data, while the central value hardly changes. 
For our joint fit using nonperturbatively corrected HISQ data at five lattice spacings we report 
the best compromise between the different contributions to the error budget as 
\begin{equation}
\alphasmZ=0.11660^{+0.00110}_{-0.00056}, 
\label{alMZ_TUMQCD}
\end{equation}
where the total, symmetric lattice error amounts to $0.00047$ for $|\bm{r}| \le 0.073\,\mathrm{fm}$. 

In~\cite{Ayala:2020odx}, a reanalysis of a subset of these data ($a=0.024\,\mathrm{fm}$, $\sqrt{8}\le \sfrac{|\bm{r}|}{a} \le 4$) was carried out that included resummation of next-to-leading ultrasoft logarithms
$\sfrac{\alphas^{4+n} \ln^n{\alphas}}{r}$, \ie\ the full $\mathrm{N^3LL}$ accuracy, and hyperasymptotic expansion, 
resulting in $\alphasmZ=0.1181(9)$. 
Concerning the central value, we have already commented that it differs from \eqref{alMZ_TUMQCD} mostly 
because of the inclusion of the subleading ultrasoft logarithms.
Concerning the error budget, it may possibly increase by including an estimate of the lattice spacing dependence and a variation of $\min{(|\bm{r}|)}$ or $\max{(|\bm{r}|)}$ to smaller values.
Nevertheless, even inside the quoted errors the result is consistent with \eqref{alMZ_TUMQCD}. 

For Wilson loops, spatial Wilson lines connecting the temporal ones entail additional, $\bm{r}/a$-dependent self-energy divergences. The static energy from Wilson loops is usually computed only for few specific geometries, and spatial link smearing is applied to suppress these divergences. 
The static energy has been obtained from Wilson loops for two different geometries $\bm{r}/a \propto (1,0,0)$ or $(1,1,0)$~\cite{Takaura:2018lpw,Takaura:2018vcy} in (2+1)-flavour QCD on lattices with M\"obius domain-wall fermions 
with a pion mass of $m_\pi \approx 300\,\mathrm{MeV}$ 
and three lattice spacings $a \in (0.04,\ldots,0.08)\,\mathrm{fm}$ 
that had been generated by the JLQCD collaboration~\cite{Kaneko:2013jla}. 
Two separate analyses were performed: a two-step analysis (I) with a continuum extrapolation at large distances 
sequentially followed by the $\alphas$ extraction, and another one-step analysis (II) using a single joint fit 
to achieve both at once. 
Both analyses relied on a particular form of operator product expansion, and used data in the range 
$0.24\,\mathrm{fm} \le |\bm{r}| \le 0.6\,\mathrm{fm}$ or $0.044\,\mathrm{fm} \le |\bm{r}| \le 0.36\,\mathrm{fm}$, respectively. 
The reported values from the two analyses are $\alphasmZ = 0.1166^{+0.0021}_{-0.0020}$ and  
$\alphasmZ = 0.1179^{+0.0015}_{-0.0014}$. They are both dominated by the estimate of the truncation errors. 
As both analyses extend far into ranges where $E_0(|\bm{r}|,a)$ is known to be sensitive to the pion mass~\cite{Weber:2018bam}, 
the authors had to include condensate terms, while reporting no significant mass dependence when assuming only 
a perturbative contribution from massive light quarks. 
The first analysis has no data for $|\bm{r}| \lesssim 0.10\,\mathrm{fm}$. 
The second analysis does have data in that region, but has to fit $\Lambda_{\MSbar}$ 
simultaneously with the lattice artifacts at small $\sfrac{\bm{r}}{a}$. 
As a result the continuum extrapolated static energy from this analysis has large errors for $|\bm{r}| \lesssim 0.10\,\mathrm{fm}$ and lies systematically below the HISQ result from Ref.~\cite{Bazavov:2019qoo} (Fig.~\ref{fig:static energy}). 

\subsubsection{Static force}

Another possibility consists of computing the force directly from the lattice, \ie\, not as the slope of the static energy. 
The force, $F$, between a static quark located in ${\bf r}$ and a static antiquark located in ${\bf 0}$ can be defined as~\cite{Vairo:2015vgb}
\begin{eqnarray}
F(r) = \partial_r E_0(r) 
      =  \lim_{T \to \infty}
     -i \frac{\left\langle \textrm{Tr}\left\{  W_{r \times T} \,\hat{\mathbf{r}} \cdot g \mathbf{E}(\mathbf{r},t^\ast)\right\} \right\rangle}{\langle \textrm{Tr}\{W_{r \times T}\} \rangle} .
\label{EQN_F}
\end{eqnarray}
The chromoelectric field $\mathbf{E}(\mathbf{r},t^\ast)$  is located at the quark line of the Wilson loop.

This definition permits to obtain the force directly from the lattice instead of 
reconstructing it, after interpolation, from the lattice data of the static energy.
The perturbative calculation of the force in continuum QCD is free of the leading renormalon and it is 
known at N$^3$LL. Therefore the force provides a clean way to extract $\alphas$.
In \cite{Brambilla:2021wqs}, we have used the multilevel algorithm to perform a preliminary quenched lattice study of
the chromoelectric insertion in a static Wilson loop given by Eq. \eqref{EQN_F} 
both with smeared Wilson loops and with Polyakov loops. The result 
is consistent with the force obtained via derivative of the static energy upon multiplication with a constant renormalization factor $Z_E$ that encodes the very slow 
convergence of lattice operators containing gluonic operators. Recently in
\cite{Leino:2021vop} we have performed the same calculation with gradient flow, which 
eliminates the necessity of $Z_E$ and makes the lattice calculation more efficient.
We plan to go to very fine lattice spacings and perform an extraction of $\Lambda_{\MS}r_0$.

\subsubsection{Static singlet free energy} 

One reason for which it is challenging to reach the fine lattice 
spacings needed for the best extraction of $\alphas$ is  that one has to simultaneously maintain the control over finite volume effects from the propagation of the lightest hadronic modes, namely, the pions. A lattice simulation at high enough temperature avoids this infrared problem, and  thus enables reaching much finer lattice spacings using smaller volumes. In \cite{Bazavov:2019qoo},  
we considered the extraction of the strong coupling 
from the singlet free energy at nonzero temperature, as it is expected
that at small distances medium effects are small.
The singlet free energy in terms of the correlation function 
of two thermal Wilson lines in Coulomb gauge is given by ($T$ is now the temperature)
\begin{equation}
F_S(r,T)=-T\ln \left(\frac{1}{N_c}\langle {\rm Tr}\left[ W(r) W^{\dagger}(0) \right]\rangle\right).
\end{equation}
At distances much smaller than the inverse temperature,  
$r T \ll 1$, we can write using potential Nonrelativistic QCD~\cite{Berwein:2017thy}
\begin{equation}
F_S(r,T)=V_s(r,\mu)+\delta F_S(r,T,\mu).
\end{equation}
The form of the US correction depends on the scale hierarchy that is now featuring also the temperature and the Debye mass $m_D \sim gT$: if $1/r \gg \alphas/r \gg T \gg m_D$ then $\mu \sim \alphas/r$ and $\delta F_S(r,T,\mu)=\delta E_{US}(\mu)+\Delta F_S(r,T)$, with $\delta E_{US}(\mu)$ being the US contribution to $E_0$ in vacuum that we already discussed.
The singlet free energy has been studied on the lattice in \mbox{Ref.}~\cite{Bazavov:2018wmo} using a wide temperature range and several lattice spacings, \ie, several temporal extents $N_{\tau}$. 
The shortest distance that we can access, due to a single lattice 
spacing on our finest lattice at \(T>0\), is \(0.00814\,{\rm fm}\). 
For our analysis the relevant data correspond to $N_{\tau}=10,~12$ and $16$, since $rT$ has to be small. 
From the analysis of \mbox{Ref.}~\cite{Bazavov:2018wmo}, we see that thermal effects are small for $rT \lesssim 0.3$. 
In particular we see that for $|\bm{r}|/a \le \sqrt{6}$ the difference between the singlet free energy at $(T>0)$ and the static energy at $(T=0)$ approaches a constant proportional to the temperature. 
No temperature effects beyond a constant can be seen in this range within the errors of the lattice results. 
Therefore we treat the finite temperature data in this range as if they were at zero 
temperature and fit them with the three-loop plus leading ultrasoft resummed result of the static energy at $T=0$.

\begin{figure}[htpb!]
\centering
\includegraphics[width=0.49\textwidth]{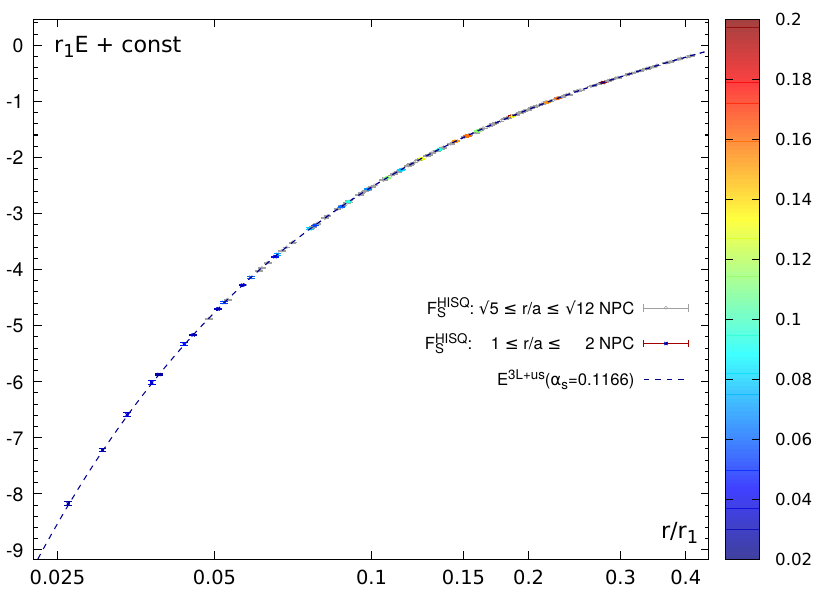}
\caption{
The nonperturbative lattice result $r_1 E(r)$ ($E\equiv E_0$) using the singlet free energy at $T>0$ and the perturbative continuum results for the static energy. 
The HISQ data~\cite{Bazavov:2019qoo} are nonperturbatively corrected (NPC, coloured bullets and gray bullets). 
The colour indicates the lattice spacing in units of the $r_1$ scale, $\sfrac{a}{r_1}$. 
The dashed line represents the {three-loop result with resummed leading ultrasoft logarithms}, corresponding 
to the result of the optimal $T=0$ analysis $\alphasmZ=0.1166$. 
Only the NPC HISQ data at $T>0$ with $\sfrac{|\bm{r}|}{a} \le 2$ (coloured) are used in the fit, but all data in the range $2 < \sfrac{|\bm{r}|}{a} \le \sqrt{12}$ (gray) are still consistent with it. 
The figure is taken from Ref.~\cite{Bazavov:2019qoo}.}
\label{fig:free energy}
\end{figure}

A drawback of this analysis is the restriction to very small $|\bm{r}|/a$, which implies that the 
nonperturbative correction (NPC) is indispensable. 
Yet the main advantage of the analysis is that at very small distances, $\max{(|\bm{r}|)} \le 0.03\,\mathrm{fm}$, 
the truncation error becomes very small.
Beyond what has been necessary at zero temperature, one must verify that $T>0$ effects do not become significant when 
including larger $|\bm{r}|/a$ or when varying the temperature $T=\sfrac{1}{aN_\tau}$ at fixed lattice spacing. 
We confirmed the statistical irrelevance in these two measures of $T>0$ effects by comparing $\sfrac{|\bm{r}|}{a} \le 2$ 
with  $\sfrac{|\bm{r}|}{a} \le 3$ and $N_\tau=12,~16$, and, where possible, $64$, and included these error estimates 
with the other lattice errors in the error budget. 
The key ingredient to making this extraction work with substantial constraining power is that a wide range in $|\bm{r}|$ can be covered using only narrow windows in $\sfrac{|\bm{r}|}{a}$ by performing a joint analysis using multiple lattice spacings $a$ that cover $|\bm{r}|$ in multiple, overlapping segments. 
For our joint fit using nonperturbatively corrected $T>0$ HISQ data at eight lattice spacings and an $|\bm{r}|$ region and set of lattice spacings nonoverlapping with the zero temperature analysis we report 
\begin{equation}
\alphasmZ=0.11638^{+0.00095}_{-0.00087}
\end{equation}
(the total, symmetric lattice error amounts to $0.00085$) for $|\bm{r}| \le 0.03\,\mathrm{fm}$~\cite{Bazavov:2019qoo}.
This result is marginally lower than the result from our zero temperature analysis. 
Remarkably, however, the zero temperature result is still compatible with the $T>0$ data over a much wider range in $\sfrac{|\bm{r}|}{a}$, a factor of sixteen in $|\bm{r}|$, and over almost a factor of ten in lattice spacings (Fig.~\ref{fig:free energy}).

\subsubsection{Outlook}

In summary, in order to reach a few permil accuracy in the extraction of $\alphas$ with the outlined methods we need:
\begin{itemize}
    \item 
    Lattice calculations of the static energy at smaller distances, ideally extrapolated to the continuum. 
    The one-loop correction should be computed in lattice perturbation theory. Based on the experience with the nonperturbative correction, the one-loop correction is expected to substantially alleviate the restriction of $\sfrac{|\bm{r}|}{a}$ (from  $\sfrac{|\bm{r}|}{a} \gtrsim 3$ to $\sfrac{|\bm{r}|}{a} \gtrsim 2$) in a field-theoretically rigorous manner. 
    \item 
    To exploit more systematically finite temperature simulations in order to go to shorter distances.
    On the lattice side there is no major impediment to further increasing the temperature and decreasing the lattice spacing.
    \item 
    To have the nonlogarithmic four-loop contribution to the static potential. This would be important to 
    verify if large numerical cancellations between soft and US contributions also happen beyond three loops.
    \item 
    To further exploit the definition of the force in terms of a chromoelectric field insertion in the static  Wilson loop in order to generate high precision quenched and unquenched lattice data at short distances and to compare with the perturbative expression at N$^3$LO or N$^3$LL.
    \item 
    To use recent high quality lattice data at 2+1+1 flavours to extract $\alphas$  from the static energy with an active charm quark~\cite{Steinbeisser:2021jgc,Brambilla:2022het}.
    To achieve precision competitive with 2+1 flavour QCD, a full three-loop calculation of the contribution to the static potential due to the finite charm quark mass may be necessary.
    \end{itemize}

\medskip

\noindent{\it Acknowledgments---} P.P. was supported by U.S. Department of Energy under Contract No.\ DE-SC0012704. J.H.W.'s research was funded by the Deutsche Forschungsgemeinschaft (DFG,
German Research Foundation) - Projektnummer 417533893/GRK2575 ``Rethinking Quantum Field Theory''.

\clearpage

\subsection{Remarks on determining \texorpdfstring{$\alphas$}{alphas} \protect\footnote{A\lowercase{uthors:} A. S. K\lowercase{ronfeld} (F\lowercase{ermilab})}}

Every method to determine the strong coupling $\alphas$ starts with an observable that depends on a short distance, $1/Q$ (or high energy $Q$).
The notion of ``short'' or ``high'' relates to other scales in the problem.
The observable can be multiplied by the appropriate power of $Q$ to obtain a dimensionless quantity, $\mathcal{R}(Q)$, which can be written
\begin{equation}
    \mathcal{R}(Q) = R(Q) + Q^{-p} S(M,Q),
    \label{eq:Ralphas}
\end{equation}
where $M$ is another (energy) scale, $M\ll Q$, the first term $R$ does not depend on $M$, and the power $p>0$ is something like
2,~4, or~1.
The separation can be justified with tools such as the operator product expansion (OPE)~\cite{Wilson:1969zs} or an effective field theory
(EFT)~\cite{Symanzik:1983dc,Symanzik:1983gh}.
(Sometimes Eq.~(\ref{eq:Ralphas}) is posited on the basis of arguments or assumptions.)
With the OPE or an EFT, the power $p$ corresponds to the dimension of an operator.
For example, $R$ might be the short-distance (or ``Wilson'') coefficient of the unit operator, and $S$ the product of another coefficient and
the matrix element of a dimension-4 operator, in which case $p=4$.

In an asymptotically free quantum field theory, such as QCD, the short-distance part can be well approximated with perturbation theory:
\begin{equation}
    R(Q) = \sum_{l=0}^\infty R_l(Q/\mu) \alphas^l(\mu) = \sum_{l=0}^\infty R_l \alphas^l(Q),
\end{equation}
where $\mu$ is a separation scale introduced by the OPE, EFT, or other consideration; in the last expression, $R_l=R_l(1)$.
Because $\mu$ is unphysical, it can be chosen, at least after thorough analysis of the scale-separation details, equal or proportional to $Q$.
At this stage, it is useful to introduce (with $\mu=Q$)
\begin{equation}
    \alpha_\mathcal{R}(Q) = \frac{\mathcal{R}(Q) - R_0}{R_1},
\end{equation}
which is known as the effective charge (or effective coupling) for the observable $\mathcal{R}$.
It may be more apt to note that $\alpha_\mathcal{R}$ is simply a physical, regularization-independent choice of renormalization scheme.

It is tempting to identify ``short distance'' with ``perturbative'' and ``long distance'' with ``nonperturbative''.
In the OPE or an EFT, however, the home for small instantons (for example) of diameter $1/Q$ is in short-distance coefficients.
Fortunately, small-instanton contributions scale with a large power, $11-2\Nf/3\ge7$ (where $\Nf$ is the number of active quark flavour), so they are small enough that neglecting
them is safer than other compromises that must be made.
Further complications in separating scales arise, such as renormalons.
In the end, the point of Eq.~(\ref{eq:Ralphas}) is that short-distance quantities depend on $Q$ logarithmically with power-law corrections.

A practical version, then, of Eq.~(\ref{eq:Ralphas}) is
\begin{equation}
    \alpha_\mathcal{R}(Q) =\sum_{l=1}^{n_\text{loop}} r_l\alphas^l(Q) + \sum_{l=1+n_\text{loop}} r_l\alphas^l(Q) + \sum_p d_p \left(M/Q\right)^p, \quad r_1=1,
    \label{eq:alphasR}
\end{equation}
presupposing a preferred renormalization scheme (such as $\MSbar$), truncating the perturbative series at $n_\text{loop}$ terms (those available from explicit
perturbative calculation), acknowledging that the remaining terms do not vanish, and allowing for several power corrections.
The renormalization group provides more information, in particular showing how to relate $\alpha_\mathcal{R}(Q)$ at one scale to
$\alpha_\mathcal{R}(Q_0)$ at a fiducial scale~$Q_0$.
Instead of using the $r_l$ to relate $\alpha_\mathcal{R}$ to $\alpha_{\MSbar}$, they can be used to convert the coefficients of the
$\beta$~function in the $\MSbar$ scheme to the $\mathcal{R}$ scheme.
Thus, perturbation theory predicts the logarithmic $Q$ dependence of $\alpha_\mathcal{R}(Q)$ without explicit reference to the ultraviolet
regularization.

Equation~(\ref{eq:alphasR}) provides a guide to controlling a determination of $\alphas$.
The quantity $\mathcal{R}$ should be something that can be measured (in a laboratory experiment) or computed (with numerical lattice QCD)
precisely.
The higher the order in $\alphas$, $n_\text{loop}$, the better.
One wants $Q$ to be as large as possible, both to reduce the power corrections and to reduce the truncation error.
Even better is a wide range of $Q$, both to verify the running of $\alpha_\mathcal{R}$ and to separate the logarithmic dependence on $Q$ from
the power-law behavior of the remainder.
An observable is better suited when $p$ can be proven, argued, or demonstrated to be large.
Another desirable feature is to have several similar quantities, especially if the power corrections are related.

The quantity $\mathcal{R}$ can be computed in lattice QCD or measured in high-energy scattering or heavy-particle decays.
Many reviews separate the two as if they are completely different objects.
Table~\ref{tab:common} compares the ingredients in the two approaches.
\begin{table}
    \centering
    \caption{Ingredients of $\alphas$ determinations, with nonperturbative ``measurements'' directly from lattice gauge theory (LGT), from the continuum limit of lattice QCD, or (to choose one example) scaling violations of the moments of structure functions in deep inelastic scattering (DIS).
    Effects beyond the Standard Model (BSM) may be present in experimental data (no one knows) and are omitted (as a rule) from the theory.\vspace{0.2cm}}
    \label{tab:common}
    \tabcolsep=4.5mm
    \begin{tabular}{l@{\qquad}c@{\quad}c@{\quad}c}
    \hline
         Ingredient & Small Wilson loops & LGT with $a\to0$ & DIS scaling violation \\
    \hline
         Obtain $\mathcal{R}(Q)$ &  \multicolumn{2}{c}{Compute from QCD Lagrangian} & Measure e-p scattering \\
         Large energy scale & $a^{-1}$ & $L^{-1}$, $2m_Q$, ... & Momentum transfer~$Q$ \\
         Scale separation & OPE & Various & OPE \\
         Perturbation theory & Lattice (NLO, maybe NNLO) & \multicolumn{2}{c}{Dimensional regularization (NNLO, N$^3$LO)} \\
         Number of quantities & Several & one or few & Several \\
         Electroweak & \multicolumn{2}{c}{Omitted by construction} & Included in data and theory \\
         BSM & \multicolumn{2}{c}{Omitted by construction} & Unknown/omitted \\
         Units in GeV & \multicolumn{2}{c}{Hadronic quantity, viz., $Q=M_\text{PDG}[(Qa)/(Ma)]_\text{lat}$} & Detector calibration \\
    \hline
    \end{tabular}
\end{table}
The scaling violations of moments of deep-inelastic structure functions are taken as a textbook example of high-energy scattering.
(Further columns could be added without much work.)
Table~\ref{tab:common} shows two classes of methods based on lattice gauge theory: small Wilson loops and every observable for which the
continuum limit is taken (including everything else discussed in the subsections below).
Even if one starts with a spacetime lattice as an ultraviolet regularization, the continuum limit is the same QCD as probed by high-energy
experiments.
In particular, the same methods for perturbation theory---based on dimensional regularization---apply and, thus, the issues related to 
truncating the series, the size of and range in $Q$, etc., are \emph{the same}.
Indeed, moments of quarkonium correlation functions can be calculated with lattice QCD or measured in $e^+e^-\to\text{hadrons}$:
the perturbative series are exactly the same.
We have here an example of a lattice-QCD method that has more in common with a high-energy-scattering method than it has with other
lattice-QCD methods.
(See Secs.~\ref{sec:latt:moments} and~\ref{sec:RelSumRulCharm} for details.)

Because of the similarities, criteria for assessing issues such as truncation of perturbation theory and the range in $Q$ should be the same for both. Ideally, the $\alphasmZ$ averaging approaches used by the PDG from the world data~\cite{ParticleDataGroup:2020ssz} and by the FLAG collaboration from lattice results~\cite{FlavourLatticeAveragingGroup:2019iem} should be more closely aligned.
Some remarks on the criteria are provided below.

In Table~\ref{tab:common}, small Wilson loops are listed separately because they are defined at the scale of the lattice spacing, \ie\ at the ultraviolet cutoff.
There are further such quantities, including the bare coupling.
They are a different object because the lattice---including details of the chosen lattice action---is present.
The lattice is like ``new physics'' at the highest energies, except that the action of the new physics is exactly known.
Determinations of~$\alphas$ from small Wilson loops warrant discussion here in order to clarify discussions in~\cite{Komijani:2020kst,DelDebbio:2021ryq,Aoki:2021kgd}.

In continuum language, a Wilson loop is a path-ordered exponential integrating $dz\cdot A$ around a closed loop~\cite{Wilson:1974sk}:
\begin{equation}
    W_P = \mathbb{P} \exp ig \oint_P dz\cdot A(z),
\end{equation}
where $A=t^aA^a$ is the gauge potential, and $\mathbb{P}$ denotes path ordering.
For small loops of linear size~$a$, this operator admits an OPE:
\begin{equation}
    W_P \doteq Z_P\left[\openone + C_P^{(FF)}a^4\alphas F^2 + C_P^{(\bar{q}q)}a^4 m\bar{q}q + \mathcal{O}(a^6) \right],
    \label{eq:loop-OPE}
\end{equation}
where the equivalence $\doteq$ is in the weak sense of matrix elements between low-energy states.
Here, $Z_P$,  $Z_PC_P^{(FF)}$, and  $Z_PC_P^{(\bar{q}q)}$, are short distance coefficients, so they can be calculated in perturbation theory.
Note that the operators that appear do not depend on the path: when taking matrix elements, the same quantities enter over and over again.
Note also the high power $4$ in the power corrections.

Equation~(\ref{eq:loop-OPE}) applies equally well in a lattice gauge theory.
Indeed, in pure gauge theory, the OPE is on a very solid footing~\cite{Bali:2014fea,Bali:2014sja}.
The short-distance coefficients must be calculated in lattice perturbation theory, which is less developed than perturbation theory with dimensional regularization.
It is instructive to show the tree-level expression for the coefficient $Z_PC_P^{(FF)}$ of a  planar Wilson loop of size $ma\times na$,
\begin{equation}
    Z_{m\times n}C_{m\times n}^{(FF)} = -\frac{\pi(mn)^2}{36},
\end{equation}
where $a$ is now the lattice spacing.
The condensate contribution to the $1\times1$ loop is 16 times smaller than that of the $2\times2$ loop.

Vacuum expectation values $\mathcal{R}_P\equiv\frac{1}{3}\langle\Re\tr W_P\rangle$ or
$\mathcal{R}'_P\equiv-\ln\frac{1}{3}\langle\Re\tr W_P\rangle$, and combinations thereof, satisfy Eq.~(\ref{eq:alphasR}).
They can be computed very precisely.
(The bare coupling, mentioned above, is known exactly.)
Fits of the precise data can include several orders beyond the $n_\text{loop}^\text{th}$ term~\cite{Mason:2005zx,Davies:2008sw,Maltman:2008bx},
which is not a shortcoming but a strength of the method.
Such fitting could be applied to any quantity $\mathcal{R}$ with per mil uncertainties,
because assuming that the higher-order terms vanish is obviously wrong.
In practice, the fits include terms that cannot be determined by the data, but the correct interpretation of these parameters is a 
marginalization over terms whose $Q$ dependence is known, even if their strength is not.
For determining $\alphas$, one is not interested in the values (and errors) of higher-order or power-law coefficients.
One is interested is how imperfect knowledge of these terms propagates to uncertainty in~$\alphasmZ$.

Incorporating the next few terms via fit parameters, with suitable priors, 
can be seen as more conservative and more robust than the popular method of
varying the scale by a factor of two up and down.
The popular method sets $\mu=sQ$ and moves $\ln s\in[-\ln2,+\ln2]$.
It picks out a one-dimensional curve in a multidimensional space, rather than allowing a data-guided exploration of the space.
Reference~\cite{DelDebbio:2021ryq} uses the scale variation method to estimate the truncation uncertainty in analyses such as those in
Refs.~\cite{Mason:2005zx,Davies:2008sw,Maltman:2008bx}, 
not discussing
that the additional fitted terms 
could absorb such a variation.

The FLAG collaboration~\cite{Aoki:2021kgd} is considering making its quality criteria stricter.
It is worth scrutinizing the criteria and asking whether they are the most apt.
One of the criteria requires $\alpha_\mathcal{R}$ to be sufficiently small.
The bare coupling of lattice QCD satisfies the 2021 criterion and probably any future one, but it has been deprecated (for well-known reasons) as a route to~$\alphas$.
Another criterion demands that the truncation error, $\alpha_\mathcal{R}^{m_\text{loop}+1}$, be smaller than the statistical (and systematic) error.
This criterion, as it stands, 
can exclude quantities that are precise enough to verify higher-order perturbative behavior by fitting.

Because small Wilson loops are defined at the scale of the lattice spacing, effects of QED and strong isospin breaking ($m_u\neq m_d$) are often very small.
The leading effect arises not in the effective $\alphas$ but in the conversion from lattice units to GeV.
The ensembles of lattice gauge fields best suited to a future study of small Wilson loops are those being used for the 
hadronic-vacuum-polarization contribution to the muon's anomalous magnetic moment, because they have the widest range in $a$, highest 
statistics, and include QED and strong isospin contributions in the determination of the lattice spacing.
A typical target is $\delta a/a\lesssim0.5\%$, leading to a $\delta\alphas/\alphas$ that is $\beta_0\alphas$ times smaller, 
or~$\lesssim0.1\%$.

\clearpage
\section{\texorpdfstring{\boldmath $\alphasmZ$}{alphasmZ} from hadronic tau decays}
\label{sec:tau}

\subsection{Determination of \texorpdfstring{$\alphas(m_\tau^2)$}{alphasmtau} from ALEPH $\tau$ decay data
\protect\footnote{A\lowercase{uthors:} A. P\lowercase{ich} (IFIC V\lowercase{al\`encia}), A. R\lowercase{odr\'iguez}-S\lowercase{\'anchez} (IJCL\lowercase{ab} O\lowercase{rsay})}}
\label{sec:tau_pich_et_al}

The inclusive distribution of the final hadrons in $\tau$ decay provides the needed information to perform a clean low-energy determination of the strong coupling~\cite{Pich:2020gzz}. The relevant dynamical quantities 
are the two-point correlation functions for the vector $V^{\mu}=\overline{u}\gamma^{\mu} d$ and axial-vector
$A^{\mu}=\overline{u} \gamma^{\mu}\gamma_5 d$ colour-singlet charged currents:
%
\begin{equation}
i \int d^{4}x\; e^{iqx} \;
\bra{0}T[\mathcal{J}^{\mu}(x)\mathcal{J}^{\nu \dagger}(0)]\ket{0}
\, =\,
(-g^{\mu\nu}q^{2}+q^{\mu}q^{\nu})\; \Pi_{\mathcal{J}}(q^{2})
+ g^{\mu\nu}q^{2}     
\;\Pi^{(0)}_{\mathcal{J}}(q^{2}) \, ,
\end{equation}
with $\mathcal{J}=V,A$.  
If the tiny up and down quark masses are neglected, $q^2 \;\Pi^{(0)}_{\mathcal{J}}(q^{2})=0$ and only the first term needs to be considered. The correlators $\Pi_{\mathcal{J}}(s)$ are analytic functions in the entire complex $s$ plane, except for a cut along the positive real axis where their imaginary parts (spectral functions) have discontinuities. This implies the exact mathematical identity~\cite{Braaten:1991qm}
\begin{equation}\label{aomega}
A^{\omega}_{\mathcal{J}}(s_{0})\;\equiv\; \int^{s_{0}}_{s_{\mathrm{th}}} \frac{ds}{s_{0}}\;\omega(s)\, \mathrm{Im}\Pi_{\mathcal{J}}(s)\; =\; \frac{i}{2}\;\oint_{|s|=s_{0}}
\frac{ds}{s_{0}}\;\omega(s)\, \Pi_{\mathcal{J}}(s)\, ,
\end{equation}
where
$\omega(s)$ is any weight function analytic in $|s|\le s_0$, $s_{\mathrm{th}}$ is the hadronic mass-squared threshold, and the complex integral in the right-hand side  runs counter-clockwise around the circle $|s|=s_{0}$.
The inclusive Cabibbo-allowed hadronic decay width of the $\tau$ just corresponds to the weight $\omega_\tau(x) = (1-x)^2 (1+2x)$ with $x=s/s_0$ and $s_0=m_\tau^2$. The measured invariant-mass distribution determines then the left-hand-side integral for $s_0\le m_\tau^2$.

For large-enough values of $s_0$, the operator product expansion (OPE)~\cite{Shifman:1978bx},
\begin{equation}\label{eq:ope}
\Pi^{\mathrm{OPE}}_{\mathcal{J}}(s) \; =\; \sum_{D}\frac{1}{(-s)^{D/2}}\sum_{\mathrm{dim} \, \mathcal{O}=D} C_{D, \mathcal{J}}(-s,\mu)\;\langle 0|\mathcal{O}(\mu)| 0\rangle
\;\equiv\; \sum_{D}\;\dfrac{\mathcal{O}_{D,\, \mathcal{J}}}{(-s)^{D/2}}\, ,
\end{equation}
can be used to expand the contour integral in inverse powers of $s_0$. The first term ($D=0$) contains the perturbative QCD contribution, which is known to $\mathcal{O}(\alphas^4)$~\cite{Baikov:2008jh}, while nonperturbative power corrections have $D\ge 4$.
The small differences between the physical values of the integrated moments $A^{\omega}_{\mathcal{J}}(s_{0})$ and their OPE approximations are known as quark-hadron duality violations. They are very efficiently minimized by taking ``pinched'' weight functions which vanish at $s=s_0$, suppressing in this way the contributions from the region near the real axis where the OPE is not valid~\cite{Braaten:1991qm,LeDiberder:1992zhd,Gonzalez-Alonso:2016ndl}.

The high sensitivity of the $\tau$ hadronic width to the strong coupling follows from four important facts:
\begin{enumerate}
\item The perturbative contribution amounts to a sizeable 20\% effect because 
$\alphas(m_\tau^2)\sim 0.3$ is large.
\item The OPE can be safely used at $s_0\sim m_\tau^2$. The weight $\omega_\tau(x)$ contains a double zero at $s=s_0$, heavily suppressing the numerical impact of duality violations.
\item Since $\omega_\tau(x) = 1-3x^2+2x^3$, the contour integral is only sensitive to OPE corrections with $D=6$ and 8, which are strongly suppressed by the corresponding powers of $m_\tau$. Moreover, these power corrections appear with opposite signs in the vector and axial-vector correlators~\cite{Braaten:1991qm,Davier:2013sfa,Pich:2016bdg}, which implies an additional numerical cancellation in the total vector\,$+$\,axial ($V+A$) decay width.
\item The opening of high-multiplicity hadronic thresholds dilutes very soon the prominent $\rho(2\pi)$ and $a_1(3\pi)$ resonances, leading to a quite flat $V+A$ spectral distribution that approaches very fast the perturbative QCD predictions.
\end{enumerate}

The small nonperturbative contributions can be extracted from data, analysing moments more sensitive to power corrections~\cite{LeDiberder:1992zhd}. The detailed analyses performed by ALEPH~\cite{ALEPH:2005qgp}, CLEO~\cite{CLEO:1995nlc} and OPAL~\cite{OPAL:1998rrm} confirmed a long time ago that these corrections are smaller than the perturbative uncertainties. The latest and more precise experimental determination of the strong coupling obtains $\alphas^{(\Nf=3)}(m_\tau^2)= 0.332\pm 0.005_{\mathrm{exp}}\pm 0.011_{\mathrm{th}}$~\cite{Davier:2013sfa}. Taking as input their measured nonperturbative correction, the strong coupling can be directly extracted from the total $\tau$ hadronic width (and/or lifetime), which results in $\alphas^{(\Nf=3)}(m_\tau^2)= 0.331\pm 0.013$~\cite{Pich:2013lsa}.\\

An exhaustive reanalysis with the updated ALEPH $\tau$ data~\cite{Davier:2013sfa} has been performed in Ref.~\cite{Pich:2016bdg}, in order to carefully assess any possible source of systematic uncertainties. A large variety of methodologies, including all previously considered strategies, have been explored, looking for potential hidden weaknesses and testing the stability of the fitted results under slight variations of the assumed inputs. The most reliable determinations of $\alphas(m_\tau^2)$, obtained with the total $V+A$ distribution, are summarized in Table~\ref{tab:AlphaTauSummary}. The dominant uncertainties are the perturbative errors associated with the unknown higher-order corrections. This is clearly illustrated in the table, which shows the results obtained under two different procedures, either performing the contour integrals with a running $\alphas(-s)$, by solving numerically the five-loop $\beta$-function equation (contour-improved perturbation theory, CIPT)~\cite{LeDiberder:1992jjr,Pivovarov:1991rh}, or naively expanding them in powers of $\alphas(m_\tau^2)$ (fixed-order perturbation theory, FOPT). FOPT generates a somewhat larger perturbative contribution and, therefore, leads to a slightly smaller fitted value of $\alphas$. Within each procedure, the perturbative error has been estimated taking a very conservative range for the fifth-order coefficient of the Adler series, $K_5=275\pm 400$, and varying the renormalization scale in the interval $\mu^2/(-s)\in (0.5,2)$. The values obtained with the two procedures are finally combined, adding quadratically half their difference as an additional systematic error. 

\begin{table}[t]
\centering
\tabcolsep=4.mm
\caption{Determinations of $\alphas^{(\Nf=3)}(m_{\tau}^2)$ from $\tau$ decay data, in the $V+A$ channel~\protect\cite{Pich:2016bdg}.\vspace{0.2cm}}
\label{tab:AlphaTauSummary}
\begin{tabular}{c|c|c|c}
\hline 
\multirow{2}{*}{Method}
 &\multicolumn{3}{c}{} \\[-9pt]
  & \multicolumn{3}{c}{$\alphas^{(\Nf=3)}(m_{\tau}^2)$}
\\[1.2pt] \cline{2-4}
& \raisebox{-2pt}{CIPT} & \raisebox{-2pt}{FOPT} & \raisebox{-1.5pt}{Average}
\\[1.2pt] \hline &&&\\[-8pt]
$\omega_{kl}(x)$ weights & $0.339 \,{}^{+\, 0.019}_{-\, 0.017}$ &
$0.319 \,{}^{+\, 0.017}_{-\, 0.015}$ & $0.329 \,{}^{+\, 0.020}_{-\, 0.018}$
\\[3pt]
$\hat\omega_{kl}(x)$ weights  & $0.338 \,{}^{+\, 0.014}_{-\, 0.012}$ &
$0.319 \,{}^{+\, 0.013}_{-\, 0.010}$ & $0.329 \,{}^{+\, 0.016}_{-\, 0.014}$
\\[3pt]
$\omega^{(2,m)}(x)$ weights  & $0.336 \,{}^{+\, 0.018}_{-\, 0.016}$ &
$0.317 \,{}^{+\, 0.015}_{-\, 0.013}$ & $0.326 \,{}^{+\, 0.018}_{-\, 0.016}$
\\[1.5pt]
$s_0$ dependence  & $0.335 \pm 0.014$ &
$0.323 \pm 0.012$ & $0.329 \pm 0.013$
\\[1.5pt]
$\omega_{a}^{(1,m)}(x)$ weights  & $0.328 \, {}^{+\, 0.014}_{-\, 0.013}$ &
$0.318 \, {}^{+\, 0.015}_{-\, 0.012}$ & $0.323 \, {}^{+\, 0.015}_{-\, 0.013}$
\\[2pt] \hline &&&\\[-9pt]
Average & $0.335 \pm 0.013$ & $0.320 \pm 0.012$ & $0.328 \pm 0.013$
\\[1pt] \hline
\end{tabular}
\end{table}

The different rows in the table correspond to different choices of pinched weights:
\begin{align}
\omega_{kl}(x) & = (1-x)^{2+k}\, x^l \, (1+2x) \, ,
\qquad\qquad & (k,l) = \{(0,0),\, &(1,0), (1,1), (1,2), (1,3)\}\, ,
\nonumber\\
\hat\omega_{kl}(x) & = (1-x)^{2+k}\, x^l \, ,
\qquad\qquad  &(k,l) = \{(0,0),\, &(1,0), (1,1), (1,2), (1,3)\}\, ,
\nonumber\\
\omega^{(2,m)}(x) & = 
1-(m + 2)\, x^{m + 1} + (m + 1)\, x^{m + 2}\, ,
&&  1\le m\le 5\, ,
\nonumber\\
\omega_{a}^{(1,m)}(x)& = (1- x^{m+1})\, \mathrm{e}^{-ax}\, ,
&&  0\le m\le 6\, .
\end{align}
In all cases, we find a high sensitivity to the strong coupling and a very small sensitivity to power corrections, which gets reflected in large uncertainties for the fitted condensates. The first set of weights was adopted by ALEPH and allows for a direct use of the measured distribution. With the five indicated moments, we have fitted $\alphas(m_\tau^2)$ and the nonperturbative corrections $\mathcal{O}_{4,6,8}$. The impact on $\alphas$ from neglected condensates of higher dimensions (the highest moment involves corrections with $D\le 16$) has been estimated including $\mathcal{O}_{10}$ in the fit and taking the difference as an additional uncertainty. Additional power corrections with $D>10$ would certainly modify the poorly determined values of the fitted condensates, specially those with higher dimensions, but they have a negligible impact on $\alphas$, compared with the errors already included. Our results (first line in the table) are in good agreement with Ref.~\cite{Davier:2013sfa}. Very similar values are obtained with the modified weights $\hat\omega_{kl}(x)$ (second line), which eliminate the highest-$D$ power correction from every moment, showing that these contributions do indeed play a minor numerical role.

Moments constructed with the optimized (double-pinched) weights $\omega^{(2,m)}(x)$ only receive power corrections with $D= 2(m+2)$ and $2(m+3)$. The third line in the table shows that they give values of $\alphas(m_\tau^2)$ in very good agreement with those in the first two lines. 
Similar results (not shown in the table) are obtained from a fit to four moments, based on the weights $\omega^{(n,0)}(x) = (1-x)^n$, with $0\le n\le 3$, which have a different sensitivity to power corrections and (for $n=0,1$) are less protected against duality violations  and experimental uncertainties.
The values of $\alphas(m_\tau^2)$, $\mathcal{O}_{2(m+2)}$ and $\mathcal{O}_{2(m+3)}$ can also be extracted from a fit to the $s_0$ dependence of a single $A^{(2,m)} (s_{0})$ moment,   above some $\hat s_0 \ge 2.0\;\mathrm{GeV}^2$. One finds a quite poor sensitivity to power corrections, as expected, but a surprising stability in the extracted values of $\alphas(m_\tau^2)$ at different $\hat s_0$. Combining the information from three different moments ($m=0,1,2$), and adding as an additional theoretical error the fluctuations with the number of fitted bins, one gets the $\alphas(m_\tau^2)$ values given in the fourth line of Table~\ref{tab:AlphaTauSummary}. 
This determination is much more sensitive to violations of quark-hadron duality because the $s_0$ dependence of consecutive bins feels the local structure of the spectral function.\footnote{
Refs.~\cite{Pich:2016bdg,Pich:2016mgv,Pich:2018jiy,Pich:2022tca} have also analysed a different strategy, advocated in some recent works (see Sec.~\ref{sec:tau_Boito_et_al}), that maximises duality violations: the experimental spectral function is fitted in the interval $\hat s_0 < s < m_\tau^2$ with an ad-hoc 4-parameter ansatz, and the resulting model is used to correct the perturbative prediction of $A^\omega_\mathcal{J} (\hat s_0)$ with $\omega(x)=1$ (no weight). This is a dangerous procedure because 1) the OPE is not valid in the real axis, 2) $\alphas$ is fixed at a very low scale $\hat s_0 = (1.2\;\mathrm{GeV})^2$ where theoretical errors are large, and 3) the subtracted duality-violation contribution is large and model dependent. Slight changes on the functional form of the assumed ansatz result in large variations of the fitted value of $\alphas(\hat s_0)$~\cite{Pich:2016bdg,Pich:2022tca} that have not been taken into account in the quoted uncertainties. 
Ref.~\cite{Pich:2022tca} provides a very detailed anatomy of this duality-violation approach to the strong coupling, exhibiting unaccounted systematic errors, some formal inconsistencies of the adopted assumptions and the tautological nature of some of the tests.
}
%
The agreement with the determinations in the other lines of the table confirms the small size of duality violations in the $V+A$ distribution above $\hat s_0$.

A completely different sensitivity to nonperturbative corrections is achieved
with the weights $\omega_{a}^{(1,m)}$. Their exponential suppression nullifies the high-$s$ region, strongly reducing the violations of quark-hadron duality, at the price of enhancing the exposition to power corrections of any dimensionality.
In a pure perturbative analysis the neglected power corrections should manifest as large instabilities of $\alphas$ under variations of $s_0$ and $a\not=0$; however, stable results are found for a broad range of both $s_0$ and $a$, which indicates that power corrections are small. The last line in the table has been obtained combining the information from seven different moments with $\omega_{a}^{(1,m)}$ ($0\le m\le 6$) weights.

Fully compatible results with slightly larger uncertainties are also obtained from the separate $V$ and $A$ distributions~\cite{Pich:2016bdg}. The excellent overall agreement among determinations obtained with a broad variety of numerical approaches that have very different sensitivities to nonperturbative corrections, and the many complementary tests successfully performed in Ref.~\cite{Pich:2016bdg},
demonstrate the robustness and reliability of the results shown in Table~\ref{tab:AlphaTauSummary}. 
The final average value 
\begin{equation}
\alphas^{(\Nf=3)}(m_{\tau}^2)\, =\, 0.328\pm 0.013
\end{equation}
is in good agreement with the results of Ref.~\cite{Davier:2013sfa} and with a recent determination (see Sec.~\ref{sec:tau_Ayala_et_al}) based on Borel--Laplace sum rules and a renormalon-motivated model~\cite{Ayala:2021yct}.
After evolution up to the scale $m_\mathrm{Z}$, the strong coupling
decreases to
\begin{equation}
\alphas^{(\Nf=5)}(m_\mathrm{Z}^2)\, =\,  0.1197\pm 0.0015\, ,
\end{equation}
in excellent agreement with the direct N${}^3$LO determination at the Z peak.

A much better control of the small nonperturbative contributions could be achieved with more precise data, specially at the highest energy bins. The high-statistics expected from Belle-II or a future TeraZ facility should make that possible. A reduction of the dominant perturbative error requires a better theoretical understanding of higher-order corrections (CIPT versus FOPT, renormalons, etc.). In the long term, an explicit calculation of the $K_5$ term in the Adler series would have a major impact.\\

\noindent \textit{Acknowledgments}---
This work has been supported by 
MCIN/AEI/10.13039/501100011033, Grant No.\ PID2020-114473GB-I00,
by the Generalitat Valenciana, Grant No.\ Prometeo/2021/071,
and by the Agence Nationale de la Recherche (ANR), Grant ANR-19-CE31-0012 (project MORA).

\subsection{The strong coupling from hadronic \texorpdfstring{$\tau$}{tau} decays: present and future
\protect\footnote{A\lowercase{uthors:} D. B\lowercase{oito} (U.\,V\lowercase{ienna \&} U.\,S\lowercase{{\~a}o} P\lowercase{aulo}), M. G\lowercase{olterman} (SF S\lowercase{tate} U\lowercase{niv.} \& IFAE B\lowercase{arcelona}), K. M\lowercase{altman} (YU T\lowercase{oronto} \& CSSM A\lowercase{delaide}), S. P\lowercase{eris} (IFAE B\lowercase{arcelona})}}
\label{sec:tau_Boito_et_al}


We review the sum-rule framework for determining the strong coupling,
$\alphas$, from inclusive spectral functions measured in hadronic $\tau$
decays. We then discuss a new inclusive vector isovector spectral function,
obtained from combining the dominant exclusive-mode spectral-function data
from ALEPH and OPAL with BaBar data for $\tau\to K^- K^0\nu_\tau$
and several CVC-converted $R$-ratio data sets for small exclusive-mode
contributions that had previously been estimated using Monte Carlo methods. We
summarize our most recent results for $\alphas$, and discuss prospects
for future improvements.

\subsubsection{Review}\label{review}

Finite-energy sum rules (FESRs) allow for the extraction of the strong
coupling at the $\tau$ mass, $\alphas(m_\tau^2)$, from inclusive vector
($V$) and/or axial ($A$) non-strange spectral functions, which can and
have been measured through hadronic $\tau$ decays~\cite{OPAL:1998rrm,ALEPH:1998rgl,ALEPH:2005qgp}. Such
sum rules have the form
\begin{equation}
\label{FESR}
I^{(w)}(s_0)\equiv\frac{1}{s_0}\int_0^{s_0} ds\,w(s)\,\rho(s)=
-\frac{1}{2\pi i s_0}\oint_{|z|=s_0}dz\,w(z)\,\Pi(z)\ ,
\end{equation}
where $\Pi(z)$ is a vacuum polarization,
$\rho(s)=\frac{1}{\pi}\mbox{Im}\,\Pi(s)$, with $s=q^2$, is the corresponding
spectral function, $s_0>0$, and $w(s)$ is typically a polynomial
in $s/s_0$. The left-hand side of Eq.~(\ref{FESR}) can be determined from
data for $s_0\le m_\tau^2$, while, for $s_0$ large enough, the right-hand
side can be represented in QCD perturbation theory, with nonperturbative
(NP) corrections. The latter are needed because of the relatively
low value of $m_\tau$.

The perturbative expansion for $\Pi(z)$ in powers of $\alphas(\mu^2)$
and $\log(-z/\mu^2)$ is known to order $\alphas^4$~\cite{Baikov:2008jh}, where
$\mu$ is the renormalization-group scale. In most of the literature,
two scale-choice prescriptions have been considered, either FOPT, in
which $\mu^2=s_0$, or CIPT~\cite{Pivovarov:1991rh,LeDiberder:1992jjr}, in which $\mu^2$ is set equal to $z$
when evaluating the perturbative contribution to the contour integral on
the right-hand side of Eq.~(\ref{FESR}).

NP corrections (away from the positive real axis, see below)
are incorporated through the operator product expansion (OPE), which
schematically takes the form
\begin{equation}
\label{OPE}
\Pi_{\mathrm{OPE}}(z)=\sum_{k=0}^\infty(-1)^k \frac{C_{2k}(z)}{z^k}\ .
\end{equation}
The $k=0$ term represents purely perturbative contributions,
the $k=1$ term perturbative mass corrections (in the non-strange
channels, $C_2$ can be set to zero because of the smallness of the up and
down quark masses); NP condensate contributions start at $k=2$.
The presence of a cut in the complex $z=q^2$ plane extending to
infinity along the positive real axis, means the OPE is not convergent;
it is, at best, an asymptotic series. It is also generally believed
that the $D=2k$ terms (for $k>1$) are related to renormalon
ambiguities in the perturbative expansion~\cite{Beneke:1998ui,Beneke:2008ad,Beneke:2012vb}.
While the higher-dimension coefficients $C_{2k}$ in principle depend
logarithmically on $z$, they are generally taken as constants,
as their $z$ dependence is suppressed by two powers of $\alphas$, in
addition to the $1/z^k$ suppression.

Recently, it was shown that the Borel sums of the FOPT and CIPT
versions of the perturbative expansion of the right-hand side of
Eq.~(\ref{FESR}) are not the same and, related, that the OPE\ (\ref{OPE})
does not reflect the renormalon ambiguities for CIPT~\cite{Hoang:2020mkw,Hoang:2021nlz},
creating a mismatch between the use of CIPT and the OPE in the
form\ (\ref{OPE}). Because of this, we will always employ
FOPT~\cite{Boito:2020xli}. The recent work of Ref.~\cite{Benitez-Rathgeb:2022yqb} (see Sec.~\ref{sec:FOCI})
paves the way for a reconciliation between the FOPT and CIPT series but,
until this is realized in practice, averaging the results
obtained using FOPT and CIPT should be avoided.

The nonconvergent, at-best-asymptotic, nature of the OPE noted
above has two implications. First, it is advisable to restrict the
use of the OPE to low orders in the expansion~(\ref{OPE}), as little
is known about where it starts to diverge (at the energies
relevant in $\tau$ decays). Second, one expects NP corrections
{\it beyond} the OPE, just as the OPE represents NP corrections
beyond perturbation theory. This phenomenon, known as the violation of
quark-hadron duality~\cite{Poggio:1975af,Blok:1997hs,Bigi:1998kc}, manifests itself physically in the
presence of overlapping resonance contributions in $\rho(s)$.
These duality violations (DVs) need to be modeled (as long as a NP
solution to QCD is not known). We have developed a theoretical
framework, based on generally accepted assumptions about QCD
in the framework of hyperasymptotics~\cite{Boito:2017cnp}, which leads to the
following large-$s$ asympotic form for the duality-violating
part of the spectral function:
\begin{equation}
\label{ansatz}
\rho_{\mathrm{DV}}(s)=e^{-\delta-\gamma s}\sin\left(\alpha+\beta s\right)
\left[1+{\cal{O}}\left(\frac{1}{\log{s}};\, \frac{1}{s};\,
\frac{1}{N_c}\right)\right]\ .
\end{equation}
With this {\it ansatz}, the sum rule~(\ref{FESR}) can be reformulated
as~\cite{Cata:2005zj}
\begin{equation}
\label{FESRDV}
\frac{1}{s_0}\int_0^{s_0} ds\,w(s)\,\rho(s)=-\frac{1}{2\pi i s_0}
\oint_{|z|=s_0}\hspace{-12pt}dz\,w(z)\,
\Pi_{\mathrm{OPE}}(z)-\frac{1}{s_0}\int_{s_0}^\infty ds\,
w(s)\, \rho_{\mathrm{DV}}(s)\ .
\end{equation}
This introduces a new set of parameters into the theory representation
of the right-hand side of Eq.~(\ref{FESR}). A separate set of DV
parameters is needed for each of the $V$ and $A$ channels.
The main assumption is that the form\ (\ref{ansatz}) can be used at
values of $s$ below $m_\tau^2$---whether this is the case can be tested
by a variety of fits to data. For a more detailed version of this very brief
review, and many more references, we refer to Ref.~\cite{Boito:2020xli}.

\begin{figure}[t]
\centering
\includegraphics[width=9cm]{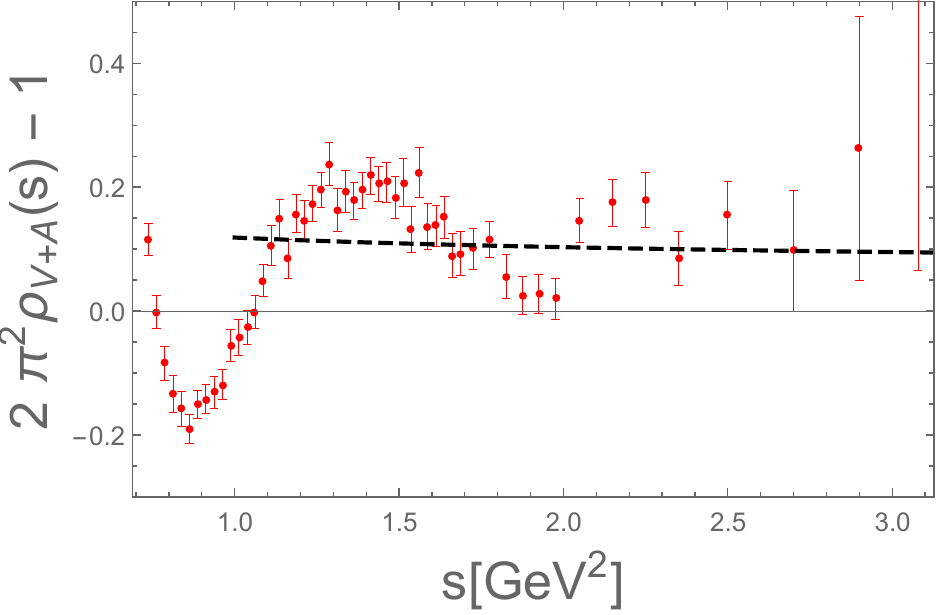}
\caption{Blow-up of the large-$s$ region of the parton-model-subtracted ALEPH $V+A$ non-strange spectral
function~\cite{Davier:2013sfa}. The dashed curve represents QCD perturbation theory.
\label{fig1}}
\end{figure}
At this point, a quick look at data is instructive. Figure~\ref{fig1}
shows a rescaled version of the sum of the non-strange $V$ and $A$
spectral functions from Ref.~\cite{Davier:2013sfa}, with the $\alphas$-independent
parton-model contribution subtracted. Resonance oscillations are immediately
evident, with an amplitude comparable in size to the $\alphas$-dependent
pQCD contribution, represented by the dashed
curve. The figure makes clear that DVs are important, though,
because of the large errors, it is difficult to tell whether oscillations
above $s=m_\tau^2$ (relevant to determining DV contributions to the sum
rules (\ref{FESRDV}) with $s_0=m_\tau^2$) will be numerically relevant
or not. Clearly, DVs need to be accounted for, and their contributions
quantitatively assessed, even in the $V+A$ channel.

Two different strategies, which we refer to as the ``truncated OPE''
(tOPE)~\cite{LeDiberder:1992zhd} and the ``DV-model'' strategies~\cite{Boito:2011qt,Boito:2012cr},
have been developed to deal with the NP contributions discussed above.

In the tOPE approach, $s_0$ is taken equal to $m_\tau^2$, and, to
suppress contributions to the contour integral from the region near the
positive real axis, and thus from DVs, weights with a double or triple zero
at $s=s_0=m_\tau^2$ are employed. The set of such multiple-zero (``pinched'')
polynomial weights typically extends up to degree $k=7$, and thus probes the
OPE up to dimension $D=2k+2=16$. The number of OPE parameters to be
fitted ($\alphas$ and the condensates $C_{D}$, $D=4,\, 6,\cdots ,\, 16$)
necessarily exceeds the number of such independent weights, and hence the
number of independent $s_0=m_\tau^2$ spectral integrals available to fix these
parameters. Some subset of the $C_D$ in principle present must thus be
set to zero by hand to leave a fit with fewer parameters than data
points. DVs are also neglected (in terms of Eq.~(\ref{ansatz}),
$\delta$ is set to $\infty$) under the assumption that the use of pinched
moments sufficiently suppressess these contributions. 

The main problem with this strategy, employed most recently in
Refs.~\cite{Davier:2013sfa,Pich:2016bdg}, is the neglect of these higher-$D$ OPE
contributions. Several tests of the tOPE strategy were carried out in
Refs.~\cite{Boito:2016oam,Boito:2019iwh}, exposing clear inconsistencies.
Such inconsistencies can arise because integrated dimension-$D$ OPE
contributions scale as $1/s_0^{D/2}$. If contributions from omitted
higher-$D$ condensates are, in fact, not numerically negligible, then
when these are absorbed into the values of the lower-$D$ OPE
parameters obtained in a fixed-$s_0$ tOPE fit, the resulting theory
representation will, in general, have an $s_0$ dependence which is
incorrect. The tOPE truncation can thus be tested for selfconsistency
by comparing spectral-integral predictions obtained using the theory
parameters obtained from the single-$s_0$ tOPE fit to the experimental
values of these same spectral integrals over a range of $s_0$ values.
Such tests, designed to take into account the impact of the very strong
correlations between (i) spectral integrals at different $s_0$, (ii) theory
integrals at different $s_0$, and (iii) fitted theory parameters and spectral
integrals, were carried out in Ref.~\cite{Boito:2019iwh}, with related tests
also carried out in Ref.~\cite{Boito:2016oam}. The $s_0$ dependence of
the theory predictions was found to be in clear disagreement with that
of the corresponding experimental spectral integral
combinations.\footnote{See, for example, Fig.~8 of Ref.~\cite{Boito:2019iwh}.}

In a recent publication \cite{Pich:2022tca}, Pich and Rodriguez-Sanchez have addressed the
relative merits of the tOPE and DV-model based approaches, commenting on 
Refs.~\cite{Boito:2016oam,Boito:2019iwh}. 
In this brief overview, there is not enough space to discuss, in sufficient
detail, a number of the ongoing misconceptions and/or mis-statements
concerning the tOPE and DV model approaches contained in both this new
work and in Sec.~\ref{sec:tau_pich_et_al} of this review.  Neither, in fact, addresses the points
raised in Ref.~\cite{Boito:2019iwh} beyond a few vague comments, while the discussion of  Ref.~\cite{Boito:2016oam} in Ref.~\cite{Pich:2022tca}  is based on a number of assumptions/prejudices about the OPE and misinterpretations of the results presented in Ref.~\cite{Boito:2016oam}. Some further discussion of these shortcomings can be found in the Mattermost forum provided for the recent
ECT$^*$ Trento $\alphas$(2022) Workshop on Precision Measurements of the
QCD Coupling Constant.\footnote{See \href{https://mattermost.web.cern.ch/alphas-2022/channels/town-square}{https://mattermost.web.cern.ch/alphas-2022/channels/town-square}.}  We leave further discussion, including of Ref.~\cite{Pich:2022tca}, to a future publication.

In the alternate DV-model strategy, recent implementations employ weights
sensitive to terms in the OPE only up to $D=8$, avoiding potential
problems with higher-$D$ OPE contributions.\footnote{Some tests were
also performed with a degree-4 weight, sensitive to $C_{10}$. We avoid
weights that project onto~$C_4$~\cite{Boito:2011qt,Boito:2012cr,Beneke:2012vb}.} DVs cannot,
in general, be ignored, and are instead modeled by Eq.~(\ref{ansatz}),
with fits performed to spectral integrals for a range of $s_0$ between a
minimum value $s_0^{min}$ and $m_\tau^2$, where $s_0^{min}$ is determined
by the quality of the fits and turns out to be $\sim 1.5$~GeV$^2$. Results
are found to be very stable against variations of $s_0^{min}$.

To determine both DV parameters and $\alphas$, the analysis should
include at least one weight with good sensitivity to DVs. The choice $w=1$,
which produces no DV suppression, is ideal for this purpose. Stability of
the DV-model approach has been tested using various combinations of weight
functions in analyses of purely $V$-channel $\tau$ data, both $V$- and
$A$-channel $\tau$ data~\cite{Boito:2014sta}, and the KNT~\cite{Keshavarzi:2018mgv,Keshavarzi:2019abf} compilation
of $R$-ratio (inclusive $e^+e^-\to\mbox{hadrons}$) data~\cite{Boito:2018yvl}.
Ref.~\cite{Boito:2020xli} contains our most recent application of this strategy.
Earlier applications, laying out the framework in more detail, can be
found in Ref.~\cite{Boito:2011qt}. The DV-model strategy uses a
range of $s_0$ as it is important to check the $s_0$ dependence of
the match between experiment and theory (left-hand and right-hand sides
of Eq.~(\ref{FESRDV}), respectively), to test the validity of
the theory representation employed for $\Pi(z)$. One also
needs to check that oscillations in the $V$ and $A$ (hence also $V+A$)
spectral functions are well represented by the {\it ansatz}\ (\ref{ansatz}),
for at least some range of $s$ below $m_\tau^2$.

\subsubsection{Data}\label{data}

Traditionally, in most papers since the appearance of the perturbative result
to order $\alphas^4$ in Ref.~\cite{Baikov:2008jh}, the $\tau$-based $\alphas$ has been
obtained from the ALEPH data~\cite{Davier:2013sfa,Pich:2016bdg,Boito:2014sta,Ayala:2021mwc,Ayala:2021yct}. However,
the value obtained from OPAL data~\cite{Boito:2011qt,Boito:2012cr} is consistent with
the ALEPH-based result~\cite{Boito:2014sta}, and it thus makes sense to combine
the ALEPH and OPAL data sets prior to integration, following the approach
now routinely employed for electroproduction data when obtaining
dispersive estimates for the leading hadronic contribution to the muon
anomalous magnetic moment. Moreover, while the bulk of the inclusive spectral
functions comes from the 2-pion and the 4-pion modes (in the $V$ channel),
and the 3-pion and 5-pion modes (in the $A$ channel), most of the residual
exclusive channels (including some of the 5-pion channels) were obtained
from Monte Carlo simulations rather than from experimental data in Refs.~\cite{OPAL:1998rrm,ALEPH:1998rgl,ALEPH:2005qgp}.
In the $V$ channel, this can now be remedied by using, in addition to BaBar
results for $\tau\to K^-K^0\nu_\tau$~\cite{BaBar:2018qry}, recent high-precision
results for higher-multiplicity, $G$-parity positive (hence $I=1$)
exclusive-mode electroproduction cross sections. CVC allows these to be
converted to the equivalent higher-multiplicity contributions to the
$I=1$, $V$ $\tau$-based spectral function.\footnote{The existence
of a Dalitz-plot-based $I=0/1$ separation of the $e^+e^-\to K\bar{K}\pi$
cross sections means CVC can also be used to determine the $K\bar{K}\pi$
contribution to the $\tau$ $V$ spectral function~\cite{Davier:2008sk}.} The
detailed construction of the
combined inclusive $I=1$, $V$ spectral function along these lines is
described in Ref.~\cite{Boito:2020xli}. With 98\% of the inclusive total
by branching fraction (BF) coming from the combined ALEPH and
OPAL 2-pion and 4-pion modes, and the support for the residual
modes lying well above the $\rho/\omega$ meson interference region,
isospin-breaking corrections, which should in principle be applied to
the CVC relations, will be safely negligible.

\begin{figure}[t]
\centering
\includegraphics[width=9cm]{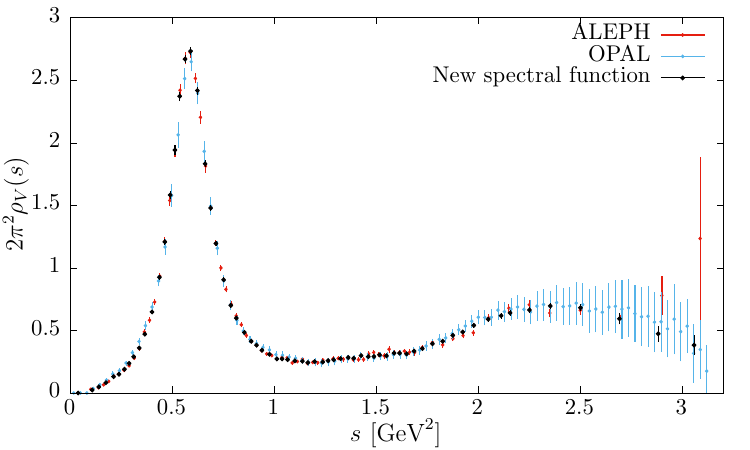}
\caption{New non-strange $V$ spectral function~\cite{Boito:2020xli}, compared with the ALEPH and OPAL spectral functions.
\label{fig2}}
\end{figure}

The new $V$ spectral function is shown in Fig.~\ref{fig2}. The
increased precision in the large-$s$ region, coming from the replacement of
Monte Carlo data with electroproduction data (which are not kinematically
limited by the $\tau$ mass) is immediately evident. A more quantitative
measure of the improved precision is provided in Table~\ref{tab1},
which shows a dramatically reduced statistical
error on the $w=1$ spectral moment $I^{(w=1)}(s_0)$ resulting from
the improved precision of the combined spectral function, compared to
the ALEPH or OPAL spectral functions, in the larger $s$ region.

\begin{table}[t]
\centering
\caption{Comparison of the left-hand side of Eq.~(\ref{FESR})  for $w=1$, at two values
of $s_0$, for the combined, ALEPH, and OPAL versions of the
$V$ spectral function. $s_0^*$ is a value close
 to $1.5$~{\rm GeV}$^2$ for each case, and $s_0^{**}$ is close
 to $2.9$~{\rm GeV}$^2$ for each case.
Note that, because the values of $s_0^*$ and $s_0^{**}$ are slightly
different for the three cases, the central values cannot be directly
compared.  The errors can, however, be compared, as they vary
slowly with $s_0$.   For details, see Ref.~\cite{Boito:2020xli}.\vspace{0.2cm}
\label{tab1}}
\tabcolsep=4.mm
\begin{tabular}{l|l|l|l}
\hline
& combined & ALEPH & OPAL \\
\hline
$s_0^*\approx 1.5$~GeV$^2$  & 0.03137(14) & 0.03145(17) & 0.03140(46) \\
$s_0^{**}\approx 2.9$~GeV$^2$ &  0.02952(29) & 0.03133(65) & 0.03030(170) \\
\hline
\end{tabular}
\end{table}

\subsubsection{Results}\label{results}

The increased precision of the new non-strange $I=1$ $V$ spectral
function constructed in Ref.~\cite{Boito:2020xli} produces an improved
determination of $\alphas(m_\tau^2)$, with the CVC-based improvement
of residual-mode contributions playing a particularly important role.
Since electroproduction data are purely vector, a similar
improvement is not possible for the axial channel. Higher precision is
thus now obtainable from $V$-channel than from $V+A$-channel analyses.\footnote{Note, however, that combined $V$- and $A$-channel
analyses of both OPAL data~\cite{Boito:2011qt,Boito:2012cr} and ALEPH
data~\cite{Boito:2014sta} produced results in very good agreement with those
of the corresponding $V$-channel analyses.}

Our result for $\alphas$ at the $\tau$-mass is~\cite{Boito:2020xli}
\begin{equation}
\label{alphastau}
\alphas(m_\tau^2)=0.3077\pm 0.0065_{\mathrm{exp}}\pm 0.0038_{\mathrm{theory}}
 \\
=0.3077\pm 0.0075\qquad(\Nf=3\ ,\ \mbox{FOPT})\ ,
\end{equation}
where the first error is the fit error, and the second an estimate
of the systematic uncertainty associated with the truncation of perturbation
theory and the use of Eq.~(\ref{ansatz}) to model DVs. For comparison,
analyses of ALEPH or OPAL data alone produce combined errors of $\pm 0.010$
and $\pm 0.018$, respectively.

Converting Eq.~(\ref{alphastau}) to the $\Nf=5$, Z-mass scale using
five-loop running and four-loop matching at the charm and bottom
thresholds, one finds
\begin{equation}
\label{alphasZ}
\alphasmZ = 0.1171\pm 0.0010\qquad(\tau\ ,\ \Nf=5\ ,\ \mbox{FOPT})\ .
\end{equation}
This is in good agreement with an independent determination from
electroproduction data using the same theoretical framework
\cite{Boito:2018yvl}:
\begin{equation}
\label{alphasEM}
\alphasmZ = 0.1158\pm 0.0022\qquad(\mbox{EM}\ ,\ \Nf=5\ ,\ \mbox{FOPT})\ .
\end{equation}
The latter determination was obtained from FESRs with
$3.25$~GeV$^2\le s_0^{min}\le 4$~GeV$^2$, significantly higher than the
interval $\sim 1.5$~GeV$^2\le s_0^{min}\le m_\tau^2$ available from
$\tau$ decays. The wider range of accessible $s_0$ allowed tests of our
model for DVs, Eq.~(\ref{ansatz}), to be performed by comparing values of
$\alphas(m_\tau^2)$ obtained from the $w=1$ FESR (which is maximally
sensitive to DV effects, and hence provides a good probe of their effect)
with and without DV contributions included. The effect found was small (of
order $0.005$)\footnote{This
systematic effect is included in the error shown in Eq.~(\ref{alphasEM}).}  for $3.25$~GeV$^2\le s_0^{min}\le 4$~GeV$^2$
but increased rapidly as $s_0^{min}$ was lowered below
$m_\tau^2$.\footnote{Recall that the dependence of Eq.~(\ref{ansatz}) on
$s$ is exponential.}

\subsubsection{Future improvements}\label{future}

The value of $\alphas$ obtained from hadronic $\tau$ decays, in
addition to being competitive, provides a particularly strong test of the
running predicted by QCD since the $\tau$ mass is much lower than the
energies at which other determinations have been carried out.
Important questions are, of course, what future data or theory
improvements might produce a reduction in the error on the current
best result, Eq.~(\ref{alphastau}), and how to test the assumed NP
behavior of the spectral function around and below the $\tau$ mass.
As we will now discuss, the best prospect for improvement appears
to be from future, more precise data for hadronic $\tau$ decays.

Improved $\tau$ data will allow for a reduction in the
experimental part of the error on $\alphas(m_\tau^2)$, currently
$\pm 0.0065$. The theory error in Eq.~(\ref{alphastau}) has two (to some
extent related) sources: the truncation of perturbation theory, and the
unavoidable need to estimate NP effects. A calculation of the
order-$\alphas^5$ term in QCD perturbation theory would be interesting,
given the sometimes slow convergence of $I^{(w)}(s_0)$ for typical
weights $w$. On the nonperturbative side, increased data
precision would allow more stringent tests, in particular of the DV
contribution from Eq.~(\ref{ansatz}). Subleading effects predicted by
Ref.~\cite{Boito:2017cnp} might become accessible, and either further
confirmation, or a breakdown of this representation of DVs would
increase our ability to probe the limits on the potential precision
of the strong-coupling determination from hadronic $\tau$ decays.

A further point, of relevance to improving tests of the reliability of
the theoretical assumptions underlying the two FESR approaches, concerns
how such tests should be carried out. In the past, overlapping experimental
spectral integral and theory integral error bands were often taken as
evidence of the compatibility of the underlying theory representation
with data. Such a conclusion is, however, statistically unjustified
since the theory errors, which result from a fit to the experimental
spectral integrals, are necessarily very strongly correlated with their
experimental counterparts. Plotting both, and interpreting their overlap
as if the two were independent runs the risk of serious double
counting. To properly test underlying theory assumptions, it is crucial
to make comparisons that account for not just correlations between the
spectral integrals at different $s_0$ and theory integrals at different
$s_0$, but also those between fitted theory and experimental integrals.

While FCC-ee would produce $\tau$-leptons copiously~\cite{FCC:2018evy,Dam:2021ibi}, the best near-future
prospect is Belle-II~\cite{Belle:2008xpe}, which has access to many more
$\tau$-leptons than were produced at LEP. Our construction
of the new inclusive $V$ spectral function, moreover, suggests a clear path
forward: {\it no fully-inclusive spectral function} needs to be obtained
from Belle-II data. Residual-mode contributions (which, though
representing only 2\% of the inclusive total by BF, are important in the
upper part of the $\tau$ kinematic range) have already been brought under
good control using CVC and electroproduction data~\cite{Boito:2020xli}. More
precise 2-pion and 4-pion exclusive-mode $\tau$ data would thus suffice
to produce a new $I=1$ $V$ spectral function with even smaller errors
than those of Ref.~\cite{Boito:2020xli}, and hence to reduce the experimental
error in Eq.~(\ref{alphastau}). Details will need to be carefully
considered. For instance, while the existing Belle unit-normalized
2-pion distribution is more precise than that of ALEPH or OPAL, the
$\tau\to\pi^-\pi^0\nu_\tau$ BF has been measured less well by
Belle, with the HFLAV~\cite{HFLAV:2019otj} value still dominated by ALEPH.
An improved, BF-normalized 2-pion distribution will thus require combining
input from different experiments. The situation is, presumably, similar
for the two 4-pion modes.\\

\noindent \textit{Acknowledgments}--- DB is supported by the S\~ao Paulo Research Foundation (FAPESP)
Grant No.\ 2021/06756-6, by CNPq Grant No.~309847/2018-4, and by the Coordena\c c\~ao de Aperfei\c coamento de Pessoal de N\'ivel Superior---Brazil (CAPES)---Finance Code 001.
MG is supported by the U.S.\ Department of Energy,
Office of Science, Office of High Energy Physics, under Award No.\
DE-SC0013682.
KM is supported by a grant from the Natural Sciences and Engineering
Research Council of Canada.
SP is supported by the Spanish Ministry of Science, Innovation
and Universities (project PID2020-112965GB-I00/AEI/10.13039/501100011033).
IFAE is partially funded by the CERCA program of the Generalitat de Catalunya.

\subsection{Extraction of \texorpdfstring{$\alphas$}{alphas} using Borel--Laplace sum rules for tau decay data
\protect\footnote{A\lowercase{uthors:} C. A\lowercase{yala} (U.\,T\lowercase{arapac\'a}), G. C\lowercase{veti\v{c}} (UTFSM, V\lowercase{alparaiso}), D. T\lowercase{eca} (UTFSM, V\lowercase{alparaiso})}}
\label{sec:tau_Ayala_et_al}

The application of double-pinched Borel--Laplace sum rules to ALEPH $\tau$-decay data is discussed. For the leading-twist ($D=0$) Adler function a renormalon-motivated extension is used, and the 5-loop coefficient is taken to be $d_4=275 \pm 63$. Two $D=6$ terms appear in the truncated OPE ($D \leq 6$) to enable cancellation of the corresponding renormalon ambiguities. Two variants of the fixed order perturbation theory, and the inverse Borel transform, are applied to the evaluation of the $D=0$ contribution. The truncation index $N_t$ is fixed by the requirement of local insensitivity of the momenta $a^{(2,0)}$ and $a^{(2,1)}$ under variation of $N_t$. The averaged value of the coupling obtained is $\alphas(m_{\tau}^2)=0.3235^{+0.0138}_{-0.0126}$ (corresponding to $\alphasmZ=0.1191 \pm 0.0016$). The theoretical uncertainties are significantly larger than the experimental ones.\\


The sum rule corresponding to the application of the Cauchy theorem to a contour  integral containing the ($u$-$d$) quark $V+A$ correlator $\Pi(Q^2)$ ($Q^2 \equiv - q^2$) and a weight function $g(Q^2)$, $\oint_{C_1+C_2} d Q^2 g(Q^2) \Pi(Q^2) = 0$, gives the sum rule
\begin{equation}
\int_0^{\sigma_{\rm m}} d \sigma g(-\sigma) \omega_{\rm exp}(\sigma)  =
\frac{1}{2 \pi}   \int_{-\pi}^{\pi}
d \phi \; {\cal D}_{\rm th}(\sigma_{\rm m} e^{i \phi}) G(\sigma_{\rm m} e^{i \phi}) ,
\label{sr}
\end{equation}
where $\sigma_{\rm m} \equiv \sigma_{\rm max}$ is the maximal used energy in the data, and $\omega(\sigma)_{\rm exp}$ is the ALEPH-measured discontinuity (spectral) function of the $(V+A)$-channel polarization function
\begin{equation}
\omega(\sigma) \equiv 2 \pi \; {\rm Im} \ \Pi(Q^2=-\sigma - i \epsilon) \ .
\label{om1}
\end{equation}
The function $g(Q^2)$ is the double-pinched Borel--Laplace weight function
\begin{equation}
g_{M^2}(Q^2) =  \left( 1 + \frac{Q^2}{\sigma_{\rm m}} \right)^2  \frac{1}{M^2} \exp \left( \frac{Q^2}{M^2} \right),
\label{g}
\end{equation}
$G(Q^2)$ is the integral of $g$
\begin{equation}
G(Q^2)= \int_{-\sigma_{\rm m}}^{Q^2} d Q^{'2} g(Q^{'2}),
\label{GQ2}
\end{equation}
and ${\cal D}_{\rm th}(Q^2)$ is the full Adler function ${\cal D}(Q^2) \equiv - 2 \pi^2 d \Pi(Q^2)/d \ln Q^2$, whose OPE truncated at dimension $D=6$ terms has the form
\begin{equation}
{\cal D}_{\rm th}(Q^2) = d(Q^2)_{D=0} + 1 + 4 \pi^2 \frac{\langle O_{4} \rangle}{ (Q^2)^2}  +   \frac{6 \pi^2}{ (Q^2)^3}  \left[ \frac{\langle O_{6}^{(2)} \rangle}{a(Q^2)} + \langle O_{6}^{(1)} \rangle \right] .
\label{DOPE}
\end{equation}
Here, $a(Q^2) \equiv \alphas(Q^2)/\pi$. The two terms of $D=6$ in the above OPE are needed to enable the cancellation of the corresponding $u=3$ IR renormalon ambiguities originating from the $D=0$ contribution $d(Q^2)_{D=0}$. The latter contribution has the perturbation expansion
\begin{equation}
d(Q^2)_{D=0, {\rm pt}}= d_0 a(\kappa Q^2) + d_1(\kappa) \; a(\kappa Q^2)^2 + \ldots + d_n(\kappa) \; a(\kappa Q^2)^{n+1} + \ldots, 
\label{dpt}
\end{equation}
where $\kappa \equiv \mu^2/Q^2$ is the renormalization scale parameter ($0 < \kappa \lesssim 1$; usually $\kappa=1$), the first four terms ($d_0=1$; $d_1$, $d_2$, $d_3$) are exactly known~\cite{Baikov:2008jh}, and for the coefficient $d_4$ [$\equiv d_4(\kappa)$ with $\kappa=1$ and $N_f=3$] we take the following values based on various specific estimates in the literature~\cite{Kataev:1995vh,Baikov:2008jh,Boito:2018rwt,Beneke:2008ad,Cvetic:2018qxs}:
\begin{equation}
d_4 = 275 \pm 63 \ .
\label{d4est}
\end{equation}
The expansion of the Borel transform of $d(Q^2)_{D=0}$ is ${\cal B}[d](u;\kappa)_{\rm ser.} = \sum_{n \geq 0} d_n(\kappa) u^n/n!/\beta_0^n$.
The extension of $d(Q^2)_{D=0}$ beyond $\sim a^5$ is performed with a renormalon-motivated model~\cite{Cvetic:2018qxs} in which the Borel transform is constructed first for an auxiliary quantity ${{\widetilde d}}(Q^2)$ of the Adler function~\cite{Cvetic:2018qxs}, resulting in the Borel transform ${\cal B}[d](u)$ having terms $\sim 1/(2-u)^{{\widetilde {\gamma}}_2}$, $1/(3-u)^{{\widetilde {\gamma}}_3+1}$, $1/(3-u)^{{\widetilde {\gamma}}_3}$ and $1/(1+u)^{1+{\overline {\gamma}}_1}$, and similar terms with lesser powers, where ${\widetilde {\gamma}}_p=1 + p \beta_1/\beta_0^2$ (${\overline {\gamma}}_p=1 - p \beta_1/\beta_0^2)$, and $\beta_0=(11 - 2 N_f/3)/4$ and $\beta_1=(1/16)(102 - 38 N_f/3)$ are the first two $\beta$-function coefficients ($N_f=3$).\footnote{In our ansatz~\cite{Cvetic:2018qxs} and notation, the effective one-loop $D=6$ anomalous dimensions  $-\gamma_{O_6}^{(1)}/\beta_0$ (appearing beside ${\widetilde {\gamma}}_3$ in the mentioned powers ${\widetilde {\gamma}}_3-\gamma_{O_6}^{(1)}/\beta_0$) were taken to be large-$\beta_0$, \ie\ $-\gamma_{O_6}^{(1)}/\beta_0=1, 0$. The work~\cite{Boito:2015joa} implies that these quantities can be evaluated beyond large-$\beta_0$, resulting in a decreasing sequence of nine numbers $-\gamma_{O_6}^{(1)}/\beta_0 \approx  -0.197; -0.247; \ldots$. It remains an open question how to extend the renormalon-motivated~\cite{Cvetic:2018qxs} model to include these results.}
This extension gives, for the choice $d_4=275.$, the coefficients of the expansion (\ref{dpt}): $d_5=3159.5$; $d_6=16136.$; $d_7=3.4079 \times 10^5$; $d_8=3.7816 \times 10^5$; $d_9=6.9944 \times 10^7$; $d_{10}=-5.8309 \times 10^8$; etc. 

The cancellation of the IR renormalon ambiguity requires: (i) for $u=2$ IR renormalon term ${\cal B}[d](u) \sim 1/(2-u)^{{\widetilde {\gamma}}_2}$, the $D=4$ OPE term of the Adler function to be of the form $1/(Q^2)^2$; (ii) for the $u=3$ IR renormalon term ${\cal B}[d](u) \sim 1/(3-u)^{{\widetilde {\gamma}}_3}$ to be of the form $1/(Q^2)^3$; (iii) and for the $u=3$ IR renormalon term ${\cal B}[d](u) \sim 1/(3-u)^{{\widetilde {\gamma}}_3+1}$ to be of the form $1/(Q^2)^3/a(Q^2)$. These three terms ($D=4, 6$) are taken into account in the OPE (\ref{DOPE}).

The $D=0$ contribution $d(Q^2)_{D=0}$ to the Adler function in the sum rule contour integral (\ref{sr}) is evaluated in three different ways. We apply two variants of fixed order perturbation theory (FOPT). In the first variant, the powers of $a(\kappa \sigma_{\rm m} e^{i \phi})^n$ are expressed as truncated Taylor series in powers of $a(\kappa \sigma_{\rm m})$ (FO). In the second variant, $d(Q^2)_{D=0}$ is expressed as the sum of the logarithmic derivatives ${\widetilde a}_n(Q^2)$ [$\propto (d/ d \ln Q^2)^{n-1} a(Q^2)$], and then ${\widetilde a}_n(\kappa \sigma_{\rm m} e^{i \phi})$ are expressed as truncated Taylor series of ${\widetilde a}_k(\kappa \sigma_{\rm m})$ (${\widetilde {\rm FO}}$). The third way of evaluation is the use of the inverse Borel transformation of $d(Q^2)_{D=0}$, where the Borel integral is evaluated with the Principal Value (PV) prescription; in the integrand, the Borel transform ${\cal B}[d](u)$ is taken as as a series consisting of the mentioned (renormalon-related) inverse powers $\sim (p-u)^k/(p-u)^{\gamma}$ ($k=0,1,\ldots$), where the series is truncated; this truncation requires for $d(\sigma_{\rm m} e^{i \phi})_{D=0}$ introduction of an additional correction polynomial  $\delta d(\sigma_{\rm m} e^{i \phi})_{D=0}^{[N_t]}$ in powers of $a(Q^2)$. In all the three methods, a truncation index $N_t$ is involved, \ie\ only the terms up to the power $a^{N_t}$ (or ${\widetilde a}_{N_t}$ in ${\widetilde {\rm FO}}$) are taken into account.

\begin{figure}[htpb!] 
\centering
\includegraphics[width=90mm]{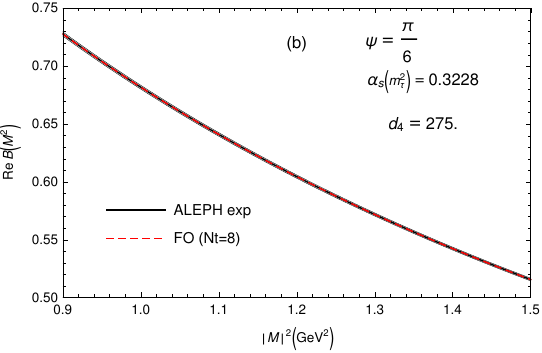}
\caption{The values of ${\rm Re} B(M^2;\sigma_{\rm m})$ along the ray $M^2=|M^2| \exp( i \Psi)$ with $\Psi=\pi/6$. The narrow grey band are the experimental predictions. The red dashed line inside the band is the result of the FOPT global fit with truncation index $N_t=8$. Similar fitting curves are obtained for the rays with $\Psi=0$ and $\Psi=\pi/4$.}
\label{fig:FigPsiPi6}
\end{figure}

We apply the Laplace-Borel sum rules, with the weight function (\ref{g}), to the ALEPH $V+A$ data with $\sigma_{\rm max} (\equiv \sigma_{\rm m})$ $=2.8 \ {\rm GeV}^2$ (\ie\ the last two bins are excluded due to large uncertainties). In the sum rule (\ref{sr}), this gives on both sides the Borel--Laplace sum rule quantity $B(M^2;\sigma_{\rm m})$. In practice, the rule is applied to the real parts only, ${\rm Re} B_{\rm exp}(M^2;\sigma_{\rm m}) = {\rm Re} B_{\rm th}(M^2;\sigma_{\rm m})$, and for the scale parameters $M^2$ along rays in the first quadrant: $M^2=|M^2| \exp(i \Psi)$ with $0 \leq \Psi < \pi/2$. We minimise (with respect to $\alphas$, $\langle O_4 \rangle$,  $\langle O_6^{(1)} \rangle$ and $\langle O_6^{(2)} \rangle$) the  following sum of squares:
\begin{equation}
\chi^2 = \sum_{\alpha=0}^n \left( \frac{ {\rm Re} B_{\rm th}(M^2_{\alpha};\sigma_{\rm m}) - {\rm Re} B_{\rm exp}(M^2_{\alpha};\sigma_{\rm m}) }{\delta_B(M^2_{\alpha})} \right)^2 ,
\label{xi2} \end{equation}
where $\{ M_{\alpha}^2 \}$ was taken as a dense set of points along the chosen rays with $\Psi=0, \pi/6, \pi/4$  and $0.9 \ {\rm GeV}^2 \leq |M_{\alpha}|^2 \leq 1.5 \ {\rm GeV}^2$. We chose 11 equidistant points along each of the three rays, and the series (\ref{xi2}) thus contains 33 terms (the fit results remain practically unchanged when the number of points is increased). In the sum (\ref{xi2}), the quantities $\delta_B(M^2_{\alpha})$ are the experimental standard deviations of ${\rm Re} B_{\rm exp}(M^2_{\alpha};\sigma_{\rm m})$, with the ALEPH covariance matrix for the $(V+A)$-channel taken into account (cf.~App.~C of~\cite{Ayala:2017tco} for more explanation). For each evaluation method (FO, ${\widetilde {\rm FO}}$, PV) and for each chosen truncation index $N_t$, the fit procedure gives us results, and the fit is usually of good quality, $\chi^2 \lesssim 10^{-3}$ (Fig.~\ref{fig:FigPsiPi6}).

The truncation index $N_t$ is then fixed by considering the first two double-pinched momenta $a^{(2,0)}(\sigma_{\rm m})$ and $a^{(2.1)}(\sigma_{\rm m})$\footnote{The weight functions for double-pinched momenta $a^{(2,n)}$ are: $g^{(2,n)}(Q^2) = ((n+3)/(n+1)) (1/\sigma_{\rm m}) (1 + Q^2/\sigma_{\rm m})^2 \sum_{k = 0}^{n} (k + 1)(-1)^k (Q^2/\sigma_{\rm m})^k$. The obtained values of $a^{(2,0)}(\sigma_{\rm m})$ and $a^{(2.1)}(\sigma_{\rm m})$ are well within the experimental band.}
and requiring local stability of their values under the variation of $N_t$.
The resulting extracted values of the coupling are
\begin{eqnarray}
\alphas(m_{\tau}^2)^{\rm (FO)} & = & 0.3228 \pm 0.0003({\rm exp})^{-0.0026}_{+0.0070}(\kappa)^{-0.0103}_{+0.0079}(d_4)^{+0.0081}_{-0.0057}(N_t)
\nonumber\\
& = &  0.3228^{+0.0134}_{-0.0121} \approx 0.323^{+0.013}_{-0.012} \qquad (N_t=8^{+2}_{-3}).
\label{BLalFO} \\
\alphas(m_{\tau}^2)^{\rm ({\widetilde {\rm FO}})} & = & 0.3209 \pm 0.0003({\rm exp})^{-0.0038}_{+0.0201}(\kappa)^{-0.0039}_{+0.0047}(d_4)^{+0.0293}_{-0.0084}(N_t)
\nonumber\\
& = &  0.3209^{+0.0359}_{-0.0100} \approx 0.321^{+0.036}_{-0.010} \qquad (N_t=5 \pm 2).
\label{BLaltFO} \\
\alphas(m_{\tau}^2)^{\rm (PV)} & = & 0.3269 \pm 0.0003({\rm exp})^{+0.0007}_{+0.0102}(\kappa)^{-0.0064}_{+0.0155}(d_4)^{+0.0092}_{-0.0006}(N_t)^{+0.0167}_{-0.0067}({\rm amb})
\nonumber \\
& = &  0.3269^{+0.0266}_{-0.0093} \approx 0.327^{+0.027}_{-0.009} \qquad (N_t=8^{+2}_{-3}).
\label{BLalPV} \\
\alphas(m_{\tau}^2)^{\rm (CI)} & = & 0.3488 \pm 0.0005({\rm exp})^{+0.0078}_{+0.0004}(\kappa) \pm 0.0000(d_4)^{-0.0027}_{+0.0119}(N_t)
\nonumber \\
& = &  0.3488^{+0.0142}_{-0.0028} \approx 0.349^{+0.014}_{-0.003}  \qquad (N_t=4^{+2}_{-1}).
\label{BLalCI}
\end{eqnarray}
The uncertainties are presented as separate terms. The variation of the renormalization scale parameter $\kappa \equiv \mu^2/Q^2$ was taken in the range $2/3 \leq \kappa \leq 2$ ($\kappa=1$ for the central values). The truncation index is $N_t=8,5,8$ for the central cases of FO, ${\widetilde {\rm FO}}$ and PV.

The (truncated) Contour Improved perturbation theory (CIPT) results are also included in the above results, for comparison. However, the truncated CIPT approach for $d(Q^2)_{D=0}$ evaluation appears to require a different type of OPE in the $D>0$ part of the contributions, because the renormalon structure and the related renormalon ambiguities are not reflected in the truncated CIPT series~\cite{Hoang:2021nlz}.
We thus include only FO, ${\widetilde {\rm FO}}$ and PV results in the average
\begin{eqnarray}
\alphas(m_{\tau}^2) &=& 0.3235^{+0.0138}_{-0.0126} \qquad ({\rm FO}+{\widetilde {\rm FO}}+{\rm PV}) 
\nonumber \\
\Rightarrow \;
\alphasmZ &=& 0.1191 \pm 0.0016.  
\label{3av} \end{eqnarray}
In Table~\ref{tab:rescomp} we compare these results with some other in the literature.
%
\renewcommand{\arraystretch}{1.3}
\begin{table}
\tabcolsep=2.mm
\centering
 \caption{$\alphas(m_{\tau}^2)$ values extracted by various groups from ALEPH $\tau$-decay data applying sum rules and other methods.\vspace{0.2cm}}
\label{tab:rescomp}
\begin{tabular}{l|l|lll|l}
\hline
group &  sum rule & FO & CI & PV & average \\
\hline
Baikov et al., 2008~\cite{Baikov:2008jh} & $a^{(2,1)}=r_{\tau}$ & $0.322 \pm 0.020$ & $0.342 \pm 0.011$ \hspace{0.3cm} & --- & $0.332 \pm 0.016$ \\
Beneke\&Jamin, 2008~\cite{Beneke:2008ad} & $a^{(2,1)}=r_{\tau}$ &  $0.320^{+0.012}_{-0.007}$ & --- \hspace{0.3cm} & $0.316 \pm 0.006$ & $0.318 \pm 0.006$ \\
Caprini, 2020~\cite{Caprini:2020lff} & $a^{(2,1)}=r_{\tau}$ &  --- & --- \hspace{0.3cm} & $0.314 \pm 0.006$ &  $0.314 \pm 0.006$ \\
Davier et al., 2013~\cite{Davier:2013sfa} & $a^{(i,j)}$ & $0.324$ & $0.341 \pm 0.008$ \hspace{0.3cm} & --- & $0.332 \pm 0.012$ \\
Pich\&R.S\'anchez, 2016~\cite{Pich:2016bdg}   &  $a^{(i,j)}$     & $0.320 \pm 0.012$ &  $0.335 \pm 0.013$ \hspace{0.3cm} & --- & $0.328 \pm 0.013$  \\
Boito et al., 2014~\cite{Boito:2014sta} & DV in $a^{(i,j)}$ & $0.296 \pm 0.010$ & $0.310 \pm 0.014$ \hspace{0.3cm}  & --- & $0.303 \pm 0.012$ \\
our prev. work, 2021~\cite{Ayala:2021mwc}  & BL ($O_6, O_8$) & $0.308 \pm 0.007$ &  --- \hspace{0.3cm}  & $0.316^{+0.008}_{-0.006}$ & $0.312 \pm 0.007$ \\
\hline 
this work, 2022 (also~\cite{Ayala:2021yct})  & BL  ($O_6^{(1)}, O_6^{(2)}$) & $0.323^{+0.013}_{-0.012}$(FO)  &  --- \hspace{0.3cm}  & $0.327^{+0.027}_{-0.009}$ & $0.324 \pm 0.013$\\
 &  &  $0.321^{+0.021}_{-0.030}$(${\widetilde {\rm FO}}$) & & & \\
\hline 
\end{tabular}
\end{table}
%

The results (\ref{3av}) can be significantly affected when the assumptions or methods are changed. For example, if we chose, instead of the central value $d_4=275.$, the upper upper bound $d_4=338.$ of Eq.~(\ref{d4est}) as the central value, the results would decrease somewhat, to $\alphas(m_{\tau}^2) \approx 0.320 \pm 0.015$ (corresponding to $\alphasmZ \approx 0.1187^{+0.0016}_{-0.0019}$), \ie\ $\delta \alphas(m^2_{\tau}) \approx -0.0004$.
If we took, instead of the two mentioned $D=6$ terms in the OPE, the simple $D=6$ and $D=8$ OPE terms [$\sim 1/(Q^2)^3$ and $\sim 1/(Q^2)^4$], the central value would decrease by about $\delta \alphas(m^2_{\tau}) \approx -0.008$. In our previous work~\cite{Ayala:2021mwc} we used the OPE with simple $D=6,8$ terms, and took for $d_4$ higher values $d_4=338 \pm 63$ than here Eq.~(\ref{d4est}).
If we took $N_t=5$ in all three methods (\ie\ no extension of Adler function beyond $d_4 a^5$), then the central value of $\alphas(m_{\tau}^2)$ in FO changes from $0.3288$ to $0.3171$, and in PV from $0.3269$ to $0.3277$ $\Rightarrow$ for the average of the three methods the central value changes from $0.3235$ to $0.3219$ (correspondingly, $\alphasmZ$ goes from $0.1191$ to $0.1189$), \ie\ $\delta \alphas(m^2_{\tau}) = -0.0016 \approx -0.002$, smaller.

According to the results (\ref{BLalFO})--(\ref{BLalPV}), Borel--Laplace sum rules indicate that the theoretical uncertainties dominate over the experimental ones. Even if maximally strong correlations were assumed among the experimental Borel--Laplace sum rule values at different $M^2_{\alpha}$, the presented experimental uncertainty of $\alphas$ would increase by a factor of less than five. Part of these theoretical uncertainties would be reduced by: (1) the calculation of the five-loop Adler function coefficient $d_4$; (2) the use of the more complicated structure of the $D=6$ OPE terms~\cite{Boito:2015joa} and the corresponding terms in the $D=0$ $u=3$ IR renormalon structure; (3) the use of a variant of the QCD coupling $a(Q^2)$ without the Landau singularities in the $D=0$ contribution, because this would allow for the resummation to all orders (no truncation) of the renormalon-motivated contribution $D=0$ and would eliminate the renormalization scale ambiguity ($\kappa$). The high precision ALEPH determination of the $\tau$ spectral function represents an important source of data for better understanding the behaviour of QCD at the limit between the perturbative and nonperturbative regimes.

\subsection{Reconciling  the fixed order  and contour improved perturbative series in hadronic \texorpdfstring{$\tau$}{tau} decays.
\protect\footnote{A\lowercase{uthors:} M. A. B\lowercase{enitez}-R\lowercase{athgeb} (U.\,V\lowercase{ienna}), D. B\lowercase{oito} (U.\,V\lowercase{ienna \&} U.\,S\lowercase{{\~a}o} P\lowercase{aulo}), A. H. H\lowercase{oang} (U.\,V\lowercase{ienna \&} E.~S\lowercase{chr\"odinger} I\lowercase{nst. for } M\lowercase{ath} P\lowercase{hys.}), M. J\lowercase{amin} (U.\,V\lowercase{ienna \&} H\lowercase{eidelberg} U\lowercase{niv.})}}
\label{sec:FOCI}

In Refs.~\cite{Benitez-Rathgeb:2022yqb,Benitez-Rathgeb:2022hfj} an approach was proposed to reconcile the long-standing discrepancy between the fixed order (FOPT) and contour improved (CIPT) pQCD expansions of hadronic spectral function moments relevant for the precise determination of $\alphas$ from  $\tau$ decays.  This is achieved by a simple change of scheme of the gluon condensate matrix element so that it becomes renormalon-free. The technical aspects of the scheme change are similar to the well-known implementation of short-distance mass schemes for massive-quark-sensitive observables.    
The scheme relies on external knowledge about the gluon condensate renormalon norm and is capable of resolving the long-standing discrepancy between  FOPT and CIPT predictions, under the assumption that the gluon condensate renormalon gives a sizeable contribution to QCD perturbative coefficients at the accessible intermediate orders. The approach is briefly outlined in the following. For details we refer to Refs.~\cite{Benitez-Rathgeb:2022yqb,Benitez-Rathgeb:2022hfj}.\\



As outlined in Sec.~\ref{sec:tau_Boito_et_al}, strong coupling determinations from hadronic $\tau$ decay data are based on 
weighted integrals of the  experimental spectral functions with an upper bound $s_0\leq m_\tau^ 2$ and to integrals of the QCD Adler function over a closed contour in the complex momentum plane. For the latter one must set the renormalization scale in $\alphas$ when performing the contour integrals. The two widely used prescriptions are based on FOPT and CIPT. 
In FOPT, the renormalization scale is fixed at $\mu^2=s_0$, which leads to a power series in $\alphas(s_0)$. In CIPT, the scale is tied to the contour's momentum variable such that logarithms related to the QCD $\beta$-function are summed, so that the resulting series is not anymore a power series in $\alphas$. These two expansion methods exhibit a sizeable discrepancy at orders $\mathcal{O}(\alphas^4)$ and $\mathcal{O}(\alphas^5)$ that is larger than the individual renormalization scale variations. As a consequence, $\alphas$ determinations based on CIPT tend to yield higher values than those based on FOPT. This discrepancy has been one of the main sources of theory uncertainty in $\alphas$ from $\tau$ decays.

In Refs.~\cite{Hoang:2020mkw,Hoang:2021nlz}, Hoang and Regner suggested that the discrepancy is of infrared (IR) origin and largely dominated by the leading IR renormalon of the massless-quark QCD Adler function, which is associated with the gluon condensate (GC).
This entails that, once this renormalon is subtracted from the Adler function's perturbation series, the truncated series of the two expansion methods should yield more consistent values already at intermediate orders, reducing the theoretical uncertainty in the extractions of $\alphas$. In Refs.~\cite{Benitez-Rathgeb:2022yqb,Benitez-Rathgeb:2022hfj} a concrete subtraction method of the gluon condensate renormalon was devised and its practical value was demonstrated. 

The subtraction method is based on two premises. First, it is assumed that the GC renormalon gives a sizeable contribution to the Adler
function's perturbative coefficients at intermediate orders ($\mathcal{O}(\alphas^3)$ to $\mathcal{O}(\alphas^5)$).
This assumption can be considered as natural, since the GC represents the leading IR sensitivity. It is also supported by all-order results in the large-$\beta_0$ limit and by renormalon models~\cite{Beneke:2008ad,Beneke:2012vb} built for the Adler function in QCD. 
Second, our framework relies on the fact that the OPE condensate corrections not only provide nonperturbative corrections but also compensate for the associated factorial renormalon growth of the perturbative series coefficients, thus leading to an unambiguous theoretical description.  The first premise is important. If it were not true, the CIPT-FOPT discrepancy problem would be unrelated to IR renormalons and no information on the norm of the GC renormalon could be gained from the known QCD corrections for the Adler function. The second premise is an established characteristics of multiloop QCD calculations in the limit of vanishing IR cutoff, where sensitivities to IR momenta are a cause of QCD perturbation series to be asymptotic. Since this is the common approach for the loop corrections for inclusive quantities based on dimensional regularization and the $\overline{\rm MS}$ scheme for the strong coupling, we call this approach to regularize IR momenta `$\overline{\rm MS}$ scheme' as well. This should not be confused with the $\overline{\rm MS}$ scheme for $\alphas$, which refers to the regularization of UV momenta. The concrete analytic expressions shown in this subsection below actually use the strong coupling in the $C$ scheme (for the concrete value $C=0$)~\cite{Boito:2016pwf}, referred to as $\bar\alphas$ below. (See the appendix of Ref.~\cite{Benitez-Rathgeb:2022yqb} for concrete formulae to obtain the results for the strong coupling in the $\overline{\rm MS}$ scheme.) 

In heavy-quark physics, it is well established that the leading (linear) IR sensitivity that arises in the pole mass renormalization scheme can be eliminated by switching to short-distance mass schemes~\cite{Hoang:2020iah,Beneke:2021lkq}. In Refs.~\cite{Benitez-Rathgeb:2022yqb,Benitez-Rathgeb:2022hfj} a similar approach is proposed for the GC matrix element.
One starts from the operator product expansion (OPE) for the massless quark Adler function. Using the $\overline{\rm MS}$ scheme (to regularize IR momenta), it can be cast in the form  (using $\mu^2=-s$ as renormalization scale for the strong coupling)
\begin{equation}
\label{eq:Dresum}
\textstyle D(s) \,=\, \sum_{n=1}^\infty \,
\bar c_{n,1} (\frac{\bar \alphas(-s)}{\pi})^n  + \frac{C_{4,0}(\bar \alphas(-s))}{s^2} \langle \bar {\cal O}_{4,0} \rangle  +
\sum_{d=6}^\infty \frac{1}{(-s)^{d/2}} \sum_i  C_{d,i}(\bar \alphas(-s)) \langle \bar {\cal O}_{d,\gamma_i}\rangle \,,
\end{equation}
where the first term is the perturbative contribution, and the $1/s$ OPE power corrections, starting from the term with $d=4$, are the nonperturbative corrections. The perturbative coefficients $\bar c_{n,1}$ are exactly known up $\mathcal{O}(\alphas^4)$~\cite{Baikov:2008jh,Herzog:2017dtz}. It is customary to
also include an estimate for the $\mathcal{O}(\alphas^5)$ coefficient $\bar c_{5,1}$ in phenomenological analyses~\cite{Boito:2014sta,Boito:2020xli,Pich:2016bdg,Ayala:2021mwc}.
The terms $\langle \bar {\cal O}_{d,\gamma_i}\rangle$ are nonperturbative vacuum matrix elements (condensates) and the $C_{d,i}$ are the respective Wilson coefficients which are a power series in $\alphas(-s)$. The bar over the operator indicates that the condensates are defined in the $\overline{\rm MS}$ scheme (to regularize IR momenta).

The leading $d=4$ OPE power correction arises from the GC and is central to this work. It can be cast in terms of the renormalization scale invariant GC matrix element  $ \langle \bar G^2\rangle$ as  ($\bar a(-s)\, \equiv\, \beta_0\,\bar \alphas(-s)/(4\pi)$, $\beta_0=11-2\Nf/3$)
\begin{eqnarray}
\label{eq:AdlerOPEGC}
\delta  D^{\rm OPE}_{4,0}(s) \, = \,
\textstyle \frac{1}{s^2} \frac{2\pi^2}{3}\,[ 1 -\frac{22}{81} \, \bar a(-s) ]\, \langle \bar G^2\rangle \,.
\end{eqnarray}
Associated with the GC OPE correction is a renormalon singularity in the Borel function of the Adler function that has the form
\begin{eqnarray}
\label{eq:AdlerBorelGC}
B_{4,0}(u) & = &
\textstyle [ 1 -\frac{22}{81} \, \bar a(-s)  ]\, \frac{N_{4,0}}{(2-u)^{1+4 \hat b_1}}\,,
\end{eqnarray}
where $N_{4,0}$ is  the  Adler function's GC renormalon norm and  $\hat b_1=\beta_1/2\beta_0^2$, with $\beta_1$ being the two-loop QCD $\beta$-function coefficient.  This GC renormalon singularity contributes to the Adler function's perturbative series coefficients $\bar c_{n,1}$ in the form
\begin{equation}
\label{eq:AdlerseriesOPEterm}
\textstyle \delta  \hat D_{4,0}(s) \, =  \,N_{4,0}\, [ 1 -\frac{22}{81} \, \bar a(-s)
+ \ldots  ]\,
{\textstyle \sum_{n=1}^\infty} \,
r_{n}^{(4,0)} \,\bar a(-s)^n\,,
\end{equation}
where the terms $r_{n}^{(4,0)}=2^{-(n+4\hat b_1)}\Gamma(n+4\hat b_1)/\Gamma(1+4\hat b_1)$ diverge factorially.
The use of the $C$-scheme~\cite{Boito:2016pwf} is convenient because the term 
$1/(2-u)^{1+4 \hat b_1}$
in $B_{4,0}(u)$ and the coefficients $r_n^{(4,0)}$ in $\delta\hat D_{4,0}(s)$ are exact, \ie\ they do not receive any further higher-order corrections. The first premise entails that $N_{4,0}$ is sufficiently sizeable, such that the series in $\delta\hat D_{4,0}(s)$ makes up for a sizeable contribution in the coefficients $\bar c_{n,1}$ at the intermediate orders relevant for phenomenological analyses. In the context of this assumption it has already been shown in Refs.~\cite{Hoang:2020mkw,Hoang:2021nlz}, that the FOPT-CIPT discrepancy can be caused by the diverging Adler function series contributions given in~$\delta\hat D_{4,0}(s)$.


Let's now turn to the construction of the renormalon-free (RF) gluon condensate scheme.  The RF scheme only deals with the GC renormalon; all other renormalons are strictly unaltered. The starting point is by imposing that the order-dependence of $\langle \bar G^2\rangle$
that compensates the factorial growth of the coefficients in Eq.~(\ref{eq:AdlerseriesOPEterm}) is made explicit, while  maintaining the generic form of the GC OPE correction in Eq.~(\ref{eq:AdlerOPEGC}). The relation between the original, order dependent, $\overline{\rm MS}$ GC matrix element and a new renormalon-free GC matrix element, $\langle G^2\rangle(R^2)$, is then
\begin{equation}
\label{eq:GCIRsubtracted}
\langle \bar G^2\rangle^{(n)}
\, \equiv \,
\langle G^2\rangle(R^2)
\,- \,
R^4 \,  {\textstyle \sum_{\ell=1}^n} \,
N_g\,
r_{\ell}^{(4,0)} \,\bar a^\ell_R\,,
\end{equation}
where $\bar a_R\equiv\beta_0\,\bar \alphas(R^2)/(4\pi)$ and
the order dependence on the left-hand side has been made explicit. The term $N_g$ is the universal GC renormalon norm, related to $N_{4,0}$ by  $N_g=3N_{4,0}/(2\pi^2)$. The condensate $\langle G^2\rangle(R^2)$ depends explicitly on the scale $R$, which sets its parametric size to be ${\cal O}(R^4)$ (instead of ${\cal O}(\Lambda_{\rm QCD}^4)$ for the $\overline{\rm MS}$ GC $\langle \bar G^2\rangle^{(n)}$). The scale $R$ plays the role of an IR factorization scale which can in principle be chosen arbitrarily. In practice, its value should be chosen smaller, but still of the order of, the typical dynamical scale of the observable of interest to avoid the appearance of (or, equivalently, to sum potentially) large logarithms. The dependence of $\langle G^2\rangle(R^2)$ on $R$ is controlled by an $R$-evolution equation~\cite{Hoang:2009yr} (see~\cite{Benitez-Rathgeb:2022yqb} for its precise form).

From a practical perspective, the $R$ dependence of $\langle G^2\rangle(R^2)$ is somewhat inconvenient. Therefore, in a second constructive step, a scale-invariant renormalon-free GC matrix element is defined. This is achieved by the addition of a function which obeys the same $R$-evolution equation as $\langle G^2\rangle(R^2)$. This is possible because the $R$-derivative of the subtraction series on the r.h.s.\ of Eq.~(\ref{eq:GCIRsubtracted}) is convergent for $n\to\infty$.
Such a function is obtained from the Borel sum of the subtraction series, defined through the principal value (PV) prescription:
\begin{eqnarray}
\label{eq:subtractclosed}
\bar c_0(R^2) & \equiv & R^4\,\,\,{\rm PV}\,   {\textstyle   \int_0^\infty \!
\frac{ {\rm d} u }{(2-u)^{1+4 \hat b_1}}  }\,e^{-\frac{u}{\bar a_R}}\,.
\end{eqnarray}
The result for $\bar c_0(R^2)$ can be given in closed analytic form~\cite{Benitez-Rathgeb:2022yqb}. Thus one can define a scale invariant renormalon-free GC matrix element, denoted as $\langle G^2\rangle^{\rm RF}$, by
\begin{eqnarray}
\label{eq:GCIRsubtracted2}
\langle G^2\rangle(R^2)
& \equiv &
\textstyle \langle G^2\rangle^{\rm RF} + N_g \, \bar c_0(R^2)\,.
\end{eqnarray}

We denote as $\langle G^2\rangle^{\rm RF}$ the GC in the `RF scheme'. The particular forms of the subtraction series in Eq.~(\ref{eq:GCIRsubtracted}) and the function $c_0(R^2)$ are particular choices to define the RF scheme, but may in principle be chosen differently. In general, the subtraction must have the correct large order behavior to eliminate the factorial growth of the coefficients $r_{n}^{(4,0)}$, but could have additional convergent contributions. This would then imply a different form for $\bar c_0(R^2)$. Moreover, one could also add an additional constant to $\bar c_0(R^2)$. The RF scheme entails that the difference between the original $\overline{\rm MS}$ GC and the new scale-independent RF GC, $\langle G^2\rangle^{\rm RF}$, is formally $\mathcal{O}(\bar a_R^{n+1})$. In this sense, the modifications to the GC in our new RF scheme are minimal.

Using Eqs.~(\ref{eq:GCIRsubtracted2}) and~(\ref{eq:AdlerOPEGC}), it is straightforward to write down the resulting perturbation series for the Adler function in the RF GC scheme. Treating the term proportional to $\bar c_0(R^2)$, which is part of the definition of $\langle G^2\rangle^{\rm RF}$, like a tree-level contribution, the perturbation series has the form: ($\bar c_{n}\equiv 4 \bar c_{n,1}/\beta_0$ for $n=1,2,\ldots$)
\begin{eqnarray}
\label{eq:invBorelDR}
\hspace{-3mm}
\hat D^{\rm RF}(s,R^2) =
\textstyle \frac{1}{s^2}\,[ 1 -\frac{22}{81}  \bar a(-s) ]\,N_{4,0}\, \bar c_0(R^2)
\, + \,
{\textstyle \sum_{n=1}^\infty } \, \bar c_{n} \,\bar a^n(-s)
\, - \,
[ 1 -\frac{22}{81} \, \bar a(-s)  ]\, N_{4,0}\,\frac{R^4}{s^2}
{\textstyle \sum_{n=1}^\infty }
r_{n}^{(4,0)}\bar a^\ell_R\,.
\end{eqnarray}
It is essential to reexpand and truncate the perturbative series terms (excluding the $\bar c_0$ term) coherently, using the strong coupling at a common renormalization scale. Only then the cancellation of the GC renormalon is realized is a consistent way. The treatment of
the `tree-level' term proportional to $\bar c_0(R^2)$ entails that the GC OPE correction retains its form of Eq.~(\ref{eq:AdlerOPEGC}) with $\langle \bar G^2\rangle$ replaced by $\langle G^2\rangle^{\rm RF}$. Furthermore, it is possible to show that
the Borel sum of the perturbation series for the RF scheme Adler function $\hat D^{\rm RF}(s,R^2)$ based on the PV prescription (as shown in Eq.~(\ref{eq:subtractclosed})) remains the same as that of the original $\overline{\rm MS}$ scheme Adler function $\hat D(s)= \sum_{n=1}^\infty \bar c_{n} \bar a^n(-s)$ independently of the value of $N_g$ (or $N_{4,0}$) and the choice for $R$. Variations of $R$ therefore vanish in the limit of larger-order truncations. The value of norm $N_g$ is an input to the RF scheme that must be supplemented independently.

\begin{figure}[htpb!]
\centering
\includegraphics[width=0.48\textwidth]{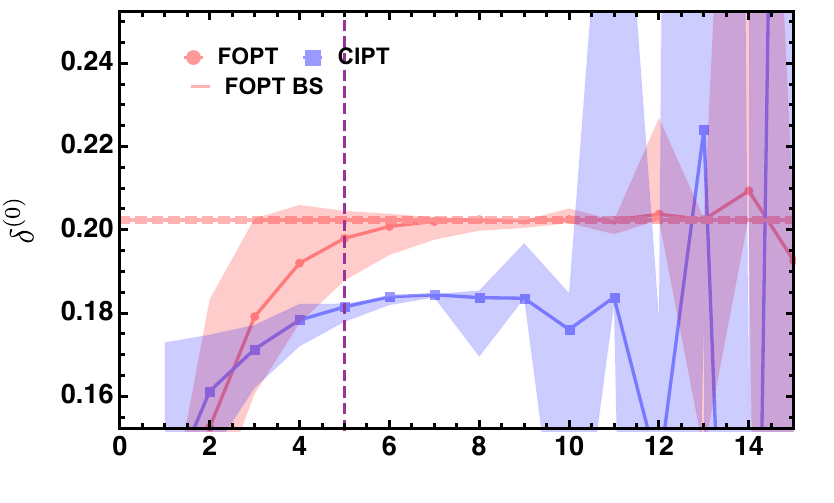}
\includegraphics[width=0.48\textwidth]{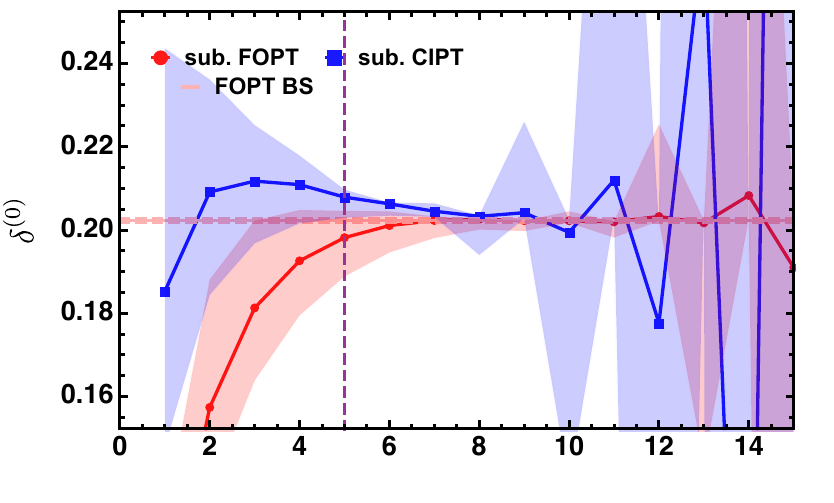}
\caption{Left: Series for the perturbative FOPT and CIPT expansions
of the $\tau$ hadronic decay width order by order  in full QCD in the $\overline{\rm MS}$ GC scheme. The orders beyond 5 are obtained from a Borel model. Renormalization scale variations are indicated by the coloured bands. Right: Corresponding series for FOPT and CIPT expansions for $\delta^{(0)}_{W}(m_\tau^2,R^2)$ in the RF GC scheme for $R=0.8\, m_\tau$. The results are shown for $\alphas(m_\tau^2)=0.315$ (in the $\overline{\rm MS}$ scheme) and $\Nf=3$. The figures are taken from Ref.~\cite{Benitez-Rathgeb:2022yqb}.}
\label{fig:ptseriesRF}
\end{figure}

In Ref.~\cite{Benitez-Rathgeb:2022yqb}, the effectiveness of the RF scheme for the perturbation series of the hadronic $\tau$ decay width has been demonstrated. In the following we quote some results given in Ref.~\cite{Benitez-Rathgeb:2022yqb}, where the known QCD corrections up to $\mathcal{O}(\alphas^4)$ have been included and an estimate for the 6-loop coefficient $\bar c_{5,1}$, consistent with the values used in the recent analyses~\cite{Boito:2014sta,Boito:2020xli,Pich:2016bdg,Ayala:2021mwc} has been employed.
In addition, corrections from beyond $\mathcal{O}(\alphas^5)$ were considered from a renormalon Borel model for the Adler function, based on the works of~\cite{Beneke:2008ad,Beneke:2012vb}, which has been shown there to be realistic within the first premise mentioned above. The model provides the concrete value $N_g=0.64$, so that the decay width series in the RF scheme can be studied in a concrete way. In the left panel of Fig.~\ref{fig:ptseriesRF}, the results for FOPT (red) and CIPT (blue) are shown in the $\overline{\rm MS}$ scheme for the GC.
The vertical dashed line indicates the order up to which current state-of-the phenomenological analyses are carried out (including the concrete perturbative coefficients up to 6-loop), and for orders beyond the series relies on the renormalon model.
The FOPT and CIPT series exhibit the well-known discrepancy at intermediate orders, which does not diminish with the successive inclusion of higher-order terms predicted by the model. In the context of the first premise, the results show that the discrepancy is systematic and  not related to missing higher orders. The two series eventually run into the divergent behavior expected for asymptotic series. The right panel displays the results in the RF GC scheme for $R= 0.8\, m_\tau$. The FOPT series in the RF scheme is almost unaltered. This can be understood, since the decay width receives only a tiny contribution from the GC through the $s$-dependence of the Wilson coefficient's 1-loop correction 
and because the dominant part of the subtraction series in FOPT remains real-valued and factors out of the contour integral. The CIPT series, however, changes dramatically. This modification is by far larger than the tiny size of the GC OPE corrections itself. 
This is because the subtraction series needs to be expanded according to the CIPT prescription such that the contour integration modifies the contributions from the subtraction series in a nontrivial way. The fact that the resulting numerical effect is so large, corroborates the finding of Refs.~\cite{Hoang:2020mkw,Hoang:2021nlz}
that the CIPT series (in the $\overline{\rm MS}$ GC scheme) is not compatible with the standard analytic form of OPE corrections shown in Eq.~(\ref{eq:Dresum}). 
Remarkably, in the RF GC scheme the discrepancy between the two series diminishes order by order and at $\mathcal{O}(\alphas^5)$ it is already significantly reduced. For even higher orders (in the context of the Borel model), the two series become fully compatible and approach essentially the same value.
The same observations are made for any other spectral function moment with a suppressed GC. At the same time, the convergence of the perturbation series for moments with an enhanced GC is substantially improved.

In Ref.~\cite{Benitez-Rathgeb:2022hfj} the practical value of the RF GC scheme in phenomenological analyses using the perturbative coefficients up to $\bar c_{5,1}$ was analyzed in detail. Is was shown that $N_g$ can be determined with a relative uncertainty of $40\%$. Following the two recent state-of-the-art strong coupling determination analyses at ${\cal O}(\alphas^5)$ of Refs.~\cite{Pich:2016bdg} (in the tOPE approach) and \cite{Boito:2020xli} (in the DV-model approach) it was demonstrated that the RF GC scheme can
successfully reconcile the extractions of $\alphas({m_\tau^2})$ based on CIPT and FOPT for both approaches to treat the nonperturbative corrections. The additional uncertainties that arise in the RF GC scheme due to variations of $R$ and the uncertainty of $N_g$ only lead to a small or moderate increase of the final uncertainty of $\alphas(m_\tau^2)$, and affect mainly the CIPT expansion results. 
The RF GC scheme thus constitutes a powerful new ingredient for future analyses of $\tau$ hadronic spectral function moments. Delicate issues such as the adequate treatment of nonperturbative corrections can now be studied without having to also deal with the CIPT-FOPT discrepancy problem.\\

\noindent \textit{Acknowledgments}--- DB and MJ would like to thank the Particle Physics Group of the University of Vienna for hospitality. We acknowledge partial support by the FWF Austrian Science Fund under the Doctoral Program ``Particles and Interactions'' No.\ W1252-N27 and under the Project No. P32383-N27. DB's work was supported by  the Coordena\c c\~ao de Aperfei\c coamento de Pessoal de N\'ivel Superior -- Brasil (CAPES) -- Finance Code 001 and by the S\~ao Paulo Research Foundation (FAPESP) Grant No.~2021/06756-6.

\clearpage
\section{\texorpdfstring{\boldmath$\alphasmZ$}{alphasmZ} from DIS and parton densities}
\label{sec:PDFs}


\subsection{\texorpdfstring{$\alphasmZ$}{alphasmZ} through scheme-invariant evolution of \texorpdfstring{N$^3$LO}{N3LO} non-singlet structure functions
\protect\footnote{A\lowercase{uthors:} J. B\lowercase{\"umlein}, M. S\lowercase{aragnese} (DESY)}}

We describe the measurement of the strong coupling constant $\alphasmZ$ based
on the scheme-invariant evolution of unpolarized and polarized flavour non-singlet 
structure functions at N$^3$LO accuracy with a precision at the subpercent level. Measurements
of this kind can be performed at future facilities such as the EIC and LHeC, provided
both proton and deuteron targets are used. The theory framework for this is already 
available. The measurement requires excellent control of the experimental systematics.\\

\maketitle



Currently, the values of the strong coupling constant $\alphasmZ$ as determined in different classes of measurement, and even inside these classes, is obtained at different values differing by several standard deviations~\cite{Proceedings:2011zvx,Alekhin:2016evh,Moch:2014tta}, 
reaching an individual accuracy of $\sim$1\%. One reason for this lies in the experimental systematics but also in different theoretical approaches being applied. We will consider the measurement of $\alphasmZ$ using deep-inelastic scattering (DIS) data. To minimize theoretical uncertainties, the measurement method has to be as simple as possible and widely free of effects that are difficult to control or are even widely unknown.
Usually one performs mixed twist-2 flavour non-singlet/singlet analyses on a wide host of DIS and other hard scattering data, which are required because, besides the flavour non-singlet parton distribution functions, those of the different sea quark and gluon distributions are needed as well. Furthermore, the range of $Q^2$ usually includes also the region in which higher-twist contributions have to be fitted in addition. Also, the treatment of heavy-quark effects varies in the different approaches, although there is no such freedom in general. Some of the cross sections used in the fits may not have the same higher-order correction as the massless DIS cross sections. The presence of all these different distribution functions will cost a large part of the statistics power to be determined and various parameter correlations, and potentially introduces theoretical biases~\cite{Alekhin:2011sk,Accardi:2016ndt}. All this complicates the measurement of the strong coupling constant in using approaches of this kind.

The situation is completely different in the case of scheme-invariant flavour non-singlet evolution of the structure functions $F_2^{\rm NS}(x,Q^2)$ or $g_1^{\rm NS}(x,Q^2)$. The latter structure function usually needs a much higher luminosity, since in addition the longitudinal polarization difference has to be carried out. To form the corresponding data sets, DIS off proton and deuteron targets has to be measured and a reliable description of the deuteron wave function effects is needed. In both cases the input distribution is a measured structure function: $F_2^{\rm NS}(x,Q_0^2)$ or $g_1^{\rm NS}(x,Q_0^2)$, with experimental error bands. Both the shape of these quantities and their experimental uncertainties can be parameterized at sufficient precision, forming the input for the one-dimensional evolution equation~\cite{Blumlein:2006be}. What remains to be determined in the fit of the data for $Q^2 > Q_0^2$ is the strong coupling constant $\alphasmZ$, the precision of which receives also contributions from the experimental uncertainty of the measured input distribution.

In the following we describe the theoretical basis of the scheme-invariant measurement of $\alphasmZ$ and illustrate a few relevant aspects numerically, following Ref.~\cite{Blumlein:2021lmf}, which may be performed in future experiments at EIC~\cite{Boer:2011fh} or the LHeC~\cite{LHeCStudyGroup:2012zhm}.
We consider the combination of structure functions
\begin{eqnarray}
F_2^{\rm NS}(x,Q^2) = F_2^p - \frac{1}{2} F_2^d = \frac{1}{6} x C_2^{\rm NS}(x,Q^2) \otimes v_3(x,Q^2),
\end{eqnarray}
where $\otimes$ denotes the Mellin convolution, $C_2^{\rm NS}(x,Q^2)$ is the corresponding unpolarized flavour non-singlet Wilson coefficient and 
\begin{eqnarray}
v_3(x,Q^2) = u_v(x,Q^2) - d_v(x,Q^2),
\end{eqnarray}
the difference of valence u- minus d-quark distributions. Nor the flavour singlet, gluon, or special sea-quark distributions enter the partonic input distribution.
However, the different flavours contribute through virtual QCD corrections, including charm and bottom quarks.

The scale evolution of $F_2^{\rm NS}(x,Q^2)$ can be described by an evolution operator $E_{\rm NS}(x,Q^2)$ 
\begin{eqnarray}
F_2^{\rm NS}(x,Q^2) = E_{\rm NS}(x,Q^2,Q_0^2; m_c, m_b) \otimes F_2^{\rm NS}(x,Q^2_0),
\end{eqnarray}
which reads in Mellin space
\begin{eqnarray}
E_{\rm NS}(Q^2,Q_0^2) & =&\left(\frac{a}{ a_0 }\right)^{-\frac{ P_0 }{2\beta_0 }}
\Biggl\{
1
+\frac{a- a_0 }{2\beta_0 ^2} \biggl\{
\Bigl[1 + a^2  C_2^{Q^2} - a_0 ^2  C_2^{Q_0^2} \Bigr] \bigl(2 \beta_0 ^2  C_1 -\beta_0   P_1
+\beta_1  P_0 \bigr)
-\frac{\bigl(a^2- a_0 ^2\bigr)}{4 \beta_0 ^3} 
\nonumber\\&&
\times \bigl(2\beta_0 ^2  C_1 -\beta_0   P_1 +\beta_1
P_0 \bigr)
\Bigl[2\beta_0 ^3 C_1 ^2+\beta_0 ^2 P_2 - \beta_0 \beta_1 P_1 + \bigl(\beta_1 ^2-\beta_0
\beta_2 \bigr)
P_0 \Bigr]
+\frac{\bigl(a^2+a  a_0 + a_0 ^2\bigr)}
{3 \beta_0 ^2} 
\nonumber\\&&
\times 
\Bigl[2\beta_0 ^4  C_1 ^3-\beta_0 ^3
P_3 +\beta_0 ^2 \beta_1 P_2 + \bigl(\beta_0 ^2 \beta_2 -\beta_0
\beta_1 ^2\bigr) P_1
+ \bigl(\beta_0 ^2 \beta_3 -2 \beta_0  \beta_1  \beta_2 +\beta_1 ^3\bigr) P_0 \Bigr]
+\frac{a- a_0 }{4 \beta_0 ^2} \bigl(2\beta_0 ^2  C_1 
\nonumber\\&&
-\beta_0   P_1 +\beta_1   P_0 \bigr)^2
+\frac{(a- a_0 )^2}{24 \beta_0 ^4} \bigl(2\beta_0 ^2  C_1 -\beta_0   P_1 +\beta_1   P_0
\bigr)^3
-\frac{a+ a_0 }{2 \beta_0 } \Bigl[2\beta_0 ^3  C_1 ^2+\beta_0 ^2  P_2 -\beta_0  \beta_1 P_1
\nonumber\\&&
+ P_0  \bigl(\beta_1 ^2
-\beta_0  \beta_2 \bigr)
 \Bigr]
\biggr\}
+a^2  C_2^{Q^2} - a_0 ^2  C_2^{Q_0^2}
- C_1  \Bigl[a^3  C_2^{Q^2} - a_0 ^3  C_2^{Q_0^2} \Bigr]
+ a^3 C_3^{Q^2} - a_0 ^3  C_3^{Q_0^2}
\Biggr\}.
\nonumber\\
\label{eq:ENS}
\end{eqnarray}
Here $a = a(Q^2)$ denotes the strong coupling constant and $a_0 = a(Q^2_0)$, $P_i$ are the non-singlet splitting functions and $C_i = c_i + h_i$ the expansion coefficients of the Wilson coefficient, where $c_i$ is the massless contribution and $h_i$ the massive contribution, where $h_1 = 0, h_2 = h_2^c + h_2^b$ and $h_3 = h_3^c + h_3^b + h_3^{cb}$, \ie\ there are mixed charm and bottom contributions from three-loops onward~\cite{Blumlein:2021lmf} for details. The dependence on the heavy-quark mass is logarithmic, with highest logarithmic powers $\ln^k(Q^2/m_h^2)$ for $h_k$. While the non-singlet splitting functions to three-loop order are known~\cite{Moch:2004pa,Blumlein:2021enk}, the four-loop non-singlet splitting function $P_3$ is not yet known, but it can be very well described by a Pad\'{e} approximation. Eight Mellin moments are known, with the earliest calculation~\cite{Baikov:2006ai}, and the presently available set~\cite{Moch:2017uml}. The massless three-loop Wilson coefficients were calculated in~\cite{Vermaseren:2005qc} and the single and double mass three-loop heavy flavour corrections in~\cite{Ablinger:2014vwa}. At two-loop order the corrections are even available for the whole kinematic region~\cite{Buza:1995ie,Blumlein:2016xcy}. Different proposals to perform scheme-invariant evolution both in the non-singlet and singlet case have been made since 1979, cf.~Refs.~[92--100] of Ref.~\cite{Blumlein:2021lmf}.

\begin{figure}[htpb!]
\centering
\includegraphics[width=0.49\textwidth]{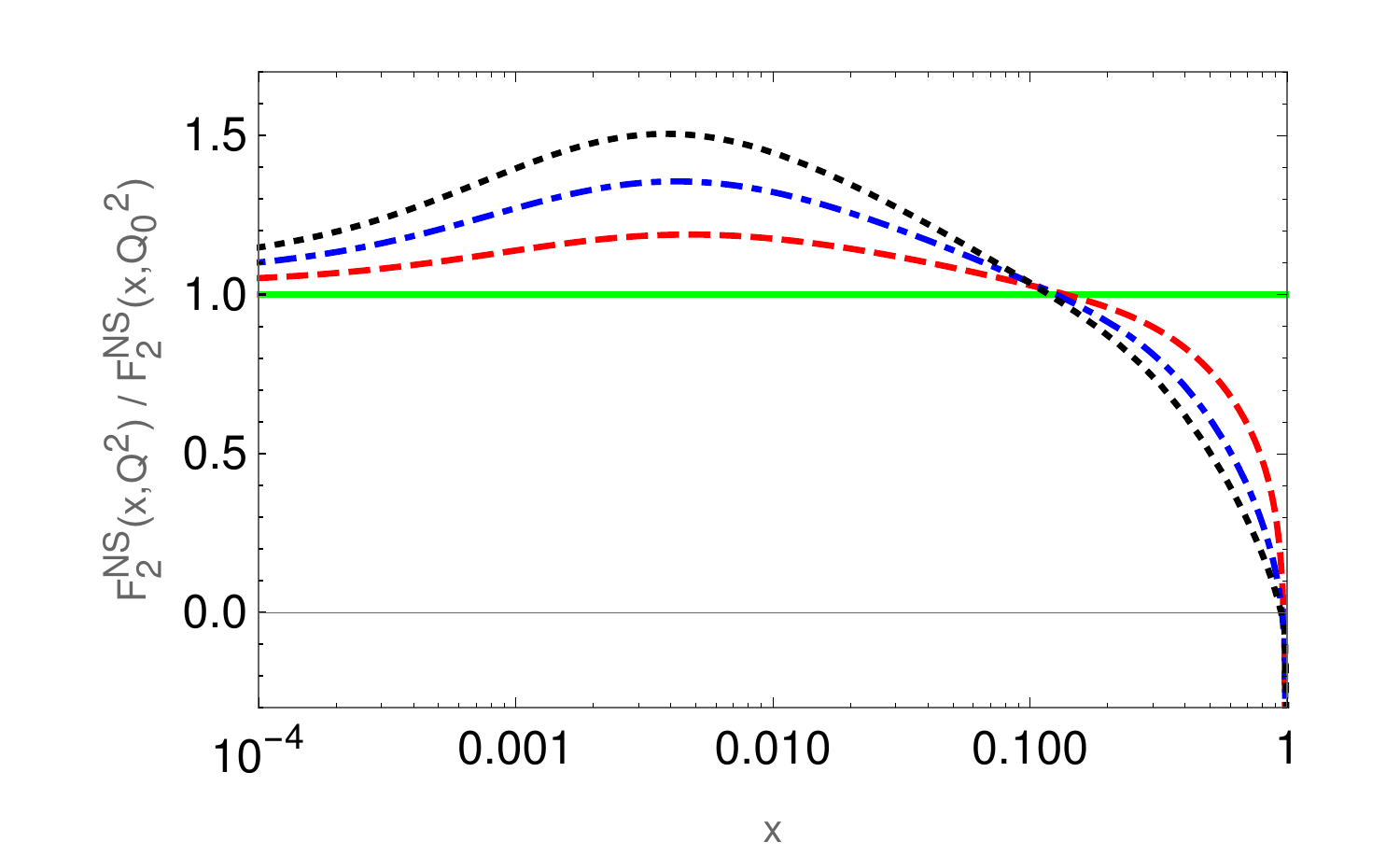}
\includegraphics[width=0.49\textwidth]{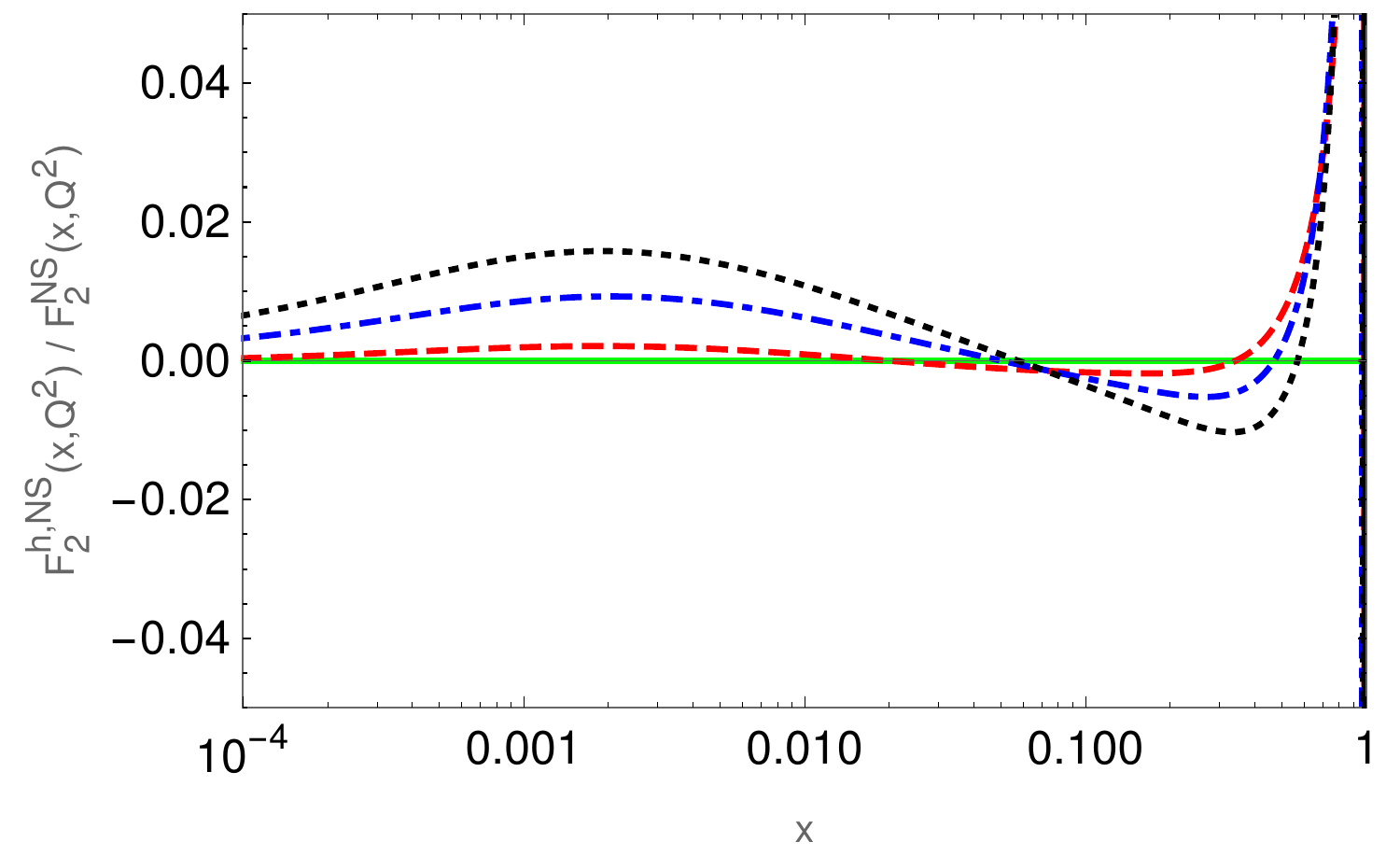}
\caption{Left: Evolution ratio of the structure function $F_2^{\rm NS}(x,Q^2)$. Right: Relative contribution of the heavy flavour corrections in $F_2^{\rm NS}(x,Q^2)$. Dashed lines: $Q^2 = 10^2$ GeV$^2$, dash-dotted lines: $Q^2 = 10^3$ GeV$^2$, dotted lines: $Q^2 = 10^4$ GeV$^2$; from Ref.~\cite{Blumlein:2021lmf}.}
\label{fig:bluemlein1}
\end{figure}


To illustrate the potential of the $\alphasmZ$ measurement using the present method, we show in Fig.~\ref{fig:bluemlein1} (left) the ratio of $F_2^{\rm NS}(x,Q^2)$ to $F_2^{\rm NS}(x,Q^2_0)$ for $Q_0^2 = 10$ GeV$^2$ up to scales $Q^2 = 10^4$~GeV$^2$. In the lower $x$ region positive corrections up to a factor of 1.5 are reached, while at large $x$ the corrections are negative. In Fig.~\ref{fig:bluemlein1} (right) we illustrate the impact of the heavy flavour corrections due to charm and bottom quark effects. In the region below $x = 0.6$ they are bound to $\pm 1.5 \%$ and grow for larger values of $x$. This shows their importance, because the future measurements will be performed at subpercent experimental accuracy. In Ref.~\cite{Blumlein:2021lmf} we also performed related studies for the case of the polarized structure function $g_1(x,Q^2)$.


A precision determination of the strong coupling constant $\alphasmZ$ requires a high luminosity
measurement of a sufficiently simple inclusive observable. The measurement must be carried out under stringent systematic control. Such a measurement would have been possible in the past, if the proposal~\cite{Guyot:1988} would have been carried out. It has not been possible at HERA, since deuterons have not been probed~\cite{Alexopoulos:2003aa} and the reconstruction of a non-singlet structure function from charged current data has not been precise enough. Given a sufficient preparation, the measurement can be carried out at the EIC using proton and deuteron targets. LHeC may also perform such a measurement, provided also deuteron data will be available and the statistics for the non-singlet measurement is high enough. The theoretical analysis method is then scheme-invariant evolution in the flavour non-singlet case for the structure function $F_2^{\rm NS}(x,Q^2)$. Both the light and heavy flavour corrections for this quantity are known at the level of the twist-2 approximation for $Q^2 \geq 10$~GeV$^2$, $W^2 \geq 15$~GeV$^2$~\cite{Alekhin:2012ig},
to measure $\alphasmZ$ at an accuracy well below the 1\% level. This is one way to decide what is the correct value of $\alphasmZ$.

To summarize, a high luminosity measurement of $F_2^p$ and $F_2^d$ at the EIC would allow to perform the N$^3$LO measurement of $\alphasmZ$ using the scheme-invariant method discussed here.\\

\noindent{\it Acknowledgments---} Support from the European Union's Horizon 2020 research and innovation programme under the Marie Sk\l{}odowska-Curie grant agreement No.\ 764850, SAGEX, is gratefully acknowledged.


\subsection{\texorpdfstring{$\alphasmZ$ from DIS large-$x$ structure function resummation}{alphasmZ from DIS large-x structure function resummation} 
\protect\footnote{A\lowercase{uthors:} A. V. K\lowercase{otikov}, V. G. K\lowercase{rivokhizhin}, B. G. S\lowercase{haikhatdenov} (JINR, D\lowercase{ubna})}}

We scrutinize the DIS $F_2(x,Q^2)$ structure functions (SFs) measured by the SLAC, NMC, and BCDMS experiments~\cite{Whitlow:1991uw,Whitlow:1990dr,Arneodo:1996qe,Benvenuti:1989rh,BCDMS:1989ggw,BCDMS:1987lzs} at NNLO accuracy in massless perturbative QCD in order to extract $\alphasmZ$. The so-called deep-inelastic scattering (DIS) scheme of the SFs~\cite{Altarelli:1978id} is considered, which leads to effective resummation of large-$x$ logarithms into the Wilson coefficient function. The study presented here is a continuation of investigations carried out in closely related papers~\cite{Shaikhatdenov:2009xd,Kotikov:2014cua,Kotikov:2016ljf,Krivokhizhin:2005pt,Krivokhizhin:2009zz}.\\

The function $F_2(x,Q^2)$ is represented as a sum of the leading twist (LT) $F_2^{\rm pQCD}(x,Q^2)$ and the twist-four terms (hereinafter, the superscripts pQCD and~LT denote the twist-two approximation with and without target-mass corrections):
\begin{equation}
F_2(x,Q^2)=F_2^{\rm pQCD}(x,Q^2)\left(1+\frac{\tilde h_4(x)}{Q^2}\right)\,.
\label{1.1}
\end{equation}

For large $x$ values, gluons do not nearly contribute and the $Q^2$ evolution of the twist-two DIS $F_2(x,Q^2)$ SF is well determined by the so-called nonsinglet (NS) part.
In this approximation, there is a direct relation between the moments of the DIS $F_2(x,Q^2)$ SF and the moments of the NS parton distribution function
${\bf f}(x,Q^2)$
\begin{equation}
M_{n}(Q^2) ~=~\int_0^1 dx x^{n-2} F^{\rm LT}_{2}(x,Q^2),~~
{\bf f}(n,Q^2) ~=~\int_0^1 dx x^{n-1}\, {\bf f}(x,Q^2) 
\label{Moments}
\end{equation}
which can be expressed as follows
\begin{equation}
M_n(Q^2) = R(f)\times C(n,a_s(Q^2))\times {\bf f}(n,Q^2)\,,
\label{3.a}
\end{equation}
where the strong coupling constant
\begin{equation}
a_s(Q^2)=\frac{\alphas(Q^2)}{4\pi} \label{as}
\end{equation}
and the Wilson coefficient function is denoted as $C(n,a_s(Q^2))$.
The constant $R(f)=1/6$
for $f=4$~\cite{Buras:1979yt}.

\subsubsection{Strong coupling constant derivation}

The strong coupling constant is found from the corresponding renormalization group equation.
At the NLO level, $a_s^{\rm NLO}(Q^2)\equiv a_{1}(Q^2)$,  the latter looks like
\begin{equation}
\label{1.coA}
\frac{1}{a_{1}(Q^2)} - \frac{1}{a_{1}(m_\mathrm{Z}^2)} +
b_1 \ln{\left[\frac{a_{1}(Q^2)}{a_{1}(m_\mathrm{Z}^2)}
\frac{(1 + b_1a_{1}(m_\mathrm{Z}^2))}
{(1 + b_1a_{1}(Q^2))}\right]}
= \beta_0 \ln{\left(\frac{Q^2}{m_\mathrm{Z}^2}\right)}\,.
\end{equation}

At NNLO level, $a_s^{\rm NNLO}(Q^2)\equiv a_{2}(Q^2)$, the strong coupling constant is derived from the following equation:
\begin{eqnarray}
\label{1.co}
\frac{1}{a_2(Q^2)} - \frac{1}{a_2(m_\mathrm{Z}^2)} &+&
b_1 \ln{\left[\frac{a_2(Q^2)}{a_2(m_\mathrm{Z}^2)}
\sqrt{\frac{1 + b_1a_2(m_\mathrm{Z}^2) + b_2a_2^2(m_\mathrm{Z}^2)}
{1 + b_1a_2(Q^2) + b_2a_2(Q^2)}}\right]} \\ \nonumber
&+& \left(b_2-\frac{b_1^2}{2}\right)\times
\Bigl(I(a_s(Q^2))- I(a_s(m_\mathrm{Z}^2))\Bigr) = \beta_0 \ln{\left(\frac{Q^2}{m_\mathrm{Z}^2}\right)}\,.
\end{eqnarray}
The expression for $I$ is
$$
I(a_s(Q^2))=
\left\{\begin{array}{c}
\displaystyle{\frac{2}{\sqrt{\Delta}}} \arctan{\displaystyle{\frac{b_1+2b_2a_2(Q^2)}{\sqrt{\Delta}}}} \hspace{1.5cm} \mbox{for}~~ f=3,4,5;~~~ \Delta>0,\\
\displaystyle{\frac{1}{\sqrt{-\Delta}}}\ln{\left[
\frac{b_1+2b_2a_2(Q^2)-\sqrt{-\Delta}}{b_1+2b_2a_2(Q^2)+\sqrt{-\Delta}}
    \right]}~~ \mbox{for}~~ f=6; \hspace{1cm} \Delta<0,
\end{array}
\right.
$$
where $\Delta=4b_2 - b_1^2$ and $b_i=\beta_i/\beta_0$ are read off from the QCD $\beta$-function:
$$
\beta(a_s) ~=~ -\beta_0 a_s^2 - \beta_1 a_s^3 - \beta_2 a_s^4 +\ldots \,.
$$

The coefficient function $C(n,a_s(Q^2))$ is then expressed in terms of the coefficients $B_j(n)$, which are exactly known
(for the odd $n$ values,
$B_j(n)$ and $Z_j(n)$ can be obtained
 by using the analytic continuation~\cite{Kazakov:1987jk,Kotikov:2005gr,Kotikov:1994mi})
\begin{equation}
C(n, a_s(Q^2)) = 1 
+ a_s(Q^2) B_{1}(n)
+ a_s^2(Q^2) B_{2}(n) + {\cal O}(a_s^3)\,.
\label{1.cf}
\end{equation}

The $Q^2$-evolution of the PDF moments can be calculated within the framework
of perturbative QCD:
\begin{equation}
\frac{{\bf f}(n,Q^2)}{{\bf f}(n,Q_0^2)}=\left[\frac{a_s(Q^2)}
{a_s(Q^2_0)}\right]^{\frac{\gamma_{0}(n)}{2\beta_0}}
\times \frac{h(n, Q^2)}{h(n, Q^2_0)}
\,, 
\label{3}
\end{equation}
where
\begin{equation}
\label{hnns}
h(n, Q^2)  = 1 + a_s(Q^2) Z_{1}(n) + a_s^2(Q^2) Z_{2}(n)
+ {\cal O}\left(a_s^3\right)\,,
\end{equation}
and
\begin{equation}
  Z_{1}(n) =
  \frac{1}{2\beta_0} \biggl[ \gamma_{1}(n) -
  \gamma_{0}(n)\, b_1\biggr]\,,~~
  Z_{2}(n)=
  \frac{1}{4\beta_0}\left[
\gamma_{2}(n)-\gamma_{1}(n)b_1 + \gamma_{0}(n)(b^2_1-b_2) \right]
+  \frac{1}{2} Z^2_{1}(n)
\,
\label{3.21}
\end{equation}
are combinations of the NLO and NNLO anomalous dimensions $\gamma_{1}(n)$ and $\gamma_{2}(n)$.

For large $n$ (this corresponds to large $x$ values), the coefficients $Z_{j}(n)\sim \ln n$ and $B_{j}(n)\sim \ln^{2j}n$.
So, the terms $\sim B_{j}(n)$ can lead to potentially large contributions and, therefore, should be resummed.

\subsubsection{Scale dependence}

We are going to consider the dependence of the results on the factorization $\mu_F$ scale caused by the truncation of a perturbative series~\cite{Shaikhatdenov:2009xd}.
This way, Eq.~(\ref{3.a}) takes the form:
\begin{equation}
M_n(Q^2) = R(f) \times \hat{C}(n, a_s(k_F Q^2))
\times {\bf f}(n,k_F Q^2)\,. \nonumber
\end{equation}

The function $\hat{C}$ is to be obtained from $C$ by modifying the r.h.s.\ of Eq.~(\ref{1.cf}) as follows:
\begin{eqnarray}
a_s(Q^2) &\to& a_s(k_F Q^2)\,,~~
B_{1}(n) \to B_{1}(n) + \frac{1}{2}\gamma_{0}(n) \ln{k_F}\,, \nonumber \\
B_{2}(n) &\to& B_{2}(n) + \frac{1}{2}\gamma_{1}(n) \ln{k_F}
+\left(\frac{1}{2}\gamma_{0} + \beta_0\right)B_{1}\ln{k_F}
+\frac{1}{8}\gamma_{0}\left(\gamma_{0} + 2\beta_0\right)\ln^2{k_F}\,.
\label{coeffun}
\end{eqnarray}

Taking a special form for the coefficient $k_F$, we can decrease contributions coming from the terms $\sim B_{j}(n)$.
To accomplish this task, we
consider
the
DIS-scheme~\cite{Altarelli:1978id},
where the NLO corrections to the Wilson coefficients are completely cancelled by changes in the factorization scale.

In the NLO case
\begin{equation}
a_s(Q^2) \to a_s(k_{\rm DIS}(n)Q^2)\equiv a_n^{\rm DIS}(Q^2),~~
k_{\rm DIS }(n)=\exp\left(\frac{-2B_{1}(n)}{\gamma_0(n)}\right) =\exp\left(\frac{-r^{\rm DIS }_{1}(n)}{\beta_0}\right) \, ,
\label{kDIS.NLO}
\end{equation}
where
\begin{equation}
r^{\rm DIS }_{1}(n)=\frac{2B_{1}(n)\beta_0}{\gamma_0}
~~~
\mbox{and}~~~
 B_1(n) \to B^{\rm DIS}_{1} =0.
\label{oBDI.NLO}
\end{equation}

The NLO coupling $a^{\rm DIS}_n(Q^2)$ obeys the following equation
\begin{equation}
\label{1.coA.DIS}
\frac{1}{a^{\rm DIS}_{n}(Q^2)} - \frac{1}{a_{1}(m_\mathrm{Z}^2)} +
b_1 \ln{\left[\frac{a^{\rm DIS}_{n}(Q^2)}{a_{1}(m_\mathrm{Z}^2)}
\frac{(1 + b_1a_{1}(m_\mathrm{Z}^2))}
     {(1 + b_1a^{\rm DIS}_{n}(Q^2))}\right]}
= \beta_0 \ln{\left(\frac{Q^2}{m_\mathrm{Z}^2}\right)}-r^{\rm DIS}_{1}(n)\,.
\end{equation}

In the NNLO case, in addition to  Eqs.~(\ref{kDIS.NLO}) and~(\ref{oBDI.NLO}), there is also
the following modification
\begin{equation}
B_2(n) \to B^{\rm DIS}_2(n)= B_2(n)-\left(\frac{1}{2}+\frac{\beta_0}{\gamma_0(n)}\right)\, B^2_1(n)-
\frac{\gamma_1(n)}{\gamma_0(n)}\, B_1(n)\, ,
\label{oBDI.NNLO}
\end{equation}
that leads to the cancellation of
the large terms $\sim \ln^4(n)$
in $B^{\rm DIS}_1(n)$.

The NNLO coupling $a^{\rm DIS}_n(Q^2)$ obeys the
equation
\begin{eqnarray}
\label{1.coB.DIS}
\frac{1}{a_n(Q^2)} - \frac{1}{a_2(m_\mathrm{Z}^2)} &+&
b_1 \ln{\left[\frac{a_n(Q^2)}{a_2(m_\mathrm{Z}^2)}
\sqrt{\frac{1 + b_1a_2(m_\mathrm{Z}^2) + b_2a_2^2(m_\mathrm{Z}^2)}
{1 + b_1a_n(Q^2) + b_2a_n(Q^2)}}\right]} \\ \nonumber
&& \hspace{-2cm} +\left(b_2-\frac{b_1^2}{2}\right)\times
\Bigl(I(a_n(Q^2))-I(a_s(m_\mathrm{Z}^2))\Bigr) = \beta_0 \ln{\left(\frac{Q^2}{m_\mathrm{Z}^2}\right)}-r^{\rm DIS}_{1}(n)\,.
 \end{eqnarray}


\subsubsection{Fit results}

Our analysis is carried out for the moments of the $F_2(x,Q^2)$ SF defined in Eq.~(\ref{Moments}). Then, for each $Q^2$, the $F_2(x,Q^2)$ SF is recovered using the Jacobi polynomial decomposition method~\cite{Parisi:1978jv,Krivokhizhin:1987rz,Krivokhizhin:1990ct}:
\begin{equation}
F_{2}(x,Q^2)=x^a(1-x)^b\sum_{n=0}^{N_\text{max}}\Theta_n ^{a,b}(x)\sum_{j=0}^{n}c_j^{(n)}(\alpha ,\beta )
M_{j+2} (Q^2)\,,
\label{2.1}
\end{equation}
where $\Theta_n^{a,b}$ are the Jacobi polynomials, $a,b$ are the parameters to be fit.
The program MINUIT~\cite{James:1975dr} is used to minimize the difference between experimental data and theoretical predictions for
the $F_2(x,Q^2)$ SF.

We use free data normalizations for various experiments. As a reference set, the most stable data of hydrogen BCDMS are used at the value of
the initial beam energy $E_0=200$ GeV.
Contrary to previous analyses~\cite{Shaikhatdenov:2009xd,Kotikov:2014cua}, the cut $Q^2\geq$ 2 GeV$^2$ is used throughout,
since for smaller $Q^2$ values Eqs.~(\ref{1.coA.DIS}) and~(\ref{1.coB.DIS}) have no real solutions.

The starting point of $Q^2$-evolution is taken at
$Q^2_0$ = 90 GeV$^2$. This value of $Q^2_0$ is close to the average values of $Q^2$ covering the corresponding data.
Based on studies done in
Ref.~\cite{Krivokhizhin:2005pt,Krivokhizhin:2009zz},
it is sufficient to take the maximum number of moments $N_\text{max} = 8$, and the cut $0.25 \leq x \leq 0.8$ is applied on the data.

We work within
the variable-flavour-number scheme (VFNS)~\cite{Shaikhatdenov:2009xd}. 
To strengthen the effect of changing the sign of twist-four corrections, we also present results 
obtained in the fixed-flavour-number scheme with $\Nf=4$.

\begin{table}[htpb!]
\centering
\tabcolsep=4.mm
\caption{Parameter values of the twist-four term in different cases obtained in the analysis of data
 (314 points: $Q^2\geq 2$ GeV$^2$) carried out within VFNS (FFNS).\vspace{0.2cm}
 \label{tab:twist4}}
\label{tab:HT}
\begin{tabular}{l|c|c|c|c}
\hline
& NLO & NLO & NNLO & NNLO  \\
$x$  & $\MSbar$ scheme & DIS scheme           &  $\MSbar$ scheme & DIS scheme \\
 &$\chi^2=246~(259)$ &  $\chi^2=238~(251)$ &$\chi^2=241~(254)$  &  $\chi^2=242(249)$    \\
&$\alphasmZ= 0.1195$ &  $\alphasmZ= 0.1177$ &$\alphasmZ= 0.1177$  &  $\alphasmZ= 0.1178$    \\
&(0.1192) &(0.1179)&(0.1170) &(0.1171)\\
\hline \hline
0.275 &$-0.245~(-0.264)$ & $-0.187~(-0.174)$ & $-0.188~(-0.204)$ & $-0.141~(-0.170)$  \\
0.35 &$-0.243~(-0.252)$ & $-0.111~(-0.134)$ & $-0.188~(-0.193)$ & $-0.133~(-0.149)$  \\
0.45 &$-0.191~(-0.187)$ & $-0.040~(-0.094)$ & $-0.172~(-0.158)$ & $-0.110~(-0.104)$  \\
0.55 &$-0.116~(0.096)$ & $-0.106~(-0.088)$ & $-0.174~(-0.137)$ & $-0.121~(-0.084)$  \\
0.65 & 0.054~(0.118) & $-0.167~(-0.094)$    & $-0.145~(-0.051)$ & $-0.223~(-0.100)$ \\
0.75 & 0.337~(0.477) &  $-0.568~(-0.442)$ & 0.115~(0.648) & $-0.587~(-0.314)$  \\
\hline
\end{tabular}
\vspace{0.5cm}
\end{table}
As one can see from Table~\ref{tab:twist4}, the central values of $\alphasmZ$ are mostly
close to each other upon taking into account total experimental and theoretical errors~\cite{Shaikhatdenov:2009xd,Kotikov:2014cua}:
\begin{equation}
\pm 0.0022~~~\mbox{(total exp. error)},~~~
\left\{\begin{array}{c}
  \displaystyle{+0.0028}\\
\displaystyle{-0.0016}
\end{array}
\right.
~~~\mbox{(theor. error)}\,.
\label{Errors}
\end{equation}
We plan to study these errors in more detail and present them in an upcoming publication.

From the table, it can also be seen that after resumming at large $x$ values (i.e. in the DIS scheme), the twist-four corrections change sign at large values of $x$. Thus,
unlike the standard analyses performed in
\cite{Kotikov:2010bm}, in this case, when twist-four corrections change the sign, (part of) the powerlike
terms can be said to be swallowed up into a QCD analytic constant~\cite{Shirkov:1997wi} just much like as it was observed at low-$x$ values~\cite{Cvetic:2009kw} in the framework of the so-called double asymptotic scaling approach~\cite{Kotikov:1998qt,Illarionov:2004nw}.

In previous papers~\cite{Kotikov:1992ht,Parente:1994bf}, where resummation at large values of $x$ was performed 
within the framework of the Grunberg approach~\cite{Grunberg:1980ja,Grunberg:1982fw}, we saw only a decrease in the twist-four contribution, 
since the terms were not studied in detail. That is why we plan to include Grunberg's approach in an
analysis similar to the present one and study in some detail the values of the twist-four corrections.


\subsection{Strong coupling \texorpdfstring{$\alphas$}{alphas} in fits of parton distributions 
\protect\footnote{A\lowercase{uthors:} S. M\lowercase{och} (H\lowercase{amburg} U\lowercase{niv.})}}

Data from deep-inelastic scattering (DIS) experiments collected during the
last few decades contributes significantly to the knowledge about the
structure of the proton and the extraction of parton distributions (PDFs).
The theory description builds on QCD factorization, which allows to express 
the structure functions in the charged-lepton- (or neutrino-) proton hard scattering 
with large momentum transfer $Q$ schematically as
\begin{equation}
  \label{eq:moch-disSF}
  F_a \,=\, f_{i}^{} \,\otimes\, c_{a, i}\, , \qquad a=2,3,L\, ,
\end{equation}
where the process-dependent coefficient functions are denoted by $c_{a, i}$
and the PDFs for a given fraction of the proton momentum $x$ by $f_{i\,}^{}(x,\mu^2)$. 
Their dependence on the (renormalization and factorization) scale $\mu$ 
is governed by the well-known evolution equations
\begin{equation}
  \label{eq:moch-evol}
  \frac{\partial}{\partial \ln \mu^2} \, f_i^{}(x,\mu^2) 
  \,=\, 
  \left[ P^{}_{ik}(\alphas(\mu^2)) \otimes f_k^{}(\mu^2) \right]\!(x) \, ,
\end{equation}
where the standard convolution is abbreviated by $\otimes$.
Both, the splitting functions $P^{}_{ik}$ and the coefficient functions by $c_{a, i}$ 
are calculable in QCD perturbation theory in an expansion 
in powers of the strong coupling constant $a_s \equiv \alphas(\mu^2)/(4\pi)$, 
\begin{eqnarray}
\label{eq:moch-Pexp}
P \, &=& 
a_s\, P^{\,(0)} 
+\, a_s^{2}\, P^{\,(1)}
+\, a_s^{3}\, P^{\,(2)} 
+\, a_s^{4}\, P^{\,(3)} 
+\, \ldots\,\, , 
\\
\label{eq:moch-Cexp}
c_{a, i} &=& 
c_{a, i}^{\,(0)} 
+\, a_s\, c_{a, i}^{\,(1)} 
+\, a_s^{2}\, c_{a, i}^{\,(2)} \,
+\, a_s^{3}\, c_{a, i}^{\,(3)} 
+\, \ldots \,\, .
\end{eqnarray}
Here, the first three terms define the NNLO predictions for DIS, which is 
currently the standard approximation~\cite{Moch:2004pa,Vogt:2004mw} for the splitting functions 
and~\cite{vanNeerven:1991nn,Zijlstra:1992qd,Zijlstra:1992kj,Moch:1999eb} for the DIS coefficient functions at this accuracy.
Form the perturbative expansion in Eqs.~(\ref{eq:moch-Pexp}) and
(\ref{eq:moch-Cexp}) as well as from the PDF evolution in Eq.~(\ref{eq:moch-evol}) 
it is obvious that there is, in general, significant correlation between the
PDFs and the value of strong coupling constant $\alphasmZ$~\cite{Accardi:2016ndt}.
In the ABM PDF fits, this information can be obtained from the (positive-definite) covariance matrix~\cite{Alekhin:2017kpj}.
  
Equation~(\ref{eq:moch-disSF}) holds up to power corrections suppressed by $1/Q^2$, 
and the accessible range of kinematics in the momentum transfer $Q$ and
Bjorken $x$ requires careful considerations of the respective kinematic regions.
Typical cuts in the application to DIS data restrict the 
invariant mass of the hadronic system $W^2 = m_p^2 + Q^2 (1-x)/x$ (with the
proton mass $m_p$) to $W^2 \ge 12.5~{\rm GeV}^2$ and $Q^2 \ge 2.5\mbox{--}10~{\rm GeV}^2$~\cite{Alekhin:2017kpj}.
Depending on these cuts, an improved theoretical description of $F_a\left(x,Q^2\right)$ 
to account for higher-twist and target-mass corrections becomes necessary.
Higher-twist corrections arise from the infinite tower of power corrections $(1/Q^2)^n$, 
in the operator product expansion (with $n=1,2,3, \dots$). 
Higher-twist terms have a physical interpretation as multiparton correlations
and modify the structure functions in Eq.~(\ref{eq:moch-disSF}) as 
\begin{eqnarray}
\label{eq:moch-ht}
F_a^{\rm ht}(x,Q^2) &=&
F_a^{\rm TMC}(x,Q^2)
+
\frac{H_a^{\tau=4}(x)}{Q^2}
\, ,
\qquad\qquad 
a \,=\, 2,T\, ,
\end{eqnarray}
where the additive terms $H_a^{\tau=4}(x)$ models the twist-four contribution 
and $F_a^{\rm TMC}$ denotes the leading-twist structure function with 
target mass corrections, which account for the finite nucleon mass~\cite{Alekhin:2017kpj}.

Another important aspect in the theoretical description of the leading-twist
structure functions from Eq.~(\ref{eq:moch-disSF}) entering in
Eq.~(\ref{eq:moch-ht}) concerns the active number of flavours and the
treatment of DIS heavy quark production~\cite{Laenen:1992zk}
A fixed-flavour number scheme and the use the standard decoupling 
relations for heavy quarks in QCD in the transition from $\Nf=3$ to $\Nf=5$ 
is justified given the currently available kinematics in $x$ and $Q$ 
for DIS charm quark data~\cite{Accardi:2016ndt}.
Variable-flavour-number schemes attempting the resummation of large
logarithms in the ratio $Q^2/m^2_h$ with the heavy-quark mass squared $m^2_h$ 
introduce additional theoretical uncertainties from the matching, 
which manifest themselves predominantly at small $Q^2$~\cite{Alekhin:2020edf}.

\begin{figure}[th!]
\begin{center}
  \includegraphics[width=0.45\textwidth]{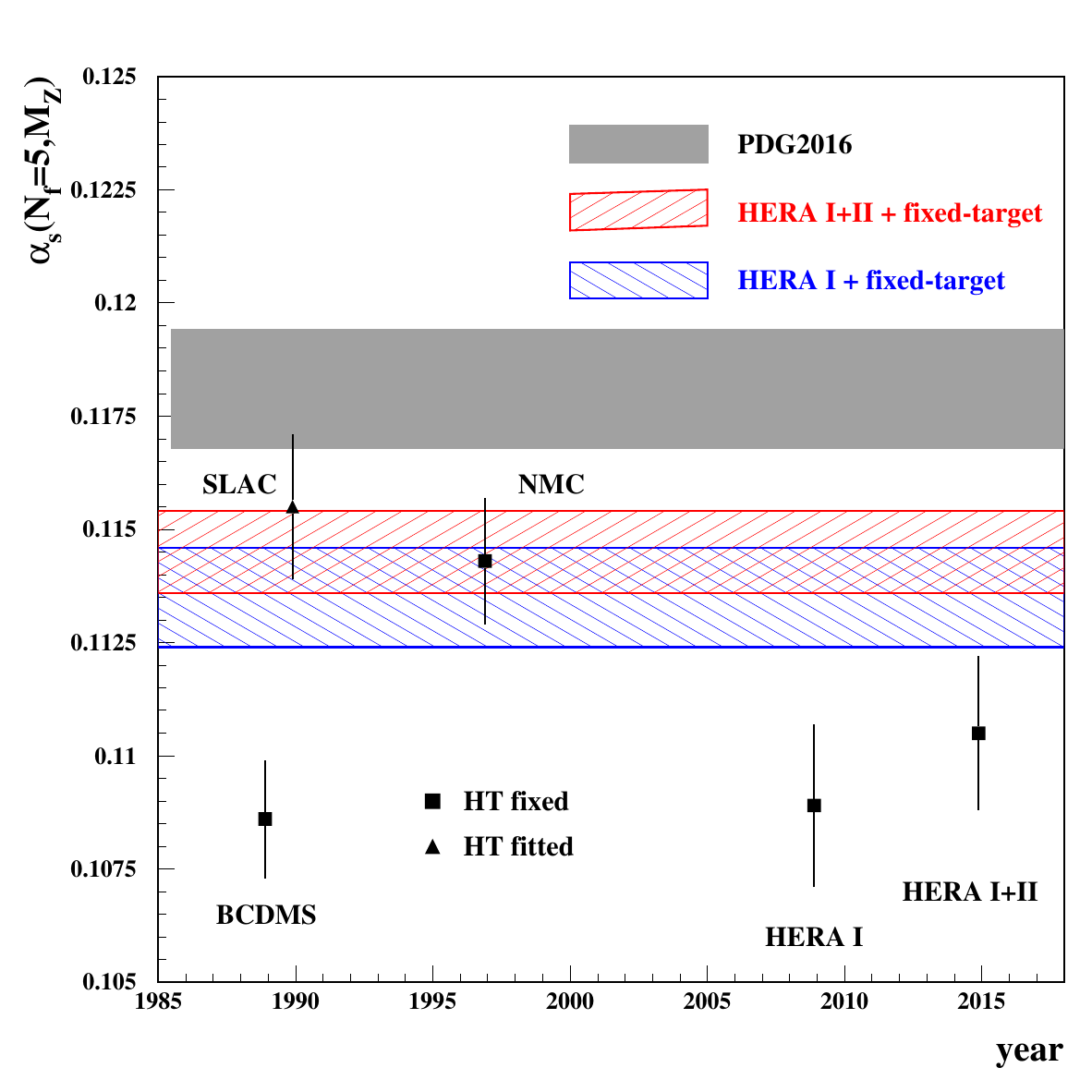}
\end{center}
\vspace*{-5mm}
  \caption{
    \label{fig:moch-alphas-history}
   Value of $\alphasmZ$ in the $\overline{\rm{MS}}$ scheme for $\Nf=5$
   flavours at NNLO in QCD preferred by individual data sets as a function of the year of their publication.
   Data from SLAC~\cite{Bodek:1979rx,Mestayer:1982ba,Whitlow:1990gk} (proton),
   BCDMS~\cite{Benvenuti:1989rh},
   NMC~\cite{Arneodo:1996qe} (proton), 
   the HERA run I~\cite{Aaron:2009aa} as well as 
   the HERA run I+II combination~\cite{Abramowicz:2015mha}
   are considered in three variants for the treatment of the higher twist
   terms defined in Eq.~(\ref{eq:moch-ht}):
   {\it (i)} the higher twist terms are set to zero (circles);
   {\it (ii)} they are fixed to the values obtained in the ABMP16
   fit from considering all data sets (squares);
   {\it (iii)} they are fitted to the individual data set
   under study (triangles).
   The bands for $\alphasmZ$ obtained by using the combination 
   of the SLAC, BCDMS and NMC samples together with those from the HERA run I 
   (left-tilted hatches) and the run I+II combination (right-tilted hatches).
   The 2016 PDG average~\cite{Patrignani:2016xqp} (shaded area) is shown for comparison.  
   Plot from Ref.~\cite{Alekhin:2017kpj}.
  }
\end{figure}

The impact of higher-twist terms and target-mass corrections in Eq.~(\ref{eq:moch-ht})
on the extracted value of $\alphasmZ$ to NNLO accuracy from world DIS data 
has been illustrated in Fig.~\ref{fig:moch-alphas-history}, where it is shown
that variants of the fits with no higher twist lead to systematically larger
values of $\alphasmZ$.
\begin{figure}[th!]
\begin{center}
\includegraphics[width=0.95\textwidth,angle=0]{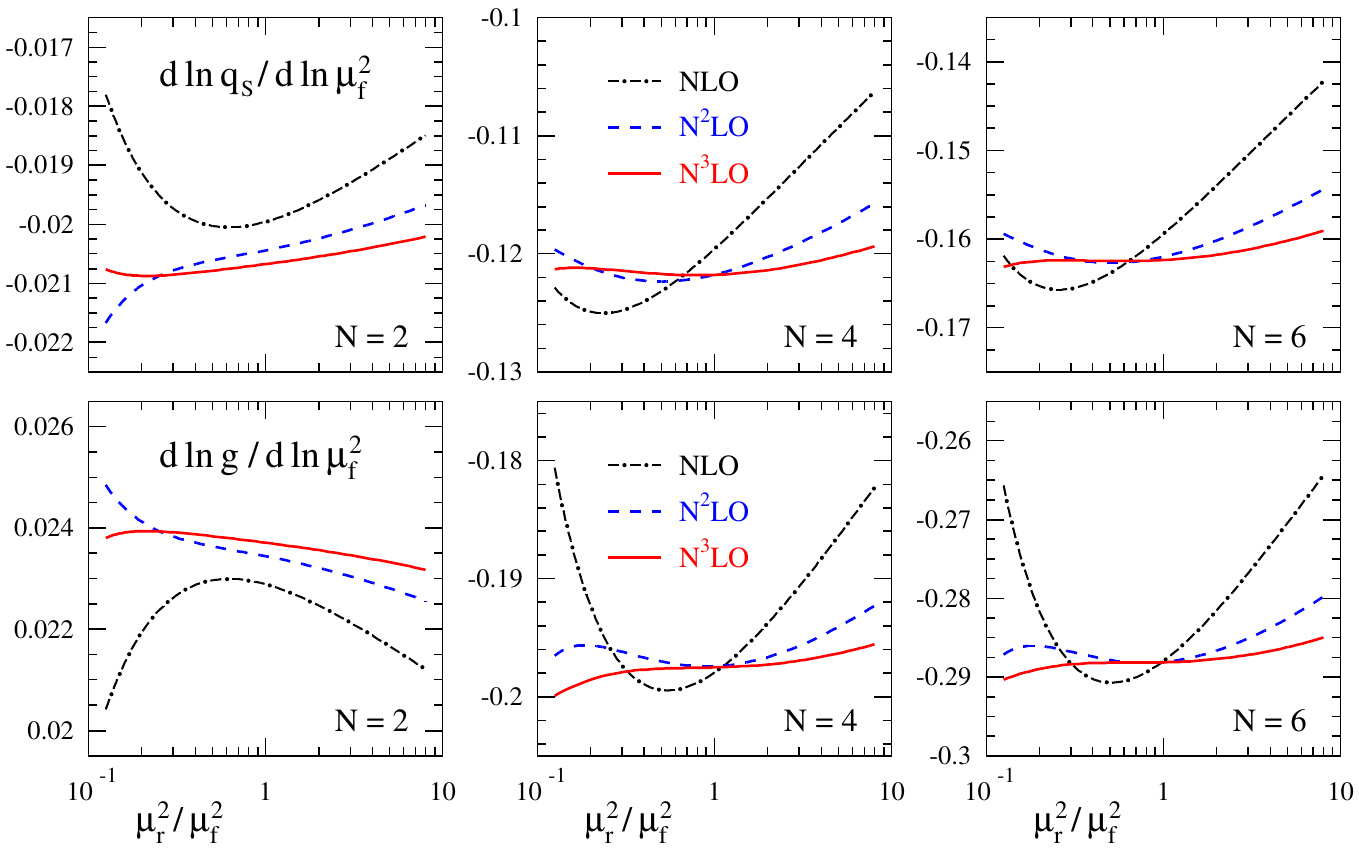}
\end{center}
\vspace*{-5mm}
\caption{
  \label{fig:moch-dqsgdlnmu}
  The dependence of the logarithmic factorization scale derivatives of
  the singlet PDFs on the renormalization scale $\mu_{r}$ at $N=2$
  (where the very small scaling violations of $q_{\rm s}$ and $g$ are
  related by the momentum sum rule), $N=4$ and $N=6$ 
  for $\alphas(\mu_{0}^{\,2}) = 0.2$ and $\Nf=4$ 
  and initial distributions of the form 
  $xq_{\rm s} = 0.6\: x^{\, -0.3} (1-x)^{3.5}\,(1 + 5.0\: x^{\, 0.8\,})$
  and 
  $xg = 1.6\: x^{\, -0.3} (1-x)^{4.5}\, (1 - 0.6\: x^{\, 0.3\,})$ 
  at the standard scale $\mu_r = \mu_f$. 
  Plot from Ref.~\cite{Moch:2021qrk}.
 }
\end{figure}
The combined fit of PDFs and $\alphasmZ$ to the world data shown in Fig.~\ref{fig:moch-alphas-history}
delivers in the ABMP16 analysis 
at NNLO in QCD the value of~\cite{Alekhin:2017kpj}  
\begin{eqnarray}
  \label{eq:asdis}
  \alphasmZ&=& 0.1145 \pm 0.0009 
  \, 
  \qquad
  \text{for}\, \Nf=5
  \, ,
\end{eqnarray}
while the full ABMP16 analysis (also including LHC data on top-quark
hadroproduction) finds the value 
\begin{eqnarray}
\label{eq:asmt}
\alphasmZ&=& 0.1147 \pm 0.0008 
  \, 
  \qquad
  \text{for}\, \Nf=5
  \, .
\end{eqnarray}

Higher-order QCD corrections beyond NNLO become important whenever high precision is needed 
for benchmark processes at the LHC and for the novel accurate DIS measurements
expected at the future Electron Ion Collider (EIC).
Then, determinations of PDFs and $\alphas$ at N$^3$LO accuracy requires
the calculation of the N$^3$LO corrections in Eqs.~(\ref{eq:moch-Pexp}) and (\ref{eq:moch-Cexp}).
Those for the DIS coefficient functions are already available~\cite{Vermaseren:2005qc,Moch:2007gx,Moch:2007rq,Moch:2008fj},
including effects of massive quarks~\cite{Ablinger:2014vwa}.
Work on the four-loop splitting functions in Eq.~(\ref{eq:moch-Pexp}) to ensure 
QCD evolution equations at N$^3$LO accuracy is ongoing~\cite{Moch:2017uml,Moch:2018wjh,Moch:2021qrk}
and the gain in the theoretical accuracy in the solution of Eq.~(\ref{eq:moch-evol})
for the evolution of the PDFs is illustrated in Fig.~\ref{fig:moch-dqsgdlnmu}
for a number of Mellin moments ($N=2$ to $N=6$) available from Ref.~\cite{Moch:2021qrk}.
With the full N$^3$LO contributions to DIS the residual theoretical
uncertainty due the scale variation and the truncation of the perturbative
series will be limited to $1\%$ in the range of parton kinematics 
relevant for the current world DIS data and the EIC.
With respect to the theory error, inclusive DIS data offer the most precise
method to measure $\alphasmZ$.



\subsection{Strong coupling determination in the CT18 global analyses
\protect\footnote{A\lowercase{uthors:} J. H\lowercase{uston} (MSU, E\lowercase{ast} L\lowercase{ansing}), P. N\lowercase{adolsky} (S\lowercase{outhern} M\lowercase{ethodist} U\lowercase{niv.}, D\lowercase{allas}), K. X\lowercase{ie} (U\lowercase{niv.\ of} P\lowercase{ittsburgh}), \lowercase{on behalf of the} CTEQ-TEA C\lowercase{ollaboration}}}

The CT18 global QCD analysis~\cite{Hou:2019efy} updates the previous CTEQ-TEA PDFs, CT14~\cite{Dulat:2015mca} and CT14HERA2~\cite{Hou:2016nqm}, adding a variety of high-precision data from the Large Hadron Collider (LHC). New measurements of single-inclusive jet production, Drell--Yan lepton pairs, top-quark pairs, as well as high transverse momentum ($\pT$) Z bosons from ATLAS, CMS and LHCb, at the center-of-mass energies of 7 and 8 TeV, provide significant additional sensitivity to the PDF determination. A new family of PDF sets, CT18, CT18A, CT18X, and CT18Z, are released at both NLO and NNLO. The nominal PDF set, CT18, is recommended for general collider phenomenology studies, while CT18A includes an additional ATLAS 7 TeV W/Z precision data set~\cite{ATLAS:2016nqi}, which is found to be in tension with other data sets in the global fit. An alternative PDF set, CT18X, has adopted an $x$-dependent DIS scale, which captures the small-$x$ behavior at low $Q^2$ and improves the QCD description of HERA DIS data. The CT18Z PDF set includes the  features of CT18A and CT18X, in addition to having a slightly larger charm pole mass, $m_c=1.4$ GeV versus $m_c=1.3$ GeV. CT18Z maximizes the difference from CT18 PDFs, but preserves a similar goodness-of-fit. More details can be found in Ref.~\cite{Hou:2019efy}.\\

\begin{figure}[htbp!]
\centering
\includegraphics[width=0.51\textwidth]{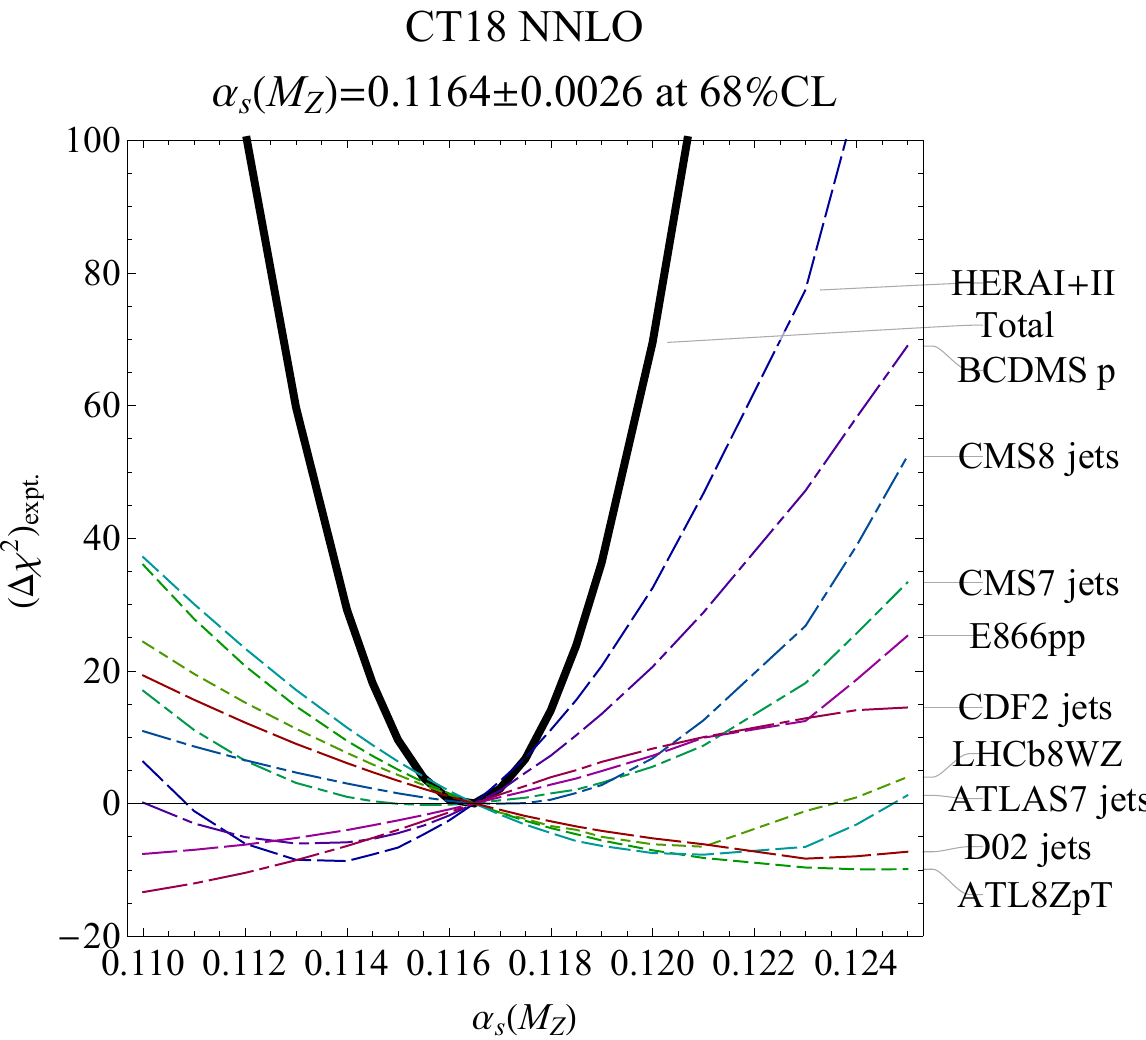}
\includegraphics[width=0.48\textwidth]{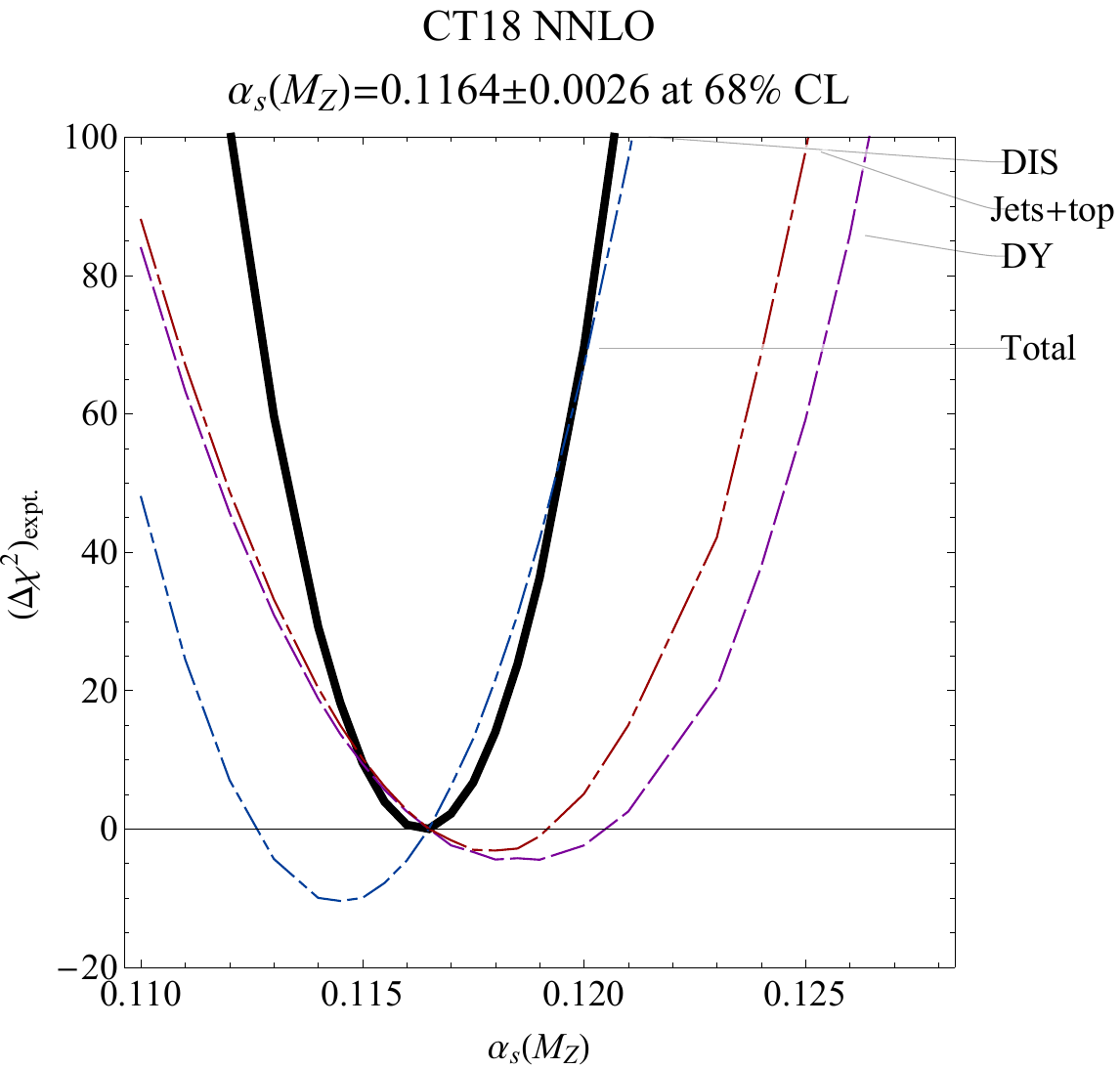}
\caption{The log-likelihood $\chi^2$ in the CT18 NNLO global PDF analysis as a function the strong coupling at scale $m_\mathrm{Z}$, plotted for the total fitted data set and the most sensitive experiments. Adapted from Ref.~\cite{Hou:2019efy}.}
\label{fig:asCT18}
\end{figure}

The final product -- the published error PDF sets such as CT18 -- takes the strong coupling at a scale $m_\mathrm{Z}$ as the world average value $\alphasmZ=0.118$~\cite{ParticleDataGroup:2018ovx}.
Alternative error PDF sets are produced with a series of fixed $\alphasmZ$ values for the use in the estimation of combined PDF+$\alphas$ uncertainties. As shown in CT10~\cite{Lai:2010nw}, a change in $\alphas$ is partially compensated by changes in the PDF parameters.  An $\alphas$ uncertainty can be defined, which quantifies the allowed variation of $\alphas$ when allowed to be freely varied in the PDF fit. 
In general, the $\alphasmZ$ sensitivity to a specific data set is introduced either through radiative corrections, such as in the Drell--Yan pair production, or through scaling violations, such as in DIS.

\begin{figure}[htbp!]
\centering
\includegraphics[width=0.8\textwidth]{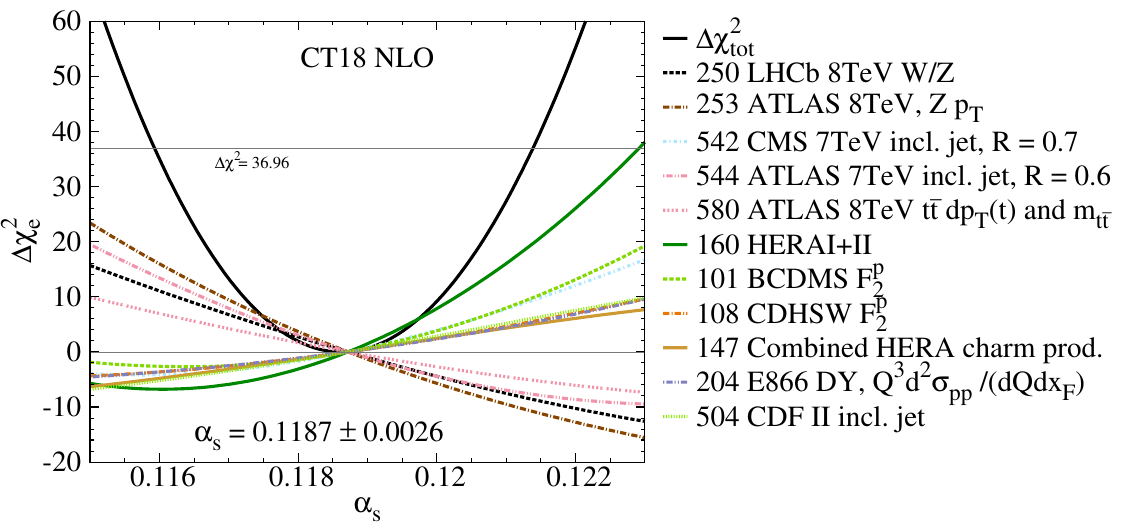}
\caption{Same as Fig.~\ref{fig:asCT18}, but now for the  $\alphasmZ$ dependence of $\chi^2$ in the CT18 NLO fit. Adapted from Ref.~\cite{Hou:2019efy}.}
\label{fig:asCT18NLO}
\end{figure}

\begin{figure}[htpb!]
\centering
\includegraphics[width=0.51\textwidth]{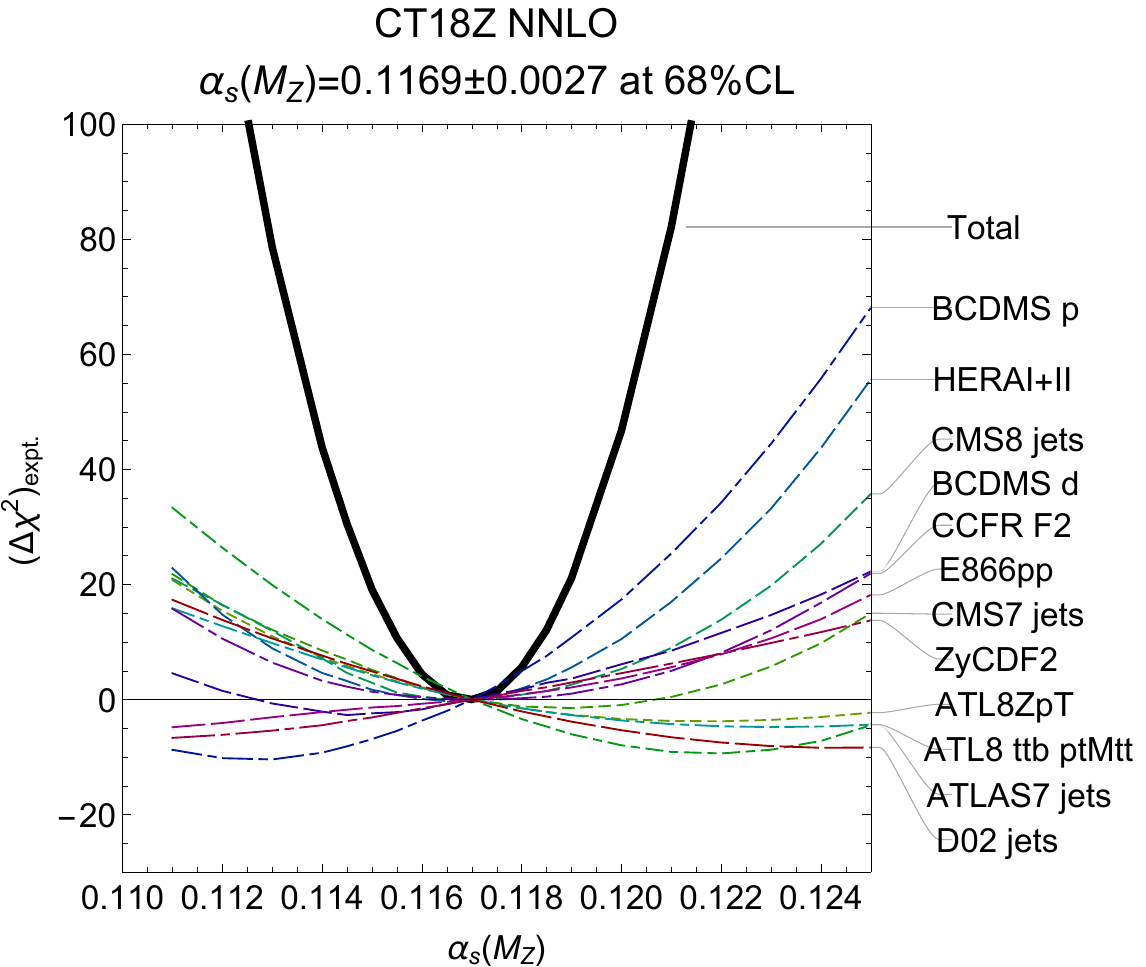}
\includegraphics[width=0.48\textwidth]{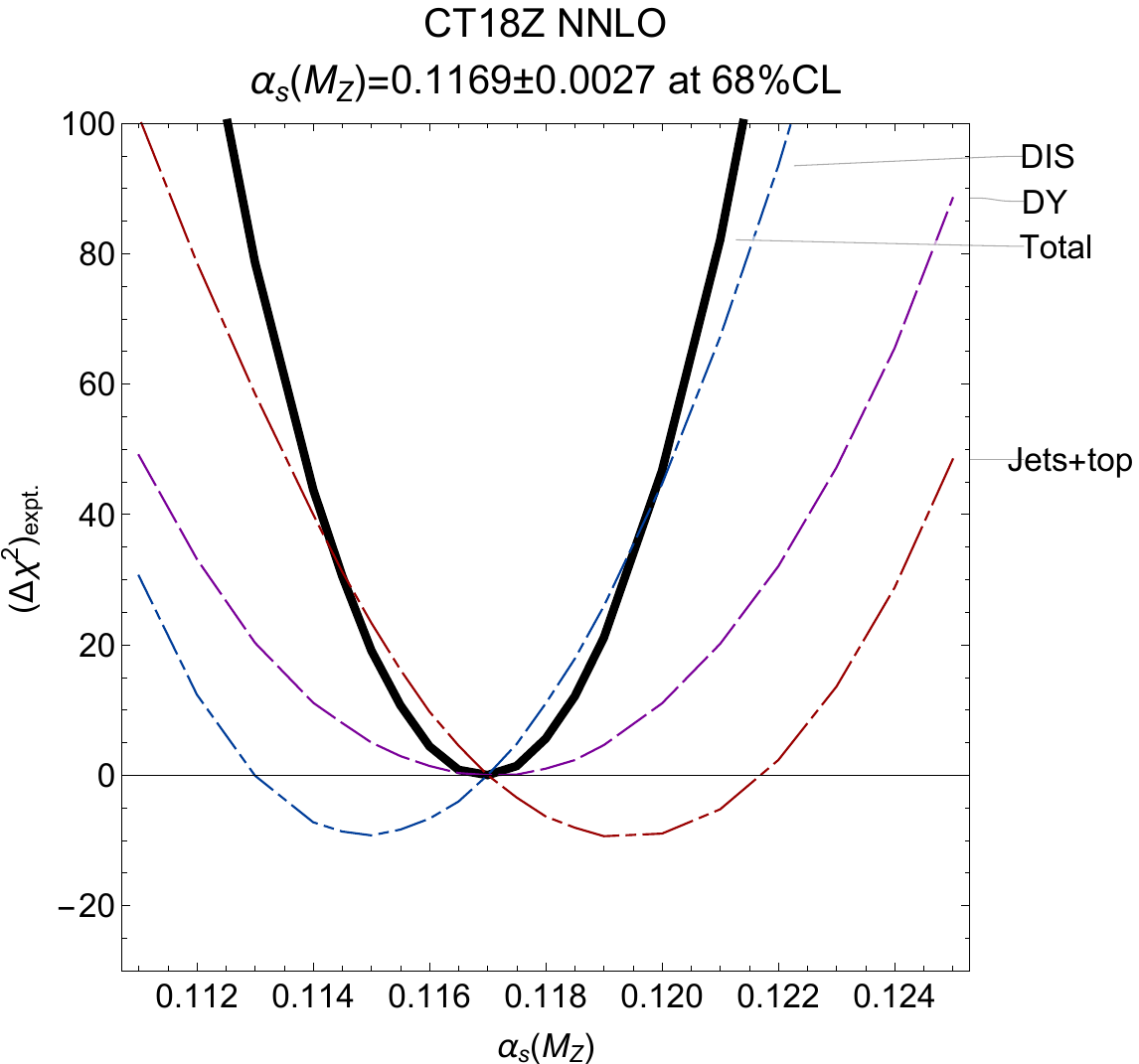}
\caption{Same as Fig.~\ref{fig:asCT18}, but for the CT18Z NNLO fit. Adapted from Ref.~\cite{Hou:2019efy}.}
\label{fig:asCT18Z}
\end{figure}

Simultaneous fits of the $\alphas$ and PDFs were also performed during the CT18 study. An optimal way to explore experimental constraints on $\alphasmZ$ is to examine the corresponding $\chi^2$ variations with the Lagrangian multiplier (LM) technique~\cite{Stump:2001gu}. In Figs.~\ref{fig:asCT18},~\ref{fig:asCT18NLO} and \ref{fig:asCT18Z}, we adapt this technique to plot a a series of curves for $\Delta\chi^2=\chi^2(\alphas)-\chi^2(\alpha_{S,0})$ for CT18 NNLO, CT18NLO, and CT18Z NNLO, respectively, where $\alpha_{S,0}$ corresponds to the global $\chi^2$ minimum in the PDF+$\alphas$ fit. 
The black solid curves are for the total $\Delta \chi^2$, and the other curves are for $\Delta \chi^2$ from the individual experiments with the highest sensitivities to $\alphas$. From the figures, it is clear that the data sets and even groups of data sets have different preferences to $\alphasmZ$ in terms of both the central values and uncertainties. 
The spread of the pulls on $\alphas$ by the data sets is broader than it would be normally expected from random fluctuations of data around theory, suggesting that a more conservative prescription for the estimation of the $\alphas$ uncertainty is necessary than the straightforward averaging over the data sets \cite{Kovarik:2019xvh}.

From the figures, we see that the HERA I and II combined DIS data~\cite{Abramowicz:2015mha} provide the strongest constraints on $\alphasmZ$ in CT18 at both NLO and NNLO. The BCDMS proton data~\cite{Benvenuti:1989rh} play a slightly stronger role in the CT18Z PDF fit.
These two DIS data sets prefer a lower value of $\alphasmZ$ than the world average,  with a value about $0.114\sim0.116$, but with larger uncertainties. On the other hand, we see that the hadron collider data, such as inclusive jet, top-quark pairs, and Drell--Yan production, pull the $\alphasmZ$  value higher, with a preferred value of $0.117\sim0.119$ for the CT18 NNLO case. The strongest pull comes ATLAS 8 TeV Z $\pT$ data~\cite{ATLAS:2015iiu}, followed by the LHCb 8 TeV W/Z measurement~\cite{LHCb:2015mad}.
We also see significant pulls from inclusive jet production at ATLAS 7 TeV~\cite{ATLAS:2014riz} and D0 Run II~\cite{D0:2008nou}, as well as the measurements of the $\pT^{t}$ and $m_{t\bar{t}}$ distributions in top-quark pair production at ATLAS 8 TeV~\cite{ATLAS:2015lsn}. In contrast to the ATLAS case, the CMS measurements of inclusive jet production at both 7 TeV~\cite{CMS:2014nvq} and 8 TeV~\cite{CMS:2016lna} prefer a smaller $\alphasmZ$, reflecting a tension between these measurements from the two experiments. 

Within the 68\% probability level, the global average values are
\begin{equation}
\alphasmZ_{\rm NNLO}=0.1164\pm0.0026,\quad \alphasmZ_{\rm NLO}=0.1187\pm0.0026,
\end{equation}
while CT18Z NNLO gives $0.1169\pm0.0027$,  all consistent with the world average~\cite{ParticleDataGroup:2018ovx}. In comparison with the CT14 global fit~\cite{Dulat:2015mca}, the CT18 data sets prefer a larger central $\alphasmZ$ with a marginally smaller uncertainty.
Note that the uncertainties quoted here are larger than those in the contemporary analyses done by other two groups, MSHT and NNPDF. This is primarily a result of a more conservative prescription for the uncertainty adopted in CT18, despite similar constraining power of the experiments included by the three groups. The CT18 prescription reflects in part some inconsistency between the fitted data sets, as discussed above. In the presence of such inconsistencies due to unidentified systematic effects, which would lead to enlarged variations in $\alphas$ when the selection of experimental data sets is varied, the final uncertainty should be generally increased \cite{Barlow:2002yb}. In addition, the CT18 uncertainty incorporates outcomes from the fits with comparable $\chi^2$ that are obtained with more than 250 alternative functional forms of the PDFs and with alternative QCD scale choices in inclusive jet and high-$\pT$ production. To achieve this, the $\alphas$ and PDF uncertainties are estimated using the tolerance chosen so as to cover a large number of candidate fits that are made with such alternative choices and pass the goodness-of-fit criteria accepted for the final fit~\cite{Kovarik:2019xvh, Guzzi:2021fre}.\\

\noindent \textit{Acknowledgments}--- The work of K.X. is supported by the U.S. Department of Energy, under grant No.\ DE-SC0007914, U.S. National Science Foundation
under Grant No.\ PHY-2112829, and in part by the PITT PACC. The work of P.N. is supported by U.S. Department of Energy under Grant
No.\ DE-SC0010129.

\subsection{NNLO \texorpdfstring{$\alphasmZ$}{alphasmZ} determination from HERA inclusive and jet data
\protect\footnote{A\lowercase{uthors:} A. M. C\lowercase{ooper}-S\lowercase{arkar} (U\lowercase{niv. of} O\lowercase{xford}), \lowercase{on behalf of} H1 \lowercase{and} ZEUS \lowercase{collaborations}}}


Deep inelastic scattering (DIS) of electrons
on protons, e-p, at centre-of-mass energies of up to $\sqrt{s} \approx 320\,$GeV
at HERA has been central to the exploration
of proton structure and quark--gluon dynamics as
described by perturbative Quantum Chromodynamics (pQCD). 
The combination of H1 and ZEUS data on inclusive e-p scattering
and the subsequent pQCD analysis, introducing the family of 
parton density functions (PDFs) known as HERAPDF2.0~\cite{Abramowicz:2015mha},
was a milestone for the exploitation of the HERA data.
The work presented here represents a completion of the 
HERAPDF2.0 family with a fit at NNLO to
HERA combined inclusive data and jet production data
published separately by the
ZEUS and H1 collaborations.
This was not possible at the time of the original introduction of HERAPDF2.0
because a treatment at NNLO of jet production in 
e-p scattering was not available then.\\

The name HERAPDF stands for a pQCD analysis within the
DGLAP formalism~\cite{Gribov:1972ri,Altarelli:1977zs,Dokshitzer:1977sg}, where predictions from pQCD are fitted to data.
These predictions
are obtained by solving the DGLAP evolution 
equations at NNLO in the $\msbar$ scheme. 
The inclusive and dijet production data which were already
used for HERAPDF2.0Jets NLO were again used for the analysis presented here.
A new data set~\cite{H1:2016goa} published by the 
H1~collaboration on jet production in low~$Q^2$ events, 
where $Q^2$ is the four-momentum-transfer squared,
was added as input to the fits.

The fits presented here were done in the same way 
as for all other members of the HERAPDF2.0 family, for full details see~\cite{ZEUS:2021sqd} and references therein.
All parameter settings were the same as for the HERAPDF2.0Jets NLO fit, with the exception of the heavy quark masses $m_c$ and $m_b$ and their uncertainty ranges, which were reevaluated using the recently published HERA combined charm and beauty data~\cite{H1:2018flt}. The default minimum $Q^2$ of data entering the fit is $Q^2 = 3.5$~GeV$^2$ and the starting scale for DGLAP evolution is $Q^2_0 = 1.9$~GeV$^2$. Note that all HERA data are at very large $W$ so that higher twist effects at small $Q^2$ and large $x$ are not relevant.

As for previous HERAPDF analyses, 
model and parametrization uncertainties were evaluated. For the present 
analysis the uncertainties on the hadronization corrections for the jet data 
were included as systematic uncertainties, $50\%$ correlated and $50\%$ 
uncorrelated, together with the experimental systematic uncertainties. 

The jet data were included in the fits at NNLO  by calculating
predictions for the jet cross sections
 using the NNLOJet~\cite{Britzger:2019kkb} extension of
NLOjet++ interfaced to the  Applfast framework 
in order to achieve the speed necessary for iterative PDF fits. The predictions were supplied with uncertainties which were also input to the fit as $50\%$ correlated and $50\%$ uncorrelated systematic uncertainties.

The NNLO analysis of the jet data was applied
 to a slightly reduced phase space compared to 
HERAPDF2.0NLOJets. All data points 
with $\sqrt{\langle \pT^2 \rangle +Q^2} \le 10$\,GeV were excluded to keep the level of scale uncertainties in the predictions for the data points to $\lesssim 10\%$. 
Six data points, the lowest $\langle \pT \rangle$ 
bin for each $Q^2$ region, were excluded from the ZEUS dijet
data set because predictions for these points were
not fully NNLO.
The trijet data which were used as input
to HERAPDF2.0NLOJets had to be excluded as their treatment at NNLO is not 
available. In addition six extra data points for H1 inclusive jet data at 
high $Q^2$ but low $\pT$, which were published more recently~\cite{H1:2016goa}, were included.

The choice of scales was also adjusted for the NNLO analysis.
At NLO, the factorization scale was chosen as 
$\mu_\mathrm{f}^2 = Q^2$,
while the renormalization scale was linked to the transverse
momenta, $\pT$, of the jets by $\mu_\mathrm{r}^2 = (Q^2 + \pT^2)/2$.
For the NNLO analysis, $\mu_\mathrm{f}^2 =\mu_\mathrm{r}^2= Q^2 + \pT^2$,
was chosen for both factorization and renormalization scales.

Jet production data are essential for the determination of
the strong coupling constant, $\alphasmZ$.
In pQCD fits to inclusive DIS data alone the value of $\alphasmZ$ is strongly correlated to
the shape of the gluon PDF.
Data on jet production cross sections provide an independent constraint
on the gluon distribution since inclusive
jet and dijet production are directly sensitive to $\alphasmZ$. 
Thus such data allow for an accurate simultaneous determination of $\alphasmZ$
and the PDFs.

The HERAPDF2.0Jets NNLO fit with free $\asmz$ and $Q^2_\mathrm{min} = 3.5$~GeV$^2$ gives a value of
\begin{eqnarray}
\nonumber
\alphasmZ =0.1156 \pm 0.0011\,\mathrm{(exp)} ^{+0.0001}_{-0.0002}\,\,\mathrm{(model/parameterization)} \pm 0.0029 \,\mathrm{(scale)}.
\end{eqnarray}
Note that the the label experimental uncertainty covers the full experimental uncertainties of the 
simultaneous PDF and free $\alphasmZ$ fit, which thus includes contributions from the PDF parametrization, the uncertainty on the hadronization corrections and the small uncertainties on the theoretical predictions. Additional model uncertainties come from variation of the input assumptions such as the the value of $Q^2_0$, the starting scale for QCD evolution, the value of the minimum $Q^2$ of data entering the fit and the values of heavy quark masses. Additional parametrization uncertainties come from the consideration of fits which have additional parameters that do not change the $\chi^2$ of the fit significantly, but which can sometimes alter PDF shape. A further cross-check on the parametrization comes from modifying the gluon parametrization such that it must remain positive definite even at very low-$x$, $x < 10^{-4}$, and $Q^2$ below $2$~GeV$^2$. The largest uncertainty comes from the scale uncertainty, as discussed later.

The HERAPDF2.0Jets NNLO fit with free $\alphasmZ$  uses
1363 data points and has a goodness-of-fit per degree of freedom of
$\chi^2/n_\text{dof} = 1614/1348 = 1.197$. This can be compared
to the $\chi^2/n_\text{dof}= 1363/1131 = 1.205$ for HERAPDF2.0 NNLO based on
inclusive data only.
The similarity of the $\chi^2/n_\text{dof}$ values indicates that 
the data on jet production do not introduce any tension. Data/fit comparisons are shown in Ref.~\cite{ZEUS:2021sqd}.

\begin{figure}
  \centering
  \includegraphics[width=0.55\textwidth]{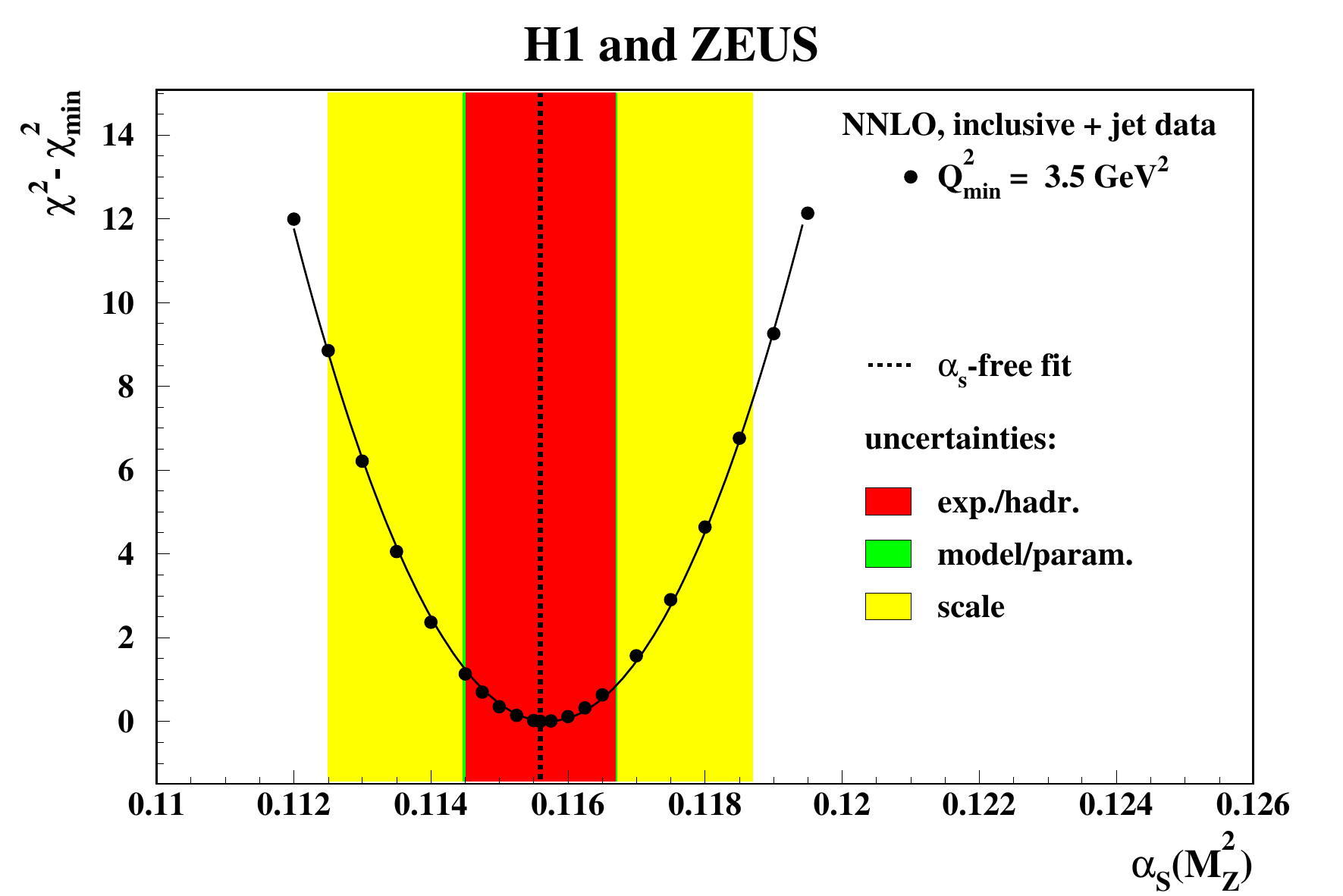}
\caption {Distribution of $\Delta \chi^2 = \chi^2 - \chi^2_\mathrm{min}$ vs.\ $\alphasmZ$ for HERAPDF2.0NNLOJets fits with fixed $\alphasmZ$ values. Experimental, model/parametrization, and scale uncertainties are illustrated. (Figure from Ref.~\cite{ZEUS:2021sqd}).
\label{fig:alphasscan}}
\end{figure}

A scan of the fit $\chi^2$ for fits with varying $\alphasmZ$ shown in Fig.~\ref{fig:alphasscan} and confirms
the value of $\alphasmZ$ and its experimental uncertainty found in the simultaneous $\alphasmZ$ and PDF fit.
The model, parameterization and scale uncertainties are also shown in the figure. The PDFs associated with this NNLO analysis are described in Ref.~\cite{ZEUS:2021sqd}, the input of the jet data reduces the gluon PDF uncertainties.

The question whether data with relatively low $Q^2$ bias the
determination of $\alphasmZ$ arises in the context of the
HERA data analysis for which low $Q^2$ is also low $x$. A treatment beyond DGLAP
may be necessary because of low-$x$ higher-twist terms, $\ln(1/x)$ terms or even parton saturation. To check for such bias
the minimum $Q^2$ entering the fit was varied to $Q^2_\mathrm{min} = 10$ and $20$~GeV$^2$. Thus the HERAPDF analysis considers the possible impact of nonperturbative and beyond DGLAP effects on the result for $\alphasmZ$. These variations do not alter the value of $\asmz$ obtained significantly, the value of $\alphasmZ$ obtained for $Q^2_\mathrm{min} = 10$~GeV$^2$ is
\begin{eqnarray}
\nonumber
\alphasmZ =0.1156 \pm 0.0011\,\mathrm{(exp)} \pm 0.0002\,\mathrm{(model/parameterization)} \pm 0.0021 \,\mathrm{(scale)}.
\end{eqnarray}
The scale uncertainty is reduced with this harder cut on $Q^2$ but the largest uncertainty still comes from the scale uncertainty.

The scale uncertainty
was evaluated by varying the renormalization and factorization 
scales by a factor of two,
both separately and simultaneously (7-point variation), 
and taking the maximal positive and negative deviations.
Scale uncertainties were assumed to be $100\%$ correlated between bins and data sets.

A strong motivation to determine
$\alphasmZ$ at NNLO was the hope to substantially reduce scale uncertainties from those determined at NLO.
However, for our previous NLO result we had treated scale uncertainties as 50\,\% correlated and 50\,\% uncorrelated.
If we repeat this treatment then the scale uncertainty for the present NNLO analysis (with $Q^2_\mathrm{min}=3.5$~GeV$^2$, comparable to the NLO analysis) would be $\pm 0.0022$, very significantly lower
than the $+0.0037,-0.0030$ previously observed for the HERAPDF2.0NLOJets
analysis. The absolute value of $\alphasmZ$ at NNLO $\alphasmZ = 0.1156$ is also lower than that observed at NLO $\alphasmZ=0.1183$. 
However, the analyses were done at different scales and with somewhat different data sets. 
In fact, if these choices were harmonized, the difference in the values of $\alphasmZ$ would be even greater: 
at NLO $\alphasmZ = 0.1186\pm 0.0014$\,(exp) and at NNLO $\alphasmZ =0.1144 \pm 0.0013$\,(exp), where the scale choices are both $Q^2 + \pT^2$, the cuts on the data sets for both orders are the harder cuts introduced for the present analysis and the H1 low $Q^2$ jet data set~\cite{H1:2016goa} is excluded (this choice is made because these data cannot be well fitted at NNLO). The main reason for the decrease in the central value at NNLO is exclusion of these low-$Q^2$ H1 inclusive and dijet data.

It is clear that the main limiting factor today is the theoretical uncertainty, which we estimate from the scale uncertainty. The decrease in scale uncertainty between NLO and NNLO gives some hope for an increase in precision at N$^3$LO. Once theoretical uncertainties are reduced, the experimental uncertainty will need to be reduced. One may hope for progress from analyses at future e-p colliders such as the EIC/LHeC/FCC-eh. Experimental accuracy on $\alphasmZ \sim 0.0001$ is projected at the LHeC~\cite{LHeC:2020van}.\\

\noindent \textit{Acknowledgments}--- AMC-S would like to thank the Leverhulme Trust.

\subsection{Determination of the strong coupling \texorpdfstring{$\alphasmZ$}{alphasmZ} in the MSHT20 NNLO PDF fit
\protect\footnote{A\lowercase{uthors:} T. C\lowercase{ridge} (UCL, L\lowercase{ondon}) \lowercase{on behalf of the} MSHT C\lowercase{ollaboration}}}


MSHT20~\cite{Bailey:2020ooq} represented a significant step forward in the global determination of PDFs within the MSHT collaboration (previously MRST/MSTW/MMHT), with more data of greater precision across more channels and more differential in nature now incorporated alongside the inclusion of full NNLO QCD theoretical predictions (and NLO EW corrections where relevant~\cite{Cridge:2021pxm}). This has been further complemented by progress on the methodological side, including in particular the extension of the PDF parameterization. The overall result was a significant improvement in our knowledge of the PDFs, including the central values and a general reduction of the PDF uncertainties. Within this context we undertook a follow-up study~\cite{Cridge:2021qfd} examining the effects of varying the strong coupling and heavy quark masses within the global fit. This updated previous work with MMHT14~\cite{Harland-Lang:2015nxa}. As a result of this new analysis we determined the preferred value of $\asmz$ and its associated uncertainty and we report on this in this brief review.\\

\begin{figure}[htpb!]
\centering
\includegraphics[width=0.49\textwidth]{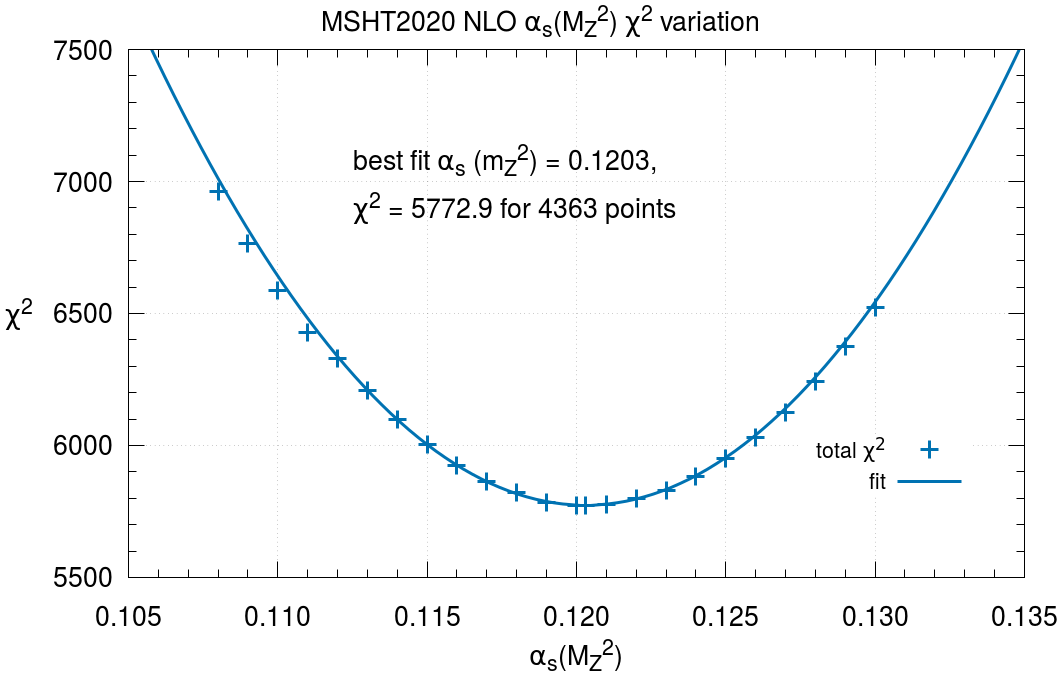}
\includegraphics[width=0.49\textwidth]{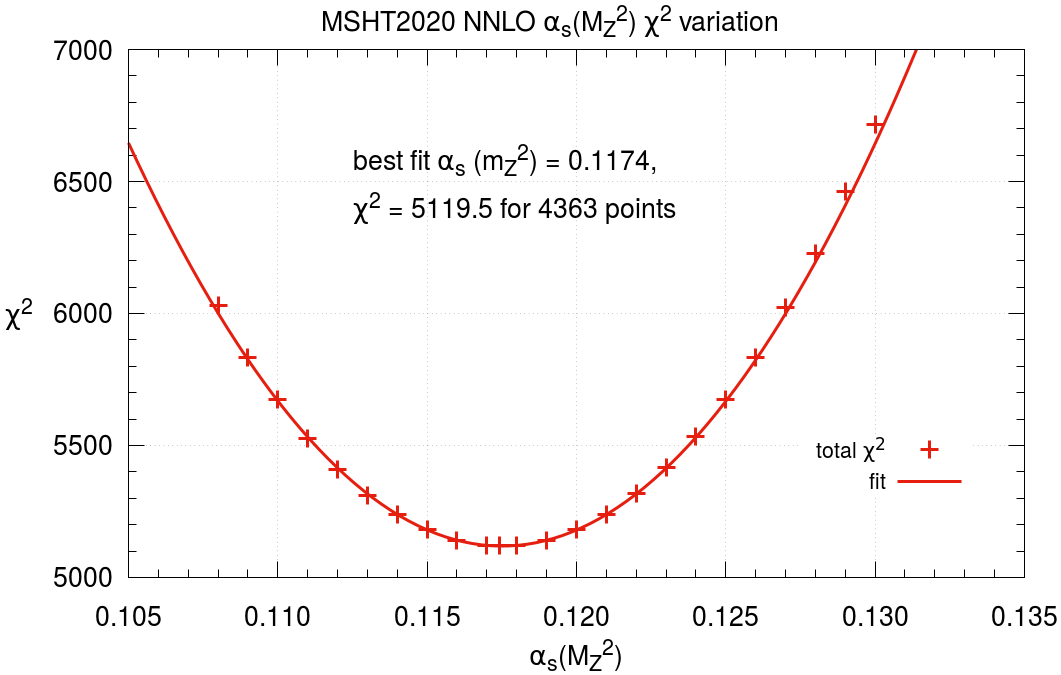}
\caption{The plots show the total $\chi^2$ as a function of $\asmz$ for NLO (left) and NNLO (right) MSHT20 fits respectively. Figure originally from \cite{Cridge:2021qfd}.
}
\label{fig:alphas_msht_global}
\end{figure}

The default PDFs within MSHT are provided at $\asmz = 0.118$ at NNLO in order to provide a common value between different PDF fitting groups which is also consistent with the world average value of $\asmz = 0.1179 \pm 0.0009$~\cite{Zyla:2020zbs}. However, $\asmz$ can also be left as a free parameter and fit alongside the PDFs, we are then able to utilize the global nature of the PDF fits to extract $\asmz$. Performing the fit with $\asmz$ free within MSHT20 at NLO and NNLO obtains the best fit values of 0.1203 and 0.1174 respectively, with the excellent quadratic $\chi^2$ profiles shown in Fig.~\ref{fig:alphas_msht_global}, demonstrating the clear sensitivity at the global fit level.

\begin{figure}[htpb!]
\centering
\includegraphics[width=0.95\textwidth]{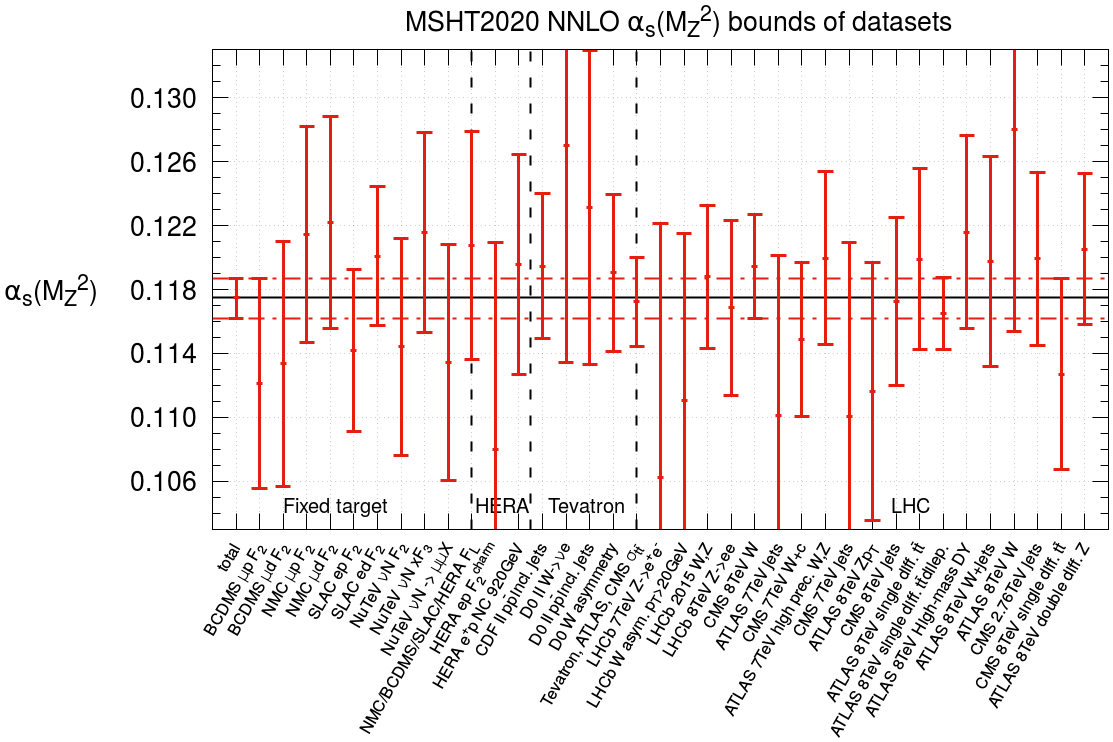}
\caption{The plot shows the value of $\asmz$ corresponding to the best fit, together
with the upper and lower $1\sigma$ constraints on $\asmz$ from the more constraining data sets at NNLO. The overall upper and lower bounds taken are given by the horizontal dashed red lines. Figure originally from \cite{Cridge:2021qfd}.
}
\label{fig:alphas_msht_bounds_nnlo}
\end{figure}

Nonetheless, performing such a fit simply extracts the best fit value of the strong coupling, in reality we wish to determine a value of the uncertainty arising on this value from the PDF fit. In order to do this, we utilize the dynamical tolerance method~\cite{Bailey:2020ooq,Harland-Lang:2014zoa,Martin:2009iq} which enables the determination of the 68\% confidence level interval for the uncertainties on $\asmz$ by obtaining a ``tolerance'' $\Delta \chi^2$ increase from the best fit value for a particular dataset. Once this is exceeded, it sets the upper and lower bound on a dataset-by-dataset level for the strong coupling. It should be noted this is the same method applied to the determination of the PDF uncertainties. Performing this across the vast array of datasets in MSHT20 one obtains bounds from each dataset on $\asmz$, the most relevant of which are presented in Fig.~\ref{fig:alphas_msht_bounds_nnlo} for the MSHT20 NNLO fit, further details can be found in~\cite{Cridge:2021qfd}.

As can be observed, different datasets may prefer significantly different values of $\asmz$ within the context of the global PDF fit and correspondingly offer different bounds. Nonetheless, several general comments can be made. The fixed target and DIS data, and in particular the BCDMS and SLAC proton data, have a preference for lower values of the strong coupling (albeit not universally) in order to slow the fall of the structure function with $Q^2$. These represent relatively clean means of determining $\asmz$ through their high $x$ nature meaning they are dominantly non-singlet. On the other hand, the HERA deep inelastic scattering data has relatively limited sensitivity compared to its large number of points in the fit, this is a reflection of its significant pull on the central values of the PDFs, meaning it naturally sits near the minimum of the profiles. The inclusive jets and Z $\pT$ data from the LHC have direct sensitivity to $\asmz$ and are seen to generally prefer lower values of the strong coupling within the fit. This is also observed for the top data, certainly at NNLO, however the differential data are dropped from the determination of the global $\asmz$ bounds due to the fixed top mass available in the theoretical prediction grids utilized, which restricts their utility for the uncertainty determination. Finally, whilst the high precision W, Z data from the LHC have only indirect sensitivity to the strong coupling, these datasets are often still able to provide bounds on $\asmz$, and are generally seen to favour higher values of $\asmz$, albeit again not universally.

Examining these bounds on $\asmz$ across the whole fit, we can then extract the global upper and lower bounds on our strong coupling determinations. At NNLO the BCDMS proton data provides the tightest upper bound, reflecting its preference for a lower $\asmz$, its $\Delta \chi^2$ rises rapidly as the strong coupling is increased and therefore bounds the upward variation of $\asmz$. On the other hand, the CMS 8~TeV W data provides the most stringent lower bound, fitting with the general trend of the W, Z data favouring larger $\asmz$ values. Nonetheless, several other datasets would provide only slightly weaker bounds including the SLAC proton data in the upwards direction and the ATLAS 8~TeV Z and high mass Drell--Yan data in the downwards direction. Overall, a symmetrized uncertainty on the strong coupling is obtained of 0.0013, corresponding to a change of $\Delta \chi^2 _{\rm global} = 17$, a similar change in fit quality to that observed for changing the PDFs by their uncertainty in an eigenvector direction. A similar analysis can be performed at NLO, for which the BCDMS proton data again provides the tightest upper bound (followed by the ATLAS 8~TeV Z $\pT$ and ATLAS 7~TeV inclusive jets data, demonstrating their preference for lower $\asmz$) whilst the LHCb 7 and 8~TeV W,Z data provides the lower bound (closely followed by the SLAC deuterium data). Overall the best fit values and uncertainties on $\asmz$ within MSHT at NLO and NNLO respectively are determined to be as follows and the NNLO result is consistent with the PDG world average\footnote{It should be noted that the uncertainties given here are experimental in nature, propagated through the PDF fit, the theoretical uncertainties are not included within the global PDF fit at this stage.}:
\begin{equation*}
\asmz_{\rm NNLO} = 0.1174 \pm 0.0013, ~~~~~~~~~~~~~ \asmz_{\rm NLO} = 0.1203 \pm 0.0015\,.
\end{equation*}

One of the benefits of performing such analyses within the context of global PDF fits, is the incorporation of correlations between the PDFs and the strong coupling, this is investigated in~\cite{Cridge:2021qfd} and the standard method of treating the $\asmz$ uncertainty as an additional eigenvector in the fit and combining it in quadrature with the PDF uncertainty to obtain the total PDF + $\asmz$ uncertainty is demonstrated. Given these correlations, there are both direct contributions to the total uncertainty on an observable from the strong coupling change, but also indirect contributions from the associated correlated changes in the PDFs. The latter can partially cancel the direct $\asmz$ contribution to the uncertainty, for example by probing the gluon below $x \lesssim 0.1$ where it is anticorrelated with the value of the strong coupling. Studies of the heavy quark masses and their correlation with $\asmz$ were also performed for $m_{\rm charm}$ and demonstrated only limited correlation, with the value of $\asmz$ extracted rising slightly with the charm quark mass utilized.

The various PDF sets determined within this work are provided for public use on the MSHT website and the LHAPDF repository \cite{Buckley:2014ana}. In particular, of relevance for this review are the sets provided at NLO and NNLO at fixed $\asmz$ in steps of 0.001 for the range $0.108 \leq \asmz \leq 0.130$, moreover the $\Delta \chi^2_{\rm global}$ in the PDF global fit at each fixed different value of $\asmz$ in these sets are provided~\cite{Cridge:2021qfd}.

Looking into the future, there are several developments which would facilitate improvements in $\asmz$ extractions from global PDF fits of deep inelastic scattering and collider data. Firstly, there are questions of the means to incorporate estimates of the theoretical uncertainties into the $\asmz$ extractions in global PDF fits (beyond relating it to (half) the difference of NLO and NNLO), this is being actively investigated by the various groups and progress is to be expected on this front. Meanwhile, the current uncertainties are limited by the experimental data available, as well as perhaps our understanding of its treatment. The measurements from the LHC in recent years of gluon-sensitive processes, such as inclusive jets and top processes have enabled improvements of our constraints on the PDFs and, in some cases also of $\asmz$. The rapid availability of grids for theoretical predictions for inclusion in PDFs for $\asmz$ sensitive processes, such as triple differential jet data, would enable further progress here and this could be complemented by the inclusion of deep inelastic scattering jet data from HERA. The ability to account for the top quark mass dependence in the top differential data would also certainly represent a step forward but is reliant on the provision of theoretical grids at different fixed values, we hope these can be provided in the short term. There is also the possibility of utilising ratios to provide particular sensitivity to $\asmz$, \eg\ the ratio of the inclusive 3- and 2-jet cross-sections~\cite{CMS:2013vbb}, in which the quantities involved are highly correlated (both experimentally and theoretically), these could further supplement the $\asmz$ sensitivity of the PDFs if included into the global fits. However, more analysis than simply providing the experimental data and corresponding theoretical predictions in a form suitable for inclusion in the PDFs is required. As precision has increased and statistical uncertainties have reduced, systematic uncertainties and their correlations have come to dominate in precision measurements. Correspondingly, increasingly global PDF fits have run into issues with such correlated systematic uncertainties and an improved understanding of such correlations would likely also benefit $\asmz$ extractions simultaneously to PDF extractions. More broadly, this study emphasizes the array of datasets with sensitivity to the strong coupling through the global PDF fits, with older fixed target data providing the tightest bounds in some places and newer LHC inclusive jet and high precision Drell--Yan data providing bounds in others. Therefore the continual evolution of the number and breadth of datasets included, along with their individual precisions, is important for further reducing these uncertainties, and we look forward to further data from the LHC and elsewhere in this regard.\\

\noindent \textit{Acknowledgments}--- T.C. would like to thank the Science and Technology Facilities Council (STFC) for support via grant awards ST/P000274/1 and ST/T000856/1.


\subsection{Strong coupling determinations from the NNPDF global analyses 
\protect\footnote{A\lowercase{uthors:} J. R\lowercase{ojo} (NIKHEF \& VU A\lowercase{msterdam}) \lowercase{on behalf of the} NNPDF C\lowercase{ollaboration}}}


\noindent
The model-independent approach based on machine
learning that is adopted by the
NNPDF Collaboration is specially suitable for
the simultaneous extraction of parton distributions
together with SM parameters, such as 
the strong coupling $\alphasmZ$, the
heavy quark masses, and the CKM
matrix elements~\cite{Ball:2009mk}
as well as BSM parameters such as Wilson coefficients in the Standard Model Effective Field Theory~\cite{Iranipour:2022iak,Greljo:2021kvv,Carrazza:2019sec}.
In this contribution, we briefly review 
studies of $\alphasmZ$ fits in the context
of the NNPDF global analyses, and outline ongoing
studies in the framework of
the NNPDF4.0 determination~\cite{Ball:2021leu,NNPDF:2021uiq}.\\

The first determination of the strong coupling
in the NNPDF framework was presented in~\cite{Lionetti:2011pw}
at NLO and updated in~\cite{Ball:2011us} to NNLO, in both cases based 
on the NNPDF2.1 dataset and fitting methodology~\cite{Ball:2011mu,Ball:2011uy}.
This dataset contained fixed-target
DIS and Drell--Yan data, HERA collider DIS, and
gauge boson and jet production measurements from
the Tevatron, and hence was devoid of LHC data.
A parabolic fit to the $\chi^2({\alphas})$ profile constructed
from a large number of 
uncorrelated Monte Carlo replicas led
to a value of the strong coupling
\begin{equation}
  \alphasmZ=0.1173 \pm 0.0007_{\rm pdf} \pm 0.0009_{\rm mhou}  \, ,
\end{equation}
at NNLO,  where the missing
higher order uncertainties (MHOUs) were estimated
using the Cacciari--Houdeau method~\cite{Bagnaschi:2014wea}.
The stability of this determination with respect
to the number of replicas was assessed, and the distribution of pulls from individual datasets
was found consistent with statistical expectations.

The extraction of $\alphasmZ$ within
the NNPDF framework was subsequently revisited in~\cite{Ball:2018iqk}, an analysis
based on the NNPDF3.1 settings~\cite{NNPDF:2017mvq}.
In order to fully account for the PDF- and data-induced
correlations on $\alphasmZ$, a new methodology denoted as the 
``correlated replica (c-replica) method'' was developed.
In this approach, one also produces variants of the global
fit for different $\alphasmZ$ values, but in a manner
that one has used correlated data- and PDF-replicas which are identical in all settings except for the value of the strong coupling itself.
The correlated replica method makes possible
a more robust estimate of the methodological
uncertainties associated to the $\alphasmZ$
extraction, at the price of a high computational cost
since the number of replicas that need to be generated
in this way is $\mathcal{O}(10^4)$.

\begin{figure}[h]
\centering
\includegraphics[width=0.99\textwidth]{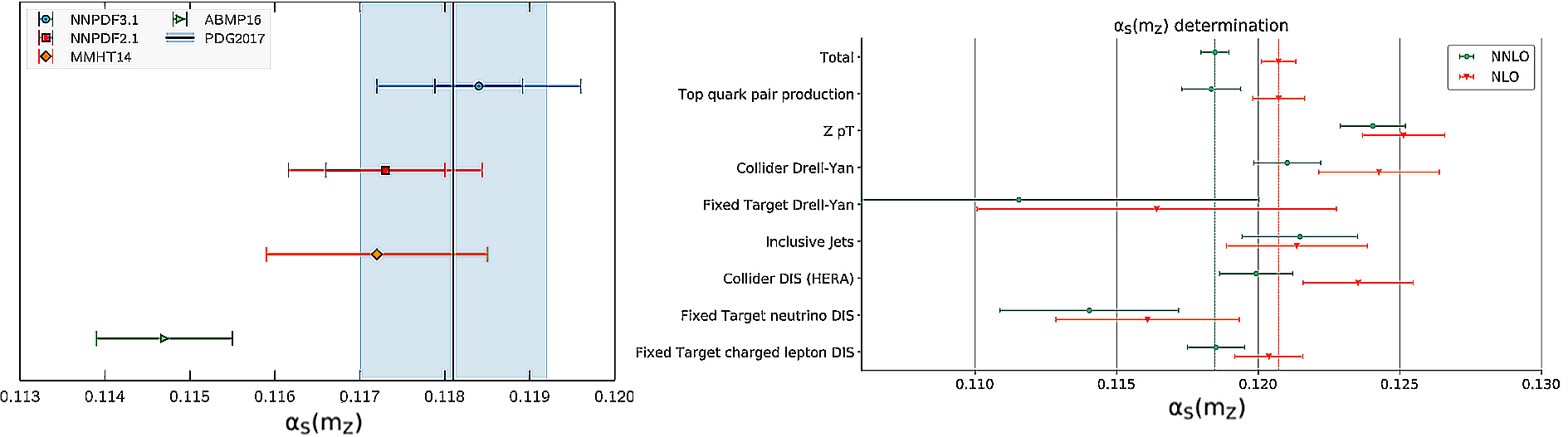}
\caption{Left: NNPDF2.1- and NNPDF3.1-based
NNLO determinations of $\alphasmZ$, compared with
the corresponding results from the MMHT14 and ABMP16 analyses
as well as to the 2017 PDG average. 
The thicker
error bars correspond to the PDF uncertainties, while
the thinner ones indicate the addition in quadrature of
PDF and MHO uncertainties.
Right: Central values and 68\% CL uncertainties
on $\alphasmZ$ in the NNPDF3.1 NLO and NNLO analyses,
comparing the baseline results with those obtained
from the partial $\chi^2_p(\alphas)$ restricted
to the specific subsets of processes indicated.
}
\label{fig:alphas_nnpdf_31fit}
\end{figure}

Figure~\ref{fig:alphas_nnpdf_31fit}~(left)
displays the  NNPDF2.1- and NNPDF3.1-based
NNLO determinations of $\alphasmZ$, compared with
the results from the corresponding MMHT14 and ABMP16 analyses
as well as to the 2017 PDG average. The thicker
error bars correspond to the PDF uncertainties, while
the thinner ones indicate the addition in quadrature of
PDF and MHO uncertainties.
The latter are estimated as half the difference
between the NNLO and NLO results.
The NNPDF3.1-based NNLO result is found to be
\begin{equation}
    \alphasmZ=0.1185 \pm 0.0005_{\rm pdf} \pm
    0.0001_{\rm meth} \pm 0.0011_{\rm mhou}  \, .
\end{equation}
One can observe the good consistency between
the two NNPDF determinations as well as among them and the MMHT14 the PDG average results.
It is also clear how, specially for the NNPDF3.1-based
analysis,
MHOUs are currently the limiting factor in strong coupling
determinations from global PDF fits.
The methodological uncertainties, associated
to \eg\ the finite number of replicas, are found
to be negligible in comparison
to both the data and the MHO uncertainties.
The role of the choice of figure of merit $\chi^2$
was also explored, showing that the frequently
used experimental definition leads to a downwards
D'Agostini bias which is avoided with
the $t_0$ definition~\cite{Ball:2009qv,Ball:2012wy}.

One issue that is frequently raised in the context
of PDF-based $\alphas$ extractions is that of the interplay
between the various types of processes that
enter the global fit, and whether or not their relative
pulls are overall consistent with the global
determination.
One possible 
strategy to gain some insight on this is provided
by repeating the $\alphasmZ$ extraction but now
using the partial, $\chi^2_p(\alphas)$, rather
than global, $\chi^2_{\rm tot}(\alphas)$, profiles.
Hence in this case $\alphasmZ$ is extracted
from a $\chi^2_p(\alphas)$ profile restricted
to the contribution of specific groups of
processes, \ie\ top quark or jet data.
It should be emphasized, that, as discussed in detail in~\cite{Forte:2020pyp}, it
would not be correct to interpret these results
as the value of $\alphas$ preferred by, say, top
or jet data: such value could
only be determined from a 
PDF fit based exclusively on this specific group of processes.
With this caveat, Fig.~\ref{fig:alphas_nnpdf_31fit}~(right) 
displays the central values and 68\% CL uncertainties
on $\alphasmZ$ in the NNPDF3.1 NLO and NNLO analyses,
comparing the baseline results with those obtained
from the partial $\chi^2_p(\alphas)$ restricted
to the specific subsets of processes listed.
One can observe that collider processes (jets, Z $\pT$, Drell--Yan) 
tend to favour a larger $\alphasmZ$ value
as compared to the global fit result, while
the opposite is found for most fixed-target
measurements.
The best-fit values from $\chi^2_p(\alphas)$ for
top quark pair production and neutral-current
DIS structure functions
are similar to the global fit result.

\begin{figure}
\centering
\includegraphics[width=0.49\textwidth]{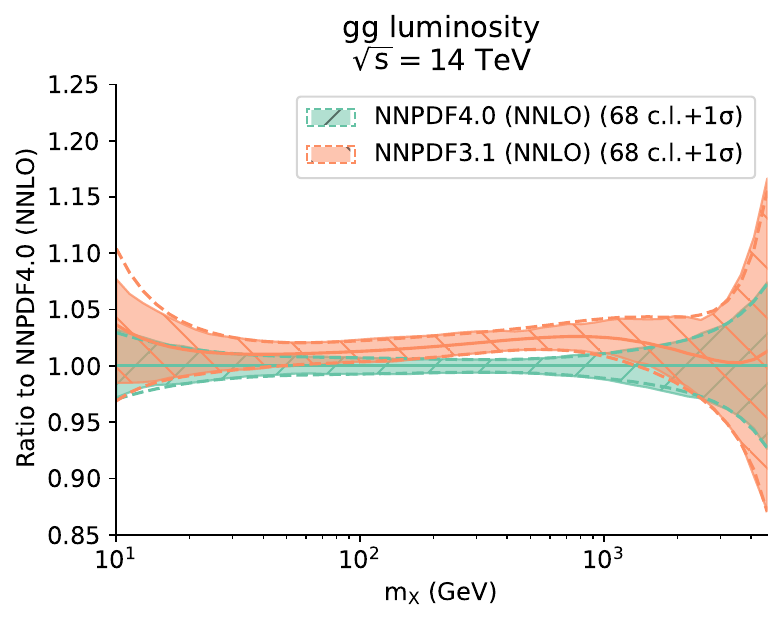}
\includegraphics[width=0.49\textwidth]{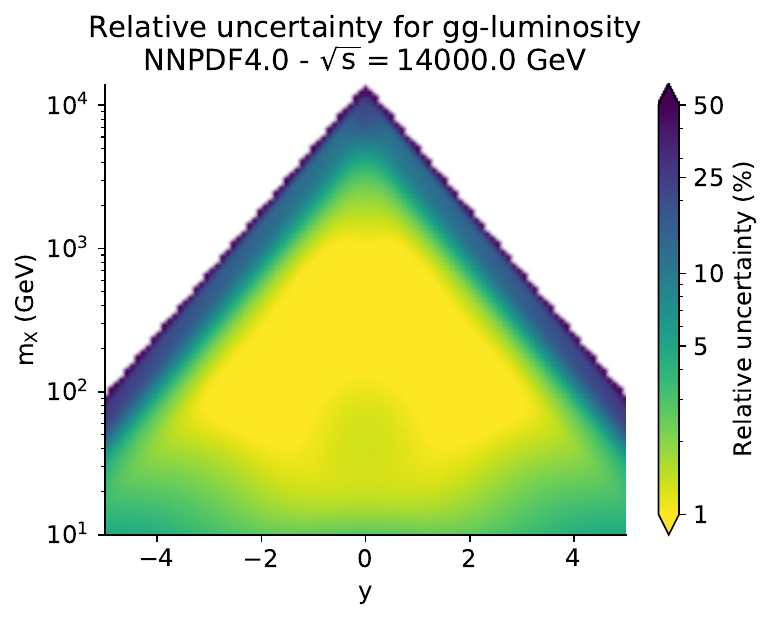}
\caption{Left: Comparison of the gluon-gluon
PDF luminosities at the LHC $\sqrt{s}=14$ TeV
as a function of the invariant mass $m_X$
between the NNPDF3.1 and NNPDF4.0 NNLO fits, highlighting the impact of the new gluon-sensitive measurements.
Right: Relative PDF uncertainties
in the double-differential gluon-gluon
luminosity as a function of $m_X$ and the rapidity $y$
in NNPDF4.0.
}
\label{fig:alphas_nnpdf_40fit}
\end{figure}

Work in progress aims to revisit
these $\alphasmZ$ extractions now based on the
recent NNPDF4.0~\cite{Ball:2021leu,NNPDF:2021uiq} analysis.
NNPDF4.0 benefits from a number of methodological
improvements as well as from the addition
of several gluon-sensitive measurements,
such as dijets~\cite{AbdulKhalek:2020jut}, direct photon production~\cite{Campbell:2018wfu},
and several new top quark production datasets, 
which in turn provide novel handles on the $\alphas$
value.
The left panel of Fig.~\ref{fig:alphas_nnpdf_40fit} compares the gluon-gluon
PDF luminosity at the LHC $\sqrt{s}=14$ TeV
between the NNPDF3.1 and NNPDF4.0 
NNLO fits, highlighting the impact of the new gluon-sensitive measurements both
in terms of uncertainty reduction as well as in the
shift in central values.
The right panel of Fig.~\ref{fig:alphas_nnpdf_40fit} displays
the percentage PDF uncertainties
in the double-differential gluon-gluon
luminosity as a function of $m_X$ and 
the rapidity $y$ in NNPDF4.0 NNLO, showing how
uncertainties are at the 1\% level for most
of the kinematic range accessible
at the LHC.
In order to fully exploit the uncertainty
reduction in the PDF determination observed in NNPDF4.0, 
it will be crucial
to assess carefully all other sources of
uncertainty that impact the $\alphas$ extraction
of global PDF fits, in particular those associated
with MHOUs, which will be tackled using the strategy laid
out in~\cite{NNPDF:2019ubu,NNPDF:2019vjt}.

Future work by the NNPDF collaboration will focus on extending 
the correlated replica method to the NNPDF4.0 fits and update 
the determination of the strong coupling constant. 
Another option that is being considered
is the recently developed {\sc\small SIMUnet} methodology~\cite{Iranipour:2022iak}, 
based on an extension of the NNPDF4.0 neural network architecture with an extra layer to simultaneously determine PDFs alongside an arbitrary number of SM
and BSM parameters from the direct joint minimization
of the $\chi^2$ figure of merit. 
The main advantage of {\sc\small SIMUnet} would be
to bypass
the need of running a very large number of MC replicas,
since the simultaneous extraction of $\alphasmZ$
together with the PDFs would only require
$N_{\rm rep}=\mathcal{O}(100)$ replicas as in standard
NNPDF fits.
Results for the determination of
$\alphasmZ$ from NNPDF4.0 using both
{\sc\small SIMUnet} and the correlated replica
method will be presented in the near future.


\subsection{Measurements of \texorpdfstring{$\alphas$}{alphas} from spin structure functions
\protect\footnote{A\lowercase{uthor:} A. D\lowercase{eur} (JLAB)}}

In this section, we discuss the extraction of $\alphas$ from nucleon spin structure data. 
The information is obtained from the cross-section asymmetry of Deep Inelastic scattering (DIS). 
The cross-section for inclusive lepton scattering is:
\begin{equation}
\frac{d\sigma}{d\Omega\ dx}=\sigma_ \text{Mott}\big[aF_1(x,Q^2)+bF_2(x,Q^2)+cg_1(x,Q^2)+dg_2(x,Q^2)\big],
\label{eq:incl-dis-xs}
\end{equation} 
where $\Omega$ is the solid angle, and $\sigma_\text{Mott}$, $a$, $b$, $c$ and $d$  are kinematic factors (see \eg~\cite{Deur:2018roz} for their explicit expressions). In particular, $\sigma_\text{Mott}$ is the Mott cross-section for scattering off a point-like target. The information on the nucleon structure is contained in the four structure functions $F_1(x,Q^2)$, $F_2(x,Q^2)$, $g_1(x,Q^2)$ and $g_2(x,Q^2)$. 
The unpolarized structure functions $F_1$ and $F_2$ are obtained with unpolarized beam and/or target and are kinematically  separated by varying $a$ and $b$. The spin structure functions $ g_1$ and $g_2$ arise when beam and target are both polarized. One accesses $ g_1$ and $g_2$ by measuring beam spin asymmetries. Varying the target polarization direction allows to separate $ g_1$ from $g_2$. 
One can extract $\alphas$ from $ g_1$ ($g_2$ is harder to measure and without simple partonic interpretation) 
by performing a global fit of its $Q^2$-evolution, see \eg~\cite{Blumlein:2010rn}. This is a thorough but
complex method, and modeled nonperturbative inputs (quark and gluon PDFs and possibly higher twists) are required.
Alternatively, the $Q^2$-evolution of the moment $\Gamma_1(Q^2) \equiv \int_0^1 g_1(x,Q^2) dx$ can be used. The method is much simpler since there is no 
$x$-dependence, and the nonperturbative inputs are the more-or-less well-measured axial charges 
$a_0$, $a_3$ and $a_8$~\cite{Deur:2018roz} (and higher-twists if low-$Q^2$ data are used). The difficulty 
is that the low-$x$ contribution to a moment cannot be measured, as it would require
infinite beam energy. Furthermore, $a_0$, which is the quark spin contribution to the nucleon spin,
is $Q^2$-dependent and in some renormalization schemes may receive a contribution from the polarized gluon PDF $\Delta G$.
Extracting $\alphas$ is further simplified when the isovector component of $\Gamma_1$ is used, 
{\it viz}, $\Gamma_1^p-\Gamma_1^n \equiv \Gamma_1^{p-n}$. The relevant axial charge is precisely measured 
($a_3=g_A=1.2762(5)$)~\cite{Zyla:2020zbs}, the $Q^2$-evolution of $\Gamma_1^{p-n}$ is known to higher order 
than $\Gamma_1^{p,n}$ or $g_1$~\cite{Kataev:1994gd,  Kataev:2005hv,  Baikov:2010je}, and there is no gluon contribution. On the other hand, the low-$x$ 
issue remains and $\Gamma_1^{p-n}$ measurements are demanding  since polarized proton and neutron
data are needed. We will focus here on using $\Gamma_1^{p-n}$. 

The Bjorken sum rule~\cite{Bjorken:1966jh, Bjorken:1969mm} generalized to finite $Q^2$ by including
QCD radiative and power corrections is~\cite{Kataev:1994gd,  Kataev:2005hv,  Baikov:2010je}:
\vspace{-0.3cm}
 \begin{eqnarray}
 \vspace{-0.2cm}
 \Gamma_1^{p-n}(Q^2) = \frac{g_A}{6} \hspace{-1mm} \bigg[1-\frac{\alphas}{\pi}
 -3.58\left(\frac{\alphas}{\pi}\right)^2
\hspace{-1mm} -20.21\left(\frac{\alphas}{\pi}\right)^{3} 
\hspace{-1mm} -175.7\left(\frac{\alphas}{\pi}\right)^{4}
\hspace{-1mm} -\sim \hspace{-1mm} 893\left(\frac{\alphas}{\pi}\right)^{5}
\hspace{-1mm} +\mathcal O(\alphas^6) \bigg]+ \nonumber \\
\frac{M^2}{Q^2}\big( a_2+4d_2 +4f_2 \big)
+\mathcal O(1/Q^{4}),
\label{eq:genBj}
\end{eqnarray}
where the series coefficient values are for $\MSbar$ and $\Nf=3$. Here, 
$M^2$ is nucleon mass,
$a_2(Q^2)$ and $d_2(Q^2)$ are 
the twist-2 and 3 target mass corrections, respectively, and $f_2(Q^2)$ is the dynamical twist-4 correction. 
The higher twist coefficients are also subject to QCD radiative corrections and thus depend on $\alphas$, \ie\ $Q^2$. 
To extract $\alphas$, one can either solve Eq.~(\ref{eq:genBj}) for $\alphas$ or fit 
the $Q^2$-dependence of Eq.~(\ref{eq:genBj}). The latter method is more accurate because it suppresses
systematic uncertainties affecting the magnitude of $\Gamma_1^{p-n}$~\cite{Altarelli:1996nm}. 
Such analysis was performed on the data of the Jefferson Lab (JLab) experiment EG1dvcs ~\cite{CLAS:2015otq, CLAS:2017qga}. 
The experiment took data during
the 6~GeV era of JLab and used the CEBAF Large Acceptance Spectrometer (CLAS) together with 
polarized NH$_3$ (for proton) and ND$_3$ (for neutron) targets and the JLab polarized electron beam. 
Since EG1dvcs was dedicated to exclusive measurements (DVCS process) it provides  inclusive data with very
high statistics. 
Only EG1dvcs data were used to avoid uncorrelated systematics between different experiments since, in contrast to 
point-to-point correlated systematics (\eg\ from polarimetries), uncorrelated errors are not suppressed in the extraction of $\alphas$.
Furthermore, since the EG1dvcs data largely dominate the world data in terms of  statistics, including other experiments would 
only increase the systematic uncertainty on the extraction without being balanced by the gains in $Q^2$ leverage and statistics.
However, one must recognize  that the availability of polarized $g_1$ data from higher energy 
experiments~\cite{E143:1998hbs, E142:1996thl, E154:1997xfa, E155:1999pwm, E155:2000qdr, SpinMuon:1993gcv, SpinMuonSMC:1997voo, SpinMuonSMC:1994met, SpinMuon:1995svc,  SpinMuonSMC:1997ixm, SpinMuonSMC:1997mkb,  COMPASS:2005xxc, COMPASS:2006mhr,  COMPASS:2010wkz, HERMES:1997hjr, HERMES:1998cbu, HERMES:2006jyl, Deur:2004ti, Deur:2008ej} is crucial for this analysis since it allows to control the low-$x$ extrapolation necessary 
to complement  $\Gamma_1^{p-n}$. 
The EG1dvcs data used cover $Q^2 > 2.32$~GeV$^2$ and are comprised of five data points.
The statistical uncertainty is negligible for this analysis. The point-to-point correlated uncertainty was separated 
from the point-to-point uncorrelated one using the {\it unbiased estimate} prescription~\cite{Schmelling:1994pz,Zyla:2020zbs}, and the uncertainties on the low-$x$ extrapolation were approximately separated into $Q^2$-dependent and 
independent parts. Thus, a large fraction of the uncertainties is suppressed in the extraction of $\alphas$. 
The dominant errors are the point-to-point uncorrelated (4.4\%) and correlated (3.3\%) uncertainties.
The others are negligible (power corrections: $<$0.1$\%$; Eq.~(\ref{eq:genBj}) truncation: 1.3\%;
truncation of the $\beta$-series  needed to evolve $\alphas$ to $m_\mathrm{Z}$: 0.1\%). 
In all the analysis provided $\alphasmZ=0.1123\pm0.0061$~\cite{Deur:2014vea} at N$^3$LO with an N$^4$LO correction
estimate, see Eq.~(\ref{eq:genBj}).

This analysis can be repeated with better precision in the future thanks to two developments.
The first is experiment EG12 using CLAS12 with the JLab 11-GeV  
beam and scheduled for June 2022. 
The expected precision is similar to EG1dvcs, but with a $Q^2$ reach of up to 6~GeV and better low-$x$ coverage.
The second is the 
EIC~\cite{Accardi:2012qut}, which will now be summarily described.

The optimal $Q^2$ range to extract 
$\alphas$ from $\Gamma_1^{p-n}$ at EIC  is $1 \lesssim Q^2 \lesssim 50$~GeV$^2$. Below, power corrections could be an issue 
and above, the unmeasured low-$x$ part of $\Gamma_1^{p-n}$ becomes large ($\gtrsim$15\%. Since for EIC there will be no higher energy data to guide the low-$x$ extrapolation, it must be kept small). 
Assuming a luminosity of $2\times 10^{33}$cm$^{-2}$s$^{-1}$ and an electron (nucleus) beam polarization of 50\% (60\%), a year of data taking yields statistical uncertainties ranging from $\Delta \Gamma_1^{p-n}=0.5\%$ for $Q^2=3$~GeV$^2$ to $0.05\%$ for $Q^2=15$~GeV$^2$.
For systematics, we assume
4\% for deuteron's nuclear corrections or negligible if the associated proton can be tagged,
100\% on the low-$x$ extrapolation,
3\% on beam polarimetries,
6\% on radiative corrections,
2.5\% on the $F_1$ structure function necessary to form $g_1$ from the measured $A_1$ asymmetry~\cite{Deur:2018roz} 
(specifically, we assumed 2\% on $F_2$ and 10\% on the $R$ ratio).
Since $g_2$ will be measured at the EIC, correcting for it to obtain $A_1$ from the experimental asymmetry $A_{||}$ 
will not add significant uncertainties. We neglected  the contamination from particle misidentification. 
Detector/trigger efficiencies, acceptance, beam currents uncertainties are negligible for asymmetry measurements.
Finally, we assumed that 60\% of the systematic is point-to-point uncorrelated, as found for the EG1dvcs data.
Fitting the simulated pseudodata yields $\Delta \alphasmZ=\pm0.0033\pm0.0005$, where the first (second) uncertainty is 
point-to-point uncorrelated (correlated). The fit was performed over
$1.5 \leq  Q^2 \leq 15$~GeV$^2$. Adding a higher $Q^2$ data point only reduces marginally the point-to-point uncorrelated
error while significantly increasing the correlated one: $\Delta \alphasmZ= \pm0.0030\pm0.0021$.
Increasing the statistics by a factor of 10 and assuming that the proton of the deuteron can be tagged yields 
$\Delta \alphasmZ=\pm0.0016\pm0.0003$.  
This simple exercise suggests that an accurate measurement of acceptable precision is possible at EIC, warranting to investigate more seriously the possibility. A more 
thorough investigation is currently being performed by the EIC structure function working group.

Another method has been used to extract $\alphas$ from  $\Gamma_1^{p-n}$ at low $Q^2$. 
It is based on the gauge-gravity duality, specifically  
LFHQCD~\cite{Brodsky:2014yha}, a model closely tied to QCD.
The Lagrangian of LFHQCD is that of QCD quantized on the light-front, a completely rigorous procedure. 
The only unknown is the interaction part of the Lagrangian. It can be determined by 
imposing the approximate chiral symmetry of QCD. 
Alternatively, one can require QCD's approximate conformal symmetry and apply de Alfaro--Fubini--Furlan 
procedure~\cite{deAlfaro:1976vlx}, or require that the light-front potential transformed to the traditional instant-form 
produces the Cornell potential in the heavy quark case~\cite{Trawinski:2014msa}.
The three methods lead to the same semiclassical potential that effectively accounts for the action of the gluons. 
Imposing upon LFHQCD a nearly exact QCD symmetry to determine the full Lagrangian 
is the same procedure as used for chiral perturbation theory to determine its Lagrangian. However, contrary to  the latter, no counterterms 
determined from phenomenology are required for LFHQCD since renormalization is unnecessary for nonperturbative frameworks. 
In fact, LFHQCD has in principle a single free parameter,  determined \eg\ from a hadron mass or
a fit to the $\Gamma_1^{p-n}(Q^2)$ data at low $Q^2$. (One free parameter is always needed to set units since 
they are arbitrary human conventions.) 
In practice, LFHQCD has a few additional parameters 
to account for hadronic higher Fock states. They are however of secondary importance and do not enter the determination of $\alphas$ discussed here.
LFHQCD provides a wide range of results that agree with experiments, including
hadron spectroscopy~\cite{Deur:2014qfa}, 
form factors~\cite{Sufian:2016hwn},
polarized and unpolarized PDFs and GPDs~\cite{deTeramond:2018ecg, deTeramond:2021lxc, Liu:2019vsn} and $\alphas(Q^2)$~\cite{Brodsky:2010ur}.
The latter is obtained by forming the effective charge~\cite{Grunberg:1980ja, Grunberg:1982fw, Deur:2016tte} characterizing $\Gamma_1^{p-n}(Q^2)$ and
calculating it with LFHQCD. The obtained effective charge, straightforwardly related to $\alphas$, 
is valid over the $0 \leq Q^2 \lesssim 2$~GeV$^2$ domain. Thus, LFHQCD
and pQCD share a domain of applicability over $1 \lesssim Q^2 \lesssim 2$~GeV$^2$. Using
the analytical expressions of $\alphas$ from LFHQCD and pQCD one can then deduce the relation between $\Lambda_s$ and 
the LFHQCD scale parameter~\cite{Deur:2016cxb, Deur:2016opc}. The procedure yields 
$\alphasmZ=0.1190\pm0.0006$ (N$^3$LO).
The uncertainty, which  is comparable to that of the best lattice results,
comes from uncertainties on the LFHQCD scale, on the chiral limit approximation and on the
truncations of the Bjorken and $\alphas$ pQCD series.

~

To summarized two determinations of $\alphas$ using the Bjorken sum $\Gamma_1^{p-n}$ were reviewed. 
The first yields $\alphas=0.1123\pm0.0061$ (N$^3$LO)~\cite{Deur:2014vea} and is based on a fit to the JLab EG1dvcs data. A similar analysis of the expected EIC data may reduce the uncertainty to $\Delta \alphas \simeq \pm0.0016$.
The second determination, $\alphas=0.1190\pm0.0006$ (N$^3$LO)~\cite{Deur:2016opc}, is obtained by matching
the pQCD expression of $\alphas$ to that from LFHQCD in the  domain of validity common to the two approaches.\\

\noindent \textit{Acknowledgments}--- This material is based upon work supported by the U.S. Department of Energy, Office of Science, Office of Nuclear Physics under contract DE-AC05-06OR23177.


\subsection{\texorpdfstring{$\alphasmZ$}{alphasmZ} from a combined NNLO analysis of normalized jet cross sections and DIS data 
\protect\footnote{A\lowercase{uthors:} D. B\lowercase{ritzger} (MPI M\lowercase{unich}) \lowercase{on behalf of the} H1 C\lowercase{ollaboration}}}

The H1 and NNLOJET Collaborations have performed a simultaneous
determination of PDFs and $\alphasmZ$ using the entire set of inclusive
neutral-current and charged-current (NC and CC) DIS data from H1 together with all
of H1's normalized inclusive jet and dijet cross
section data~\cite{H1:2017bml}.
The methodology of the analysis follows closely the well-established
formalism of PDF determinations at HERA, as it was previously employed for HERAPDF-style or
H1PDF-style fits~\cite{H1:2009pze,H1:2012qti,Abramowicz:2015mha}, but additionally
$\alphasmZ$ is further considered as a free parameter in the fit.\\

The $\alphas$ sensitivity in the analysis is obtained by exploiting H1's
normalized inclusive jet and dijet cross sections that were published
previously~\cite{H1:2007xjj,H1:2014cbm,H1:2016goa}.
These jet measurements were performed double-differentially as
functions of $Q^2$ and $\pT$ (where $\pT$ denotes the transverse
momenta of a single jet for inclusive jets, and the average $\pT$ of
the dijet system).
The data were taken during the HERA-I~\cite{H1:2007xjj} and HERA-II~\cite{H1:2014cbm,H1:2016goa}
running periods and for $Q^2$ values lower than
100\,GeV$^2$~\cite{H1:2016goa} or above~\cite{H1:2007xjj,H1:2014cbm},
and jets were measured in the Breit frame in the range of about
$4.5<\pT<50$\,GeV.
Differently than absolute jet cross sections, normalized jet cross
are measured as the ratio of absolute jet cross sections 
to the bin-integrated inclusive NC DIS cross section in the
respective $Q^2$ interval.
Thus, some experimental uncertainties are reduced or cancel (like normalization
uncertainties), and since the inclusive DIS data themselves are used in the fit, the
correlations of the uncertainties between different data sets are
correctly treated in the fit.
It is also hoped that the PDF-dependence cancels out to some extent.
Since jet cross sections in the Breit frame are proportional to
$\mathcal{O}(\alphas^1)$ in leading-order pQCD, while inclusive DIS is
of $\mathcal{O}(\alphas^0)$, the $\alphas$ dependence is preserved but smaller
experimental uncertainties are achieved, when compared to absolute jet
cross sections.

The selected inclusive DIS data comprise the full set of H1's
inclusive NC and CC DIS data, and makes particularly use of the
combined low-$Q^2$ data from HERA-I~\cite{H1:2010fzx} and
high-$Q^2$ data with polarized lepton beams from HERA-II~\cite{H1:2012qti}.
The $x$-dependence of five orthogonal PDFs ($g$, $\bar{u}$, $\bar{d}$,
$\bar{U}$, $\bar{D}$) are parameterized at a
scale of $\sqrt{1.9}$\,GeV using a common HERAPDF-style
parameterization.
The inclusive DIS data is restricted to $Q^2>12\,$GeV$^2$, to reduce
effects from heavy-quark masses and low-$x$ effects~\cite{Ball:2017otu}.
Since the $\alphas$-sensitivity arises from the matrix elements of the jet
data, the $\alphas$ results are rather insensitive to the exact
parameterization of the PDFs.
The PDF evolution and structure function calculation is performed
using the program QCDNUM~\cite{Botje:2010ay} in the ZM-VFNS scheme.

The jet data are confronted for the first time with NNLO
predictions~\cite{Currie:2016ytq,Currie:2017tpe} from
NNLOJET~\cite{Gehrmann:2018szu} with five massless flavours and interfaced to
fastNLO~\cite{Britzger:2012bs}, and the denominator of the normalized
jet cross sections is calculated in NNLO using QCDNUM.
Similarly as inclusive DIS data, also jet data are restricted to
scales that are two times higher than the $b$-quark mass in order to
reduce the impact from heavy quark masses.
The renormalization and factorization scales are identified with
$\mu^2=Q^2+\pT^2$.
The NNLO predictions are corrected for hadronization effects using
multiplicative correction factors, which were derived using the MC event
generators Djangoh~\cite{Charchula:1994kf} and Rapgap~\cite{Jung:1993gf}.

The fit minimizes a $\chi^2$ quantity based on log-normal probability
distribution functions.
At the minimum $\chi^2/n_\text{dof}=1.0$ for $n_\text{dof}=1516$
degrees of freedom is found, which is an excellent data-to-theory agreement.
The value of $\alphasmZ$ is determined in the PDF+$\alphasmZ$ fit to~\cite{H1:2017bml}
\begin{equation}
  \alphasmZ=0.1147\,(11)_{\rm exp,had,PDF}\,(2)_{\rm mod}\,(3)_{\rm par}\,(23)_{\rm scale}\,,
  \label{eq:H1PDF}
\end{equation}
where the first uncertainty denotes the \emph{fit} uncertainty, the
second and third component is obtained from variations of model (mod)
or parameterization (par) assumptions of the PDF.
The last uncertainty denotes scale uncertainties which are obtained
from repeating the fit with scale-factors of 0.5 and 2.
The $\alphas$ value is found to be consistent with a determination
from non-normalized jet cross sections (Sec.~\ref{sec:ep}).
The resulting PDF is denoted H1PDF2017nnlo and is found to be somewhat 
compatible with NNPDF3.1~\cite{NNPDF:2017mvq} and in good consistency
with NNPDF3.1sx~\cite{Ball:2017otu}.

In a PDF+$\alphasmZ$ fit to inclusive DIS data alone, large
uncertainties in $\alphasmZ$ are obtained, and it was shown that the
inclusion of normalized jet data reduces the correlation of $\alphasmZ$
and the gluon density in the PDF+$\alphas$ fit~\cite{H1:2017bml}.
It can be concluded that the sensitivity to $\alphasmZ$ arises
exclusively from the jet data, but not from the inclusive DIS data,
which is also seen from the small (mod) and (par) uncertainties
in Eq.~(\ref{eq:H1PDF}).

A recent reanalysis of the selected H1's normalized jet cross
section data from the H1PDF2017nnlo analysis by a somewhat different author group~\cite{ZEUS:2021sqd} exploits the same NNLO
predictions.
Differences are in the selection of the DIS data, and the addition of further non-normalized DIS jet data.
That analysis finds a consistent
result for $\alphasmZ$, equal-sized experimental uncertainties, and
also scale uncertainties of comparable size  (albeit a bit higher).
It is concluded, that the superior sensitivity of H1's normalized jet
data to $\alphasmZ$ makes the $\alphasmZ$ result from the PDF+$\alphas$ fit rather
insensitive to changes in the inclusive DIS data, or to
the addition of further, absolute, jet cross sections from HERA.
The scope of Ref.~\cite{ZEUS:2021sqd} focuses on different QCD aspects, like
PDFs, but it does not supersede the original $\alphas$ analysis~\cite{H1:2017bml}, since it does not employs improved predictions, or achieves smaller uncertainties in $\alphasmZ$. Therefore, the present $\alphas$ from HERA (normalized) jet data from PDF+$\alphas$ analyses remains is quoted in Ref.~\cite{H1:2017bml}, and Eq.~\eqref{eq:H1PDF}. 
Future uncertainty reductions can be achieved with improved predictions, like
N$^3$LO, or by adding normalized jet data from more experiments.


\clearpage
\section{\texorpdfstring{\boldmath$\alphasmZ$}{alphasmZ} from electroweak data}
\label{sec:EW}

The present QCD coupling world average~\cite{Zyla:2020zbs} contains an ``Electroweak precision fit'' category where an $\alphasmZ = 0.1208 \pm 0.0028$ value is determined from the average of two results: (i) a global fit of SM electroweak and Higgs observables with the QCD coupling constant left as single free parameter, yielding $\alphasmZ = 0.1194 \pm 0.0029$~\cite{Haller:2018nnx}, and (ii) a fit of three hadronic pseudoobservables measured at LEP and SLC in $\epem$ collisions at the Z mass pole~\cite{Erler:2019hds}, resulting in $\alphasmZ = 0.1221 \pm 0.0027$. More recently, Ref.~\cite{dEnterria:2020cpv} has improved both Z-boson-based determinations by incorporating higher-order theoretical corrections~\cite{Dubovyk:2018rlg,Dubovyk:2019szj,Chen:2020xzx} and an experimental update of the LEP luminosity corrections~\cite{Voutsinas:2019hwu}. In addition, it has been shown that similar future high-precision measurement of W boson hadronic decays~\cite{dEnterria:2016rbf,dEnterria:2020cpv} will also provide competitive $\alphasmZ$ determinations. The results of these two latest $\alphasmZ$ extractions from Z and W boson decays are summarized below.

\subsection{Strong coupling from electroweak boson decays at N$^3$LO accuracy
\protect\footnote{A\lowercase{uthors:} D. \lowercase{d'}E\lowercase{nterria} (CERN)}}
\label{sec:EWB}

The $\alphasmZ$ value can be neatly extracted from various electroweak-boson hadronic pseudoobservables that can be accurately measured in high-energy $\epem$ collisions, such as: 
\begin{itemize}
\item The W and Z hadronic widths, computable theoretically through the expression
\begin{eqnarray}
\Gamma^\mathrm{had}_\mathrm{W,Z}(Q) =  \Gamma^{^\mathrm{Born}}_\mathrm{W,Z} \left(\! 1 +\sum^{4}_{i=1} a_i(Q)\left(\!\frac{\alphas(Q)}{\pi}\right)^i\!\!+ \mathcal{O}(\alphas^5)+\delta_{_\mathrm{EW}}+\delta_\mathrm{mix}+\delta_\mathrm{np}\right)\!,\label{eq:Gamma_alphas}
\end{eqnarray}
where the Born width $\Gamma^{^\mathrm{Born}}_\mathrm{W,Z}=f(G_\mathrm{F},\,m_\mathrm{W,Z}^3,\,N_\mathrm{C};\mathrm{\sum |V_{ij}|^2})$ depends on the Fermi constant $G_\mathrm{F}$, the cube of the EW boson mass, the number of colours $N_\mathrm{C}$, and, in the W case, also on the sum of CKM matrix elements $\rm|V_{ij}|^2$. The $a_i(Q)$ and $\delta_\mathrm{_{EW},mix,np}$ terms are, respectively, higher-order pQCD, EW, mixed, and non-pQCD corrections discussed below. Since the total W and Z widths ---given by the sum of hadronic and leptonic partial widths $\Gamma^\mathrm{tot}_\mathrm{W,Z} = \Gamma^\mathrm{had}_\mathrm{W,Z} + \Gamma^\mathrm{lep}_\mathrm{W,Z}$--- have smaller experimental uncertainties than the hadronic ones alone, and since $\Gamma^\mathrm{lep}_\mathrm{W,Z}$ can be both accurately measured and computed, the value of $\Gamma^\mathrm{tot}_\mathrm{W,Z}$ is often directly used to determine $\alphasmZ$.
\item The ratio of W, Z hadronic-to-leptonic widths, defined theoretically as
\begin{eqnarray}
\mathrm{R}_\mathrm{W,Z}(Q) = \frac{\Gamma^\mathrm{had}_\mathrm{W,Z}(Q)}{\Gamma^\mathrm{lep}_\mathrm{W,Z}(Q)} = \mathrm{R}_\mathrm{W,Z}^\mathrm{EW} \left(1 + \sum^{4}_{i=1} a_i(Q)\left(\frac{\alphas(Q)}{\pi}\right)^i\!\!+\mathcal{O}(\alphas^5)+\delta_\mathrm{mix}
+\delta_\mathrm{np}\right)\!,
\label{eq:R_alphas}
\end{eqnarray}
where the $\mathrm{R}_\mathrm{W,Z}^\mathrm{EW}=f(\alpha,\alpha^2,\dots)$ prefactor, which depends on the fine structure constant $\alpha$, accounts for the purely electroweak dependence of the calculation. Experimentally, in the W boson case the denominator of the $R_\mathrm{W}$ ratio represents the {\it sum} of all leptonic decays, which can be accurately determined from the ratio of hadronic over leptonic decay branching ratios: $R_\mathrm{W}=\mathcal{B}_\mathrm{W}^\mathrm{had}/\mathcal{B}_\mathrm{W}^\mathrm{lep} = 2.0684 \pm 0.0254$~\cite{Zyla:2020zbs}. 
However, in the Z boson case the denominator of $R_\mathrm{Z}$ (often labeled $R^0_\ell$) is the average width over the three charged lepton species, \ie\ $R_\mathrm{Z} = \Gamma_\mathrm{Z}^\mathrm{had}/\Gamma_\mathrm{Z}^\mathrm{\ell} = 20.767 \pm 0.025$~\cite{Zyla:2020zbs} with $\Gamma_\mathrm{Z}^\mathrm{\ell} = \frac{1}{3}(\Gamma_\mathrm{Z}^\mathrm{e} + \Gamma_\mathrm{Z}^\mathrm{\mu} + \Gamma_\mathrm{Z}^\mathrm{\tau})$, which can be more precisely measured.

\item In the Z boson case, the hadronic cross section at the resonance peak in $\epem$ collisions, theoretically given by
\begin{equation}
\so = \frac{12 \pi}{m_\mathrm{Z}} \cdot \frac{\Gamma_\mathrm{Z}^{e}\,\Gamma_\mathrm{Z}^\mathrm{had}}{(\Gamma^\mathrm{tot}_\mathrm{Z})^2}\,,
\label{eq:sigma0Z}
\end{equation}
 where $\Gamma_\mathrm{Z}^{e}$ is its electronic width. This quantity is also useful because $\so$ can be measured with small experimental uncertainties in the $\epem \to \mathrm{Z} \to \mathrm{hadrons}$ process, independently of any $\Gamma_\mathrm{Z}^\mathrm{e,had,tot}$ widths appearing in the theoretical equation.
\end{itemize}

In Ref.~\cite{dEnterria:2020cpv}, two new determinations of the QCD coupling constant at the Z pole have been carried out from detailed comparisons of inclusive W and Z hadronic decays data to state-of-the-art pQCD calculations at next-to-next-to-next-to-leading order (N$^{3}$LO) accuracy, incorporating the latest experimental and theoretical developments. In the W boson case, the total width computed at N$^{3}$LO has been used for the first time in the extraction. For the Z boson pseudoobservables, the N$^{3}$LO results have been complemented with the full two- and partial three-loop EW corrections recently made available, and the experimental values have been updated to account for newly estimated LEP luminosity biases. A combined reanalysis of the Z boson data yields $\alphasmZ = 0.1203 \pm 0.0028$, with a 2.3\% uncertainty reduced by about 7\% compared to the previous state-of-the-art. From the combined W boson data, a value of $\alphasmZ = 0.107 \pm 0.035$ is extracted, with still large experimental uncertainties but also reduced compared to previous works. The levels of theoretical and parametric precision required in the context of QCD coupling determinations with permil uncertainties from high-statistics W and Z boson samples expected at future $\epem$ colliders, such as the FCC-ee, are discussed.

\subsubsection{\texorpdfstring{$\alphasmZ$}{alphasmZ} from W boson decays}

\begin{figure}[htpb!]
\centering
\includegraphics[width=0.49\textwidth]{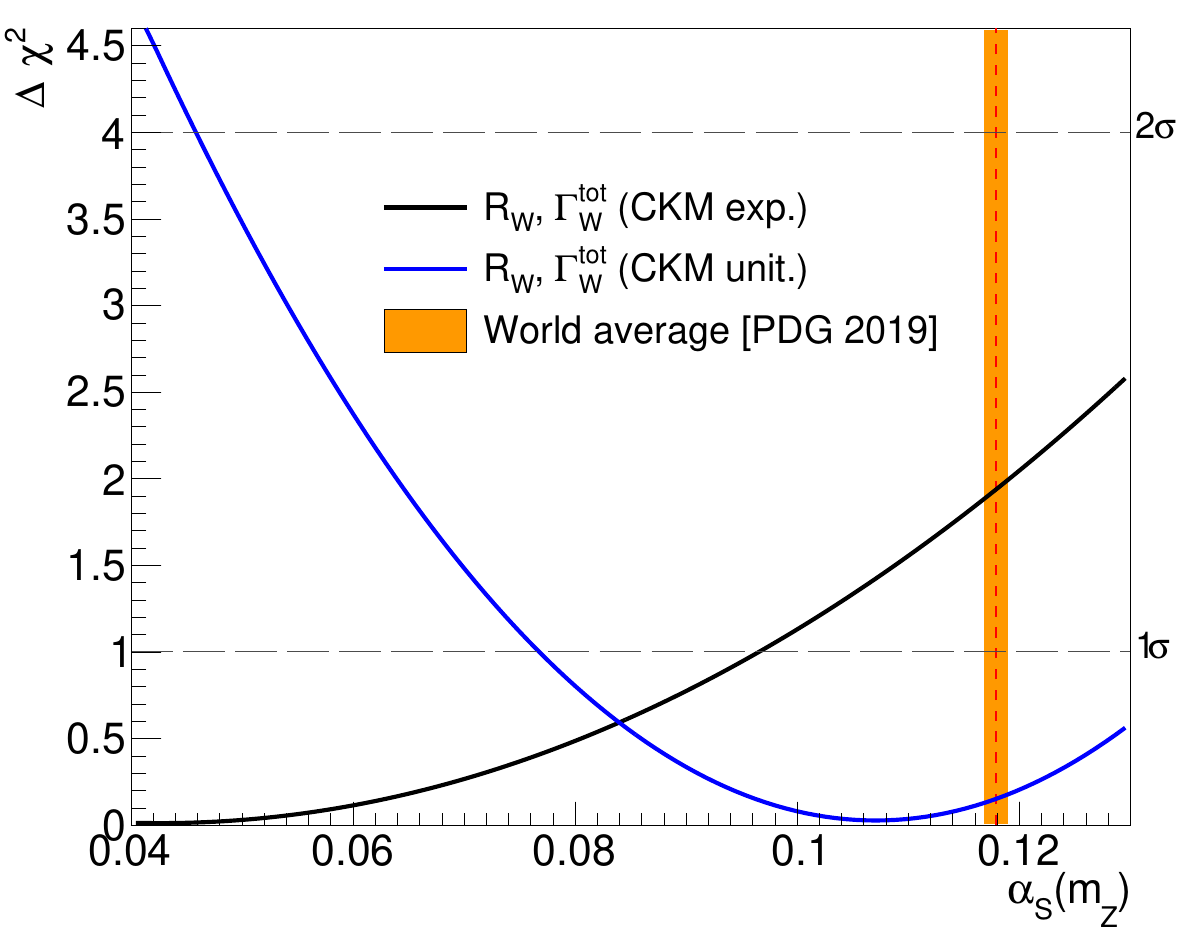}
\includegraphics[width=0.49\textwidth]{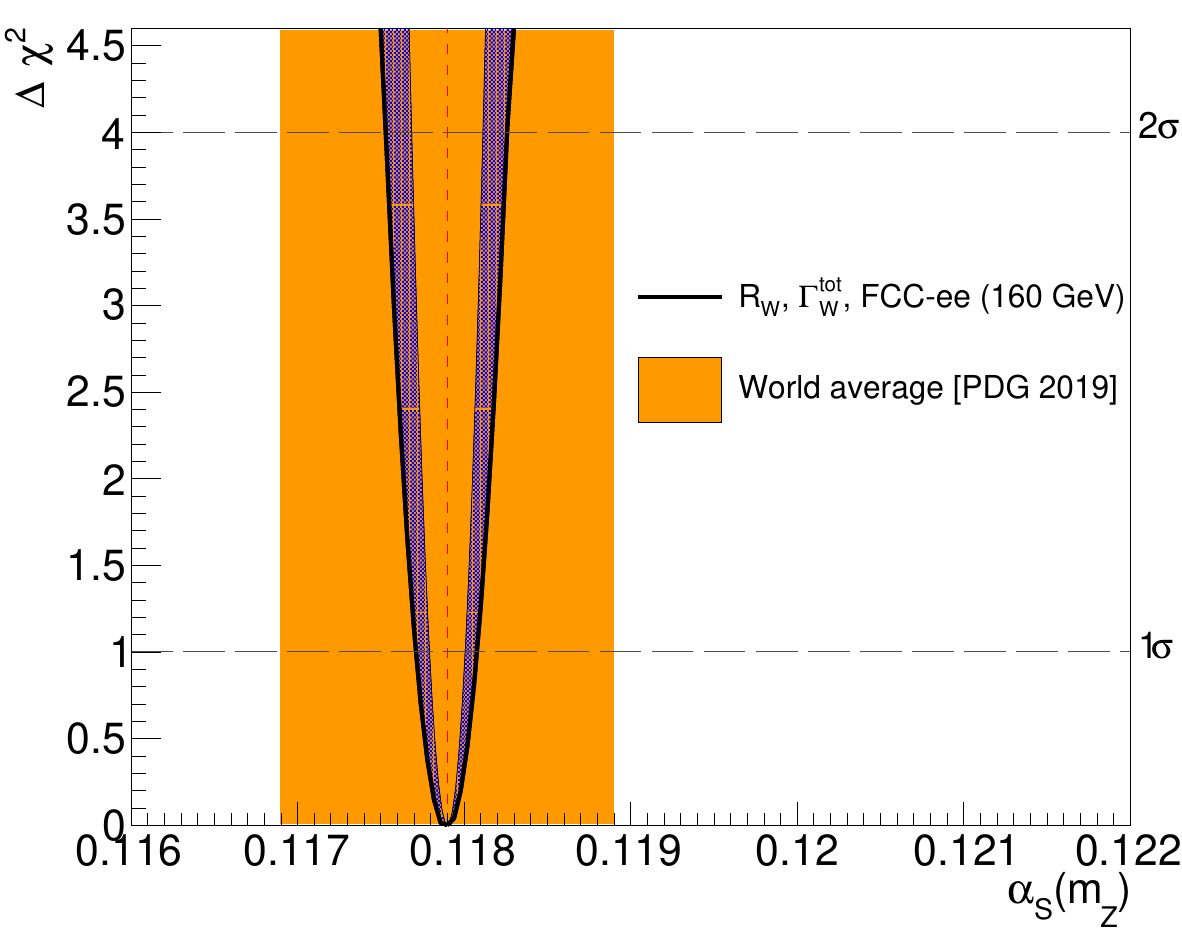}
\caption{$\Delta\chi^2$ fit profiles of the $\alphasmZ$ determined from the combined N$^3$LO analysis of the total W width and $R_\mathrm{W}$ ratio, compared to the 2019 $\alphasmZ$ world average (vertical orange band). Left: Extraction with the present W data assuming (blue curve) or not (black curve) CKM unitarity. Right: Extraction expected at the FCC-ee, with the total (experimental, parametric, and theoretical, added in quadrature) uncertainties (outer parabola) and with the experimental uncertainties alone (inner parabola).}
\label{fig:alphas_W}
\end{figure}

The state-of-the-art analytic unintegrated expressions for the leptonic and hadronic W boson decay widths~\cite{Denner:1991kt,Kara:2013dua} have been integrated out, and convenient parametrizations of all quantities of interest ($\Gamma^\mathrm{lep,had,tot}_\mathrm{W}$ and $R_\mathrm{W}$) have been derived for the subsequent $\alphasmZ$ fitting procedure. The parametrizations of all the W boson observables include the full-N$^3$LO QCD $\mathcal{O}(\alphas^4)$, the leading EW ${\cal O}(\alpha)$, and mixed QCD+EW ${\cal O}(\alphas\alpha)$ corrections. The value of $\alphasmZ$ is then obtained by a combined fit of the theoretical expressions for $\Gamma^\mathrm{tot}_\mathrm{W}$ and $R_\mathrm{W}$ to the experimental data: $\Gamma_\mathrm{W}^\mathrm{tot} = 2085 \pm 42$ and $R_\mathrm{W} = 2.0684 \pm 0.0254$ (combining the three leptonic decays, assuming lepton universality). The relative experimental uncertainties of $\Gamma^\mathrm{tot}_\mathrm{W}$ and $R_\mathrm{W}$ are $2\%$ and $1.2\%$, respectively, and combining both observables in the fit, assuming them to be fully uncorrelated~\cite{ParticleDataGroup:2018ovx,ALEPH:2013dgf}, provides some improvement in the final $\alphasmZ$ precision. The derived $\alphasmZ$ values are tabulated in Table~\ref{tab:alphas_W}, and the corresponding goodness-of-fit $\Delta\chi^2$ scans are plotted in Fig.~\ref{fig:alphas_W} (left). Without imposing CKM unitarity, the fitted QCD coupling constant is left basically unconstrained: $\alphasmZ = 0.042 \pm 0.052$, due to the large dominant parametric uncertainties of the theoretical $\Gamma_\mathrm{W}^\mathrm{tot}$ and $R_\mathrm{W}$ calculations. Imposing unitary of the CKM matrix leads to an extraction with $\sim$30\% uncertainty of experimental origin. The obtained value of $\alphasmZ = 0.107 \pm 0.035$ (with comparatively negligible parametric and theoretical uncertainties) is obviously, within its large uncertainties, in perfect accord with the current world average (orange band in Fig.~\ref{fig:alphas_W}). With respect to the previous NNLO extraction, $\alphasmZ = 0.117 \pm 0.042$ based on $R_\mathrm{W}$ alone~\cite{dEnterria:2016rbf}, our new calculation leads to a $\sim$10\% relative improvement in the experimental (as well as more accurate N$^3$LO theoretical and parametric) uncertainties.

\begin{table}[htpb!]
\centering
\caption{Values of $\alphasmZ$ extracted from the combined $\Gamma_\mathrm{W}^\mathrm{tot}$ and $R_\mathrm{W}$ measurements compared to the corresponding N$^3$LO theoretical calculations, assuming or not CKM unitarity, with the breakdown of propagated experimental, parametric, and theoretical uncertainties. The last row lists the $\alphasmZ$ result expected in $\epem$ collisions at the FCC-ee.\vspace{0.2cm}\label{tab:alphas_W}} 
\tabcolsep=4.5mm
\begin{tabular}{lccccccc}\hline
W boson observables & $\alphasmZ$   &   \multicolumn{3}{c}{$\alphasmZ$ uncertainties} \\
     &  & exp.  & param. &  theor.       \\\hline
$\Gamma_\mathrm{W}^\mathrm{tot}$, $R_\mathrm{W}$ (exp. CKM) & $0.042 \pm 0.052$ &$\pm 0.027$ & $\pm  0.045$ & $(\pm 0.0014)$  \\
$\Gamma_\mathrm{W}^\mathrm{tot}$, $R_\mathrm{W}$ (CKM unit.) & $0.107 \pm 0.035$ &$\pm 0.035$ & $(\pm 0.0002)$ & $(\pm 0.0016)$  \\\hline
$\Gamma_\mathrm{W}^\mathrm{tot}$, $R_\mathrm{W}$ (FCC-ee, CKM unit.) & $0.11790 \pm 0.00023$ & $\pm 0.00012$ & $\pm 0.00004$ & $\pm 0.00019$ \\
\hline
\end{tabular}
\end{table}

Achieving a truly competitive $\alphasmZ$ determination from the W decay data, with propagated experimental uncertainties reduced by a factor of $\times$30 at least (\ie\ below the 1\% level), requires much larger data samples than those collected in $\epem$ collisions at LEP-2 (about 40\,000 events). This can be reached at a future machine such as the FCC-ee, where the total W width can be accurately measured through a threshold $\epem\to\mathrm{W^{+}W^{-}}$ scan around $\sqrts = 2\MW$ center-of-mass energies~\cite{FCC:2018evy}, and where the $\RW$ ratio will benefit from the huge sample of $5 \cdot 10^8$ W bosons collected (about 12\,000 times larger at LEP-2) .
Without parallel progress in the measurements of $\Vcs$, $\Vcd$, and $\MW$, the parametric uncertainty would then largely dominate the precision of any $\alphasmZ$ extraction, as it is the case today when CKM unitarity is not enforced. However, both CKM elements will be also accurately determined at the FCC-ee by exploiting (i) the huge $\calO(10^{12})$ and clean samples of charmed mesons available in runs at the Z pole, and (ii) an experimental precision of 0.5 (1.2)~MeV for the W mass (width) within reach with 12~ab$^{-1}$ accumulated at the W\,W production threshold.

To assess the ultimate precision achievable from W-boson data, we run a combined $\alphasmZ$ fit with our N$^3$LO setup employing the following experimental values expected at the FCC-ee of the W observables and all other relevant parameters: (i) $\Gamma^\mathrm{tot}_\mathrm{W} = 2089.5 \pm 1.2$~MeV (to be compared to $2085 \pm 42$ today) and (ii) $R_\mathrm{W} = 2.07570 \pm 0.00008$ (instead of the current $2.068 \pm 0.0025$ value), (iii) CKM unitarity (or, equivalently, $\Vcs$ and $\Vcd$  uncertainties at the level of $10^{-5}$), and (iv) a W mass with $\MW = 80.3800 \pm 0.0005$~GeV precision, leads to $\sim$0.1\% uncertainties in $\alphasmZ$ (last row of Table~\ref{tab:alphas_W}). At such high level of experimental and parametric precision, the present propagated theoretical uncertainties would be about ten times larger than the former, although theory improvements are also expected in the coming years~\cite{Proceedings:2019vxr}. The theoretical effort should focus at computing the missing two- and three-loop ${\cal O}(\alpha^2,\alpha^3)$ EW, N$^4$LO QCD, as well as the mixed QCD+EW $\mathcal{O}(\alpha\alphas^2,\,\alpha\alphas^3,\,\alpha^2\alphas)$ corrections, which are all of about the same size and yield today a relative theoretical uncertainty of $\sim3.5\cdot10^{-4}$ in the W boson observables. With a factor of 10 reduction of the theory uncertainties, a final QCD coupling extraction at the FCC-ee with a 2-permil total uncertainty is possible: $\alphasmZ = 0.11790 \pm 0.00012_\mathrm{exp} \pm 0.00004_\mathrm{par} \pm 0.00019_\mathrm{th}$ (Table~\ref{tab:alphas_W}, last row), where the central value quoted is arbitrarily set at the world average.
Figure~\ref{fig:alphas_W} (right) shows the corresponding $\Delta\chi^2$ parabola for the $\alphasmZ$ determination expected at the FCC-ee compared to the 2019 world average (orange band), with the dashed band covering the range between taking into account all uncertainties (outer curve) or only experimental uncertainties (inner curve).

\subsubsection{\texorpdfstring{$\alphasmZ$}{alphasmZ} from Z boson decays}

For the Z boson case, we incorporate into the \gfitter~v2.2 code~\cite{Haller:2018nnx} all parametrizations of the partial and total Z widths, including the full two-loop ${\cal O}(\alpha^2)$ electroweak terms given in Ref.~\cite{Dubovyk:2018rlg}, plus the leading fermionic three-loop ${\cal O}(\alpha^3)$ EW corrections of Ref.~\cite{Chen:2020xzx}. These 2-loop (leading fermionic 3-loop) EW corrections increase the theoretical predictions for $\GZ$ by $+0.83$~MeV ($+0.33$~MeV), $R_\mathrm{Z}$ by $+0.0186$ ($-0.0037$), and $\so$ by $+1$~pb. The theoretical errors from missing higher-order ${\cal O}(\alpha^3)$ EW, ${\cal O}(\alpha^5)$ QCD, and ${\cal O}(\alpha\alphas^2,\,\alpha\alphas^3,\,\alpha^2\alphas)$ mixed corrections estimated in Refs.~\cite{Dubovyk:2018rlg,Chen:2020xzx} are of relative size of $\sim$1.5$\cdot10^{-4}$ for $\GZ$ and $\so$, and  $\sim$2.5$\cdot10^{-4}$ for $\Rlz$, \ie\ they are about a factor of two better than the corresponding theoretical calculations for the W boson pseudoobservables, as expected since the EW accuracy of the latter is only ${\cal O}(\alpha)$ today.

On the experimental data, new studies~\cite{Voutsinas:2019hwu,Janot:2019oyi} have updated the LEP luminosity corrections at and off the resonance peak that modify the PDG results for the Z boson pseudoobservables $\Gamma_\mathrm{Z}^\mathrm{tot}$ and $\so$. The change in $\Gamma_\mathrm{Z}^\mathrm{tot}$ is of $+0.012\%$. The impact on $\so$ is the largest of all pseudoobservables, with a $0.144\%$ reduction of the hadronic cross section at the Z peak that brings the data very close to the theoretical prediction. The central $\rm R_\mathrm{Z}$ ratios have not changed, but an extra precision digit is added now. The experimental uncertainties today ($\sim$0.1\%) are about a factor of four larger than their theoretical or parametric counterparts ($\sim$0.025\%). Matching the uncertainties of the current theory state-of-the-art calls for higher precision measurements in $\epem$ collisions at the Z pole with, at least, 20 times larger data samples than those collected at LEP.

\begin{figure}[htpb!]
\centering
\includegraphics[width=0.49\textwidth]{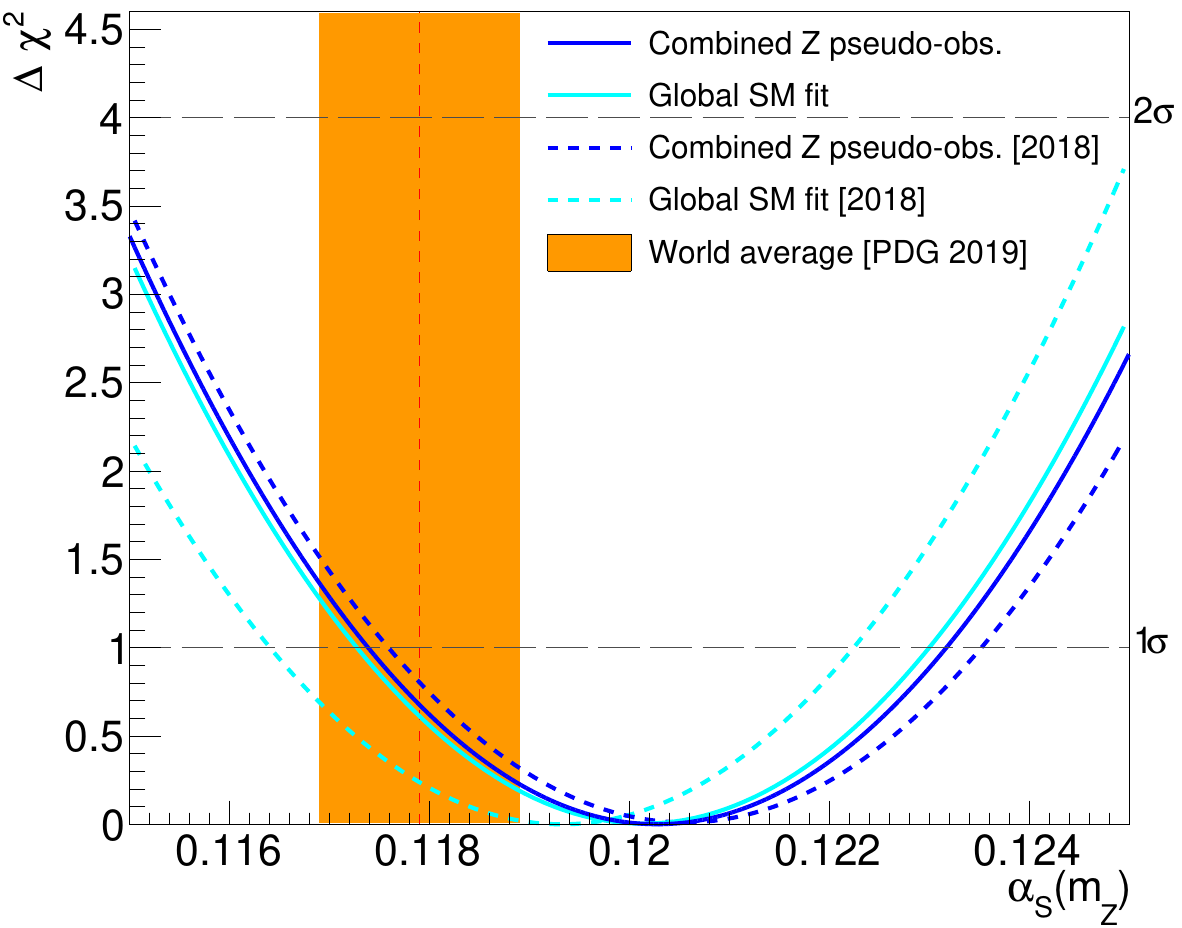}
\includegraphics[width=0.49\textwidth]{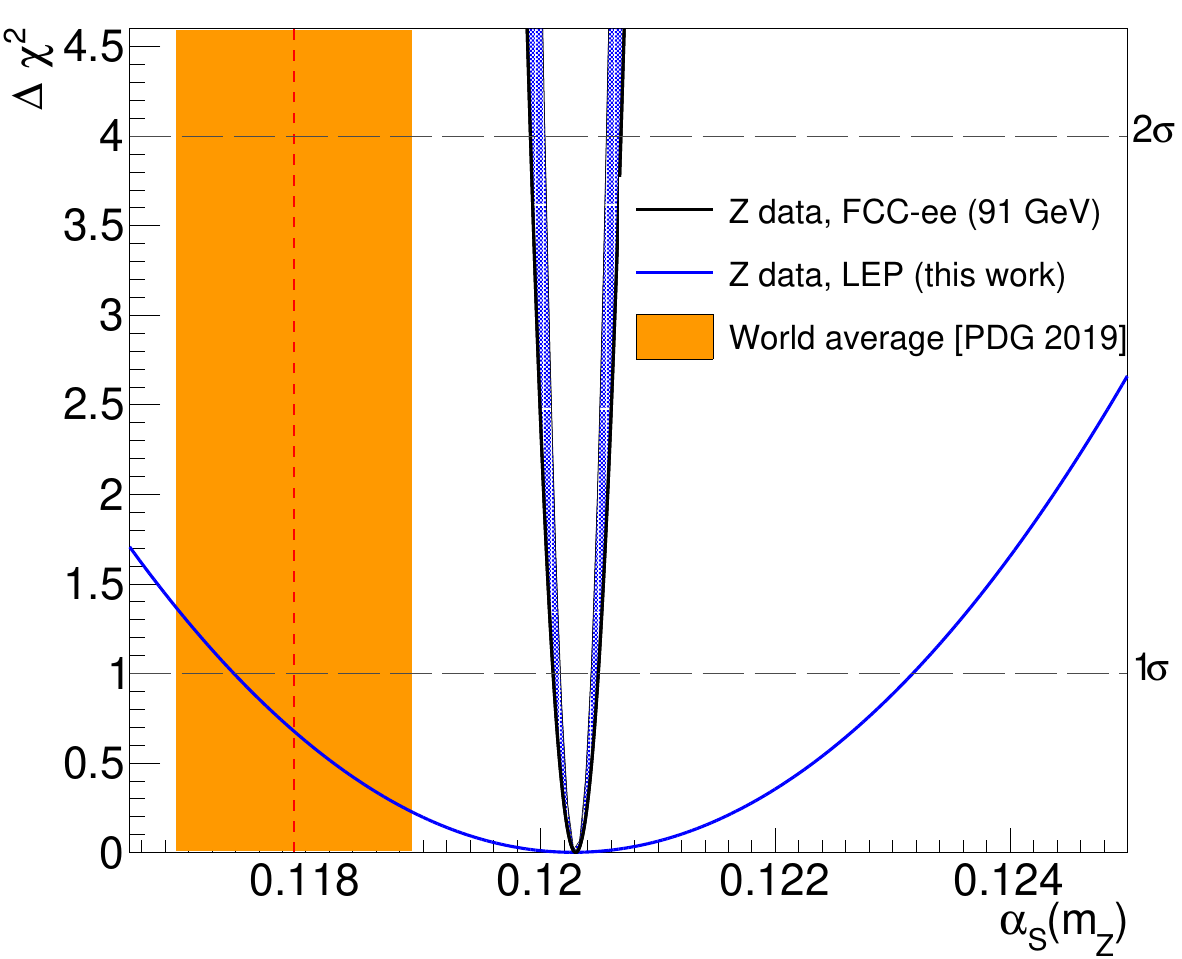}
\caption{ $\Delta\chi^2$ fit profiles of $\alphasmZ$ determined from the combined Z boson pseudoobservables analysis and/or the global SM fit compared to the 2019 world average (orange band). Left: Current results (solid lines) compared to the previous 2018 fit (dashed lines). Right: Extraction expected at the FCC-ee --with central value arbitrarily set to $\alphasmZ = 0.12030$ and total (experimental, parametric, and theoretical, added in quadrature) uncertainties (outer parabola) and experimental uncertainties alone (inner parabola)-- compared to the present one from the combined Z data (blue line).}
\label{fig:alphas_Z}
\end{figure}

The extraction of $\alphasmZ$ is carried out with 1-D scans of this variable as a free parameter using single 
and combined observables with our updated version of \gfitter~2.2. The results from these fits are listed in Table~\ref{tab:alphas_Z}, and the corresponding $\Delta\chi^2$ profiles are plotted in Fig.~\ref{fig:alphas_Z}.
The solid lines represent the results of the present improved calculations and data, whereas the dashed lines are those obtained with \gfitter\ in 2018~\cite{Haller:2018nnx}. All new QCD couplings are clustered around $\alphasmZ = 0.1200$, whereas previously the extraction from $\so$ was about 2$\sigma$ lower (and also had larger uncertainties) than the average of the three, and that from $\Rlz$ was 1$\sigma$ above it. Among $\alphasmZ$ extractions, the most precise is that from $\Rlz$ (3.4\% uncertainty), followed by the ones from $\GZ$ (3.9\% uncertainty) and $\so$ (5.6\% uncertainty). The precision did not change appreciably compared to the previous $\GZ$ and $\Rlz$ results, but the extraction from the hadronic Z cross section has been improved by about 20\% thanks mostly to the updated LEP data. When combining various Z observables, their associated correlation matrix is used in the fit. Table~\ref{tab:alphas_Z} lists all results with their propagated uncertainties broken down into experimental, parametric, and theoretical sources. Our final values, $\alphasmZ = 0.1203 \pm 0.0028$ from the combined Z boson data, and $\alphasmZ = 0.1202 \pm 0.0028$ from the full SM fit, and the PDG electroweak fit result ($\alphasmZ = 0.1203 \pm 0.0028$)~\cite{Zyla:2020zbs} are all virtually identical now.

\begin{table}[htpb!]
\centering
\caption{Values of $\alphasmZ$ determined at N$^3$LO accuracy from $\Gamma_\mathrm{Z}^\mathrm{tot}$, $R_\mathrm{Z}$, and $\so$ individually, combined, as well as from a global SM fit, with propagated experimental, parametric, and theoretical uncertainties broken down. The last two rows list the expected values at the FCC-ee from all Z pseudoobservables combined and from the corresponding SM~fit.\vspace{0.2cm}\label{tab:alphas_Z}}
\tabcolsep=4.5mm
\begin{tabular}{lcccc}\hline
Observable & $\alphasmZ$   & \multicolumn{3}{c}{uncertainties}\\
     &     & exp.  &  param. & theor.      \\\hline
$\Gamma_\mathrm{Z}^\mathrm{tot}$ & $0.1192 \pm 0.0047$ & $\pm0.0046$ & $\pm0.0005$ & $\pm0.0008$ \\ 
$R_\mathrm{Z}$ & $0.1207 \pm 0.0041$ & $\pm0.0041$ & $\pm0.0001$ & $\pm0.0009$ \\ 
$\so$ & $0.1206 \pm 0.0068$ & $\pm0.0067$ & $\pm0.0004$ & $\pm0.0012$ \\ 
All above combined & $0.1203 \pm 0.0029$ & $\pm0.0029$ & $\pm0.0002$ & $\pm0.0008$ \\ 
Global SM fit & $0.1202 \pm 0.0028$ & $\pm0.0028$ & $\pm0.0002$ & $\pm0.0008$ \\ \hline
All combined (FCC-ee)  & $0.12030 \pm 0.00026$ & $\pm0.00013$ & $\pm0.00005$ & $\pm 0.00022$\\ 
Global SM fit (FCC-ee) & $0.12020 \pm 0.00026$ & $\pm0.00013$ & $\pm0.00005$ & $\pm 0.00022$\\\hline
\end{tabular}
\end{table}

At the FCC-ee, combining the $3\cdot10^{12}$ Z bosons decaying hadronically 
at the Z pole, and the $\mathcal{O}$(tens of keV)-accurate $\sqrts$ calibration obtained using resonant depolarization~\cite{Blondel:2021zix}, will provide measurements with unparalleled precision. The statistical uncertainties in the Z mass and width, today of $\pm$1.2~MeV and $\pm$2~MeV (dominated by the LEP beam energy calibration), will be reduced to below $\pm$4~keV and $\pm$7~keV respectively. 
Similarly, the statistical uncertainty in $\RZexp$ will be negligible and the measurement in the Z\,$\to\mu^+\mu^-$ final state alone, yielding an experimental precision of 0.001 from the knowledge of the detector acceptance, will suffice to reach an absolute (relative) uncertainty of 0.001 ($5\cdot10^{-5}$) on the ratio of the hadronic-to-leptonic partial Z widths. Thus, accounting for the dominant experimental systematic uncertainties at the FCC-ee, we expect: $\delta \MZ$~=~0.025--0.1~MeV, $\delta \GZ = 0.1$~MeV, $\delta\so = 4.0$~pb, and $\delta \Rlz  = 10^{-3}$ relative uncertainties~\cite{FCC:2018evy}. In addition, the QED coupling at the Z peak will be measured with a precision of $\delta \alpha  = 3\cdot10^{-5}$~\cite{Janot:2015gjr}, thereby also reducing the corresponding propagated parametric uncertainties. Implementing the latter uncertainties into our updated \gfitter\ setup, namely taking $\GZ = 2495.2 \pm 0.1$~MeV, $\so = 41\,494\pm 4$~pb, and $\Rlz = 20.7500 \pm 0.0010$, as well as $\MZ =  91.18760 \pm 0.00001$~GeV, and $\delta \alpha_{\mathrm{had}}^{(5)}(\MZ) = 0.0275300 \pm 0.0000009$, 
we derive the results listed in the last two rows of Table~\ref{tab:alphas_Z} where, the central $\alphasmZ$ value is arbitrarily set at the current SM global fit extraction.
The final uncertainties in the QCD coupling constant are reduced to the $\sim$0.1\% level, namely about three times smaller than the propagated theoretical uncertainties today. Theoretical developments in the years to come should further bring down the latter by a factor of four~\cite{Proceedings:2019vxr}. A final QCD coupling constant extraction at the FCC-ee with a two-permil total uncertainty is thereby reachable: $\alphasmZ = 0.12030 \pm 0.00013_\mathrm{exp} \pm 0.00005_\mathrm{par} \pm 0.00022_\mathrm{th}$ (Table~\ref{tab:alphas_Z}). Figure~\ref{fig:alphas_Z} (right) shows the $\Delta\chi^2$ parabola for the $\alphasmZ$ extraction from the Z boson data (or from the SM fit that is almost identical) expected at the FCC-ee (with the central value arbitrarily set to its present result), compared to the same extraction today (blue parabola) and to the 2019 world average (orange band). 
The large improvement, by more than a factor of ten, in the FCC-ee extraction of $\alphasmZ$ from the Z boson data (and its comparison to the similar extraction from the W boson, Fig.~\ref{fig:alphas_W} right) will enable searches for small deviations from the SM predictions that could signal the presence of new physics contributions. 

\clearpage
\section{\texorpdfstring{\boldmath$\alphasmZ$}{alphasmZ} from hadronic final-states in $\epem$ collisions}
\label{sec:ee}

\subsection{Hadronic vacuum polarization function, $R$-ratio, and the strong coupling
\protect\footnote{A\lowercase{uthor:} A. V. N\lowercase{esterenko} (JINR, D\lowercase{ubna})}}
\label{sec:ee_Rratio}


The process of electron-positron annihilation into hadrons plays a 
distinctive role in elementary particle physics. This is primarily 
caused by the remarkable fact that its theoretical description 
requires no phenomenological models of hadronization, which forms 
the experimentally detected final-state particles. In~turn, this 
feature makes it possible to extract the key parameters of the theory 
from pertinent experimental data in a model-independent way, 
thereby constituting one of the cleanest methods of their 
determination.

In the theoretical studies of $\epem$~annihilation into 
hadrons, one needs to operate with the physical observable that 
depends on the timelike kinematic variable (namely, the 
center-of-mass energy squared~\mbox{$s=q^2>0$}), whereas such basic 
tools as the perturbative technique and the renormalization group~(RG) 
method are directly applicable only to quantities that depend 
on the spacelike kinematic variable~\mbox{$Q^2 = -q^2 > 0$}. The 
proper description of the strong interactions in the timelike domain 
substantially relies on the corresponding dispersion relations, 
which provide the physically consistent way to interrelate the 
``timelike'' experimentally measurable observables (such~as the 
\mbox{$R$-ratio} of $\epem$~annihilation into hadrons) 
with the ``spacelike'' theoretically computable quantities (such~as 
the hadronic vacuum polarization function and the Adler function), 
see, \eg, Ref.~\cite{Nesterenko:2016pmx} and references therein for 
details.\\

The hadronic vacuum polarization function~$\Pi(q^2)$ is defined as 
the scalar part of the hadronic vacuum polarization tensor
\begin{equation}
\label{AVN:P_Def}
\Pi_{\mu\nu}(q^2) = i\!\int\!d^4x\,e^{i q x} \bigl\langle 0 \bigl|\,
T\!\left\{J_{\mu}(x)\, J_{\nu}(0)\right\} \bigr| 0 \bigr\rangle =
\frac{i}{12\pi^2} (q_{\mu}q_{\nu} - g_{\mu\nu}q^2) \Pi(q^2),
\qquad
q^2<0.
\end{equation}
The physical kinematic restrictions on the process on hand define
the location of the cut of the function~$\Pi(q^2)$ in the complex
\mbox{$q^2$-plane}, that enables one to write down 
the corresponding dispersion relation. In~particular, as discussed 
in, \eg, Ref.~\cite{Feynman:1973xc}, for the processes involving 
final state hadrons the function~$\Pi(q^2)$~(\ref{AVN:P_Def}) 
possesses the only cut along the positive semiaxis of real~$q^2$ 
starting at the hadronic production threshold~$q^2 \ge 4 m_{\pi}^{2}$, 
that~implies
\begin{equation}
\label{AVN:P_Disp}
\Pi(q^{2}) - \Pi(q_{0}^{2}) = (q^2 - q_0^2) 
\!\int\limits_{4 m_{\pi}^{2}}^{\infty}\!
\frac{R(\sigma)}{(\sigma-q^2)(\sigma-q_0^2)}\, d\sigma,
\end{equation}
where
\begin{equation}
\label{AVN:R_Def}
R(s) = \frac{1}{2 \pi i} \lim_{\varepsilon \to 0_{+}}
\Bigl[\Pi(s+i\varepsilon) - \Pi(s-i\varepsilon)\Bigr]\! =
\frac{1}{\pi}\,{\rm Im} \!\lim_{\varepsilon \to 0_{+}}\!
\Pi(s+i\varepsilon).
\end{equation}
The function~$R(s)$~(\ref{AVN:R_Def}) is commonly identified with 
the so-called $R$-ratio of $\epem$~annihilation into hadrons 
$R(s) = \sigma(e^{+}e^{-} \!\to \text{hadrons}; s)/\sigma(e^{+}e^{-} 
\!\to \mu^{+}\mu^{-}; s)$. For the practical purposes it proves to 
be convenient to deal with the Adler function~\cite{Adler:1974gd}
\begin{equation}
\label{AVN:Adler_Def}
D(Q^2) = - \frac{d\, \Pi(-Q^2)}{d \ln Q^2},
\qquad
Q^2 = -q^2 > 0.
\end{equation}
The corresponding dispersion relation~\cite{Adler:1974gd}
\begin{equation}
\label{AVN:Adler_Disp}
D(Q^2) = Q^2 \int\limits_{4 m_{\pi}^{2}}^{\infty}
\frac{R(\sigma)}{(\sigma+Q^2)^2}\, d\sigma
\end{equation}
directly follows from Eqs.~(\ref{AVN:P_Disp}) and~(\ref{AVN:Adler_Def}) and
enables one to extract the experimental prediction for the Adler
function from the respective data on the~$R$-ratio. In~turn,
the theoretical expression for the function~$R(s)$ can be obtained by
integrating Eq.~(\ref{AVN:Adler_Def}) in finite limits, 
namely~\cite{Radyushkin:1982kg, Krasnikov:1982fx}
\begin{equation}
\label{AVN:R_Disp2}
R(s) =  \frac{1}{2 \pi i} \lim_{\varepsilon \to 0_{+}}
\int\limits_{s + i \varepsilon}^{s - i \varepsilon}
D(-\zeta)\,\frac{d \zeta}{\zeta},
\end{equation}
where the integration contour in the complex $\zeta$-plane lies in the region 
of analyticity of the integrand. At~the same time, Eq.~(\ref{AVN:Adler_Def}) 
additionally provides the relation that expresses the hadronic vacuum 
polarization function in terms of the Adler function, 
specifically~\cite{Moorhouse:1976qq, Pennington:1981cw, 
Pennington:1983rz, Pivovarov:1991bi}
\begin{equation}
\label{AVN:P_Disp2}
\Pi(-Q^2) - \Pi(-Q_0^2) = - \int\limits_{Q_0^2}^{Q^2} D(\zeta)
\frac{d \zeta}{\zeta},
\end{equation}
where $Q^{2}=-q^{2}>0$ and~$Q_{0}^{2}=-q_{0}^{2}>0$ denote, respectively, 
the spacelike kinematic variable and the subtraction point.

Basically, Eqs.~(\ref{AVN:P_Disp})--(\ref{AVN:P_Disp2}) constitute the complete set 
of relations, which mutually express the functions~$\Pi(q^2)$, $D(Q^2)$, and~$R(s)$ 
in terms of each other. The~derivation of the foregoing relations, being based only 
on the kinematics of the process on hand, requires neither additional approximations
nor model-dependent phenomenological assumptions. It~is worthwhile to note also 
that the dispersion relations~(\ref{AVN:P_Disp})--(\ref{AVN:P_Disp2}) impose 
a number of stringent physical intrinsically nonperturbative constraints on the 
functions~$\Pi(q^2)$, $D(Q^2)$, and~$R(s)$, that should definitely be accounted for 
when one reaches the limits of applicability of the perturbative approach, see, in~particular, 
Refs.~\cite{Nesterenko:2016pmx, Nesterenko:2013vja, Nesterenko:2014txa} for a detailed 
discussion of this issue. The~nonperturbative aspects of the strong interactions will 
be disregarded hereinafter and the massless limit~$m_{\pi}=0$ will be assumed in~what 
follows.\\

To~calculate the $R$-ratio of $\epem$~annihilation into hadrons one usually
starts with the perturbative expression for the hadronic vacuum polarization function
\begin{equation}
\label{AVN:PpertDef}
\Pi^{(\ell)}\bigl(q^2,\mu^2,a^{}_{{\rm s}}\bigr) =
\sum_{j=0}^{\ell}\Bigl[a^{(\ell)}_{{\rm s}}(\mu^2)\Bigr]^{j}
\sum_{k=0}^{j+1}\Pi_{j,k}\,\ln^{k}\!\!\left(\frac{\mu^2}{-q^2}\right)\!,
\qquad
q^2 \to -\infty.
\end{equation}
In this equation $\ell$~specifies the loop level, $q^2<0$~denotes the 
spacelike kinematic variable, $\mu^2>0$~is the renormalization scale, 
$\alpha_{{\rm s}}=g^{2}/(4\pi)$ stands for the strong coupling,
$a^{(\ell)}_{{\rm s}}(\mu^2) =
\alpha^{(\ell)}_{{\rm s}}(\mu^2)\,\beta_{0}/(4\pi)$, 
$\beta_{0} = 11 - 2\Nf/3$ denotes the one-loop $\beta$~function
perturbative expansion coefficient, $\Nf$~is the number of
active flavours, and the~common prefactor $N_{\text{c}}\sum_{f=1}^{\Nf} Q_{f}^{2}$ is
omitted throughout, where \mbox{$N_{\text{c}}=3$}~stands for the number of colours and
$Q_{f}$~denotes the electric charge of $f$-th quark. In~particular, at the one-loop 
level ($\ell=1$) Eq.~(\ref{AVN:PpertDef}) reads
\begin{equation}
\label{AVN:Ppert1L}
\Pi^{(1)}\bigl(q^2,\mu^2,a_{{\rm s}}\bigr) = 
\frac{5}{3} + \ln\biggl(\frac{\mu^2}{-q^2}\biggr)\!
+ a^{(1)}_{{\rm s}}(\mu^2)\biggl(\frac{4}{\beta_{0}}\biggr)\! \biggl[
\frac{55}{12} - 4\zeta(3) + \ln\biggl(\frac{\mu^2}{-q^2}\biggr)\! \biggr],
\qquad
q^2 \to -\infty.
\end{equation}
As~mentioned earlier, in practice it is convenient to employ the Adler
function~(\ref{AVN:Adler_Def}), which, contrary to the hadronic vacuum 
polarization function~(\ref{AVN:P_Def}), is an RG-invariant quantity.
Specifically, Eqs.~(\ref{AVN:Adler_Def}) and~(\ref{AVN:PpertDef}) imply 
that at the $\ell$-loop level the perturbative expression for the Adler
function~reads
\begin{equation}
\label{AVN:DpertMu}
D^{(\ell)}\bigl(Q^2,\mu^2,a_{{\rm s}}\bigr) =
\sum_{j=0}^{\ell}\Bigl[a^{(\ell)}_{{\rm s}}(\mu^2)\Bigr]^{j}
\sum_{k=0}^{j+1}k\Pi_{j,k}\,\ln^{k-1}\!\!\left(\frac{\mu^2}{Q^2}\right)\!,
\qquad
Q^2 \to \infty.
\end{equation}
Note that the hadronic vacuum polarization function~(\ref{AVN:P_Def}) and 
the Adler function~(\ref{AVN:Adler_Def}) satisfy, respectively, 
the inhomogeneous and homogeneous RG~equations, which, in~turn, enable 
one to express the higher-order coefficients~$\Pi_{j,k}$ 
entering Eqs.~(\ref{AVN:PpertDef}) and~(\ref{AVN:DpertMu}) in terms of 
the lower-order ones. In~particular, such RG~relations have been 
obtained at the first few loop levels in Refs.~\cite{Baikov:2009uw, 
Baikov:2012zm, Beneke:2008ad}, at the higher loop levels 
in~Ref.~\cite{Nesterenko:2019rag}, and at an arbitrary loop level
(in~a compact recurrent and unfolded explicit forms)
in~Ref.~\cite{Nesterenko:2020nol}.

The native choice of the renormalization scale~$\mu^2=Q^2$ (that amounts to the 
RG~summation in the spacelike domain) casts the Adler function~(\ref{AVN:DpertMu}) 
to a well-known form ($\Pi_{0,1}=1$)
\begin{equation}
\label{AVN:AdlerPert}
D^{(\ell)}(Q^2) =
\sum_{j=0}^{\ell}\Pi_{j,1}\Bigl[a^{(\ell)}_{{\rm s}}(Q^2)\Bigr]^{j} =
1 + d^{(\ell)}(Q^2),
\qquad
d^{(\ell)}(Q^2) = \!\sum_{j=1}^{\ell} d_{j}\!
\left[a^{(\ell)}_{{\rm s}}(Q^2)\right]^{j}\!,
\qquad
\Pi_{j,1}=d_{j}.
\end{equation}
It~is worthwhile to note here that a~general choice of the renormalization 
scale~$\mu^2=c\,Q^2$ (with~$c \neq 1$ being a~positive constant) retains in 
the resulting expression for the Adler function~$D^{(\ell)}(Q^2)$ all the 
terms proportional to the higher-order coefficients~$\Pi_{j,k}$ appearing 
on the right-hand side of~Eq.~(\ref{AVN:DpertMu}). The function 
$a^{(\ell)}_{{\rm s}}(Q^2)$ entering Eq.~(\ref{AVN:AdlerPert}) 
can be represented as the double sum
\begin{equation}
\label{AVN:AItGen}
a^{(\ell)}_{{\rm s}}(Q^2) =
\sum_{n=1}^{\ell}\sum_{m=0}^{n-1} b^{m}_{n}\,
\frac{\ln^{m}(\ln z)}{\ln^n z},
\qquad
z=\frac{Q^2}{\Lambda^2},
\end{equation}
where $\Lambda$ is the QCD scale parameter and~$b^{m}_{n}$ (the 
integer superscript~$m$ is not to be confused with a respective power) 
stands for the combination of the $\beta$~function perturbative 
expansion coefficients (in~particular, \mbox{$b^{0}_{1}=1$}, 
$b^{0}_{2}=0$, $b^{1}_{2}=-\beta_{1}/\beta_{0}^{2}$, see,~\eg, 
Appendix~A of Ref.~\cite{Nesterenko:2016pmx}). For~example, at the 
one-loop level ($\ell=1$) the Adler function~(\ref{AVN:AdlerPert}) 
takes a quite simple form
\begin{equation}
\label{AVN:Dpert1L}
D^{(1)}(Q^2) = 1 + d_{1}a^{(1)}_{{\rm s}}(Q^2),
\qquad
a^{(1)}_{{\rm s}}(Q^2) = \frac{1}{\ln(Q^2/\Lambda^2)},
\qquad
d_{1} = \frac{4}{\beta_{0}}.
\end{equation}
The Adler function perturbative expansion coefficients~$d_{j}$
entering Eq.~(\ref{AVN:AdlerPert}) were calculated up to the 
fourth order in the strong coupling ($1 \le j \le 4$), see 
Refs.~\cite{Baikov:2008jh, Baikov:2010je, Baikov:2012zn} and 
references therein, whereas the $\beta$~function perturbative 
expansion coefficients~$\beta_{j}$ appearing in Eq.~(\ref{AVN:AItGen}) 
are available up to the five-loop level ($0 \le j \le 4$), see 
Refs.~\cite{Baikov:2016tgj, Herzog:2017ohr, Luthe:2017ttg, 
Chetyrkin:2017bjc} and references therein.

In~turn, the dependence of the hadronic vacuum polarization function~(\ref{AVN:PpertDef}) 
on the renormalization scale can be eliminated in the following~way (see, in particular, 
Refs.~\cite{Moorhouse:1976qq, Pennington:1981cw, Pennington:1983rz}). Namely, for~this 
purpose one first calculates the corresponding Adler function~(\ref{AVN:Adler_Def}), 
then performs the RG~summation, and then employs the relation~(\ref{AVN:P_Disp2}), that 
yields at the one-loop level~\cite{Moorhouse:1976qq, Pennington:1981cw, Pennington:1983rz, 
Pivovarov:1991bi, Nesterenko:2014txa}
\begin{equation}
\label{AVN:DP1L}
\Pi^{(1)}(-Q^2) - \Pi^{(1)}(-Q_0^2) = -\ln\!\biggl(\frac{Q^2}{Q_0^2}\biggr)
- d_{1}\ln\!\Biggl[\frac{a^{(1)}_{{\rm s}}(Q_{0}^{2})}{a^{(1)}_{{\rm s}}(Q^{2})}\Biggr]\!,
\qquad
Q^{2}=-q^{2}>0,
\qquad
Q_{0}^{2}=-q_{0}^{2}>0,
\end{equation}
where~$a^{(1)}_{{\rm s}}(Q^{2})$ and~$d_{1}$ are given in~Eq.~(\ref{AVN:Dpert1L}). 
At~an arbitrary loop level the corresponding expression for the hadronic vacuum
polarization function has been obtained in Ref.~\cite{Nesterenko:2019rag}:
\begin{equation}
\label{AVN:DPHL}
\Pi^{(\ell)}(-Q^{2}) - \Pi^{(\ell)}(-Q_{0}^{2}) = 
-\ln\!\left(\frac{Q^{2}}{Q_{0}^{2}}\right) + 
\sum_{j=1}^{\ell}d_{j} \Bigl[
p^{(\ell)}_{j}(Q^{2}) - p^{(\ell)}_{j}(Q_{0}^{2})\Bigr],
\end{equation}
where
\begin{equation}
p^{(\ell)}_{j}(Q^{2}) =
\sum_{n_{1}=1}^{\ell}\ldots\sum_{n_{j}=1}^{\ell}
\sum_{m_{1}=0}^{n_{1}-1}\ldots\sum_{m_{j}=0}^{n_{j}-1}
\Biggl(\prod_{i=1}^{j}b^{m_{i}}_{n_{i}}\Biggr)
J\mbox{{\tiny$\!$}}\Biggl(\!Q^{2},\sum_{i=1}^{j}n_{i},\sum_{i=1}^{j}m_{i}\!\Biggr),
\end{equation}
\begin{equation}
J(Q^{2},n,m) =
\begin{cases}
\displaystyle
-\frac{\ln^{m+1}(\ln z)}{m+1},&\qquad  \text{if $\, n=1$},\\[4mm]
\displaystyle
\sum_{k=0}^{m}\frac{m!}{k!}(n-1)^{k-m-1}\,\frac{\ln^{k}(\ln z)}{\ln^{n-1}z},
&\qquad \text{if $\, n \ge 2$},
\end{cases}
\end{equation}
$z=Q^{2}/\Lambda^{2}$, and the coefficients~$d_{j}$ and~$b^{m}_{n}$ 
are specified in Eqs.~(\ref{AVN:AdlerPert}) and~(\ref{AVN:AItGen}), 
respectively, see Ref.~\cite{Nesterenko:2019rag} for the details.\\

At~this point there are two equivalent methods to calculate the $R$-ratio of 
$\epem$~annihilation into hadrons. Specifically, the first one consists in applying 
the relation~(\ref{AVN:R_Disp2}) to the Adler function~$D(Q^2)$~(\ref{AVN:AdlerPert}).
This method eventually leads to an integral representation for the function~$R(s)$, 
which involves the so-called spectral function, the latter being the discontinuity 
of the strong correction to the Adler function~$d(Q^2)$~(\ref{AVN:AdlerPert}) across 
the physical cut~$Q^2<0$. In~turn, the second method to calculate the $R$-ratio 
consists in applying the relation~(\ref{AVN:R_Def}) to the hadronic vacuum polarization 
function~(\ref{AVN:DPHL}). At~the one-loop level the $R$-ratio of $\epem$~annihilation 
into hadrons takes the following form (note that this expression first appeared 
in Ref.~\cite{Schrempp:1980cz} and only afterwards was derived in 
Refs.~\cite{Radyushkin:1982kg, Pivovarov:1991bi, Jones:1995rd, Milton:1996fc, Milton:1998wi}):
\begin{equation}
\label{AVN:Rprop1L}
R^{(1)}(s) = 1 + d_{1}A^{(1)}_{{\rm TL},1}(s),
\qquad
A^{(1)}_{{\rm TL},1}(s) = \frac{1}{2} -
\frac{1}{\pi}\arctan\biggl(\frac{\ln w}{\pi}\biggr)\mbox{{\tiny$\!$}},
\qquad
w=\frac{s}{\Lambda^2},
\end{equation}
where the coefficient~$d_{1}$ is given in~Eq.~(\ref{AVN:Dpert1L}) and it is 
assumed that~$\arctan(x)$ is a monotonic nondecreasing function of its argument 
[namely, $-\pi/2 \le \arctan(x) \le \pi/2$ for $-\infty < x < \infty$].

Basically, the first method of calculation of the $R$-ratio becomes somewhat inconvenient 
at the higher-loop levels. In~particular, since the required spectral function 
turns out to be rather cumbrous beyond the one-loop level, its integration can, 
in~general, be performed only by making use of numerical methods. Moreover, at 
the higher-loop levels the explicit calculation of the spectral function represents 
a rather demanding task, whereas its numerical evaluation needs a lot of computation 
resources and essentially slows down the overall computation process. Nonetheless,
the required spectral function has explicitly been calculated at the first few loop
levels in, \eg, Refs.~\cite{Solovtsov:1999in, Nesterenko:2010je, Nesterenko:2011aa} 
and at an arbitrary loop level in Refs.~\cite{Nesterenko:2016pmx, Nesterenko:2017wpb},
that, in turn, facilitates the numerical calculation of the $R$–ratio within the 
first method.

The explicit expression for the $R$-ratio of $\epem$~annihilation into hadrons, 
which properly accounts for all the effects due to continuation of the spacelike 
perturbative results into the timelike domain, has recently been obtained at an 
arbitrary loop level within the second method in Ref.~\cite{Nesterenko:2019rag},
namely
\begin{equation}
\label{AVN:Rprop}
R^{(\ell)}(s) = 1 + r^{(\ell)}(s), 
\qquad
r^{(\ell)}(s) = \sum\limits_{j=1}^{\ell} d_{j}\,
A^{(\ell)}_{{\rm TL},j}(s).
\end{equation}
In this equation~$d_{j}$ stand for the Adler function perturbative 
expansion coefficients~(\ref{AVN:AdlerPert}),
\begin{equation}
A^{(\ell)}_{{\rm TL},j}(s) =
\sum_{n_{1}=1}^{\ell}\ldots\sum_{n_{j}=1}^{\ell}
\sum_{m_{1}=0}^{n_{1}-1}\ldots\sum_{m_{j}=0}^{n_{j}-1}
\Biggl(\prod_{i=1}^{j}b^{m_{i}}_{n_{i}}\Biggr)\mbox{{\tiny$\,$}}
T\mbox{{\tiny$\!$}}\Biggl(\!s,\sum_{i=1}^{j}n_{i},\sum_{i=1}^{j}m_{i}\!\Biggr)
\end{equation}
denotes the $\ell$-loop $j$-th~order ``timelike'' effective expansion
function (which constitutes the corresponding continuation of the $j$-th 
power of $\ell$-loop function~$\bigl[a^{(\ell)}_{{\rm s}}(Q^2)\bigr]^{j}$ 
from spacelike into the timelike domain), the coefficients~$b^{m}_{n}$ are specified 
in~Eq.~(\ref{AVN:AItGen}),
\begin{equation}
T(s,n,m) =
\begin{cases}
\displaystyle -V_{0}^{1}(s),& \text{if $\, n=1 \,$ and $\, m=0$},\\[2.5mm]
\displaystyle \sum_{k=0}^{m}\frac{m!}{k!}(n-1)^{k-m-1}\,V_{n-1}^{k}(s),
\quad& \text{if $\, n \ge 2$},
\end{cases}
\end{equation}
\begin{equation}
V_{n}^{m}(s) =
\begin{cases}
0,\quad& \text{if $\, n=0 \,$ and $\, m = 0$}, \\[1.25mm]
v_{0}^{m}(s),\quad& \text{if $\, n=0 \,$ and $\, m \ge 1$}, \\[1.25mm]
v_{n}^{0}(s),& \text{if $\, n \ge 1 \,$ and $\, m=0$}, \\[1.25mm]
v_{n}^{0}(s) u_{0}^{m}(s) +
u_{n}^{0}(s) v_{0}^{m}(s),\quad& \text{if $\, n \ge 1 \,$ and $\, m \ge 1$},
\end{cases}
\end{equation}
\begin{align}
v_{0}^{m}(s) & = \!\sum\limits_{k=0}^{K(m)}\!\binom{m}{2k+1} (-1)^{k+1}\pi^{2k}
\Bigl[L_{1}(y)\Bigr]^{m-2k-1}\, \Bigl[L_{2}(y)\Bigr]^{2k+1},
\\[1.5mm]
v_{n}^{0}(s) & = \frac{1}{(y^2 + \pi^2)^{n}}
\sum_{k=0}^{K(n)} \!\binom{n}{2k+1} (-1)^{k} \pi^{2k} y^{n-2k-1},
\end{align}
\begin{align}
u_{0}^{m}(s) & = \!\sum\limits_{k=0}^{K(m+1)}\!\binom{m}{2k} (-1)^{k}\pi^{2k}
\Bigl[L_{1}(y)\Bigr]^{m-2k}\, \Bigl[L_{2}(y)\Bigr]^{2k},
\\[1.5mm]
u_{n}^{0}(s) & = \frac{1}{(y^2 + \pi^2)^{n}}
\sum_{k=0}^{K(n+1)} \!\binom{n}{2k} (-1)^{k} \pi^{2k} y^{n-2k},
\end{align}
\begin{equation}
\label{AVN:KDef}
L_{1}(y) = \ln\!\sqrt{y^{2}+\pi^{2}},
\quad
L_{2}(y) = \frac{1}{2} - \frac{1}{\pi}\arctan\biggl(\frac{y}{\pi}\biggr),
\quad
\binom{n}{m} = \frac{n!}{m!\,(n-m)!},
\quad
K(n) = \frac{n-2}{2} + \frac{n \;\mbox{mod}\; 2}{2},
\end{equation}
$(n \;\mbox{mod}\; m)$ denotes the remainder on division of~$n$ by~$m$,
and~$y = \ln(s/\Lambda^2)$, see~Refs.~\cite{Nesterenko:2017wpb, 
Nesterenko:2019rag} for the details.\\

It~is worthwhile to note that a commonly employed way of calculation
of the $R$-ratio of $\epem$~annihilation into hadrons, being different
from the two equivalent methods described above, leads to an incomplete
result for the function~$R(s)$. In~particular, here one applies the
relation~(\ref{AVN:R_Def}) directly to Eq.~(\ref{AVN:PpertDef}) and
then assigns the renormalization scale~\mbox{$\mu^2=|s|$} (that
factually amounts to an incomplete RG~summation in the timelike domain, 
see, \eg, Refs.~\cite{Moorhouse:1976qq, Pennington:1981cw, 
Pennington:1983rz, Pivovarov:1991bi, Nesterenko:2017wpb, Nesterenko:2019rag,
Nesterenko:2020nol} and references therein), that yields~\cite{ParticleDataGroup:2020ssz}
\begin{equation}
\label{AVN:Rappr}
R^{(\ell)}_{{\rm appr}}(s) = 1 + r^{(\ell)}_{{\rm appr}}(s),
\qquad
r^{(\ell)}_{{\rm appr}}(s) =
\sum_{j=1}^{\ell} r_{j} \Bigl[a^{(\ell)}_{{\rm s}}(|s|)\Bigr]^{j},
\qquad
r_{j} = d_{j} - \delta_{j}.
\end{equation}
In this equation the function $a^{(\ell)}_{{\rm s}}(|s|)$ is given
in~Eq.~(\ref{AVN:AItGen}), $d_{j}$ stand for the Adler function perturbative 
expansion coefficients~(\ref{AVN:AdlerPert}), and $\delta_{j}$~embody the
contributions of the so-called~$\pi^2$-terms, which play a significant role 
in the studies of the process on~hand~\cite{Bjorken:1989xw}. 
At~the first two-loop levels the coefficients~$\delta_{j}$~(\ref{AVN:Rappr})
vanish ($\delta_{1} = \delta_{2} = 0$) and the first several non-vanishing
coefficients~$\delta_{j}$ were reported in Refs.~\cite{Bjorken:1989xw,
Kataev:1995vh, Prosperi:2006hx, Nesterenko:2016pmx, Nesterenko:2017wpb,
Nesterenko:2019rag}. The explicit expression for the 
coefficients~$\delta_{j}$~(\ref{AVN:Rappr}) has recently been obtained 
at an arbitrary loop level in~Ref.~\cite{Nesterenko:2020nol}:
\begin{align}
\delta_{j} = & -\sum_{k=1}^{K(j)}
\frac{(-2\pi^{2})^{k}}{(2k+1)!}
\sum_{i_{1}=2(k-1)+1}^{j-2}
\underbrace{%
\sum_{i_{2}=2(k-2)+1}^{i_{1}-2}   \ldots
\sum_{i_{n}=2(k-n)+1}^{i_{n-1}-2} \ldots
\sum_{i_{k}=1}^{i_{k-1}-2}%
}_{\mbox{{\small$(k-1)$~sums}}}
(j+i_{1})i_{1} \times
\nonumber \\[0.5mm] & \times
\underbrace{%
(i_{1}+i_{2})i_{2} \times \ldots \times
(i_{n-1}+i_{n})i_{n} \times \ldots \times
(i_{k-1}+i_{k})i_{k}%
}_{\mbox{{\small$(k-1)$~products}}}
\times
\nonumber \\[0.5mm] & \times
\mathfrak{B}_{j-i_{1}-2}
\underbrace{%
\mathfrak{B}_{i_{1}-i_{2}-2} \ldots
\mathfrak{B}_{i_{n-1}-i_{n}-2} \ldots
\mathfrak{B}_{i_{k-1}-i_{k}-2}%
}_{\mbox{{\small$(k-1)$~terms}}}
d_{i_{k}},
\quad
\mathfrak{B}_{n} = \frac{1}{4} \sum_{i=0}^{n} B_{i}\,B_{n-i},
\quad
B_{i} = \frac{\beta_{i}}{\beta_{0}^{i+1}},
\quad
j \ge 3,
\end{align}
where the coefficients~$d_{j}$ are specified in Eq.~(\ref{AVN:AdlerPert}),
$\beta_{j}$~denote the $\beta$~function perturbative expansion coefficients, and
the function~$K(n)$ is defined in Eq.~(\ref{AVN:KDef}), see Ref.~\cite{Nesterenko:2020nol} 
for the details.

It is necessary to emphasize that, as argued in, \eg, 
Refs.~\cite{Moorhouse:1976qq, Pennington:1981cw, Pennington:1983rz, 
Pivovarov:1991bi}, the effects due to continuation of the spacelike 
perturbative results into the timelike domain are only partially 
accounted for in Eq.~(\ref{AVN:Rappr}) by the coefficients~$\delta_j$, 
whereas the ignorance (complete or partial) of such effects may yield 
misleading results. In~particular, it was shown~\cite{Nesterenko:2017wpb, 
Nesterenko:2019rag, Nesterenko:2020nol} that the 
approximation~$R_{{\rm appr}}(s)$~(\ref{AVN:Rappr}) factually constitutes 
the truncated re-expansion of the proper expression~$R(s)$~(\ref{AVN:Rprop}) 
at high energies and the validity range of such re-expansion is strictly 
limited to~$\sqrt{s} > \Lambda\exp(\pi/2) \simeq 4.81\,\Lambda$. Moreover, 
the contribution of a given order to the proper expression for the 
\mbox{$R$-ratio}~(\ref{AVN:Rprop}) appears to be redistributed over the 
higher-order terms in its approximate form~(\ref{AVN:Rappr}), thereby 
substantially amplifying them. In turn, this makes the loop convergence of 
a commonly employed approximation of the $R$-ratio~(\ref{AVN:Rappr}) much 
worse than that of its proper form~(\ref{AVN:Rprop}) and increases the 
resulting theoretical uncertainty of the strong coupling and the QCD scale 
parameter associated with the higher-loop perturbative corrections 
disregarded in Eq.~(\ref{AVN:Rappr}). Basically, the aforementioned 
truncation of the re-expansion of the proper expression for the 
$R$-ratio~(\ref{AVN:Rprop}) neglects all the higher-order $\pi^2$-terms 
in Eq.~(\ref{AVN:Rappr}), though the latter may not necessarily be small 
enough to be safely neglected. Specifically, it~was shown that the 
higher-order $\pi^2$-terms omitted in a commonly employed 
approximation~$R_{{\rm appr}}(s)$~(\ref{AVN:Rappr}) can produce a considerable 
effect on the determination of the strong coupling and the QCD scale parameter 
from the experimental data on the $R$-ratio, see Refs.~\cite{Nesterenko:2017wpb, 
Nesterenko:2019rag, Nesterenko:2020nol} and references therein for the details.


\subsection{ \texorpdfstring{$\alphasmZ$}{alphasmZ} from soft parton fragmentation functions
\protect\footnote{A\lowercase{uthors:} R.~P\lowercase{erez-}R\lowercase{amos} (DRII-IPSA \& LPTHE, P\lowercase{aris}), D. \lowercase{d'}E\lowercase{nterria} (CERN)}}

\label{sec:FFs}

We summarize a derivation of the QCD coupling $\alphas$ from the energy evolution of the moments of the parton-to-hadron fragmentation functions (FFs) at low hadron Feynman momentum fraction $z$. In Refs.~\cite{Perez-Ramos:2013eba,dEnterria:2014lmi,dEnterria:2014xmb,Perez-Ramos:2014pua,dEnterria:2015ljg}, the energy evolution of the moments of the parton-to-hadron FFs were computed up to approximate next-next-to-leading-order (NNLO$^\star$) fixed-order including next-to-next-to-leading-log (NNLL) resummation corrections. A fit to the corresponding experimental jet data from $\epem$ and deep-inelastic e$^\pm,\nu$-p collisions, to the NNLO$^*$+NNLL predictions yields $\alphasmZ = 0.1205 \pm 0.0010\,\mathrm{(exp)}\pm 0.0022$\,(th), in good agreement with the current $\alphas$ world average. Forthcoming prospects based on full-NNLO calculations are discussed.\\

\begin{figure}[!htpb]
\centering
\includegraphics[width=0.51\linewidth]{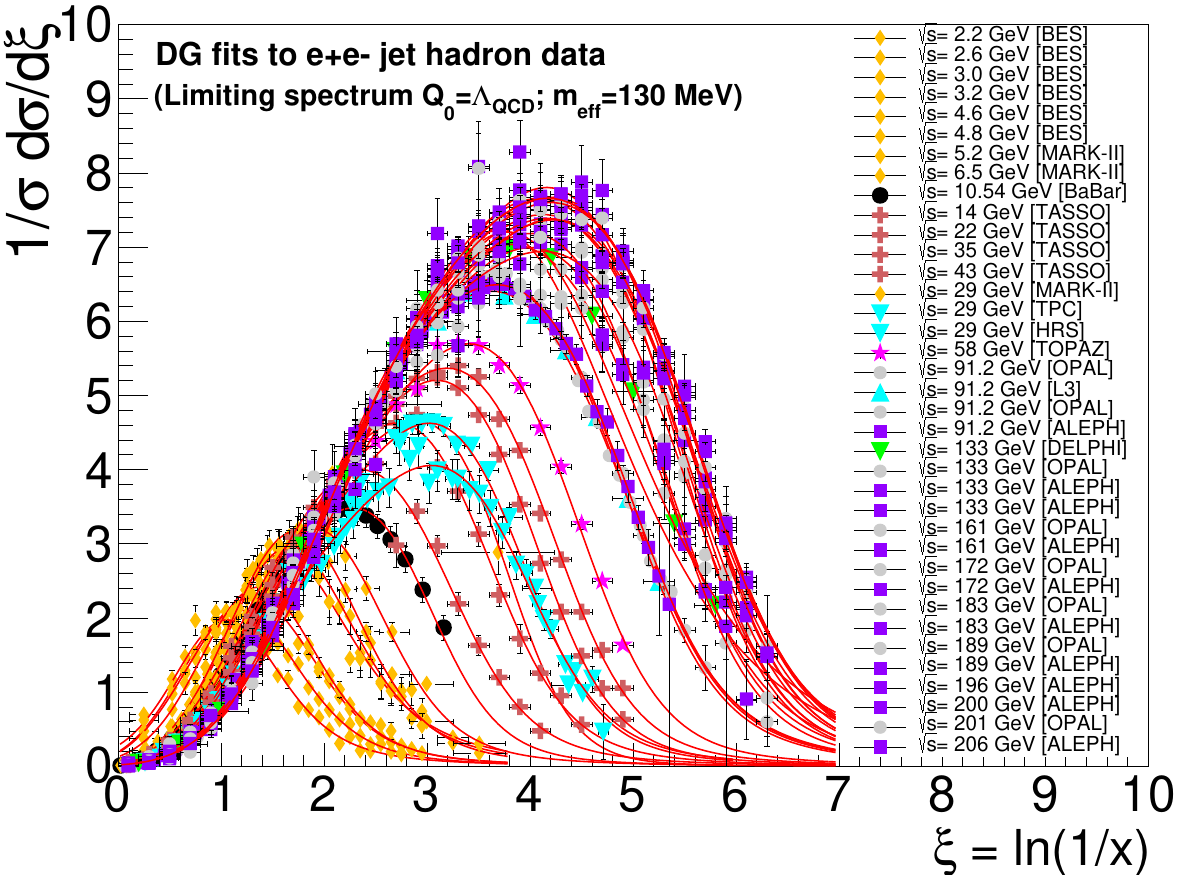}
\includegraphics[width=0.48\columnwidth]{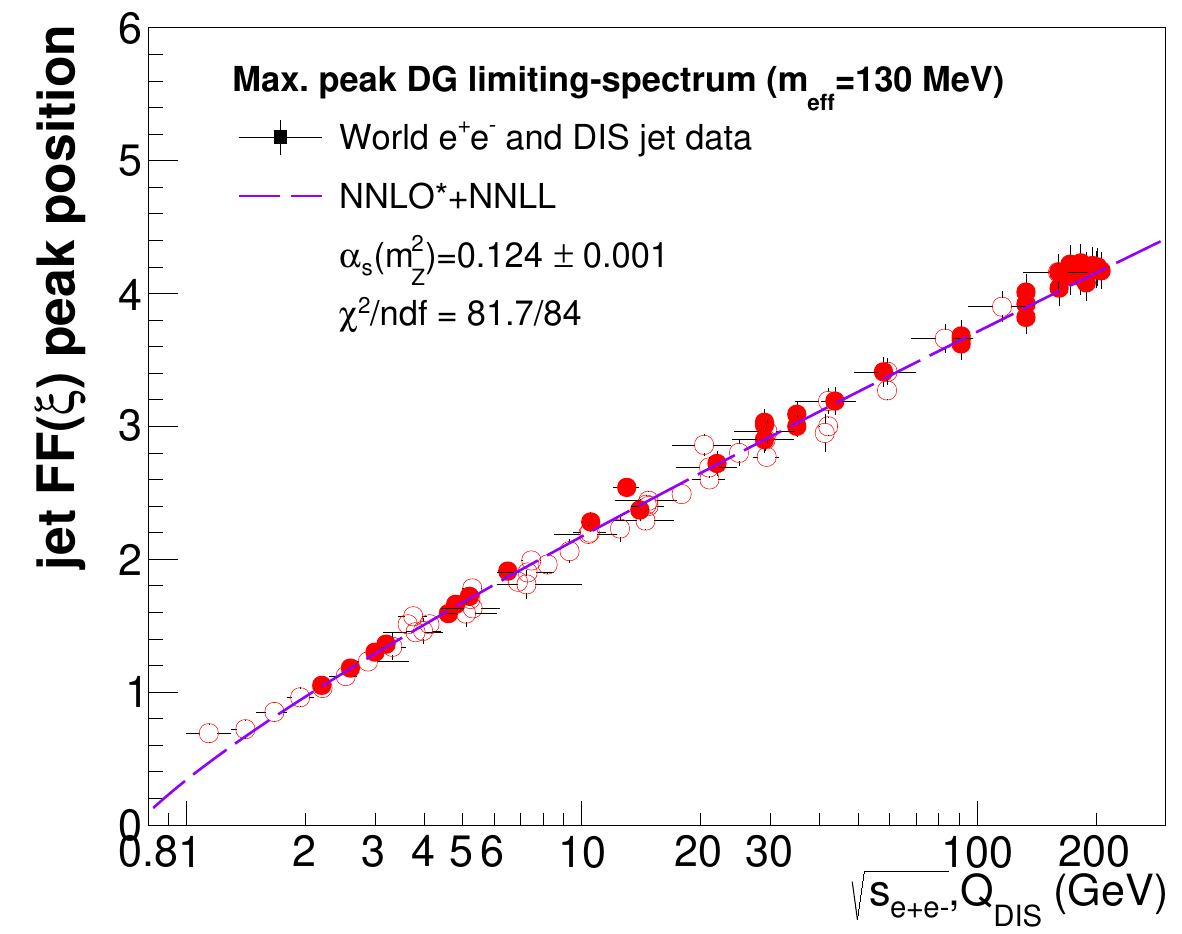}
\caption[]{Left: HBP charged-hadron spectra in jets as a function of $\xi=\ln(1/z)$ measured in $\epem$ at $\sqrts\approx$~2--200~GeV, fitted to a DG distribution. Right: Energy evolution of the peak position of the DG measured in $\epem$ and DIS data (open and closed symbols) fitted to the NNLO$^\star$+NNLL predictions (dashed curve).
\label{fig:FFs}
}
\end{figure}

The conversion of a quark and gluon (collectively called partons) into a final jet of hadrons is driven by soft and collinear gluon bremsstrahlung~\cite{Dokshitzer:1982ia} followed by the final transformation into hadrons of the last partons produced in the QCD shower  at non-perturbative scales approaching $\lqcd$. 
The distribution of hadrons inside a jet is described by its fragmentation function, $D_{\rm a\to h}(z,Q)$, that encodes the probability that an initial parton $a$ eventually fragments into a hadron $h$ carrying a fraction $z=p_{\rm hadron}/p_{\rm parton}$ of the parent parton's momentum. Starting with a parton at a given $\delta$-function energy $Q$, its evolution to any other lower energy scale $Q'$ is driven by a branching process of parton radiation and splitting, $a\to b\,c$, that can be perturbatively computed. At large $z\gtrsim 0.1$ one uses the DGLAP evolution equations~\cite{Gribov:1972ri,Altarelli:1977zs,Dokshitzer:1977sg}, whereas the Modified Leading Logarithmic Approximation (MLLA)~\cite{Dokshitzer:1991ej}, resumming soft (along with hard) and collinear logs, provides an appropriate theoretical framework at small $z$. In this latter approach, one writes the FF as a function of the log of the inverse of $z$, \ie\ $\xi=\ln(1/z)$, in order to describe the region of low hadron momenta that dominates the jet fragments. Due to colour coherence and interference in gluon radiation (angular ordering), not the softest partons but those with intermediate energies multiply most effectively in QCD cascades, leading to a final FF with a typical ``hump-backed plateau'' (HBP) shape as a function of $\xi$ (Fig.~\ref{fig:FFs}, left). Such a shape can be perfectly reproduced by a distorted Gaussian (DG,~\cite{Fong:1990nt})
parametrized in terms of the hadron multiplicity ${\cal N}$ (giving the integral, and thereby the normalization, of the DG), the mean peak position $\bar\xi$, the dispersion $\sigma$, the skewness $s$, and kurtosis $k$ of the distribution.

In Refs.~\cite{dEnterria:2014lmi}, we described a new approach that solves the set of integro-differential equations for the FF evolution combining both DGLAP and MLLA corrections. This is done by expressing the Mellin-transformed hadron distribution in terms of the anomalous dimension $\gamma$: $D\simeq C(\alphas(t))\exp\left[\int^t \gamma(\alphas(t')) dt\right]$ where $t=\ln Q$ is the ``time'' evolution variable in QCD parton showers.  The analysis leads to a series in half powers of $\alphas$: 
$\gamma\sim {\cal O}(\alphas^{^{1/2}})+{\cal O}(\alphas)+{\cal O}(\alphas^{^{3/2}})+{\cal O}(\alphas^{^2})+{\cal O}(\alphas^{^{5/2}})+\cdots$, where integer powers of $\alphas$ correspond to fixed-order corrections, and half-integer terms can be identified with increasingly accurate resummations of soft and collinear logarithms.
The full set of NLO ${\cal O}(\alphas^2)$ terms for the anomalous dimension, including the one-loop splitting functions $P_{ac}^{(1)}$ and the two-loop running of $\alphas$, plus a fraction of the ${\cal O}(\alphas^{^{5/2}})$ terms, coming from the NNLO expression of $\alphas$ have been computed~\cite{DdERPR}. Upon inverse-Mellin transformation, one can derive the analytical expressions for the energy evolution of the FF, and its associated moments, as a function of $Y = \ln(E/\lqcd)$, for an initial parton energy $E$, down to a shower cutoff scale $Q_0>\lqcd$ for $N_f=3,4,5$ quark flavours. By introducing $\lambda=\ln(Q_0/\lqcd)$, the resulting formulas for the energy evolution of the moments depend on $\lqcd$ as a {\it single} free parameter. Simpler expressions can be further obtained for $Q_0\to\lqcd$ (limiting spectrum) motivated by the ``local parton-hadron duality'' hypothesis for inclusive-enough observables. Thus, by fitting to the distorted Gaussian, the  measured HBP at various energies, one can determine $\alphas$ from the corresponding jet energy-dependence of the FF moments ${\cal N}$, $\bar\xi$, $\sigma$, $s$, and $k$.

In the phenomenological analysis, we first start by fitting to the DG all existing jet FF data measured in $\epem$ and e$^\pm,\nu$-p collisions over $\sqrts\approx$~1--200~GeV, and thereby derive the corresponding FF moments at each jet energy~\cite{Perez-Ramos:2019zxf}. The overall normalization of the HBP spectrum (${\cal K}_{\rm ch}$), which determines the average charged-hadron multiplicity of the jet (\ie\ the zeroth moment of the FF), is an extra free parameter in the DG fit which, nonetheless, plays no role 
in the finally derived $\lqcd$ value since the latter is {\it solely} dependent on the relative evolution of the multiplicity, and not on its absolute value at any given jet energy. The impact of finite hadron-mass effects in the DG fit are taken into account through a rescaling of the theoretical (massless) parton momenta with an effective mass $\meff\approx m_{\pi}$. Varying such effective mass from zero to a few hundred MeV, results in small propagated uncertainties into the final extracted $\lqcd$ (and associated $\alphasmZ$) value, as discussed in Refs.~\cite{dEnterria:2014lmi}.

Once the energy evolution of all FF moments has been obtained from the individual experimental measurements, one can perform a combined fit of them as a function of the original parton energy. In the case of $\epem$ collisions, the latter corresponds to half the centre-of-mass energy $\sqrts/2$ whereas, for DIS, the invariant four-momentum transfer $Q_{_{\rm DIS}}$ is used. The experimental and theoretical evolutions of the hadron multiplicity and FF peak position as a function of jet energy are shown in~\cite{Perez-Ramos:2019zxf}. The hadron multiplicities measured in DIS jets appear somewhat smaller (especially at high energy) than those from $\epem$ collisions, due to limitations in the FF measurement only in half (current Breit) e$^\pm$-p hemisphere and/or in the determination of the relevant $Q$ scale~\cite{dEnterria:2014lmi}. The NNLO$^\star$+NNLL limiting-spectrum ($\lambda = 0$) predictions for $N_f = 5$ active quark flavours (the moments of the lowest-$\sqrts$ data have a few-percent correction applied in order to account for the slightly different ($N_f = 3,4$) evolutions below the charm and bottom production thresholds), leaving $\lqcd$ as a free parameter, reproduce very well the data. Fit results for the rest of the FF moments can be found in~\cite{dEnterria:2014lmi}.  Figure~\ref{fig:FFs} (right) shows the result of the peak position fit. Among FF moments, the peak position $\xi_{\rm max}$ appears as the most ``robust'' for the determination of $\lqcd$, being relatively insensitive to most of the uncertainties associated with the extraction method (DG fits, energy evolution fits, finite-mass corrections, \dots) as well as to higher-order corrections~\cite{Perez-Ramos:2019zxf}. Finally, the energy evolution of all or a fraction of HBP moments can be simultaneously fitted to the corresponding theoretical predictions in the limiting-spectrum case with just $\lqcd$ as single free parameter.



The QCD coupling obtained from the combined fit of the multiplicity and peak position is $\alphasmZ = 0.1205 \pm 0.0010\,(\mathrm{exp})\,\pm 0.0022\,(\mathrm{theo})$, where the first uncertainty includes all experimentally-related sources discussed in Refs.~\cite{dEnterria:2014lmi}, and the second one is a theoretical scale uncertainty derived at NLO by stopping the parton evolution of the FFs at $Q_{_{0}} = 1$~GeV rather than at the limiting spectrum value $Q_{_{0}} = \lqcd$. 
Our extracted $\alphasmZ$ value is consistent with all other NNLO results from the latest PDG compilation~\cite{Zyla:2020zbs}, as well as with other determinations with a lower degree of theoretical accuracy~\cite{dEnterria:2018cye}. The precision of our result ($\pm2\%$) is competitive with the other extractions, with a totally different set of experimental and theoretical uncertainties.\\

\paragraph*{Outlook:}
Given the robustness of the observables chosen, energy evolution of FF moments, the purely experimental uncertainties of our $\alphasmZ$ extraction are small, $\sim$0.8\%, and will be eventually negligible with the large jet data samples, orders of magnitude larger than those at LEP, expected to be collected at future $\epem$ machines~\cite{Proceedings:2017ocd,FCC:2018evy}. Thus, the main source of imprecision is of theoretical origin. The main theoretical challenges of the approach presented here are to match the MLLA anomalous dimension to the $\MSbar$ anomalous dimension and to reach full-NNLO pQCD accuracy. It is known that the MLLA anomalous dimension obtained within the massive gluon (MG) regularization scheme~\cite{Mueller:1982cq} turns out to be inconsistent with the expected N$^4$LL+NNLO result in the $\MSbar$ scheme. For instance, the inconsistency between NNLO DLs terms to the anomalous dimension calculated in~\cite{Mueller:1982cq} and those calculated in~\cite{Mitov:2006ic} can be explained from the use of different regularization and factorization schemes~\cite{Albino:2011si}. More recently, it has been noticed that the main difference is entirely due to running coupling effects being truncated in a quite severe way~\cite{Neill:2020tzl}. In addition, in order to extract a coupling constant with NNLO fixed-order accuracy from the moments of the FFs, the diagonalization of the matrix elements (\ie\ the splitting functions) beyond the ${\cal O}(\alphas^{5/2})$ order is a crucial step forwards. The diagonalization method to be adopted was considered a major technical challenge since the splitting functions do not commute beyond leading order. However, it becomes now possible thanks to the recent work by Kotikov and Teryaev~\cite{Kotikov:2020ukv}. To conclude, the implementation of the approach used in~\cite{Kotikov:2020ukv} and the change from the MG regularization scheme to the $\MSbar$ scheme may make it possible to complete the present programme on the full-NNLO extraction of $\alphasmZ$ from the moments of FFs. 

\subsection{Power corrections to event-shape distributions and impact on \texorpdfstring{$\alphas$}{alphas} extractions
\protect\footnote{A\lowercase{uthors:} P. F. M\lowercase{onni} (CERN), P. N\lowercase{ason} (MPI M\lowercase{unich}, INFN \& U\lowercase{niv.} M\lowercase{ilano})}}
\label{sec:ee_Cparam}

In this brief note we discuss some recent developments on the fits of
$\alphas$ from event-shape distributions at lepton colliders.
These determinations rely on the fact that event shape variables that
vanish in the two-jet limit are directly sensitive to the QCD
$q\bar{q}g$ vertex, and thus arguably offer one of the simplest
frameworks to extract the strong coupling constant.
The present status of these measurements as reported in the Particle
Data Group (PDG)~\cite{Zyla:2020zbs,ParticleDataGroup:2018ovx} is not very
satisfactory.
In particular, determinations in which nonperturbative corrections
are estimated via analytic models tend to give values for the strong
coupling which are systematically smaller than those obtained using
Monte Carlo generators to correct for nonperturbative effects.
%
%
Specifically, some of the most precise $\alphasmZ$ determinations obtained
with analytical nonperturbative models
($\alphasmZ = 0.1135\pm0.0010$~\cite{Abbate:2010xh} from fitting thrust
data, and $\alphasmZ = 0.1123\pm0.0015$~\cite{Hoang:2015hka} from
$C$-parameter data), are in tension with the world average of
$0.1179 \pm0.0010$ and from other individual precise extractions, such
as $0.1185\pm 0.0008$ from lattice step scaling~\cite{Bruno:2017gxd}
and $0.1188\pm 0.0013$ from jet rates \cite{Verbytskyi:2019zhh}.
A similar extraction, that of Ref.~\cite{Gehrmann:2012sc}, uses the
thrust distribution and an analytical hadronization model and returns
a value for $\alphasmZ$ compatible with those of
Refs.~\cite{Abbate:2010xh,Hoang:2015hka}, although with larger
uncertainties.\\

A feature common to analytic approaches to nonperturbative
corrections (see
\eg\ Refs.~\cite{Manohar:1994kq,Webber:1994cp,Korchemsky:1994is,Korchemsky:1999kt,Dokshitzer:1995zt,Dokshitzer:1995qm,Nason:1995np,Dasgupta:1996ki,Dokshitzer:1997iz,Dasgupta:1999mb,Beneke:1997sr,Gardi:2003iv,Davison:2009wzs,Abbate:2010xh,Gehrmann:2012sc,Hoang:2015hka})
is that their application to event shapes relies upon a power series
in $1/Q$ (with $Q$ being the centre-of-mass energy of the $\epem$
collision), of which only the leading (linear) term is retained.
This linear power correction is proportional to a nonperturbative
parameter (that is extracted together with $\alphas$) via a
calculable, observable-dependent coefficient.
Furthermore, this coefficient is commonly calculated in the $2$-jet
limit, in general supplementing a Sudakov resummation of logarithmic
corrections assuming that the above coefficient remains constant
across the fit range, which instead covers both $2$- and $3$-jet
configurations (see
\eg~\cite{Gardi:2003iv,Davison:2009wzs,Abbate:2010xh,Gehrmann:2012sc,Hoang:2015hka}).
This assumption has recently been questioned in
Refs.~\cite{Luisoni:2020efy,Caola:2021kzt}, where the first
calculations of the nonperturbative correction in three-jet
configurations have appeared.\\

In this note, we briefly discuss the calculation of the leading power
correction reported in Refs.~\cite{Luisoni:2020efy,Caola:2021kzt}. We
start by recalling how the calculation is performed in the two-jet
limit, and then we outline how it can be performed in the symmetric
three-jet limit, as well as in a generic three-jet configuration.
Finally, we present some phenomenological considerations on the impact
on $\alphas$ fits, and discuss future developments.

\subsubsection{Definition of the observable}
\label{sec:obs}
Here we limit our discussion to the $C$-parameter, but analogous
considerations apply to other shape variables such as
thrust~\cite{Farhi:1977sg}.
The $C$-parameter variable for a hadronic final state in $\epem$
annihilation is defined as follows~\cite{Ellis:1980wv},
\begin{equation}
  \label{eq:C-from-eigenvalues}
  C = 3 \left(\lambda_1\lambda_2 + \lambda_2\lambda_3 + \lambda_3\lambda_1\right),
\end{equation}
in terms of the eigenvalues $\lambda_i$ of the linearized momentum
tensor $\Theta^{\alpha\beta}$ \cite{Parisi:1978eg,Donoghue:1979vi},
\begin{equation}
  \label{eq:momentum tensor}
  \Theta^{\alpha\beta} = \frac1{\sum_i |\vec p_i|}
  \sum_i \frac{{\vec p}_i^{\,\alpha} {\vec p}_i^{\,\beta}}{|\vec p_i|}\,,
\end{equation}
where $|\vec p_i|$ is the modulus of the three momentum of particle
$i$ and ${\vec p}_i^{\,\alpha}$ is its momentum component along 
spatial dimension $\alpha$ ($\alpha = 1,2,3$).
In events where all particles are massless, this can also be written
as
\begin{align}
C &= 3 - \frac{3}{2}\sum_{i,j} \frac{(p_i\cdot p_j)^2}{(p_i\cdot Q)
    (p_j\cdot Q)}=\frac{3}{8}\sum_{i,j} x_i x_j \sin^2\theta_{ij}\,,
\label{eq:Cpar}
\end{align}
where $Q$ is the centre-of-mass energy, $p_i$ denotes the
four-momentum of particle $i$, $x_i = 2 (p_i\cdot Q)/Q^2$, and
$\theta_{ij}$ is the angle between particles $i$ and $j$.
We consider the calculation of the linear power correction in the
context of the so-called dispersive model~\cite{Dokshitzer:1995qm},
which postulates that the leading power correction to the cumulative
distribution of the event shape, \ie\
\begin{equation}
\Sigma(C) \equiv \int_0^C d C^\prime \frac{d\sigma}{d C^\prime }\,,
\end{equation}
is due to the radiation of a soft and nonperturbative system with
gluon quantum numbers, \ie\ a \textit{gluer}.
Starting from a Born configuration of final state momenta
$\{\tilde{p}\}$ (\eg\ at the leading order
$\{\tilde{p}\} \equiv \{\tilde{p}_q,\tilde{p}_{\bar q}\}$ in the
two-jet limit and
$\{\tilde{p}\} \equiv \{\tilde{p}_q,\tilde{p}_{\bar q},\tilde{p}_g\}$
in the three-jet limit), $\Sigma(C)$ schematically reads
\begin{equation}
\label{eq:pre-master}
\Sigma(C)  = \sigma -\left\{\int d\sigma (\{\tilde{p}\})
  \Theta(C(\{\tilde{p}\})-C) + \int d\sigma(\{p\},k) \Theta(C(\{p\},k) -C)\right\}\,,
\end{equation}
where $\sigma$ is the total cross section (in which the radiation of
the gluer $k$ does not generate any linear
correction~\cite{Dokshitzer:1995qm}), $d\sigma (\{\tilde{p}\})$
includes also the contribution in which the gluer is virtual, and
$d\sigma(\{p\},k) $ encodes the one in which the gluer is real. In the
latter term, we have denoted with $(\{p\},k)$ the set of hard momenta
$\{p\}$ ($\neq \{\tilde{p}\}$ due to kinematic recoil) and the
collective set of soft particles ($k$) constituting the gluer system.

\subsubsection{Schematic illustration of the calculation}
\label{sec:calculation}
We now focus our discussion on the two-jet limit, \ie\
$\{\tilde{p}\} \equiv \{\tilde{p}_q,\tilde{p}_{\bar q}\}$.
Here the value of $C$ approaches zero (\ie\ $C(\{\tilde{p}\}) = 0$) as
\begin{equation}
 \label{eq:scaling}
C(\{p\},k) \sim {\cal O}(k)\,.
\end{equation}
Quadratic corrections to the above equation (\eg\ due to the recoil of
the hard partons against $k$) can be safely ignored if one is
interested in the computation of the linear power correction
${\cal O}(1/Q)$.
In order to single out the linear contribution ${\cal O}(1/Q)$, we can
therefore recast Eq.~\eqref{eq:pre-master} as
\begin{align}
\label{eq:master}
\Sigma(C)  &= \sigma - \left\{\int \left[ d\sigma (\{\tilde{p}\})+ d\sigma(\{p\},k) \right]
  \Theta(C(\{\tilde{p}\})-C) \right.\notag\\
&\qquad \left. + \int d\sigma(\{p\},k) \left[\Theta(C(\{p\},k)-C)-\Theta(C(\{\tilde{p}\})-C)\right]\right\}\,.
\end{align}
The first line in the r.h.s.\ of Eq.~\eqref{eq:master} reduces to the
total cross section, which is well known to be free from linear
corrections.
Therefore, the second line in the r.h.s.\ of Eq.~\eqref{eq:master} is
the only source of linear power corrections.
Because of the linear suppression~\eqref{eq:scaling}, this can be
computed in the soft approximation along the lines of
Refs.~\cite{Dokshitzer:1995zt,Catani:1998sf,Dokshitzer:1998pt}.\\
The $C$ parameter is special in that it has two singular points which
feature a scaling of the type~\eqref{eq:scaling}. One of them is the
two-jet limit discussed in the previous section (\ie\ $C=0$) and the
second is the symmetric three-jet configuration (\ie\ $C=3/4$) at
which the distribution features a \textit{Sudakov
  shoulder}~\cite{Catani:1997xc,Catani:1998sf}.
Due to this property, the value of the $C$ parameter is nearly
constant (\ie\ up to quadratic corrections) near the symmetric
three-jet limit, and the same considerations used for the two-jet case
apply also here.
One can therefore apply Eq.~\eqref{eq:master} by considering only the
contribution of the second line and restricting the calculation to the
soft approximation as done in Ref.~\cite{Luisoni:2020efy}.

In between the two- and symmetric three-jet configurations (\ie\ $0< C
< 3/4$), the simplifications derived in the above sections do not
apply trivially.
To simplify the discussion, let us consider for simplicity the
radiation of the gluer $k$ off a single colour dipole (the final
result is simply obtained by summing over all possible dipoles).
Let us assume the existence of a mapping $p \equiv p(\{\tilde{p}\},k)$
that is collinear safe in the limit where the system $k$ is collinear
to the ends of the radiating dipole, and such that in the soft limit
one has (schematically)
\begin{equation}
\label{eq:mapping}
p(\{\tilde{p}\},k) = \tilde{p} + {\cal M}(\{\tilde{p}\}) \sum_i k_i + {\cal O}(k^2)\,,
\end{equation}
where the sum runs over all constituents $k_i$ of the gluer, and the
tensor ${\cal M}$ only depends on $\{\tilde{p}\}$.
The master formula~\eqref{eq:master} can be rewritten as
\begin{align}
\label{eq:master-generic}
\Sigma(C)  &= \sigma - \left\{\int \left[ d\sigma (\{\tilde{p}\})+ d\sigma(\{p(\{\tilde{p}\},k)\},k) \right]
  \Theta(C(\{\tilde{p}\})-C) \right.\notag\\
&\qquad \left. + \int d\sigma(\{p(\{\tilde{p}\},k)\},k) \left[\Theta(C(\{p(\{\tilde{p}\},k)\},k)-C)-\Theta(C(\{\tilde{p}\})-C)\right]\right\}\,.
\end{align}
The first line of Eq.~\eqref{eq:master-generic} represents a cross
section that is integrated inclusively over $k$ at fixed
$\{\tilde{p}\}$ according to a mapping of the form~\eqref{eq:mapping}.
According to the finding of Ref.~\cite{Caola:2021kzt}, such cross
section is free of linear power corrections, and therefore the only
linear contribution comes from the second line as in the previous
cases.
Moreover, the expression in squared brackets of the second line of
Eq.~\eqref{eq:master-generic} is suppressed in the soft limit so that
one can once again compute the linear contribution in the soft
approximation.

\subsubsection{Results and impact on $\alphas$ fits}
\label{sec:res}
One finds that the leading power correction can be
parametrized as
\begin{equation}
\Sigma(C) = \Sigma^{\rm pert}\left(C -\zeta(C)\frac{\alpha_0(\mu_I^2)}{Q}\Delta^{\rm NP}\right)\,,
\end{equation}
where the quantity $\Delta^{\rm NP}$ does not depend on $C$, and
$\Sigma^{\rm pert}$ denotes the perturbative cumulative distribution.
The parameter $\alpha_0(\mu_I^2)$ is a nonperturbative quantity
related to the mean value of the strong coupling constant in a
physical scheme at scales smaller than
$\mu_I$~\cite{Dokshitzer:1995qm}. It is commonly extracted from fits
to experimental data together with $\alphas$.
The entire dependence of the leading power correction on $C$ is
encoded in the function $\zeta(C)$, which can be extracted directly
from the calculations outlined in the above sections.
Specifically, Ref.~\cite{Luisoni:2020efy} obtains that
$\zeta(3/4)/\zeta(0)\simeq 0.476$, \ie\ the nonperturbative
correction in the three-jet limit is about a factor of two smaller
than the result in the two-jet limit.

In order to study the intermediate region $0 < C < 3/4$,
Ref.~\cite{Luisoni:2020efy} considered a set of possible functional
forms for the function $\zeta(C)$, with which a fit of the strong
coupling constant from a differential distribution of the $C$
parameter was performed.
This relies on a NNLO+NNLL perturbative calculation obtained with the
results of Refs.~\cite{Gorishnii:1990vf,Banfi:2014sua,DelDuca:2016csb}
and experimental data from
Refs.~\cite{Heister:2003aj,MovillaFernandez:1998ys} (we refer to
Ref.~\cite{Luisoni:2020efy} for the technical details).
This study reveals that the variation of the functional form of
$\zeta(C)$ can impact the extracted value of the strong coupling at
the $\sim 4\%$ level. In particular, the standard assumption
$\zeta(C) = \zeta(0)$ used in past extractions leads to the following
values of $\alphasmZ$ and $\alpha_0(\mu_I^2)$
\begin{equation}
\alphasmZ = 0.1121^{+0.0024}_{-0.0016} \,,\quad
\alpha_0(\mu_I^2) = 0.53^{+0.07}_{-0.05}\,,\notag
\end{equation}
and agrees well with that of Ref.~\cite{Hoang:2015hka} (albeit with
larger uncertainties, in part due to the use of NNLL+NNLO theory in
Ref.~\cite{Luisoni:2020efy} rather than
N$^3$LL+NNLO~\cite{Hoang:2014wka} as in Ref.~\cite{Hoang:2015hka}).
Conversely, several of the assumed $\zeta(C)$ scalings lead to a
$\chi^2$ value that is the same as, or smaller than, that for the fit
within the $\zeta(C) = \zeta(0)$ assumption.
In particular, one of the considered models returns
\begin{equation}
\alphasmZ = 0.1163^{+0.0028}_{-0.0018} \,,\quad \alpha_0(\mu_I^2) = 0.51^{+0.06}_{-0.04}\,,\notag
\end{equation}
with a $\chi^2$ that is similar to that of the previous fit. This
corresponds to a potential additional uncertainty in $\alphasmZ$
due to nonperturbative corrections of about $3.7\%$. With this
uncertainty taken into account, the extracted values of $\alphas$
become compatible with the world average $\alphasmZ = 0.1179 \pm 0.0010$~\cite{PDGQCD2019}.

\begin{figure}[htpb!]
  \centering
 \includegraphics[width=0.5\linewidth]{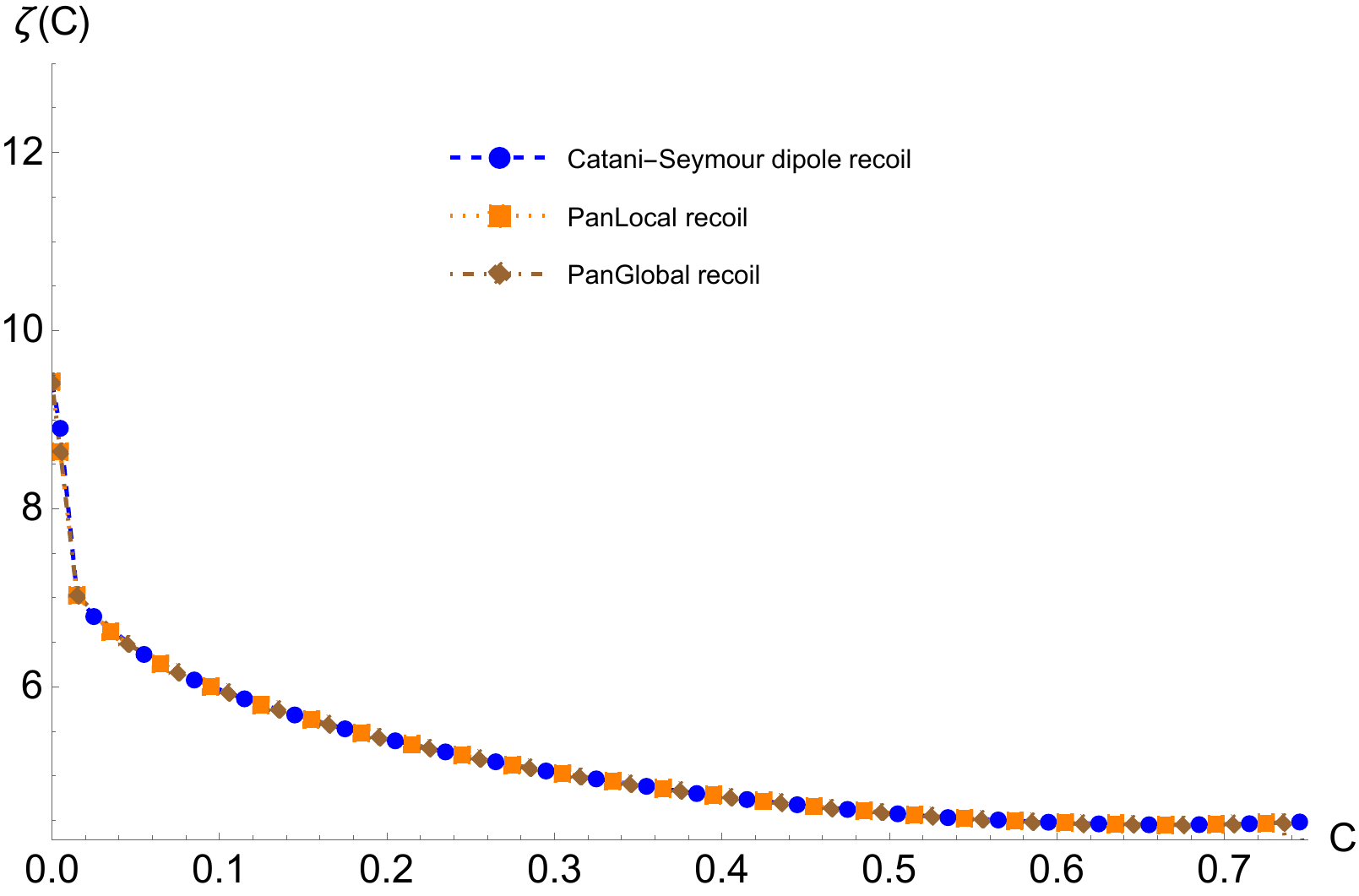}
 \caption{Fixed order calculation of $\zeta(C)$ with different recoil
   schemes~\cite{Luisoni:2020efy}.}
 \label{fig:fixed-order}
\end{figure}

With the above conclusions, an important question is whether one can
calculate $\zeta(C)$ between the two-jet and three-jet limits.
As discussed in Sec.~\ref{sec:calculation}, Ref.~\cite{Caola:2021kzt}
demonstrates that such a calculation can be performed in the soft
approximation provided certain regularity
conditions~\eqref{eq:mapping} are met by the kinematic map used to
share the recoil due to the radiation of the gluer among the three
hard particles in the event.
Figure~\ref{fig:fixed-order} shows the results of a calculation of this
type performed in Ref.~\cite{Luisoni:2020efy}, where it was observed
that the displayed $\zeta(C)$ scaling obtained with three commonly
used recoil schemes~\cite{Catani:1996vz,Dasgupta:2020fwr} yielded
identical results.
Later, in Ref.~\cite{Caola:2021kzt} it was shown that (in a simplified
theoretical framework, described in detail in that reference) one
obtains the correct power correction if the recoil scheme satisfies
scalings of the kind given in Eq.~\eqref{eq:mapping}. Specifically, as
shown in Ref.~\cite{future}, the three recoil schemes considered in
Fig.~\ref{fig:fixed-order} do satisfy an appropriate scaling.

A precise extraction of $\alphas$ with the above $\zeta(C)$ profile
and further in depth studies of the interplay of the nonperturbative
corrections discussed in this note with perturbative aspects of the
calculation at higher orders are still necessary in order to make
conclusive statements on the impact of the developments discussed in
this note.
Furthermore, similar studies for other event shape observables such as
thrust (also addressed in Ref.~\cite{Caola:2021kzt}) will be essential
to shed more light on the current discrepancies among the different
event-shape based extractions of $\alphas$ considered in the world
average~\cite{PDGQCD2019}.



\subsection{The strong coupling from groomed event shapes
\protect\footnote{A\lowercase{uthors:} S. M\lowercase{arzani} (INFN \& U\lowercase{niv.} G\lowercase{enova}), D. R\lowercase{eichelt} (D\lowercase{urham}), S. S\lowercase{chumann} (U\lowercase{niv.} G\lowercase{\"ottingen}), G. S\lowercase{oyez} (CEA-S\lowercase{aclay})}}

Event-shape variables such as thrust have been measured with high precision by the LEP experiments. These observables have played and continue to play an important role in the determination of the value of the strong coupling, see for 
instance~\cite{Bethke:2009ehn, Dissertori:2009ik, Dissertori:2009qa, OPAL:2011aa, Schieck:2012mp, Davison:2009wzs, Abbate:2010xh, Gehrmann:2012sc, Abbate:2012jh}, as well as the discussion in this report.
While event shapes can be calculated with astonishing precision within perturbative QCD, both at fixed-order and resummed levels, nonperturbative contributions in particular due to hadronization can be sizable and thus affect the ultimate precision achievable.
In the context of jet-substructure analyses of final states produced in hadronic collisions the soft-drop grooming technique offers a handle to reduce nonperturbative corrections~\cite{Larkoski:2014wba}. This idea can be transferred to (global) event shapes in lepton~\cite{Baron:2018nfz} and hadron~\cite{Baron:2020xoi} collisions. With a possibly reduced impact of nonperturbative contributions, this can offer an improved precision in extractions of $\alphaS$. In Ref.~\cite{Marzani:2019evv} this has been explored for the soft-drop version of thrust in $\epem$ collisions at $\sqrt{s}=m_\mathrm{Z}$. \\

The soft-drop variant of thrust (and similarly for other event shapes) is defined by
first determining the conventional thrust axis and accordingly separating a given event into two hemispheres. The soft-drop procedure is then applied to both hemispheres independently. Soft-drop thrust is calculated based on the remaining
constituents of both hemispheres, according to
\begin{equation}\label{eq:sd-thrust}
\tau_{\text{SD}}=\frac{\sum_{i\in\mathcal{E}_{\text{SD}}}\left|\vec{p}_i\right|}{\sum_{i\in\mathcal{E}}\left|\vec{p}_i\right|}\left[1-\frac{\sum_{i\in\mathcal{H}^{L}_{\text{SD}}}\left|\vec{n}_L\cdot\vec{p}_i\right|+\sum_{i\in\mathcal{H}^{R}_{\text{SD}}}\left|\vec{n}_R\cdot\vec{p}_i\right|}{\sum_{i\in\mathcal{E}_{\text{SD}}}\left|\vec{p}_i\right|}\right],
\end{equation}
where $\vec{n}_{L/R}$ denote the axes for the left and the right hemispheres, respectively, and the sums extend over all particles in the full event ($\mathcal{E}$), the soft-dropped event ($\mathcal{E}_{\text{SD}}$) or the left and right hemispheres ($\mathcal{H}^{L/R}$). Thereby, the $\epem$ version of soft drop operates on the hemispheres defined by the thrust axis, reclusters them using the Cambridge--Aachen algorithm and discards soft subjets failing the criterion  
\begin{equation}\label{eq:sd-ee}
\frac{\text{min}(E_i,E_j)}{E_i+E_j}>z_\text{cut}(1-\cos\theta_{ij})^{\beta/2}\,.
\end{equation}
While the parameter $z_\text{cut}$ determines how stringent the cut on the subjet energies is, $\beta$ provides an angular suppression to grooming.

\begin{figure}[htb!]
\centering
\includegraphics[width=0.49\textwidth]{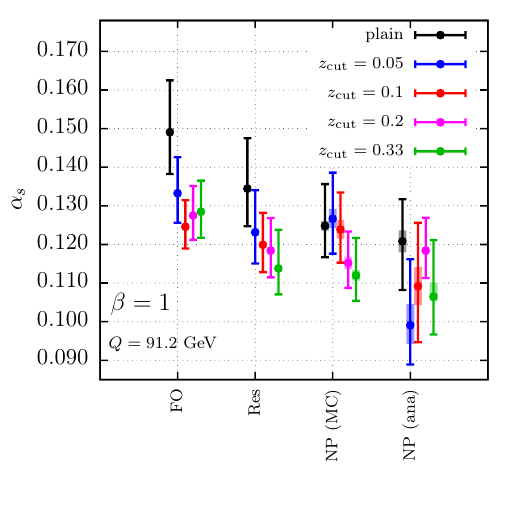}
\includegraphics[width=0.49\textwidth]{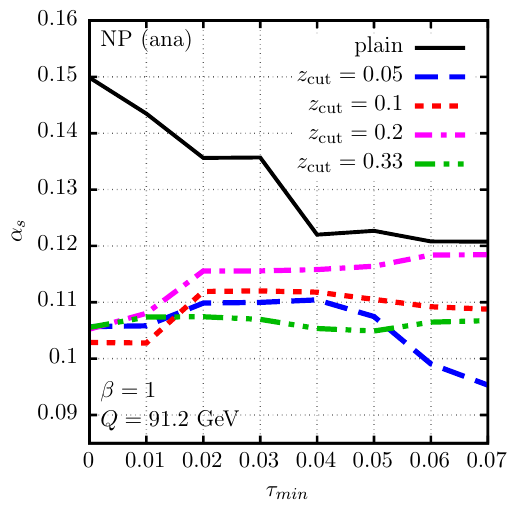}
\caption{Results of~\cite{Marzani:2019evv} illustrating the advantages of using soft-drop thrust in fits of $\alphaS$. Both are based on fits to pseudodata generated with \sherpa. Left: Dependence of the results on the effects included in the calculation, starting from only fixed order, \ie\ NLO QCD, on the left and including effects of resummation, corresponding to NLO+NLL$^\prime$ accuracy, and upon including nonperturbative effects in two different approaches. Right: Dependence of 
the $\alphaS$ best-fit value on the range of (soft-drop) thrust included in the fit.}
\label{fig:sdas}
\end{figure}

Though originally designed to reduce nonperturbative effects like what is
usually accounted for in multiple parton interaction (MPI) and underlying event
(UE) simulations at hadron colliders~\cite{Buckley:2011ms}, soft drop has been shown 
to also significantly reduce the impact of hadronization corrections in event shapes
and jet observables at lepton colliders~\cite{Baron:2018nfz}. The viability
of using soft-drop thrust in fits of $\alphas$ was studied in detail
in~\cite{Marzani:2019evv}. Theoretical predictions computed at NLO+NLL$^\prime$ accuracy were employed in fits to hadron-level pseudodata generated with \sherpa~\cite{Sherpa:2019gpd} based on merging the NLO pQCD matrix elements for $\epem\to 2,3,4,5$ partons with the dipole shower~\cite{Schumann:2007mg}, using a nominal value of $\alphasmZ=0.117$, and the cluster-fragmentation model~\cite{Winter:2003tt}. The main results of this study are summarized in Fig.~\ref{fig:sdas}.

Firstly, the results obtained from fitting $\alphaS$ in the NLO+NLL$^\prime$ prediction for the soft-drop thrust distribution depend less strongly on nonperturbative effects than for plain thrust when using the same Monte Carlo (MC) pseudodata. This is illustrated in the left plot of Fig.~\ref{fig:sdas}. While the best-fit $\alphaS$ values obtained from plain thrust change significantly when going from using fixed-order (FO) results to including resummation effects (Res) and again when including nonperturbative (NP) effects, either extracted from MC simulations or from an analytic model (ana). This change is reduced when considering soft-drop thrust, shown in Fig.~\ref{fig:sdas} for $\beta=1$ and a variety of $\zcut$ values. Note, in Ref.~\cite{Marzani:2019evv} also the cases $\beta=0$ and $\beta=2$ have been considered.

Secondly, for soft-drop thrust the value of $\alphaS$ obtained from the fits is less affected by the choice of the considered observable range. This is illustrated in the right plot of Fig.~\ref{fig:sdas}. The minimal observable value $\tau_\text{min}$ included in the fit is lowered from the default value $\tau_\text{min} = 0.07$. The upper boundary is thereby kept fixed at $\tau_\text{max}=0.25$. For plain thrust, this results in dramatic changes due to increasing and severe hadronization corrections. Consequently, for plain thrust this additional observable range, \ie\ phase-space region, cannot be reliably used in an $\alphaS$ extraction. However, the results obtained from soft-drop thrust, with the various choices of $\zcut$, are significantly more stable and the fit range could be extended into those areas.

From an experimental point of view, an obvious problem is that soft-drop event shapes have not been measured by the LEP experiments. However, their data are preserved and have recently been reanalysed for other observables~\cite{Fischer:2015pqa,Kluth:2017xai,Chen:2021iyj, Chen:2021uws}. Thus, we believe, it would be  very interesting to measure soft-drop event shapes,  as well as broader families of (groomed) substructure observables like angularities or energy correlation functions, on archived LEP data and use them to perform novel fits of the strong coupling.
On the other hand, soft-drop groomed event shapes can also be studied in hadron collisions~\cite{Baron:2020xoi}, so analogous analyses could be performed at the currently running LHC experiments. The same comment applies to analyses of soft-drop groomed jet substructure observables, see for example~\cite{Caletti:2021oor, CMS:2021iwu, ALICE:2021njq, Caletti:2021ysv, Zhu:2021xjn, Reichelt:2021svh,Makris:2021drz} for recent theoretical and experimental results. However, it should be noted that the study in~\cite{Proceedings:2018jsb} identified a strong correlation between $\alphaS$ and the overall structure of the hard collision, in particular the fraction of quarks and gluons that enter the calculation, as one of the main challenges for extractions of $\alphaS$ from jet-shape measurements in proton-proton collisions. Thus, in this context, it would be important to find observables that can break this degeneracy.  

On the theory side, we also anticipate that higher-accuracy calculations will be necessary to be competitive with fits using traditional event shapes. At fixed order, NNLO accurate predictions are in principle available~\cite{Kardos:2018kth}. 
Furthermore, resummed predictions for soft-drop observables have been obtained at NNLL~\cite{Frye:2016aiz} and some results are available even at N$^3$LL~\cite{Kardos:2020gty}.  Regarding all-orders soft-gluon effects, a crucial ingredient is an improved understanding of the transition region where emissions soft enough to be groomed become important for the distribution. The resummed calculations like the ones used in~\cite{Marzani:2019evv} are performed in the strict $\tau\ll\zcut\ll 1$ limit. However, judging from the ranges where $\alphaS$ has been extracted from LEP data of the thrust distribution, and assuming a traditional choice of $\zcut \sim {\cal{O}}(0.1)$, it seems however inevitable that a fit would also rely on the region $\tau \lesssim \zcut$ or even $\tau\sim\zcut$. Extending calculations to this region has been discussed recently in~\cite{Benkendorfer:2021unv}. With this, NNLL accuracy appears to be achievable over the whole range, however still limited to the assumption $\zcut\ll 1$. A better understanding of those power corrections in $\zcut$ is still a missing ingredient to date.

Finally, despite the fact that soft drop successfully reduces the sensitivity to nonperturbative effects, calculations aimed at precision determinations of the strong coupling cannot neglect them. In this context, significant improvement has been recently achieved, both in the analytic and MC approaches. Using effective field theory techniques, a more detailed understanding of power corrections due to hadronization in the presence of soft-drop grooming has been achieved in Refs.~\cite{Hoang:2019ceu,Pathak:2020iue}. The Monte Carlo approach has also been improved thanks to the implementation of so-called transfer matrices~\cite{Reichelt:2021svh}. This method allows us to better model the effect that the hadronization process has on the event kinematics, although it was shown in~\cite{Marzani:2019evv} to not significantly alter the effect of nonperturbative corrections in the range of soft-drop thrust used for the central fits, like the ones in the left plot of Fig.~\ref{fig:sdas}. However, outside of this range, differences might be sizable. 
In this context, it would be interesting to compare this improved numerical model with the aforementioned first-principle analysis. 



\clearpage
\section{\texorpdfstring{\boldmath$\alphasmZ$}{alphasmZ} from hadronic final-states in e-p and p-p collisions}
\label{sec:ep_pp}




\subsection{\texorpdfstring{$\alphas$}{alphas} from jet-production cross sections in neutral-current DIS using NNLO predictions
\protect\footnote{A\lowercase{uthors:} D. B\lowercase{ritzger} (MPI M\lowercase{unich})}}
\label{sec:ep}
The measurements of jet-production cross sections in
neutral-current deep-inelastic scattering (NC DIS) are performed at HERA in the
Breit frame of reference and provide clean and precise measurements.
The Breit frame is defined as a brick-wall frame, where in
leading-order NC DIS the incoming parton, literally, bounces back from
the photon wall.
Consequently, once the outgoing partons have significant transverse
momenta, the process is described by a $2\to 2$  photon-parton
scattering process in pQCD and is proportional to
$\mathcal{O}(\alphas)$ in leading order.
Progress in the antenna subtraction formalism enabled to
perform predictions for single-inclusive jet and dijet production
cross sections up to next-to-next-to-leading order (NNLO) in
pQCD~\cite{Currie:2016ytq,Currie:2017tpe}.
Several measurements of inclusive jet and dijet cross section
measurements from the H1 and ZEUS collaborations, from different
run-periods and different kinematic ranges in $Q^2$, were already
exploited for the determination of $\alphas$.

~ 
\paragraph{Methodology}
The value of $\alphasmZ$ is determined from HERA jet cross sections in
a $\chi^2$-minimization procedure of NNLO predictions to data.
Following the application of the factorization theorem, the pQCD
predictions for jet cross sections are
\begin{equation}
  \sigma = \sum_{k=q,\bar{q},g}\int \text{d}xf_k(x)\hat{\sigma}_k(x) \cdot c_\text{had}\,,
\end{equation}
where $f_k$ denotes the parton-distribution functions,
$\hat{\sigma}$ the partonic NNLO cross section, and $c_\text{had}$
nonperturbative correction factors which account for hadronization
effects.
Both components at the hard scale, $f_k$ and $\hat{\sigma}_k$, exhibit
a sensitivity to $\alphasmZ$, which can directly be seen when calculating
the partial derivative
$\tfrac{\partial\sigma}{\partial\alphasmZ}$.
In order to account for both $\alphas$-sensitive terms, a
constant \emph{starting scale} $\mu_0$ is introduced, similarly as in
DGLAP-based PDF-fits, and consequently the PDFs $f_k$ are defined
through their $x$-dependence at $\mu_0$ and \emph{evolved} to the
factorization scale using the DGLAP formalism, where also the
$\alphas$-sensitivity in the DGLAP kernels is exploited in the fitting procedure.
The evolution starting scale is chosen to be $\mu_0=20$~GeV. In a
study by H1, it was observed that the extracted values of
$\alphasmZ$ are rather insensitive to the exact choice of the starting
scale, and that the predominant sensitivity to $\alphasmZ$ arises from
the NNLO coefficients $\hat\sigma$.
The value of $\alphasmZ$ is then determined by minimizing the $\chi^2$
expression based on log-normal probability distribution functions
\begin{equation}
  \chi^2= \bm{r}^T(V_\text{exp} + V_\text{had} +
  V_\text{PDF})^{-1}\bm{r}
  ~~~\text{using}~~~
  r_i=\log\sigma_{\text{data},i} - \log\sigma_{\text{pred.},i}\,,
\end{equation}
where the covariance matrices are calculated from relative
uncertainties of the data ($V_\text{exp}$), the hadronization factors ($V_\text{had}$) and the
PDFs ($V_\text{PDF}$) and the cross sections $\sigma_{\text{data},i}$
and $\sigma_{\text{pred.},i}$ refer to data and the NNLO predictions
(pred.) in a bin $i$, respectively.
The PDF uncertainties are calculated from the eigenvectors or replicas
of a given PDF set from an external analysis. Note that, by including the
PDF uncertainties in the $\chi^2$ expression, the fit exploits the same
degrees of freedom as the respective PDF fit, and the PDFs become
\emph{profiled} in the minimization procedure. This procedure is de-facto
equivalent to adding the HERA jet to the PDF fit, while contrary the
$\alphasmZ$ inference exploits exclusively the jet data and thus
ensures a theoretically and experimentally cleaner determination
of $\alphasmZ$ than PDF fits, which mix different predictions
and processes.

The NNLO predictions are obtained using the program
NNLOJET~\cite{Currie:2017tpe}, which is
interfaced~\cite{Britzger:2019kkb} to 
fastNLO~\cite{Britzger:2012bs} and Applgrid~\cite{Carli:2010rw} to
enable repeated calculations with differing values 
for $\alphasmZ$ and differing PDFs. The DGLAP evolution is done with
QCDNUM~\cite{Botje:2010ay} or Apfelxx~\cite{Bertone:2017gds}, and the hadronization
correction factors are provided by the experimental collaborations
together with the data and are commonly determined using the MC event generators
Djangoh~\cite{Charchula:1994kf} or Rapgap~\cite{Jung:1993gf}.
The PDFs are obtained from NNPDF3.1~\cite{NNPDF:2017mvq} and further PDF sets
are
studied~\cite{Alekhin:2017kpj,Dulat:2015mca,Harland-Lang:2014zoa,Abramowicz:2015mha}
and are used to define a so-called PDFset uncertainty.
The renormalization and factorization scales are identified with
$\mu^2=Q^2+\pT^2$, where $\pT$ is identified with the single-jet
transverse momentum in case of inclusive jets, and with the average
$\pT$ of the two leading jets in case of dijet cross sections.

~ 

\paragraph{$\alphas$ from single-jet inclusive cross sections}
The value of $\alphasmZ$ was determined from inclusive jet cross
sections from previously published data by H1 in Ref.~\cite{H1:2017bml}, and
from H1 and ZEUS inclusive jet cross sections in Ref.~\cite{Britzger:2019kkb}.
Both experiments employ the $k_t$ jet-algorithm with a distance
measure $R=1.0$ and provide double-differential cross section data as
functions of $Q^2$ and jet transverse momenta $\pT^\text{jet}$.
The combined analysis of the H1 data using NNLO predictions exploits
five independent cross section measurements~\cite{H1:2000bqr,H1:2009pze,H1:2010mgp,H1:2014cbm,H1:2016goa} in the kinematic range of
$5<Q^2<15\,000$~GeV$^2$ and $4.5<\pT^\text{jet}<50$~GeV and 
yields a value of
$\alphasmZ=0.1157\,(10)_\text{exp}\,(36)_\text{th}$~\cite{H1:2017bml}, where
the first uncertainty comprises experimental uncertainties, and the
second collects theoretical uncertainties from PDFs, hadronization, and
missing higher orders.
The analysis of ZEUS inclusive jet cross sections exploits two data
sets from the HERA-I running period~\cite{ZEUS:2002nms,ZEUS:2006xvn} in the kinematic range
$120<Q^2<20\,000~$GeV$^2$ and yields
$\alphasmZ=0.1227\,(21)_\text{exp}\,(19)_\text{th}$~\cite{Britzger:2019kkb}.
Since H1 provides more data, and data at lower scales $\mu$, their
experimental uncertainty is smaller as compared to the ZEUS result,
while contrary the highly-sensitive low-scale data results in increased
scale uncertainties.

\begin{figure}[thbp!]                           
  \centering
  \includegraphics[width=0.40\textwidth,valign=t]{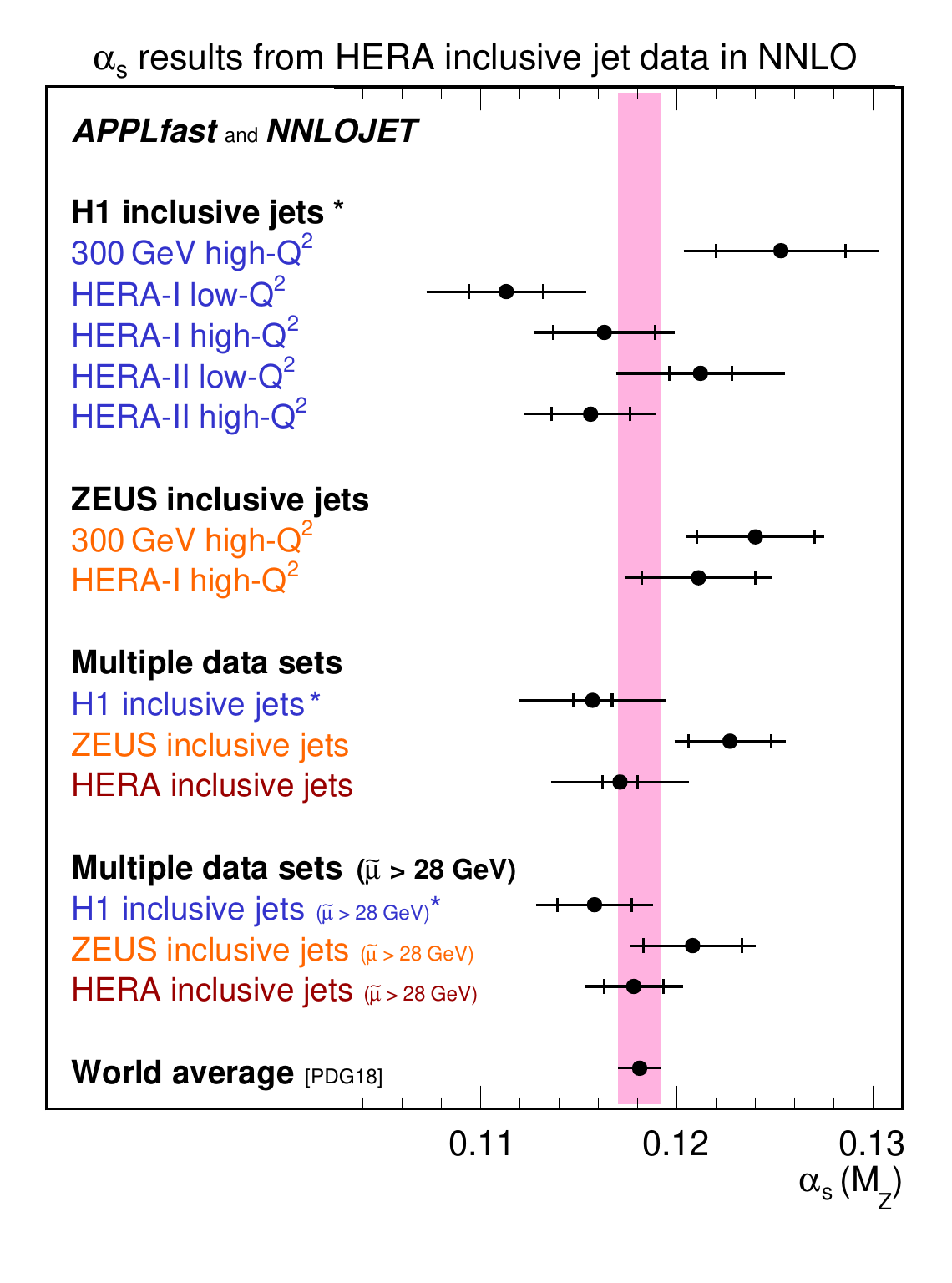}
  \hspace{0.08\textwidth}
  \includegraphics[width=0.40\textwidth,valign=t]{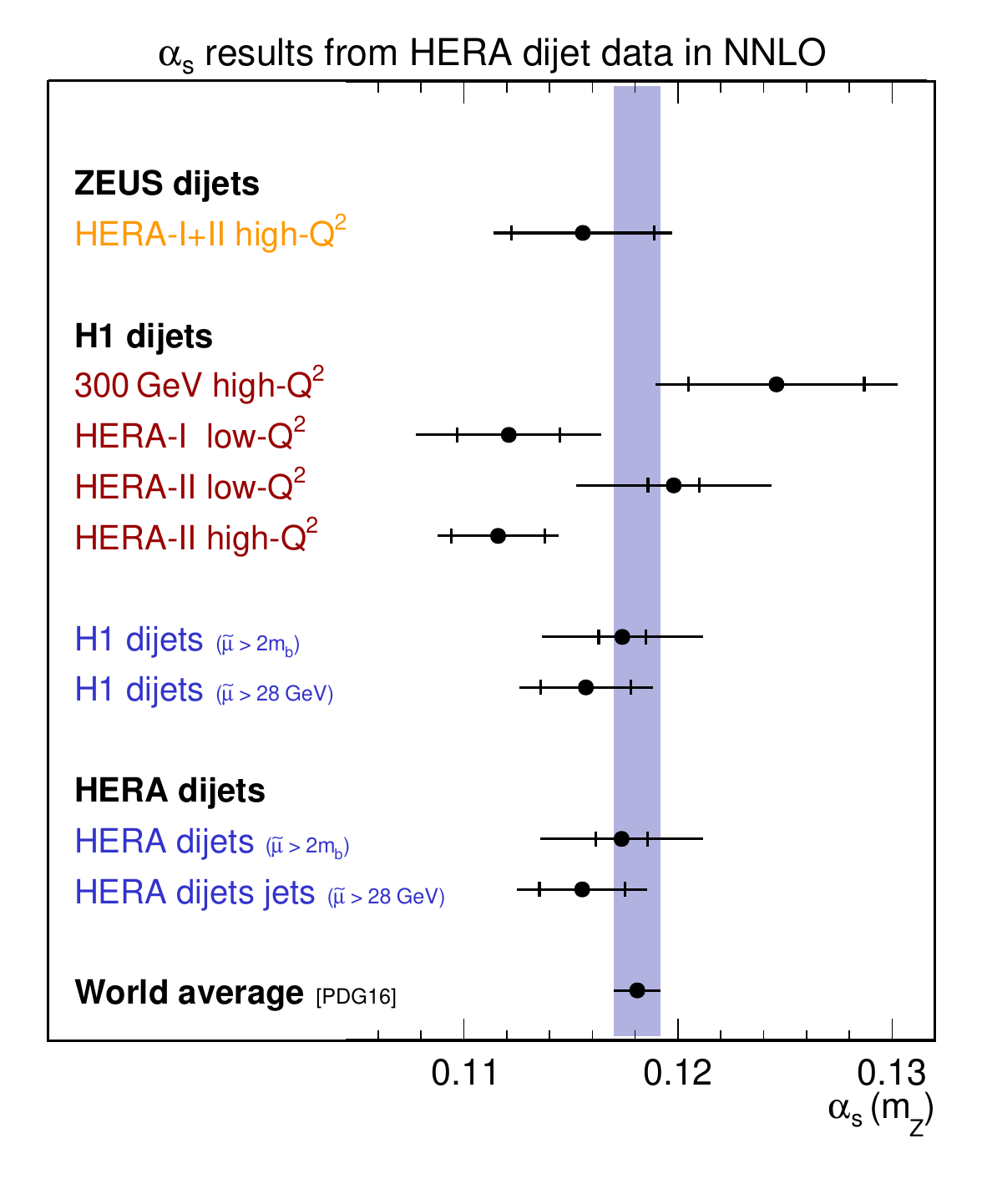}
  \caption{Strong coupling $\alphasmZ$ determined 
    from HERA inclusive jet (left) and dijet (right) cross section measurements
    using NNLO pQCD predictions (left).
    The vertical bands indicate the 2019 world average.
  }
  \label{fig:HERAjets}
\end{figure}

The combined analysis of H1 and ZEUS inclusive jet data yields the most precise
result of~\cite{Britzger:2019kkb}
\begin{equation}
  \alphasmZ = 0.1178\,(15)_\text{exp}\,(21)_\text{th}\,,
\end{equation}
which is obtained by restricting the selected data to $\mu>28$~GeV (Fig.~\ref{fig:HERAjets}, left).
The NNLO predictions provide an excellent description of the data in
all aspects and the value of $\chi^2/n_\text{dof}$ is found to be
$79.2/104$.
The dominant uncertainty arises from scale variations of the NNLO
predictions, while several PDF-related uncertainties are found to be
(negligibly) small.
Future improved predictions may exploit these low-$\mu$ data and may
achieve total uncertainties of less than $\pm0.002$.

~ 

\paragraph{$\alphas$ from di-jet cross sections}
A determination of $\alphasmZ$ from dijet cross sections in NC DIS was performed by H1.
Four previously published double-differential dijet cross section measurements as a function of $Q^2$ and the average transverse momenta of the two leading jets in the Breit frame
$\langle \pT \rangle$ were analyzed using NNLO pQCD predictions.
The data were taken during different HERA run-periods,
at different e$^\pm$-p center-of-mass energies and different kinematic
regions in $Q^2$.
From all four dijet data sets together a value of~\cite{H1:2017bml}
\begin{equation}
  \alphasmZ = 0.1157\,(22)_\text{exp}\,(23)_\text{th}
\end{equation}
is determined, and the fit exhibits an excellent quality with
$\chi^2/n_\text{dof}=31.6/43$.
While the above value is restricted to data with $\mu>28$~GeV, a
determination from all dijet data yields
$\alphasmZ=0.1174\,(10)_\text{exp}\,(36)_\text{th}$ (Fig.~\ref{fig:HERAjets}, right).
Hence, the experimental uncertainty would significantly be reduced, but
contrary the theoretical uncertainties increase overly, which is due to enhanced sensitivity to
$\alphasmZ$ of data at lower scales and thus increased scale uncertainties.


An analysis of dijet cross section measurements from ZEUS exploits a single
double-differential measurement~\cite{ZEUS:2010vyw}, as a function
of $Q^2$ and $\langle \pT \rangle$, where data were taken during HERA-I and HERA-II.
However, its sensitivity to $\alphasmZ$ is somewhat reduced,
since the most sensitive data at low $\langle \pT \rangle$ have
to be omitted since the definition of the dijet observable causes instabilities
of the pQCD predictions because of symmetric cuts on the two jets~\cite{ZEUS:2010vyw,Currie:2017tpe}. The ZEUS dijet cross sections yield a value
\begin{equation}
  \alphasmZ = 0.1156\,(34)_\text{exp}\,(25)_\text{th}\,.
\end{equation}
Due to the limited sensitivity of the ZEUS dijet cross sections to $\alphasmZ$, a
combined analysis with H1 data does not improve over the H1 only
result.
For the future, a reanalysis of these data would be desirable, such
that they can have a relevant impact on $\alphasmZ$, and the analysis
would profit from uncorrelated experimental uncertainties of the two independent
experiments and high statistics.

~ 
\paragraph{$\alphas$ from jet cross sections of the H1 Collaboration}
The H1 Collaboration performed a determination of $\alphasmZ$ from
inclusive jet and dijet data simultaneously using NNLO predictions.
Although these observables are highly statistically correlated, this 
was made possible by a simultaneous analysis of inclusive jet and
dijet data, where the statistical uncertainties and their correlations
were measured as well~\cite{H1:2014cbm,H1:2016goa}.
The analysis further includes inclusive jet data from H1 from HERA-I
and when restricting the data to $\mu>28$~GeV the result yields a value of~\cite{H1:2017bml}
\begin{equation}
  \alphasmZ = 0.1166\,(19)_\text{exp}\,(24)_\text{th}\,.
\end{equation}
Since the inclusive jet and dijet data are highly correlated, the
result improves only moderately over the result from the respective
inclusive jet data alone.
The smallest experimental uncertainty is achieved with relaxed cut on
$\mu$ and yields an experimental uncertainty of
$\delta\alphas=\pm0.0009$, which motivates future improved predictions
to underbid the experimental precision.

\subsubsection{The running of \texorpdfstring{$\alphas$}{alphas} from HERA jet cross sections}

Measurements of jet cross sections at HERA can be employed to test the
running of the strong coupling, because these measurements cover a
wide kinematic range.
Once the renormalization scale $\mu_r$ is identified with final-state
observables, every single cross section measurement of the
double-differential data sets covers a well-defined range in $\mu_r$.
Hence, determinations of $\alphasmZ$ from selected data points with similar
values of $\mu_r$ provide a determination of $\alphas$, where the
validity of the renormalization group equation (RGE) is employed only within a limited $\mu_r$ range. 
The value of $\alphasmZ$ can be translated to $\alphas(\mu_r^2)$, using
a representative value of $\mu_r$ of the selected data sets, and
multiple measurements at different $\mu_r$ provide a test of the
running of $\alphas$. 

\begin{figure}[thbp!]                           
  \centering                                      
  \includegraphics[width=0.40\textwidth]{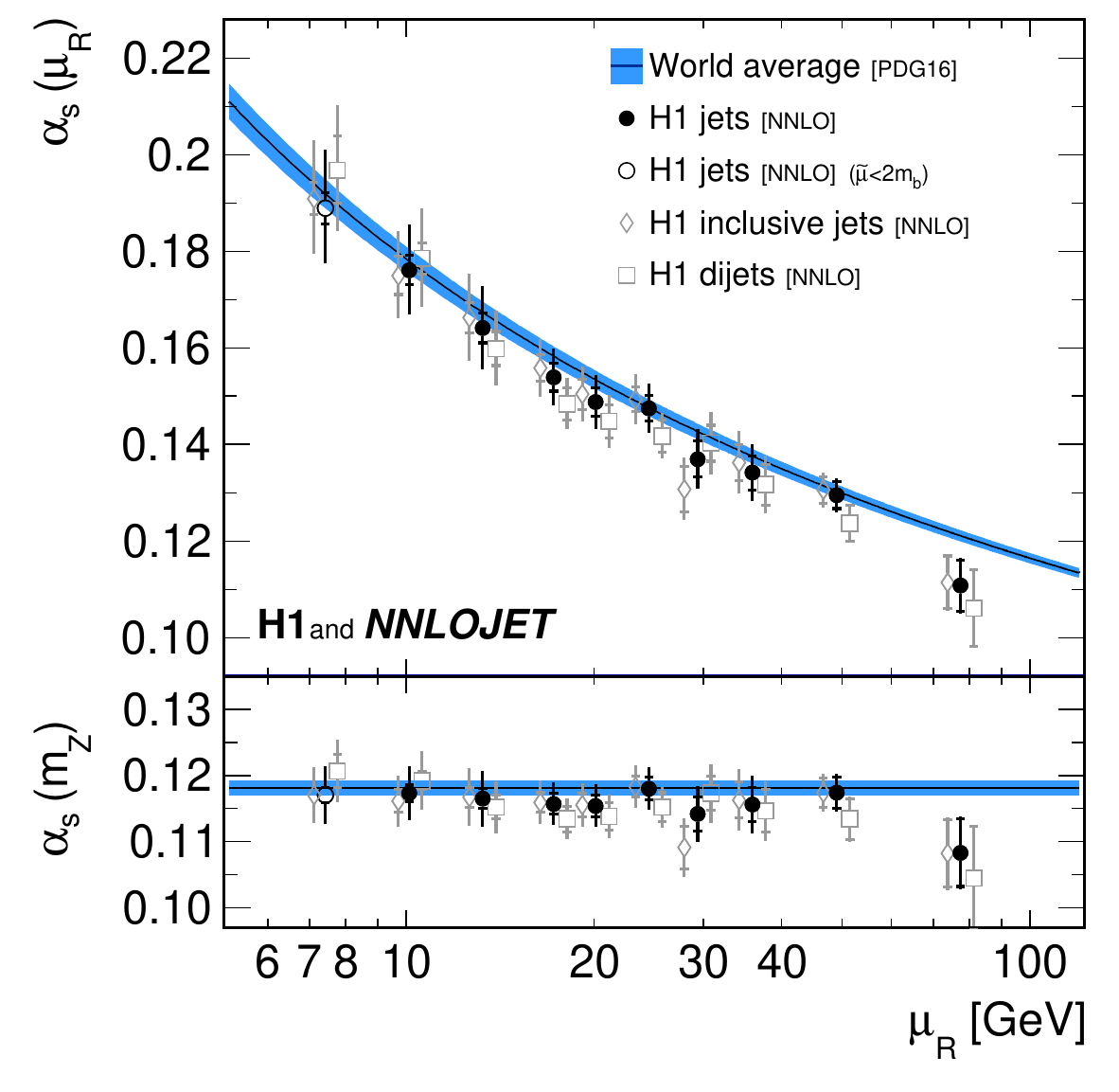}
  \hspace{0.08\textwidth}
  \includegraphics[width=0.40\textwidth]{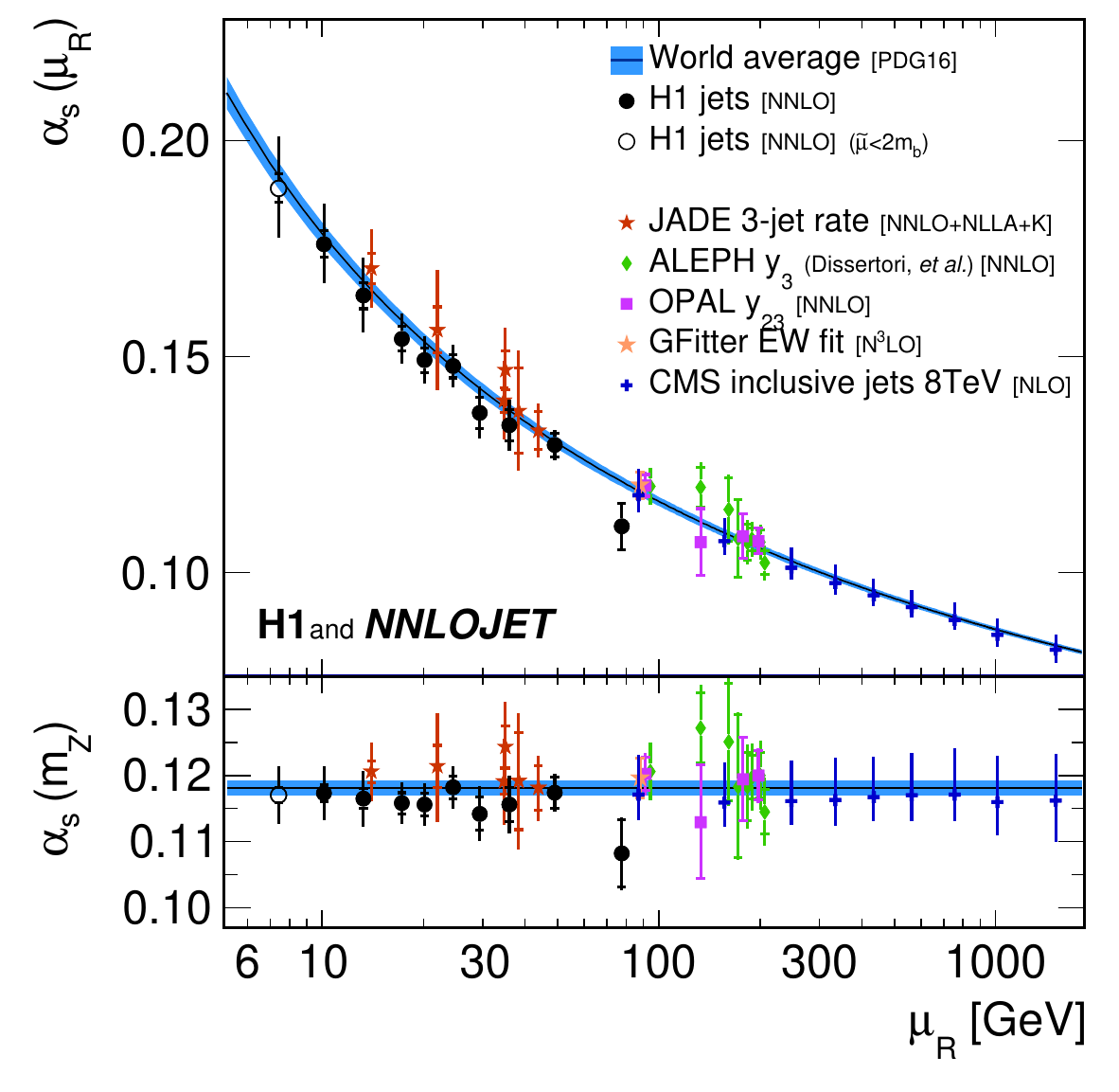}
  \caption{
    Left: Tests of the running of $\alphas$ using NNLO pQCD
    predictions together with H1 inclusive jet data,  dijet data, or
    both together.
    The lower panel displays $\alphasmZ$ as obtained in the fit to
    selected data points, and the upper panel the respective value of
    $\alphas(\mu_r^2)$ for a representative value of $\mu_r$. The blue
    shaded band shows the expectation when assuming the world average
    value from PDG 2016.
    Right: A comparison of $\alphasmZ$ and corresponding
    $\alphas(\mu_r^2)$ from different
    measurements~\cite{Baak:2014ora,OPAL:2011aa,Schieck:2012mp,Dissertori:2007xa,CMS:2016lna}. 
    Both figures taken from Ref.~\cite{H1:2017bml}.
  }
  \label{fig:runningDIS}
\end{figure}                                    
The running of the strong coupling constant is tested using NNLO
predictions together with
inclusive jet and dijet data from H1~\cite{H1:2017bml}, or solely inclusive jet data but
from H1 and ZEUS were exploited~\cite{Britzger:2019kkb}, respectively.
The $\overline{\text{MS}}$ renormalization scheme with 5 active
flavours is used.
Results from Ref.~\cite{H1:2017bml} are displayed in
Fig.~\ref{fig:runningDIS} and compared to the world average value and
other determinations.
Good consistency between results from inclusive jet and from dijet
cross sections is observed, and all results are in good agreement with
the world average value and the expectation from the RGE.

The HERA jet data are capable of testing the running of $\alphas$ in
the range from about 7 to 90~GeV with a considerable precision of
about 2.5 to 4\%.

\subsubsection{Further processes}

In the recent years, the determination of $\alphasmZ$ from e-p
collision data focused mainly on inclusive jet and dijet
data from HERA.
This is because the recently achieved NNLO
calculations~\cite{Currie:2016ytq,Currie:2017tpe} provided a significant improvement over
previously available NLO predictions~\cite{H1:2017bml}, and since H1 provided 
a comprehensive set of jet measurements from the HERA-II running period~\cite{H1:2014cbm,H1:2016goa}.
However, many further observables and final states can be exploited for $\alphasmZ$,
either once HERA data are further analyzed or when theoretical
advancements are achieved.

Some examples, where future improvements could be possible, would be
three-jet cross sections, heavy-flavour cross sections, event-shape
observables, jet substructure observables or observables in photoproduction.
As an example, three-jet cross sections from measurements
by H1 were proven to provide small experimental uncertainties in
$\alphas$ of about 1\%~\cite{H1:2014cbm,H1:2016goa}, but only
NLO~\cite{Nagy:2001xb}  
predictions are available and thus largely limit the precision, and no
corresponding measurement from ZEUS is published either.
Similarly, in photoproduction, a precision measurement of inclusive jet cross sections by ZEUS exhibits high
sensitivity to $\alphas$~\cite{ZEUS:2012pcn}, but no corresponding
measurement from H1 is available, and no theoretical advancements were
achieved for this process since then.
The measurements of various event-shape observables in NC DIS by
H1~\cite{H1:2005zsk} and ZEUS~\cite{ZEUS:2006vwm} from HERA-I data
proved a significant sensitivity to $\alphas$. 
However, although recent theoretical improvements were
achieved~\cite{Gehrmann:2019hwf} and new calculation techniques and
observables~\cite{Kang:2013lga,Kang:2013nha,Kang:2015swk}, or
substructure~\cite{Makris:2021drz} could be
studied, no suitable measurement of any such observable was published by the
HERA experiments from HERA-II data (though, note the ongoing work in
Ref.~\cite{Hessler:2021usr}).
A recent measurement of lepton-jet correlation observables in NC DIS~\cite{H1:2021wkz}, where the jets are 
defined in the laboratory rest-frame, may provide high sensitivity to $\alphasmZ$, and
also theoretical predictions with considerable precision are available.

\subsubsection{Future prospects}

Future electron-proton collider experiments provide many opportunities
for precision determinations of $\alphas$.
At lower center-of-mass energies, the electron-ion collider in the USA
(EIC)~\cite{Accardi:2012qut,AbdulKhalek:2021gbh} and in China
(EicC)~\cite{Anderle:2021wcy} will provide new high-luminosity data.
Early studies investigate a measurement of the 1-jettiness global event
shape observable and prospect a determination of $\alphasmZ$ at a level
of a few percent~\cite{AbdulKhalek:2021gbh}.

\begin{figure}[thbp!]                           
  \centering                                      
  \includegraphics[width=0.42\textwidth]{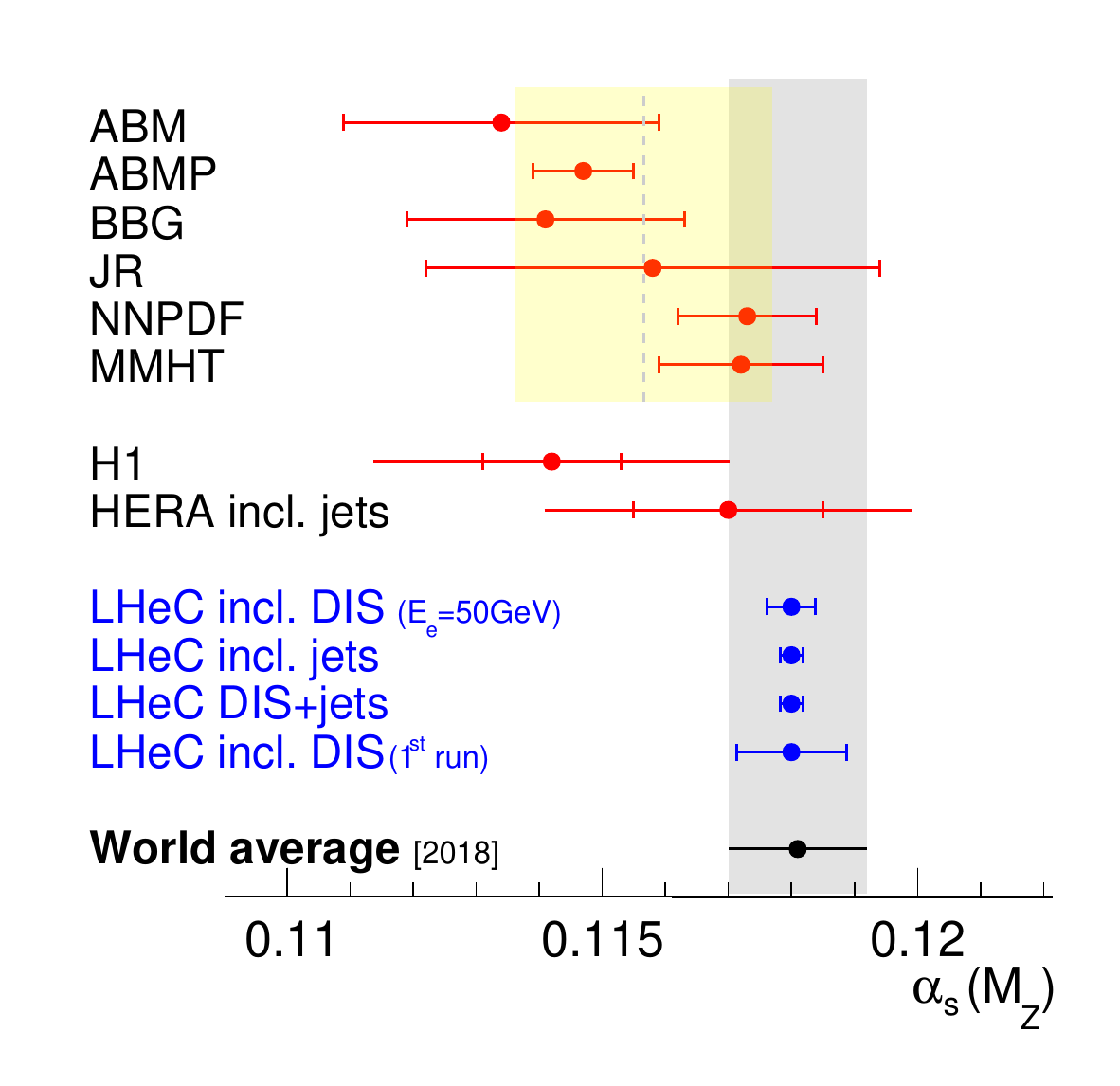}
  \hspace{0.08\textwidth}
  \includegraphics[width=0.38\textwidth]{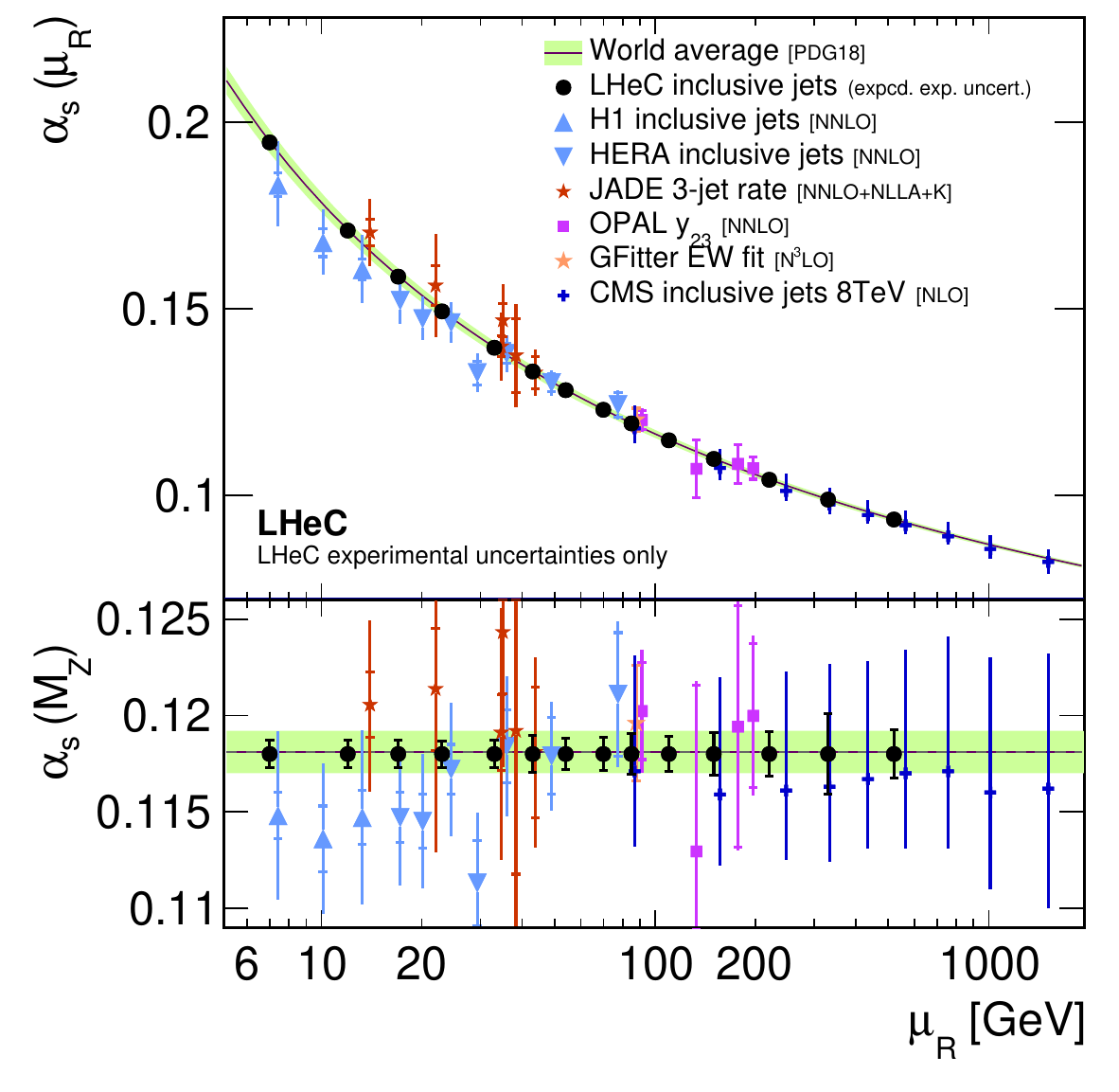}
  \caption{
    Left: Comparison of prospected determination of $\alphasmZ$ from
    inclusive DIS data at the LHeC and 
    (global) PDF fits.
    Right:
    An illustration of the prospected experimental uncertainties in a
    study of the running of $\alphas$ from inclusive jet cross
    sections at the proposed LHeC experiment.
    Both figs.\ taken from Ref.~\cite{LHeC:2020van}.
  }
  \label{fig:lhec}
\end{figure}

The proposed Large-Hadron-electron-Collider experiment at CERN
(LHeC)~\cite{LHeCStudyGroup:2012zhm,LHeC:2020van} will provide
e$^\pm$-p collision data at a center-of-mass energy of 1.3~TeV and hence the measurements of
hadronic final state observables cover a considerably larger range
than it was possible at HERA.
These data will provide new precision measurements of inclusive NC and
CC DIS cross section data.
Because of an excellent detector acceptance and high-luminosity, also
the high-$x$ region will be measured with high precision.
This will
provide the opportunity to determine $\alphas$ from inclusive DIS data
alone, something that was not possible with HERA data, and an
experimental uncertainty of~\cite{LHeC:2020van}
\begin{equation}
  \delta\alphasmZ = \pm0.00022\,\text{(exp+PDF)}\,,
\end{equation}
can be expected in combined determination of PDFs and $\alphasmZ$.
The prospected uncertainties are compared with recent determinations
in global PDF fits in Fig.~\ref{fig:lhec}.

From a simulation of inclusive jet cross section data with a complete
set of systematic uncertainties, a determination of $\alphasmZ$ with
uncertainties of
\begin{equation}
  \delta\alphasmZ = \pm0.00013\,\text{(exp)}~\pm0.00010\,\text{(PDF)}\,,
\end{equation}
where the first uncertainties accounts for experimental uncertainties
only and the second for PDF uncertainties, is expected in the LHeC/HL-LHC era.
While experimental uncertainties can be well estimated, it is not
quite possible to estimate the size of theoretical uncertainties reliably.

Similarly as at HERA, the running of $\alphas$ can be tested because
of the dynamic scale of the jet data. Prospects for scale-dependent
determinations of $\alphasmZ$ (and corresponding values
$\alphas(\mu_r^2)$) are displayed in Fig.~\ref{fig:lhec}.
It is observed that, from LHeC inclusive jet cross sections, the
running can be tested in the range from a few GeV up to about 600~GeV
with permille precision.
These measurements will become an indispensable experimental
confirmation of its validity, which will allow to combine the low-scale
$\alphas$ determinations from $\tau$-decays or lattice QCD with those
at the electroweak scale.

\subsection{\texorpdfstring{$\alphasmZ$}{alphasmZ} from inclusive W and Z cross sections in p-p collisions
\protect\footnote{A\lowercase{uthors:} D. \lowercase{d'}E\lowercase{nterria} (CERN)}}
\label{sec:pp_sigma_WZ}

A new determination of $\alphasmZ$ has been recently proposed~\cite{dEnterria:2019aat,CMS:2019oeb} based on comparing the fiducial cross sections of electroweak (EW) boson production in p-p collisions at the LHC, $\mathrm{pp\to W,\,Z\,+\,X}$ (with the EW bosons decaying into clean final states with electrons or muons) to the corresponding theoretical cross sections computed at next-to-next-to-leading-order (NNLO) accuracy. The method employed follows a similar approach to that used to derive $\alphasmZ$ from inclusive top-quark pair cross sections~\cite{Klijnsma:2017eqp}.
Such an extraction exploits the fact that, first, there are many W and Z boson cross sections available and that those are the most precisely measured ones at the LHC (with typical $\pm0.5\%$ statistical uncertainties and $\pm2\%$ systematic uncertainties, dominated by the integrated luminosity), and, second, that their theoretical values can also be precisely computed (with scale and PDF uncertainties amounting to 0.5--1\% and 2–-4\%, respectively). Theoretically, the cross sections can be derived from the convolution of parton densities $f_i(x_i,\mu_F)$ (evaluated at fractional momentum $x_i$ and factorization scale $\mu_F$) and elementary parton-parton cross sections (evaluated at renormalization scale $\mu_R$) written as an expansion in the QCD coupling,
\begin{equation}
\!\! \sigma_{\rm pp\to W,Z+X} =\! \int\!\!\int \mathrm{d}x_1 \mathrm{d}x_2\,f_1(x_1,\mu_F)f_2(x_2,\mu_F)\, \left[\hat{\sigma}_{\textsc{lo}}+ \alphas(\mu_R)\hat{\sigma}_{\textsc{nlo}} +\alphas^2(\mu_R)\hat{\sigma}_{\textsc{nnlo}} + \cdots \right].
\end{equation}
Although the bulk of the cross section, given by $\hat{\sigma}_{\textsc{lo}}$, is a pure EW quantity, the contributions of NLO and NNLO higher-order pQCD corrections increase the overall $\sigma_{\rm W,Z}$ value~\cite{Anastasiou:2003ds,Catani:2009sm,Gavin:2010az,Campbell:2010ff,Boughezal:2016wmq}, and provide the dependence on $\alphasmZ$ that allows extracting this parameter from a combined data-theory comparison. The size of the higher-order corrections, encoded in the so-called K-factor given by the ratio of NNLO to LO cross sections, amounts to $\rm K = \sigma_{_{\rm NNLO}}/\sigma_{_{\rm LO}}\approx$~1.22, 1.33, and 1.29 in the typical ATLAS, CMS, and LHCb fiducial acceptance for \Wpm\ and Z final states, respectively. Such a result indicates that indeed W and Z boson production in p-p collisions is sensitive to $\alphasmZ$, through $\sim$25\% higher-order matrix-elements direct contributions to their total cross sections. 

Up to the year 2019, there were $12+9+7=28$ W$^\pm$and Z fiducial cross sections measured in p-p collisions by CMS, LHCb, and ATLAS, respectively, that have been exploited in~\cite{dEnterria:2019aat,CMS:2019oeb} to determine $\alphasmZ$ by comparing them to the corresponding NNLO theoretical predictions computed with \mcfm~v.8~\cite{Campbell:2010ff,Boughezal:2016wmq} for a variety of PDF sets (CT14~\cite{Dulat:2015mca}, HERAPDF2.0~\cite{Abramowicz:2015mha}, MMHT14~\cite{Harland-Lang:2014zoa}, and NNPDF3.0~\cite{NNPDF:2014otw}) and $\alphasmZ = 0.115$--0.121 values. The absolute W$^\pm$and Z cross section data exploited here were not used by any of these PDF sets in their global fits to extract the parton densities themselves, so there is no ``double counting'' of the same data samples in the $\alphasmZ$ determination. The final numerical accuracy of the calculations is 0.2–0.6\%, comparable to $\sim$1\% differences found with alternative NNLO
calculators such as FEWZ~\cite{Gavin:2010az} or DYNNLO~\cite{Catani:2009sm}. In the theoretical cross sections, we included corrections (mostly negative, of a few percent size) due to EW and photon-induced production processes evaluated at NLO accuracy with \textsc{mcsanc}~v.1.01~\cite{Bondarenko:2013nu}. 

\begin{figure}[htbp!]
\centering
\includegraphics[width=0.49\textwidth]{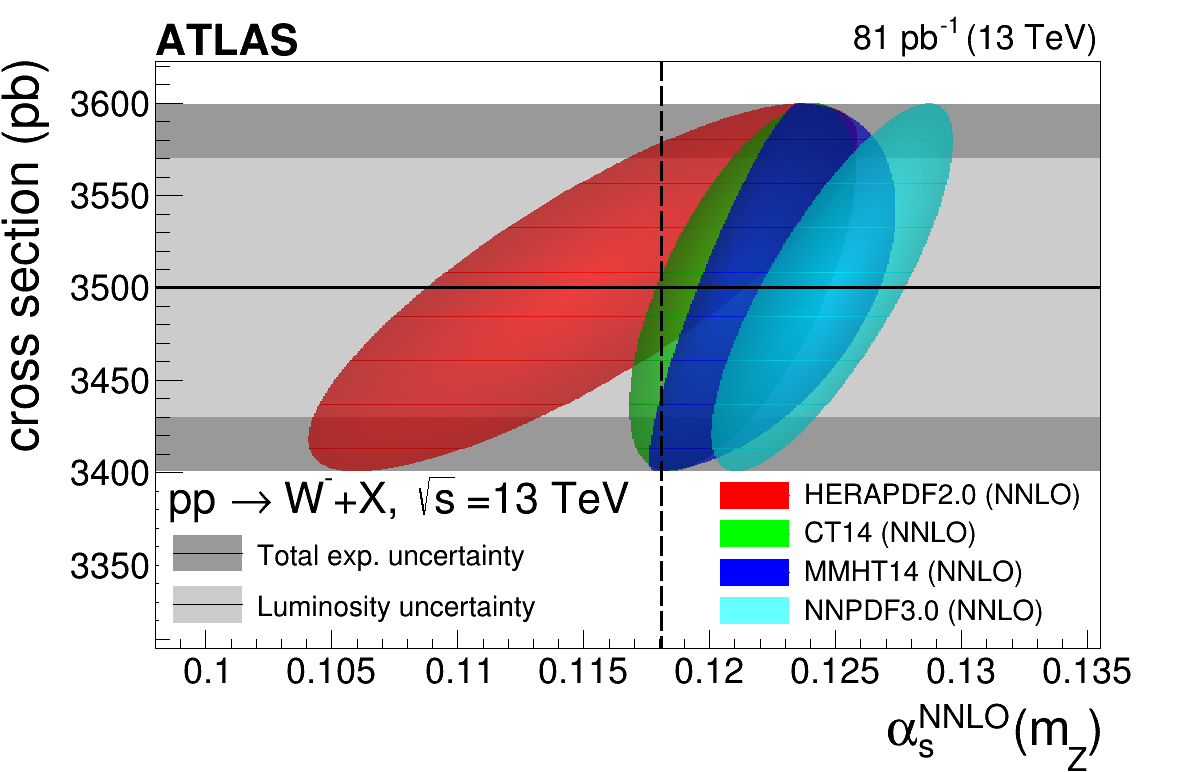}
\includegraphics[width=0.49\textwidth]{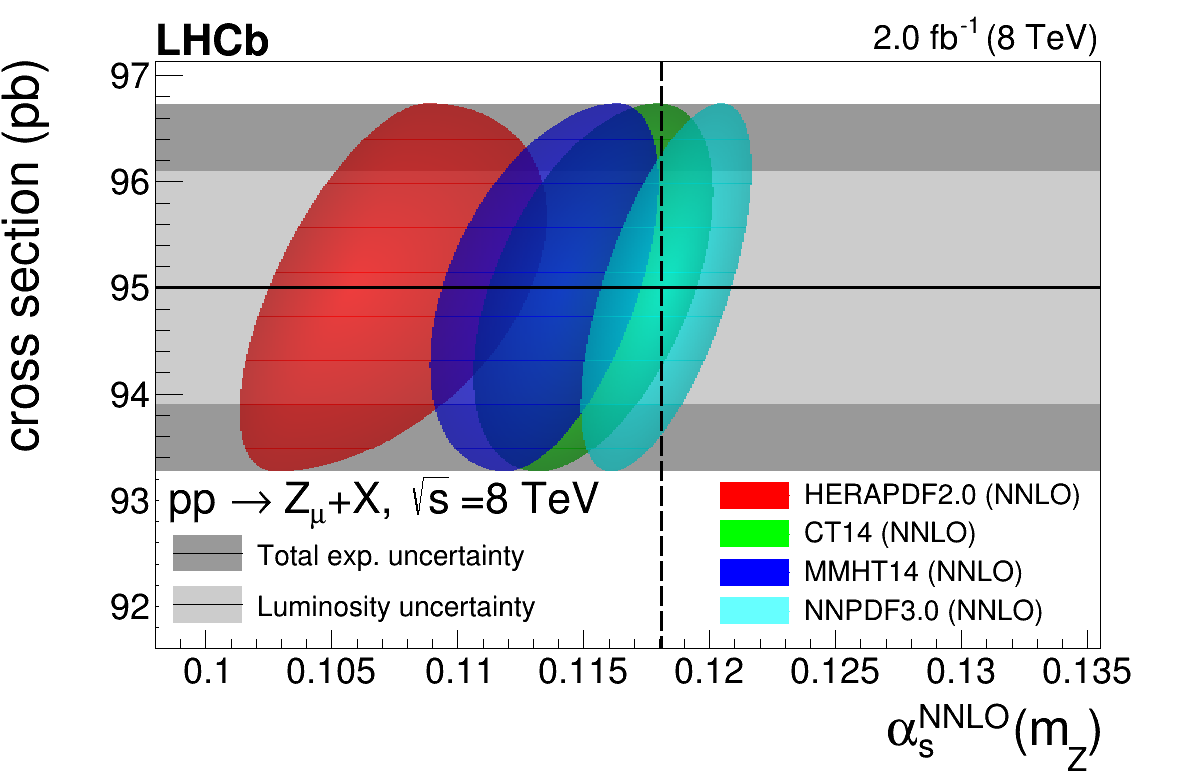}
\caption{Examples of data-theory comparisons for EW fiducial cross sections in p-p collisions measured by ATLAS (W$^+$ bosons at $\sqrts = 13$~TeV, left)~\cite{ATLAS:2016fij} and LHCb (Z boson at $\sqrts = 8$~TeV, right)~\cite{LHCb:2015mad}. The experimental cross sections are plotted as horizontal black lines with outer darker (inner grey) bands indicating their total (integrated luminosity) uncertainties. The theoretical predictions are computed for each PDF set as a function of $\alphasmZ$, and combined with the experimental results into Jpdfs shown as filled ellipses. The vertical dashed line indicates the expected predictions for $\alphasmZ = 0.118$.
\label{fig:ellipses}}
\end{figure}

Figure~\ref{fig:ellipses} shows examples of experimental ATLAS (left) and LHCb (right) cross sections (horizontal lines and bands) compared to the corresponding theoretical predictions per PDF as a function of $\alphasmZ$ (coloured ellipses). 
From the computed cross sections, a linear dependence of $\sigma^\mathrm{th}_\mathrm{W,Z}$ on $\alphasmZ$ is derived, and the filled ellipses are constructed to represent the contours of the joint probability density functions (Jpdfs) of the theoretical and experimental results, with a width representing a two-dimensional one standard-deviation obtained from the product of both probability densities for each PDF. The uncertainty in the theoretical cross sections is given by the quadratic sum of its associated PDF, scale, and numerical uncertainties. 

The predictions are consistent with data within uncertainties but not systematically for the same $\alphasmZ$ value (in particular, HERAPDF2.0 results do not always overlap with any of the others within the 1 std.-dev.\ region). For a fixed $\alphasmZ$ value, HERAPDF2.0 (NNPDF3.0) predict larger (smaller) cross sections. Namely, HERAPDF2.0 (NNPDF3.0) favour systematically smaller (larger) $\alphasmZ$ values, whereas MMHT14 and CT14 predictions are in between and less scattered over the 28 measurements considered. The predictions derived with HERAPDF2.0 (MMHT14) always feature the smallest (largest) slope, \ie\ HERAPDF2.0 (MMHT14) cross sections are the least (most) sensitive to underlying $\alphasmZ$ variations. With the 28 fiducial cross sections computed, we find that for the baseline QCD coupling constant value of $\alphasmZ = 0.118$ of all PDF sets, the data-theory accord is overall better for the predictions calculated with CT14 and MMHT14 (goodness-of-fit per degree of freedom, $\chi^2/n_\text{dof} \approx 1$) than those obtained with the HERAPDF2.0 and NNPDF3.0 sets ($\chi^2/n_\text{dof}\approx 2.1$).

The cross sections calculated with different $\alphasmZ$ values are fitted (using $\chi^2$-minimization) to a first-order polynomial and the corresponding slope $k$ is extracted for each PDF and measurement. 
The value of $\alphasmZ$ preferred by each individual measurement is determined by the crossing point of the fitted linear theoretical curve with the experimental horizontal line. All the individual 28 $\alphasmZ$ derived per PDF set, and the correlation matrices associated with all their uncertainties, are given as inputs of the \textsc{convino} v1.2~\cite{Kieseler:2017kxl} program employed to determine the final best estimate of all combined values. 
Table~\ref{tab:final_alphas_perPDF} lists the $\alphasmZ$ values, along with the uncertainty breakdowns from every source, determined for each PDF set through the combination of the 28 individual determinations. 
The total symmetrized uncertainties amount to $\sim$1.4\% for CT14, $\sim$2.1\% for HERAPDF2.0, $\sim$1.3\% for MMHT14 and $\sim$1.6\% for NNPDF3.0. The last column of Table~\ref{tab:final_alphas_perPDF} lists the ($\chi^2/n_\text{dof}$) of the final single combined result compared to the 28 individual $\alphasmZ$ extractions. 

\begin{table}[htbp!]
\centering
\tabcolsep=2.mm
\caption{Strong coupling constant $\alphasmZ$ values extracted per PDF set by combining all the individual results obtained for each W$^\pm$ and Z boson production cross section measurements, listed along with their propagated total and individual uncertainties. The last column tabulates the 
$\chi^2/n_\text{dof}$ of the final single combined result compared to the 28 individual $\alphasmZ$ determinations.\vspace{0.2cm}
\label{tab:final_alphas_perPDF}}
\renewcommand*{\arraystretch}{1.2}
\begin{tabular}{lcccccccc}\hline
PDF & $\alphasmZ$  & $\delta\stat$ & $\delta\lum$ & $\delta\syst$ & $\delta$(PDF) & $\delta$(scale) & $\delta\statt$ & $\chi^2/n_\text{dof}$ \\\hline
CT14       & $0.1172^{+0.0015}_{-0.0017}$ & 0.0003 & 0.0005 & 0.0006 &\,$^{+0.0011}_{-0.0013}$ & 0.0006 & 0.0003 & 23.5/27 \\
HERAPDF2.0 & $0.1097^{+0.0022}_{-0.0023}$ & 0.0004 & 0.0009 & 0.0009 & \,$^{+0.0015}_{-0.0016}$ & 0.0007 & 0.0005 & 27.0/27 \\
MMHT14     & $0.1188^{+0.0019}_{-0.0013}$ & 0.0002 & 0.0008 & 0.0003 & \,$^{+0.0015}_{-0.0007}$ & 0.0007 & 0.0002 & 19.3/27 \\
NNPDF3.0   & $0.1160 \pm 0.0018$ & 0.0006 & 0.0004 & 0.0005 & 0.0013 & 0.0006 & 0.0007 & 56.9/27 \\\hline
\end{tabular}
\end{table}

\begin{figure}[htbp!]
\centering
\includegraphics[width=0.58\textwidth]{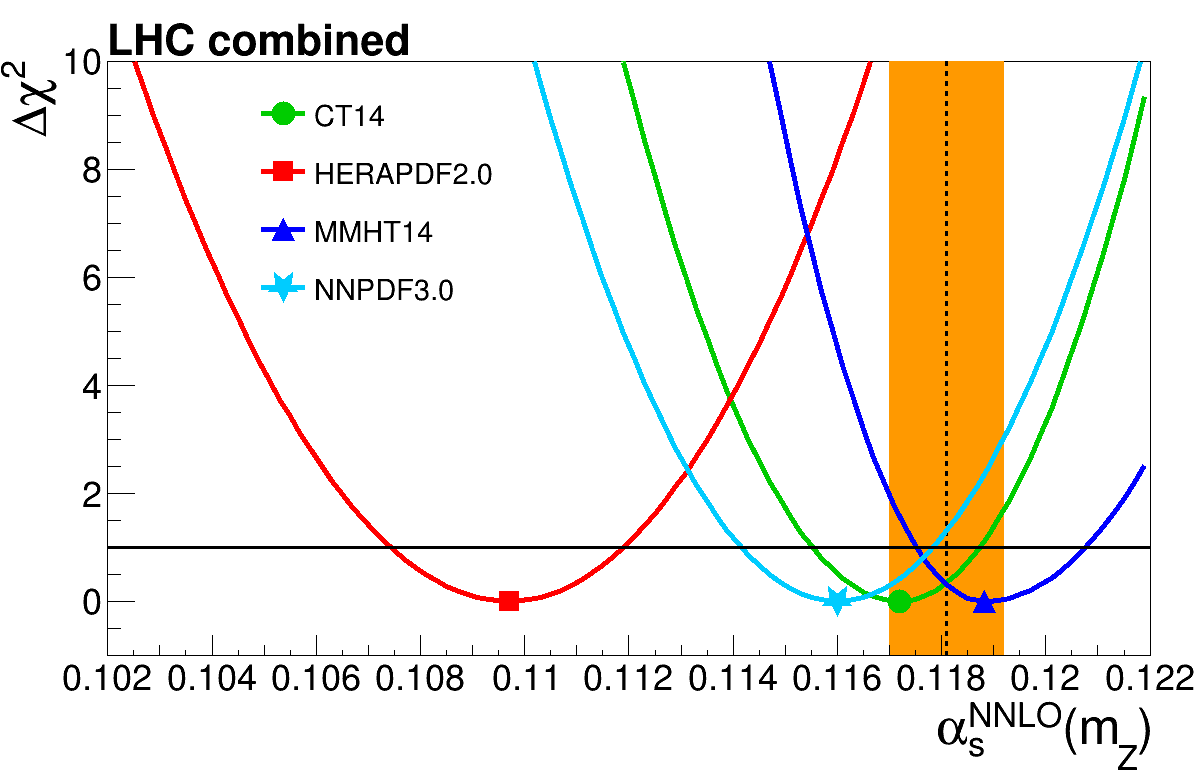}
\includegraphics[width=0.41\textwidth]{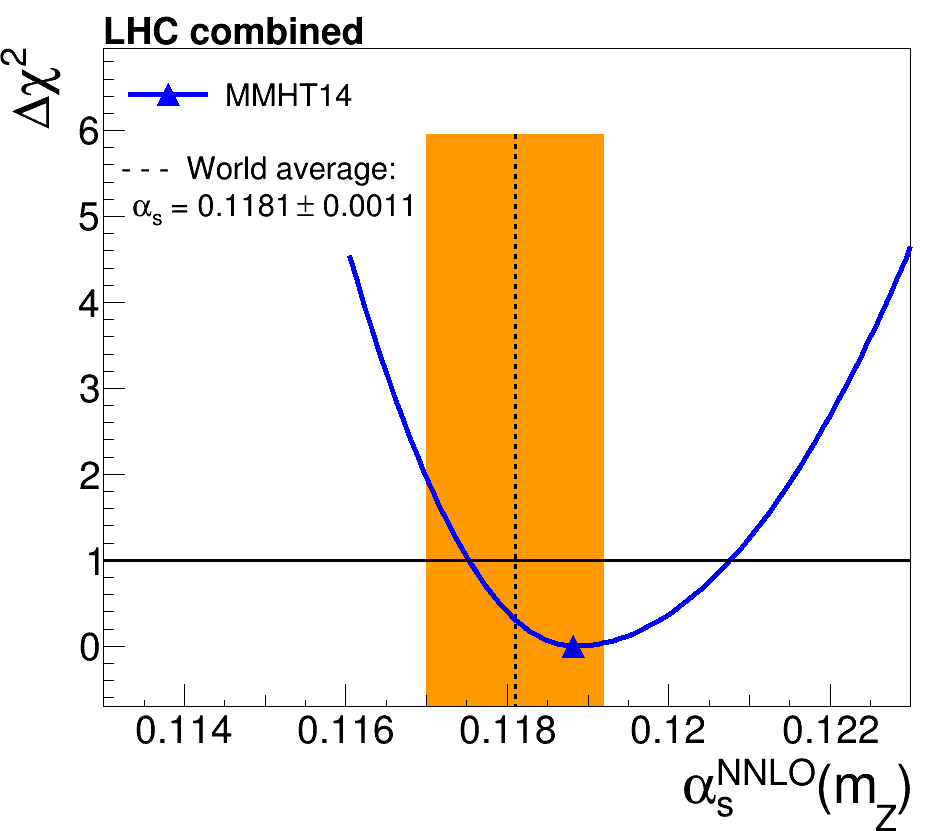}
\caption{Final $\alphasmZ$ values determined from the analysis of the EW boson inclusive cross sections at the LHC using the CT14, HERAPDF2.0, MMHT14, and NNPDF3.0 PDF sets (left), and $\alphasmZ$ extraction from the MMHT14 PDF alone (right), compared to the 2018 world average (vertical orange band).
\label{fig:final_alphas_per_PDF}}
\end{figure}

The final $\alphasmZ$ values determined for each individual PDF are plotted in Fig.~\ref{fig:final_alphas_per_PDF} (left) compared with the 2018 world average of $\alphasmZ = 0.1181 \pm 0.0011$ (orange band)~\cite{ParticleDataGroup:2018ovx}. The (asymmetric) parabolas are constructed to have a minimum at the combined value and are fitted to go through $\Delta\chi^2=1$ (horizontal black lines) at the one std.\,deviation  uncertainties quoted in Table~\ref{tab:final_alphas_perPDF}.
The robustness and stability of the final $\alphasmZ$ determination per PDF is cross-checked by varying key
experimental and theoretical ingredients and uncertainties. For this purpose, the {\sc convino} combination is redone as follows: (i) for a fraction of the data subsets (ATLAS, CMS, or LHCb alone; or for 7, 8, 13 TeV c.m.\ energies only), (ii) varying the correlation factors of the PDF/scale uncertainties between 0 and 1, (iii) shifting the central values of the computed cross sections by $\pm 1\sigma$ of the theoretical uncertainty prior to combination, and (iv) adding $\pm 1\%$ uncorrelated theoretical uncertainty (to cover differences among NNLO calculators). The results of such tests indicate that the MMHT14 result is the most stable against any variations in the analysis, whereas a few larger-than-1-standard-deviation changes appear in some cases for the results of the other PDF sets.
The preferred QCD coupling value extracted from this study is that of MMHT14, $\alphasmZ=0.1188 \pm 0.0016$ plotted in Fig.~\ref{fig:final_alphas_per_PDF} (right), because (i) it features the largest sensitivity (slope) of $\sigma_\mathrm{W,Z}$ to $\alphas$, (ii) it shows the lowest $\chi^2/n_\text{dof}$ of the final single combined result compared to the 28 individual $\alphasmZ$ extractions, (iii) it has the smallest (symmetrized) propagated uncertainties, and (iv) it is the most stable against any data or theory analysis variations.

This study confirms that the total inclusive \Wpm\ and Z boson cross sections at hadron colliders are new promising observables that can provide useful constraints on the value of the QCD coupling constant, and that can eventually help improve the precision of the $\alphasmZ$ world average. The recent availability of N$^3$LO codes~\cite{Duhr:2020sdp,Camarda:2021ict} for the calculation of inclusive \Wpm\ and Z boson production cross sections, with one extra higher degree of theoretical accuracy compared to the one used here, will allow for further reductions of the propagated scale uncertainties, provided that PDF uncertainties are available at the same level of pQCD accuracy. Such theoretical developments, combined with upcoming EW boson measurements at the LHC with $\mathcal{O}($1\%) experimental uncertainties, mostly thanks to further reduced integrated luminosity uncertainties, will enable future $\alphasmZ$ extractions with propagated uncertainties below the 1\% level.

\subsection{\texorpdfstring{$\alphasmZ$}{alphasmZ} from the transverse-momentum distribution of Z bosons
\protect\footnote{A\lowercase{uthors:} S. C\lowercase{amarda} (CERN), M. S\lowercase{chott} (JGU M\lowercase{ainz})}}
\label{sec:pp_zpt}

A new determination of $\asmz$, using QCD resummed theory predictions and based on
a semi-inclusive (\ie\ radiation inhibited) observable at hadron-hadron colliders, has been recently proposed in Ref.~\cite{Camarda:2022qdg}.
The strong-coupling constant $\alphasmZ$ is determined from the transverse-momentum distribution of Z bosons measured at $\sqrts = 1.96$~TeV with the CDF experiment, using predictions based on $q_\mathrm{t}$-resummation at N$^3$LO+N$^3$LL accuracy, as implemented in the \texttt{DYTurbo}~\cite{Camarda:2019zyx,Camarda:2021ict}
program. The measurement is performed through a simultaneous fit of $\alphasmZ$ and the nonperturbative Sudakov form factor.
This measurement has all the
desirable features for a precise determination of
$\alphasmZ$: large observable’s
sensitivity to $\alphasmZ$ compared to the experimental precision;
high accuracy of the theoretical prediction; small size of
nonperturbative QCD effects.\\

Measuring $\alphasmZ$, or equivalently $\Lambda_{\textrm{QCD}}^{\MSbar}$,
from semi-inclusive Drell--Yan cross sections was first proposed in
Ref.~\cite{Catani:1990rr}, by using Monte Carlo parton showers to determine
$\Lambda_{\textrm{QCD}}^{\textrm{MC}}$ and later convert it to
$\Lambda_{\textrm{QCD}}^{\MSbar}$. The conversion is based on resummation
arguments showing that a set of universal QCD corrections can be
absorbed in coherent parton showers by applying a simple rescaling,
the so-called Catani-Marchesini-Webber (CMW) rescaling.

The Z-boson transverse-momentum distribution at small transverse
momentum is one of such semi-inclusive observables. The recoil of Z
bosons produced in hadron collisions is mainly due to QCD
initial-state radiation, and the Sudakov form factor is responsible for the
existence of a Sudakov peak in the distribution, at
transverse-momentum values of approximately 4 GeV. The position of the
peak is sensitive to the value of the strong-coupling constant.

For the measurement of $\alphasmZ$ from the Z-boson transverse-momentum
distribution it is necessary to rely on fast computing codes which
allow the calculation of variations in the input parameters with small
numerical uncertainties. To this end, the \texttt{DYTurbo} program was used.

The CDF measurement of Z-boson transverse-momentum
distribution~\cite{CDF:2012brb} at the Tevatron collider is ideal
for testing the extraction of $\alphasmZ$ with {\texttt{DYTurbo}}
predictions. This measurement was performed with the angular coefficients
technique, which allows extrapolating the cross section to full-lepton
phase space with small theoretical uncertainties.
The full-lepton phase space cross section allows fast predictions and
avoid any theoretical uncertainties on the modelling of the Z-boson polarization.
Another advantage of this measurement with respect to similar
measurements performed at the LHC is the fact that Tevatron is a
proton-antiproton collider, and the Z-boson production has reduced
contribution from heavy-flavour-initiated processes compared to
proton-proton collisions at the LHC.

The CDF measurement is performed in the electron channel, with central
($|\eta^e| < 1.1$) and forward ($1.2 < |\eta^e| < 2.8$) electrons,
allowing a coverage up to Z-boson rapidity of $y = 2.8$, and a small
extrapolation to the full rapidity range $|y_{\textrm{max}}| \approx 3.1$
of Z-boson production at $\sqrts = 1.96$~TeV. The data sample is
characterized by low values of the average number of interactions per
bunch crossing, and by good electron resolution, at the level of
1~GeV for central electrons, and 1.5~GeV for forward electrons. The
good resolution allows fine transverse-momentum binning (0.5~GeV)
while keeping the bin-to-bin correlations smaller than 30\%.

The nonperturbative QCD corrections to the Z-boson
transverse-momentum distribution are modelled by including a
nonperturbative term in the Sudakov form factor: $S(b) \to S(b) \cdot
S_{\textrm{NP}}(b)$. The general form of $S_{\textrm{NP}}(b)$ is mass and centre-of-mass
energy dependent~\cite{Landry:2002ix}.
However, at fixed invariant mass $q = m_Z$, and for one value of centre-of-mass
energy, the form of $S_{\textrm{NP}}(b)$ can be simplified to depend on a single parameter $g$:
$S_{\textrm{NP}}(b) = \exp ( -g \cdot b^2 )$.
The nonperturbative parameter $g$ is generally determined from the
data, and its value depends on the chosen prescription to avoid the
Landau pole in the impact-parameter b-space, which
corresponds to a divergence of the Sudakov form factor.
The divergence is avoided by using the
so-called $b_\star$ prescription, which freezes $b$ at a given value
$b_{\textrm{lim}}$:  $b \to b_\star = \frac{b}{1+b^2/b_{\textrm{lim}}^2}$. In this
analysis $b_{\textrm{lim}}$ is set to the value of 3~GeV$^{-1}$.

\begin{figure}[htbp!]
\begin{center}
\resizebox{0.48\textwidth}{!}{\includegraphics[width=0.99\columnwidth]{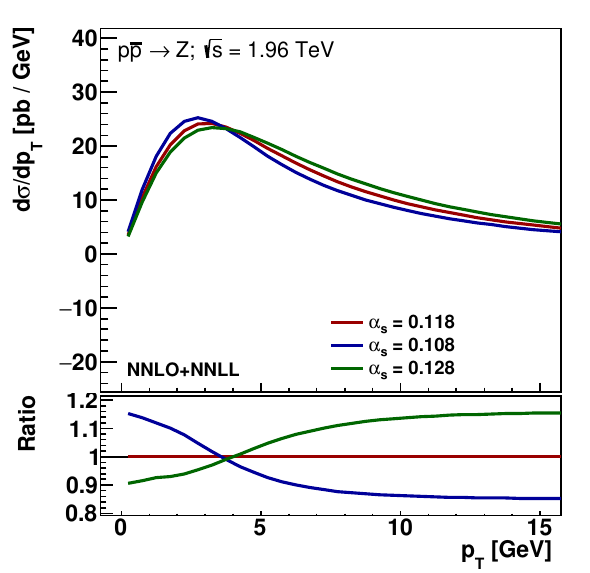}}
\resizebox{0.48\textwidth}{!}{\includegraphics[width=0.99\columnwidth]{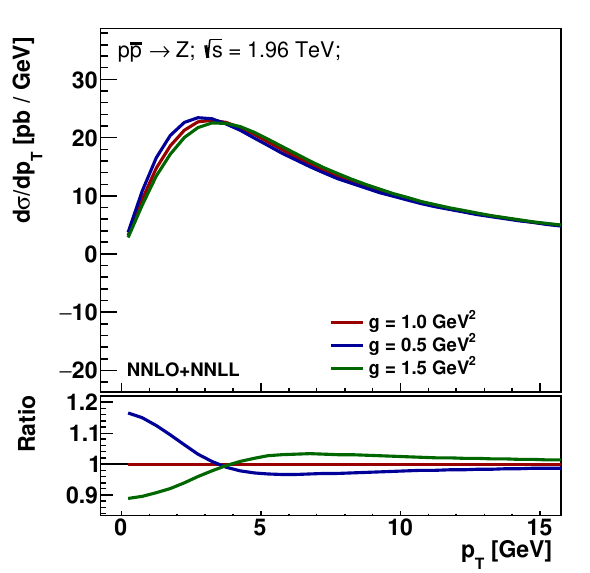}}
\caption{Sensitivity of the Z-boson transverse-momentum distribution to $\alphasmZ$ (left) and to the nonperturbative QCD parameter $g$ (right).
\label{fig:sens}}
\end{center}
\end{figure}

The sensitivity of the Z-boson transverse-momentum distribution to
$\alphasmZ$ mainly comes from the position of the Sudakov
peak, and is related to the average recoil scale $\langle p_{\textrm{T}} \rangle \approx 10$~GeV.
The sensitivity of the Z-boson transverse-momentum distribution to $g$ also comes from the
position of the Sudakov peak. However, the scale of the
nonperturbative smearing governed by $g$ corresponds to the value of primordial $k_\mathrm{T}$. Typical
values of $g \approx 0.6$~GeV$^2$ corresponds to a primordial $k_\mathrm{T}$ of
approximately 1.5~GeV. It is possible to disentangle the perturbative
contribution to the Sudakov form factor, governed by $\alphasmZ$,
from the nonperturbative one, determined by $g$, thanks to their
different scale, as shown in Fig.~\ref{fig:sens}.

The statistical analysis leading to the determination of
$\alphasmZ$ is performed by interfacing \textsc{DYTurbo} to
xFitter~\cite{Alekhin:2014irh}. The agreement between data and predictions is
assessed by means of a $\chi^2$ function, which includes experimental
and PDFs theoretical uncertainties~\cite{HERAFitterdevelopersTeam:2015cre}. The nonperturbative form factor
is added as unconstrained nuisance parameter in the $\chi^2$
definition, \ie\ it is left free in the fit. The fit to the data is
performed in the region of transverse momentum $p_{\textrm{T}} < 30$~GeV by minimising the $\chi^2$ as a
function of $\alphasmZ$, with $\alphas$ variations as provided in
LHAPDF.

The corrections to the Z-boson transverse-momentum distribution due
to QED initial-state radiation are estimated with \textsc{Pythia8} and the AZ tune~\cite{ATLAS:2014alx}
of the parton shower parameters, and
applied as multiplicative corrections. They are the level of
1\%, and are responsible for a shift in the measured value of
$\alphasmZ$ of $\delta\alphasmZ = -0.0004$.

\begin{figure}[htp!]
\centering
\includegraphics[width=0.45\textwidth]{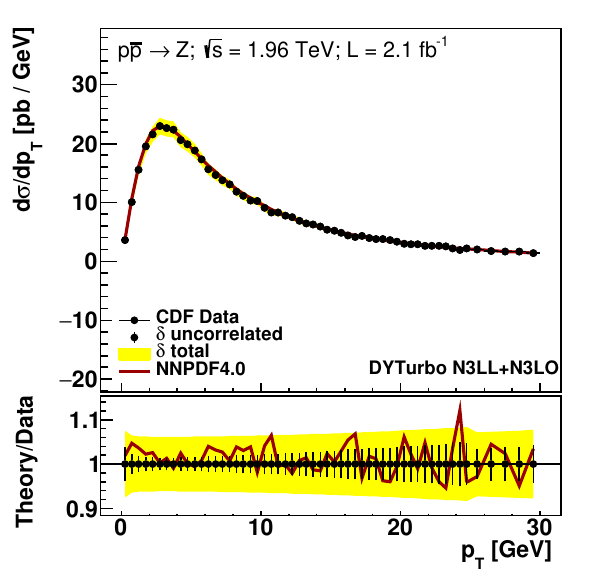}
\includegraphics[width=0.54\textwidth]{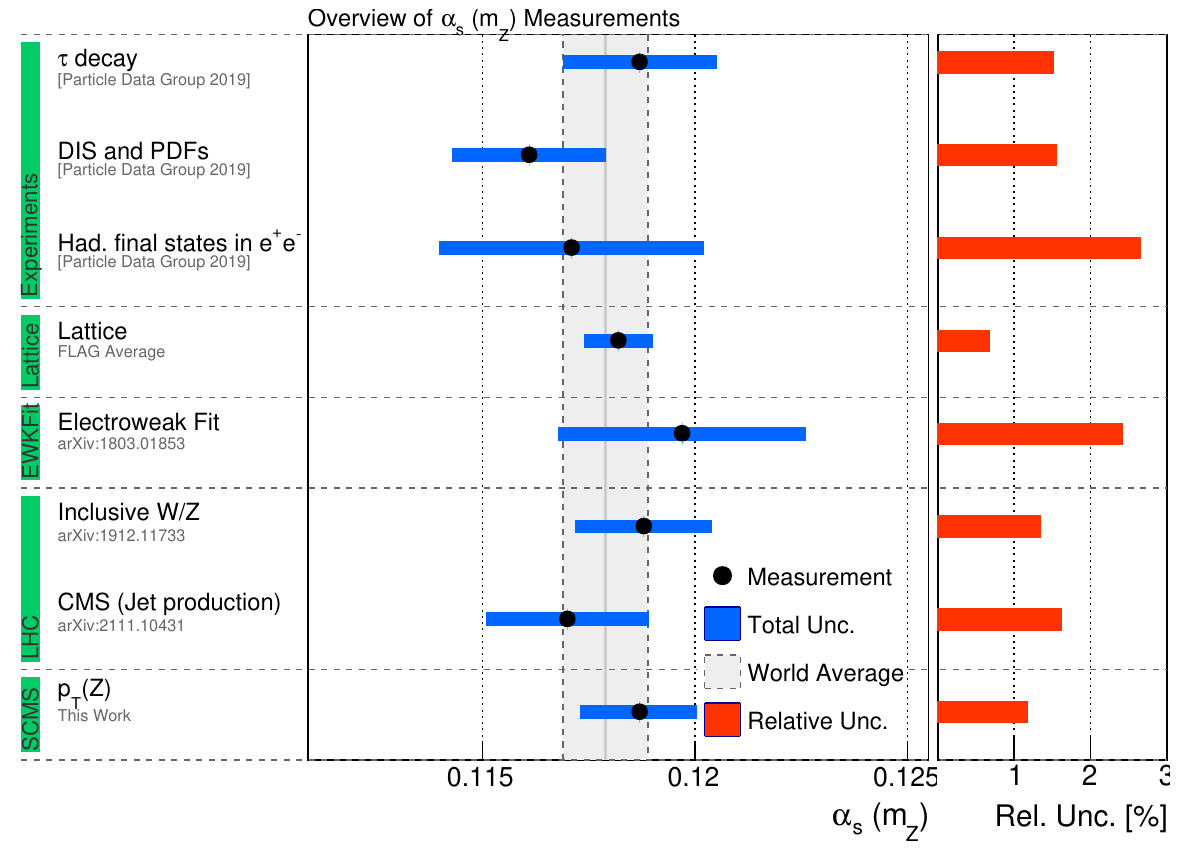}
\caption{Left: Comparison of the N$^3$LO+N$^3$LL \texttt{DYTurbo} prediction to the measured Z-boson transverse-momentum distribution. Right: Comparison of the $\alphasmZ$ derived here to the results of other categories of the world-average value.\label{fig:posfit_overview}}
\end{figure}

The determination of $\alphasmZ$ with the NNLO NNPDF4.0 PDF set~\cite{Ball:2021leu}
yields $\alphasmZ = 0.1187$, with a statistical uncertainty of
$\pm 0.0007$, a systematic experimental uncertainty
of $\pm 0.0001$, and a PDF uncertainty of
$\pm 0.0004$. The value of $g$ determined in the fit
is $g = 0.66 \pm 0.05$~GeV$^2$, and the value of the $\chi^2$ function
at minimum is 41 per 53 degrees of freedom. Various alternative NNLO
PDF sets are considered: CT18~\cite{Hou:2019efy}, CT18Z, MSHT20~\cite{Bailey:2020ooq},
HERAPDF2.0~\cite{Abramowicz:2015mha}, and ABMP16~\cite{Alekhin:2017kpj}. The determined values of
$\alphasmZ$ range from a minimum of 0.1178 with the ABMP16 PDF
set to a maximum of 0.1192 with the CT18Z PDF set. The midpoint value
in this range of $\alphasmZ = 0.1185$ is considered as nominal
result, and the PDF envelope of $\pm 0.0007$ as an additional source of uncertainty.
Missing higher order uncertainties are estimated
through independent variations of $\mu_R$, $\mu_F$ and $Q$ in the
range $m_{\ell\ell}/2 \leq \{ \mu_R, \mu_F, Q \} \leq 2m_{\ell\ell}$
with the constraints $0.5\leq \{ \mu_F/\mu_R, Q/\mu_R, Q/\mu_F \}\leq
2$, leading to 14 variations.
The determined values of $\alphasmZ$ range from a minimum of
0.1177 to a maximum of 0.1193, yielding a scale-variation envelope of
$\pm 0.0008$.
Alternative fits with a value of $b_{\textrm{lim}} = 2$~GeV$^{-1}$ in
the $b_*$ regularization procedure and with the minimal prescription
yields an uncertainty of $^{+0.0006}_{-0.0004}$. A fit in which the NNPDF4.0 PDF set is evolved
with a variable-flavour number scheme yields $\delta\alphasmZ = -0.0002$.
The stability of the results upon variations of the fit range is
tested by performing fits in the regions of Z-boson transverse
momentum $p_\mathrm{T} < 20$~GeV and $p_\mathrm{T} < 40$~GeV. The spread in the
determined values of $\alphasmZ$ is at the level of
$\pm 0.0001$ and is not considered as a source of
uncertainty. Since the region $20 < p_\mathrm{T} < 40$~GeV is sensitive to the
matching of the resummed cross section to the fixed order prediction,
this test provides a strong confirmation that the missing
O($\alphas^{3}$) contributions to the asymptotic term and the
$V$+jet finite-order cross section are negligible for this
analysis. The measured value of the strong-coupling constant is
$\alphasmZ = 0.1185^{+0.0015}_{-0.0014}$, with a statistical
uncertainty of $\pm 0.0007$, an experimental systematic uncertainty of
$\pm 0.0001$, a PDF uncertainty of $\pm 0.0008$, missing higher order
uncertainties of $\pm 0.0008$, and additional theory uncertainties of
$^{+0.0007}_{-0.0004}$.
The post-fit predictions are compared to the measured
Z-boson transverse-momentum distribution in Fig.~\ref{fig:posfit_overview} (left).

We make some observations on this determination of $\alphasmZ$.
Contrary to other hadron collider observables, the Z-boson
transverse-momentum distribution in the Sudakov region is not included
in PDF fits, therefore this determination does not have any issue of correlations with existing PDF sets. The PDF uncertainties are estimated with a conservative
approach, including the envelope of six different PDF sets. Missing
higher order uncertainties are estimated with the standard approach of
computing an envelope of scale variations.
The model for nonperturbative QCD effects based on a Gaussian form factor is simple but effective, as it describes the data very well. Generous uncertainties on this model are estimated with variations of the Landau pole prescription.
The measured value of $\alphasmZ$ has a relative uncertainty of
$1.2\%$, and is compatible with other determinations and with the
world-average value, as illustrated in Fig.~\ref{fig:posfit_overview} (right).

Finally, we outline our personal wish-list for experimental and theoretical developments.
LHC measurement of Z-boson transverse-momentum distributions are 
significantly more precise than measurements at the Tevatron.
By analysing LHC data it is likely to reach a few $10^{-4}$ 
experimental uncertainty on $\alphasmZ$ with the LHC Run-2 and Run-3 data samples.
The main experimental limitation will be the lepton momentum/energy scale, currently known at $\sim 10^{-3}$. Improving the lepton scale to $10^{-4}$ will help to reach high precision on $\alphasmZ$.
Measurements of Drell--Yan transverse-momentum distributions at high mass will 
bring further sensitivity to $\alphasmZ$, but resolution may be a limiting factor.
Precise measurements of Drell--Yan transverse-momentum at low and intermediate masses
will help reducing the nonperturbative uncertainties.

From the point of view of theory predictions, the analysis will clearly benefit
from even higher order predictions. The N$^4$LL accuracy is likely to be possible
in the near future, and approximate N$^4$LL' could also be on reach in the coming years.
The required ingredients for N$^4$LL are the 5-loop cusp anomalous dimension, the 4-loop rapidity anomalous dimension. For N$^4$LL’ also N$^3$LO PDFs, N$^4$LO TMD, and the 4-loop quark form factor are needed.
In order to make full usage of the extremely precise LHC measurements, it is needed to have an improved heavy-flavour treatment, with variable-flavour number scheme and/or massive corrections. The inclusion of joint $q_\mathrm{T}$/small-$x$ resummation~\cite{Marzani:2015oyb} may also be relevant or even required.
The precision of the measurement will greatly benefit from first-principle understanding of nonperturbative corrections. As a last remark, the availability of public software for Z+jet predictions at NNLO would help the inclusion of currently missing O($\alphas^3$) terms in the matching to fixed order.
In conclusion, prospects to reach subpercent precision in the next 5--10 years mostly rely on theory developments.

\textbf{\subsection{Crucial aspects of PDF fits relevant for  \texorpdfstring{$\alphasmZ$}{alphasmZ} determination
\protect\footnote{A\lowercase{uthor:} F. G\lowercase{iuli} (CERN)}}
\label{sec:pp_ATLASpdf21}}

Parton densities uncertainties play an important role in $\alphasmZ$ determinations from hadron collider measurements. A novel analysis at NNLO accuracy for the determination of a new set of proton PDFs using diverse measurements in p-p collisions at $\sqrt{s}$ = 7, 8 and 13~TeV performed by the ATLAS experiment at the LHC, and combined with DIS data from e-p collisions at the HERA collider, has been presented in Ref.~\cite{ATLAS:2021vod} with the resulting set of PDFs called ATLASpdf21. In this analysis, particular attention is paid to the correlation of systematic uncertainties within and among the various ATLAS data sets and to the inclusion of theoretical scale uncertainties, two crucial aspects of PDF fits if an ultimate precision below $\mathcal{O}(1\%)$ is sought on PDFs determination.

\begin{figure}[t!]
\centering
\includegraphics[width=0.48\textwidth]{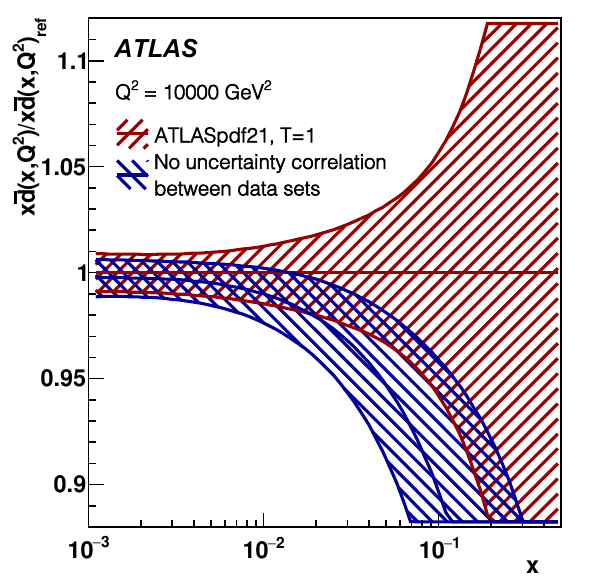}
\includegraphics[width=0.48\textwidth]{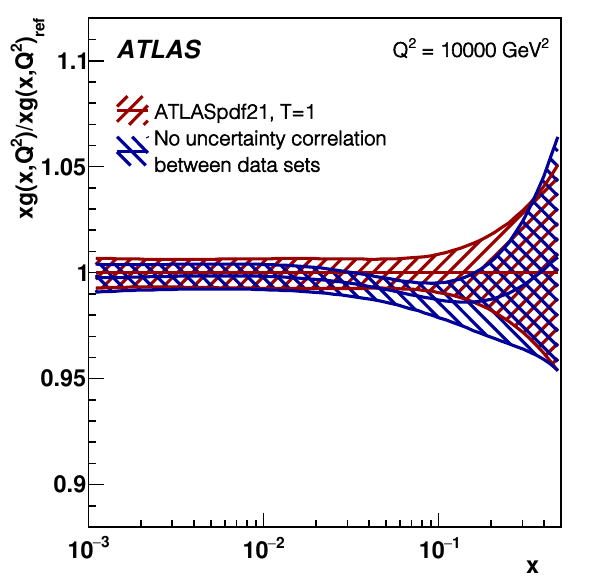}
\caption{Difference in the gluon and the $x\bar{d}$ PDFs shown in ratio to the ATLASpdf21 (default) PDFs at scale $Q^{2}$ = 10$^{4}$~GeV$^{2}$. The default (red) analysis applies the full correlation of specified systematic sources among the data sets which use jet data, and the alternative (blue) analysis does not apply any correlation of systematics sources (apart from the integrated luminosities) among the data sets.}
\label{fig:atlas_corr}
\end{figure}

Specifically, the correlations of various systematic sources have been considered between different analyses that use jet data: $\ttbar$ data in the lepton+jets channel~\cite{ATLAS:2015lsn,ATLAS:2019hxz}, W/Z\,$+$\,jets data~\cite{ATLAS:2017irc,ATLAS:2019bsa}, and inclusive jet data~\cite{ATLAS:2017kux}. The difference in the resulting $x\bar{d}$ and gluon $xg$ PDFs, when such correlations among the input data sets are considered and when they are not, is shown in Fig.~\ref{fig:atlas_corr}. Such differences are still visible at LHC energy scales, indeed this figure is made for the scale $Q^2=10^{4}$~GeV$^2$ to illustrate so.

It is visible how correlations of sources of systematic uncertainty both within and among data sets need to be carefully considered in PDF fits and although the difference between the resulting PDFs is not large in the best-known kinematic region (namely $0.01 < x < 0.1$, corresponding to mass scales $\sim$100 GeV $\to$ 1 TeV at the LHC) it can nevertheless be large enough to have an impact. In the less well-known regions, at smaller and larger mass scales, the impact can be considerably greater.

Another important aspect to be considered is the inclusion of theoretical scale uncertainties, which are evaluated as follows. The $K$-factors are evaluated for separate changes of the renormalization ($\mu_{R}$) and factorization ($\mu_{F}$) scales by factors of 2 and 0.5. The magnitude of the $K$-factor difference is symmetrized as $(K[\mu_{R}(2)]-K[\mu_{R}(0.5)])/2$ and ($K[\mu_{F}(2)]-K[\mu_{F}(0.5)])/2$ and its sign is preserved as positive if the upward variation of $\mu_{R}$ or $\mu_{F}$ makes the $K$-factor increase, and negative if it makes the $K$-factor decrease. In the ATLASpdf21 analysis, for the ATLAS W and Z$/\gamma^{*}$ inclusive data sets at both 7 and 8 TeV, both the total experimental uncertainty and the scale uncertainties approach $\sim 0.5\%$, so the scale uncertainties are considered as additional theoretical uncertainties (the effect of scale uncertainties for the other data sets entering in the fit was studied and was found to be negligible). Due to the similarity of the W and Z processes, both the $\mu_{R}$ and $\mu_{F}$ scales are considered correlated within the W, Z data sets at 7 TeV and between the W and Z data sets at 8 TeV. They are also considered to be correlated between the W and Z data sets at 7 and 8 TeV for the central fit.

\begin{figure}[t!]
 \centering
 \includegraphics[width=0.48\textwidth]{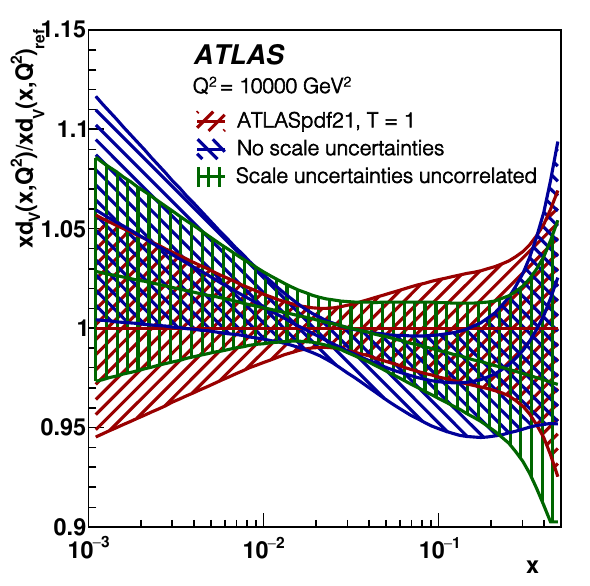}
 \includegraphics[width=0.48\textwidth]{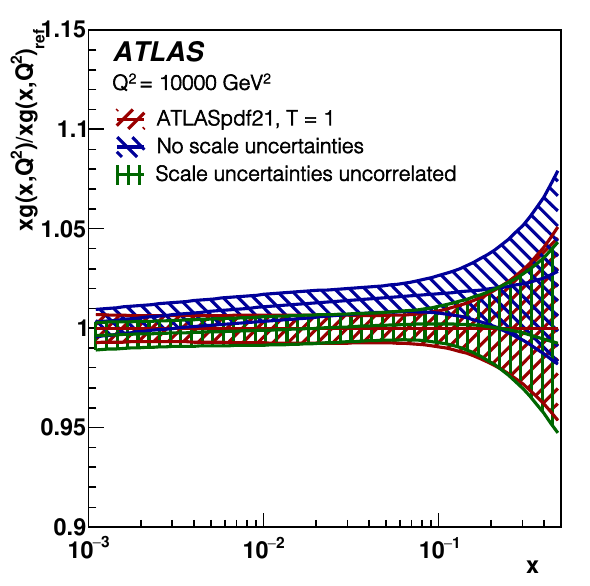}
 \caption{ATLASpdf21, showing the ratios of a fit not including theoretical scale uncertainties in the inclusive W, Z data to the central fit which does include these uncertainties, at the scale $Q^{2}$ = 10$^{4}$~GeV$^{2}$, for $xd_{v}$ (eft) and $xg$ (right).}
\label{fig:atlas_scales}
\end{figure}

Different choices for the treatment of the scale uncertainties in inclusive W, Z data are considered, and two alternative cases are considered:
\begin{itemize}
\item the scale uncertainties are not correlated between the 7 and 8 TeV data,
\item scale uncertainties are not applied at all.
\end{itemize}
Figure~\ref{fig:atlas_scales} shows the results of fits for these two cases, compared with the central fit, shown as a ratio at a scale $Q^{2}$ = 10$^{4}$~GeV$^{2}$, relevant for LHC physics. The uncertainties are very similar in size. The differences between the shapes of the PDFs are not large, but they can be important if the desired accuracy of the PDFs is $\mathcal{O}(1\%)$. The difference between the cases where the scale uncertainties are applied as being correlated or uncorrelated between the 7 and 8 TeV inclusive W, Z data sets is shown by the green line in these figures and it can be seen that it is generally a smaller effect.

The two aforedescribed effects should be taken into account properly when performing future precision data analyses, such as the measurement of $m_\mathrm{W}$, $\sin^{2}\theta_\mathrm{W}$ or $\alphasmZ$, because fits to these quantities can be very sensitive to (even) small changes.

\subsection{Exact fixed-order pQCD predictions for cross section ratios
\protect\footnote{A\lowercase{uthors:} L. S\lowercase{awyer}, C. W\lowercase{aits}, M. W\lowercase{obisch} (L\lowercase{ouisana} T\lowercase{ech} U\lowercase{niv.})}}
\label{sec:sigma_ratios}

In the standard approach, pQCD predictions
for ratios of cross sections are computed
as the ratio of the fixed-order predictions
for the numerator and the denominator.
Beyond the lowest order in the perturbative expansion,
this result does, however,
not correspond to a fixed-order prediction for the ratio.
This contribution describes how exact fixed-order results
for ratios of arbitrary cross sections can be obtained.
Differences between the standard approach and
the exact fixed-order results
should be regarded as uncertainties of the predictions,
related to possible higher-order contributions
to the perturbative expansion.
While this idea was motivated by our group's $\alphas$ determinations
from ratios of jet cross sections in hadron collisions
(for which pQCD predictions are currently available up to NLO),
the method can be applied at any order pQCD, for any processes,
and arbitrary cross sections.
The full details of the method are documented in Ref.~\cite{Sawyer:2021bzn}. 
Here, we describe the general method for the computation
of the exact fixed-order results and provide specific results
for the NLO and NNLO cases.
For various multijet cross section ratios in hadron collisions
that were previously used in $\alphaS$ determinations,
we compare the NLO pQCD predictions for the two methods,
and study how they describe the experimental data.\\

A measurable quantity $R = \frac{\sigma_n}{\sigma_d}$
is defined as the ratio of two cross sections $\sigma_n$ and $\sigma_d$.
It is assumed that the
quantity $R$ is defined in bins of an energy
or transverse momentum related variable $p$
which is defined for both $\sigma_n$ and $\sigma_d$.
At a fixed value (or in a given bin) of $p$,
the quantity $R$ is given by 
$R(p) = \frac{\sigma_n(p)}{\sigma_d(p)}$,
and it is assumed that in a pQCD calculation
the renormalization scale $\mu_\mathrm{r}$
can be related to $p$ by the same simple function
(like $\mu_\mathrm{r} = p$ or $\mu_\mathrm{r} = p/2$) for both, $\sigma_n$ and $\sigma_d$.
In other words, in a given bin of $p$, the ratio $R$ is probing
$\alphas$ and the pQCD matrix elements for $\sigma_n$ and $\sigma_d$
at the same $\mu_\mathrm{r}$.
For the sake of brevity, the dependence on $p$ is
omitted in the following.

In phenomenological analyses of experimental data,
the LO, NLO, and NNLO pQCD prediction for $R$ are usually computed 
from the ratios of the corresponding pQCD predictions for the numerator and denominator, $\sigma_n$ and $\sigma_d$, as
\begin{equation} 
  R_\mathrm{LO} = \frac{\sigma_{n,\rm LO}}{\sigma_{d,\rm LO}} 
  \qquad   \mbox{and} \qquad 
  R_\mathrm{NLO} = \frac{\sigma_{n,\rm NLO}}{\sigma_{d,\rm NLO}} 
   \qquad   \mbox{and} \qquad 
   R_\mathrm{NNLO} = \frac{\sigma_{n,\rm NNLO}}{\sigma_{d,\rm NNLO}} \, .
  \label{eq:standard}
\end{equation}
While the LO pQCD prediction for $R$ is uniquely defined,
the higher-order pQCD predictions can be obtained in different ways.
In the following, we refer to the above results
as the ``standard'' approach.
Note that, beyond LO, these ratios are not 
exact fixed-order pQCD results for the quantity $R$.

We write the perturbative expansion for $\sigma_n$ as
$\sigma_n = \sigma_{n,\rm LO} \cdot (1 + k_{n,1} + k_{n,2} + ...)$
where $k_{n,1}$ is related to the NLO correction 
($k_{n,1} = \frac{\sigma_{n,\rm NLO} - \sigma_{n,\rm LO}}{\sigma_{n,\rm LO}}$) 
and $k_{n,2}$ to the NNLO correction 
($k_{n,2} = \frac{\sigma_{n,\rm NNLO} - \sigma_{n,\rm NLO}}{\sigma_{n,\rm LO}}$).
The variables $k_{n,i}$ for $i \ge 3$ (corresponding to corrections beyond NNLO) 
are defined correspondingly, and also the corresponding variables $k_{d,i}$ $i \ge 1$
for the denominator  $\sigma_d$.
The ratio $R$ is then given by
$$
R = R_\mathrm{LO} \cdot \left( 1 + k_{n,1} + k_{n,2} + \cdots \right)
   \cdot
   \left(  1 + k_{d,1} + k_{d,2} + \cdots \right)^{-1} \, .
$$
To obtain an exact fixed-order result for $R$,  
the second parenthesis is expanded in a Taylor series
$(1+x)^{-1} = 1 -x +x^2 -x^3 +\cdots$
with $x = k_{d,1} + k_{d,2} + \cdots$.
The terms of this series are multiplied with the terms
in the left parenthesis,
and the resulting products of $k_{n,i}$ and $k_{d,j}$ are sorted in powers of $\alphas$.
The infinite series is then truncated at the corresponding order
at which $\sigma_n$ and $\sigma_d$ were
computed.
This is the general procedure by which exact fixed-order results for $R$
are obtained at arbitrary orders in pQCD.
At NLO, (with $x = k_{d,1}$ in the Taylor series) this yields
\begin{equation}
  R_\mathrm{NLO} = R_\mathrm{LO} \cdot (1 + k_{n,1} - k_{d,1} )  \, ,
\end{equation}
and at NNLO (with $x = k_{d,1} + k_{d,2}$ in the Taylor series) 
\begin{equation}
  R_\mathrm{NNLO} = R_\mathrm{LO} \cdot \left[ 1 + (k_{n,1} - k_{d,1})
    + (k_{n,2} - k_{d,2} -  k_{d,1} (k_{n,1}- k_{d,1})
    \right]   \, .
\end{equation}
In the following discussion, 
these results are referred to as the ``fixed-order'' results for $R$.

The obtained formulas for the ``standard'' and ``fixed-order'' expressions
are now used to compute NLO pQCD predictions for selected quantities
which are then compared to each other and
to the results from experimental measurements.
For this purpose, we focus on five measurements of different multijet
cross section ratios at the CERN LHC (in p-p collisions at
$\sqrt{s} = 7$ and 8~TeV) and the Fermilab Tevatron Collider
(in p-$\bar{\rm p}$ collisions at $\sqrt{s} = 1.96$~TeV).
These include measurements of the quantities  
$R_{3/2}$, $R_{\Delta \phi}$, and $R_{\Delta R}$,
which are different ratios of three-jet and two-jet production
processes.
The theoretical predictions for the ratios at NLO
are obtained from the
LO and NLO pQCD results for the two-jet and three-jet cross section
calculations, which are computed using \textsc{nlojet}~\cite{Nagy:2001fj,Nagy:2003tz}
with fastNLO~\cite{Kluge:2006xs,Britzger:2012bs}.
The proton PDFs are taken from the results of the CT18
global analysis~\cite{Hou:2019efy}.
The renormalization, $\mu_\mathrm{r}$, and factorization scales, $\mu_\mathrm{f}$,
are set to the same values
as used in the experimental publications of the measurement results,
either to one of the relevant jet $\pT$ variables, or to half of the
total jet $\pT$ sum, $\HT/2$.
The uncertainty of the pQCD results due to the $\mu_\mathrm{r,f}$ dependence
is computed from independent variations of $\mu_\mathrm{r}$ and $\mu_\mathrm{f}$
by factors of 0.5--2 around the nominal choices.
The corresponding range of variations is referred to as ``scale dependence''.
Correction factors, to account for non-perturbative contributions
are taken from
the estimates that were obtained in the experimental analyses.
PDF uncertainties are not relevant for the following discussions
and have not been evaluated.
The computations are referred to
as ``fixed-order'' results and ``standard'' method, respectively.

\begin{figure}[htpb!]
\centering
\includegraphics[width=0.32\textwidth]{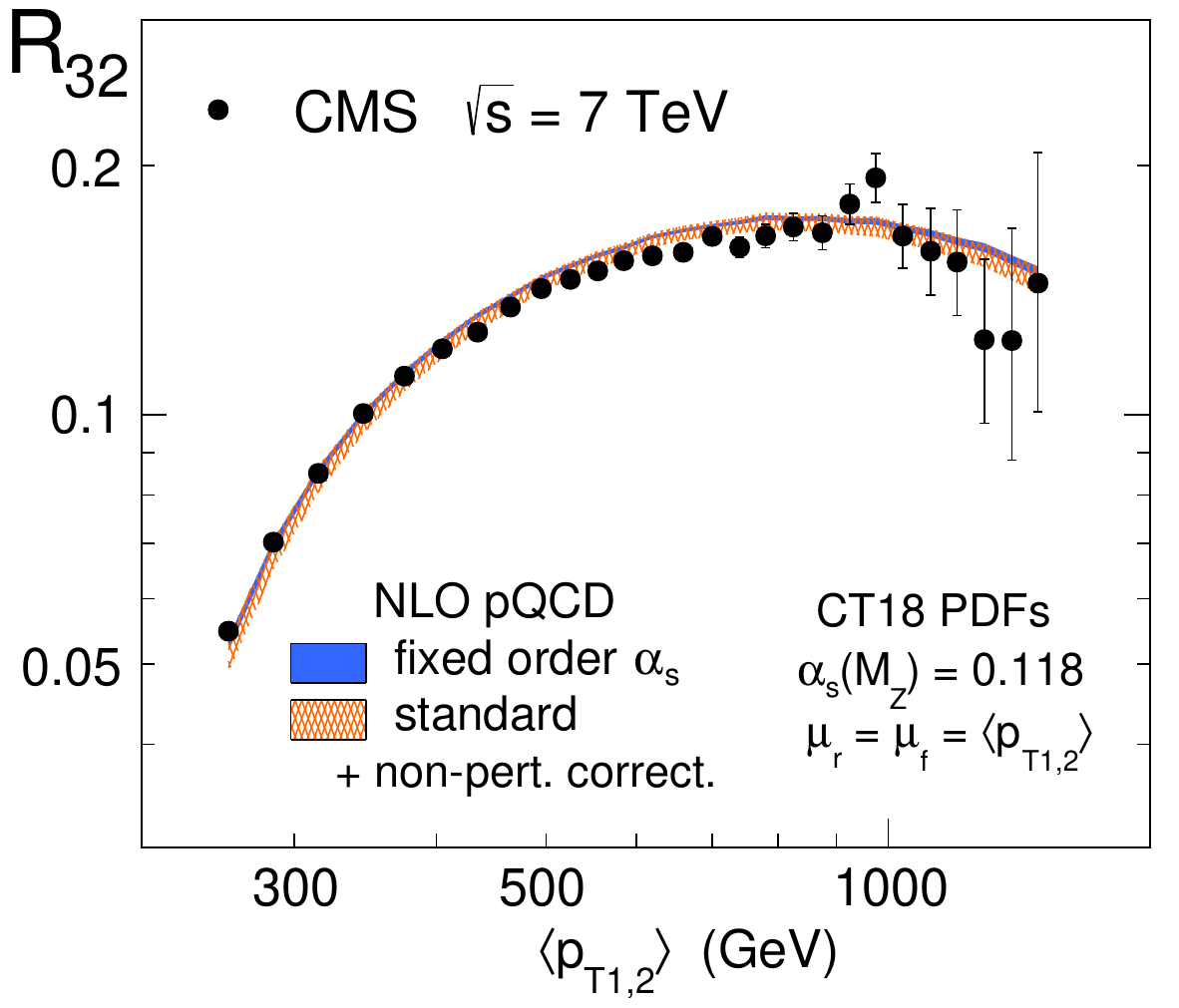}
\includegraphics[width=0.65\textwidth]{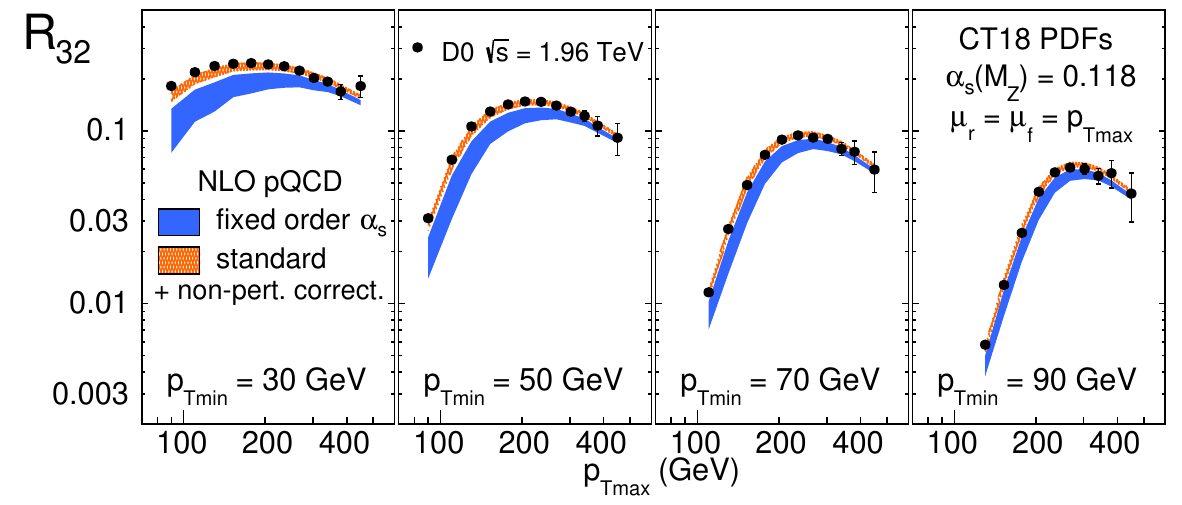}
\caption{The multijet cross section ratio $R_{3/2}$, measured in p-p collisions at $\sqrt{s}=7$~TeV as a function of $\langle p_{\mathrm{T}1,2} \rangle$ in the CMS experiment~\cite{CMS:2013vbb}, and 
in p-$\bar{\rm p}$ collisions at $\sqrt{s}=1.96$~TeV  as a function of $\ptmax$ in the D0 experiment~\cite{D0:2012dcq}.
Two sets of pQCD predictions, corrected for non-perturbative contributions, are compared to the data: the fixed-order results for $R_{3/2}$, and the results from the ``standard'' approach, computed from the ratio of the fixed-order results for the two cross sections. The shaded areas represent the ranges of the scale dependencies of the calculations.}
\label{fig:ratio-cms-d0}
\end{figure}

%
The CMS Collaboration has measured the ratio
of the inclusive three-jet and two-jet cross sections, $R_{3/2}$,
for jets with $\pT > 150$~GeV and rapidities of $|y|<2.5$~\cite{CMS:2013vbb}.
The results are published as a function of the
average transverse momentum
of the two leading jets in the event,
$\langle p_{\mathrm{T}1,2} \rangle$,
over the range
$0.42 < \langle p_{\mathrm{T}1,2} \rangle <1.39$~TeV,
as displayed in Fig.~\ref{fig:ratio-cms-d0} (left). 
%
%
Another measurement of the ratio $R_{3/2}$
was made by the D0 Collaboration
for jets with rapidities $|y|<2.4$
and for various lower jet $\pT$ requirements, $\pTmin$.
The results are published as a function
of the leading jet $\pT$, $\pTmax$, 
over the range $80 < \pTmax < 500$~GeV
and for four $\pTmin$ choices from 30 to 90~GeV,
as displayed in Fig.~\ref{fig:ratio-cms-d0} (right). 
The results of the fixed-order and ``standard'' calculations
for the CMS and D0 measurements are compared to the data, with their error bands
representing the range of their respective scale dependences.
For the CMS results, both methods results are in agreement, and both describe the data
equally well.
For the D0 data, the fixed-order calculation predicts lower values everywhere
and the scale uncertainty bands of the two calculations do either
not, or hardly, overlap.
Only towards larger $\pTmin$ and larger $\pTmax$, the uncertainty bands
get closer.

\begin{figure}[htpb!]
\centering
\includegraphics[width=0.49\textwidth]{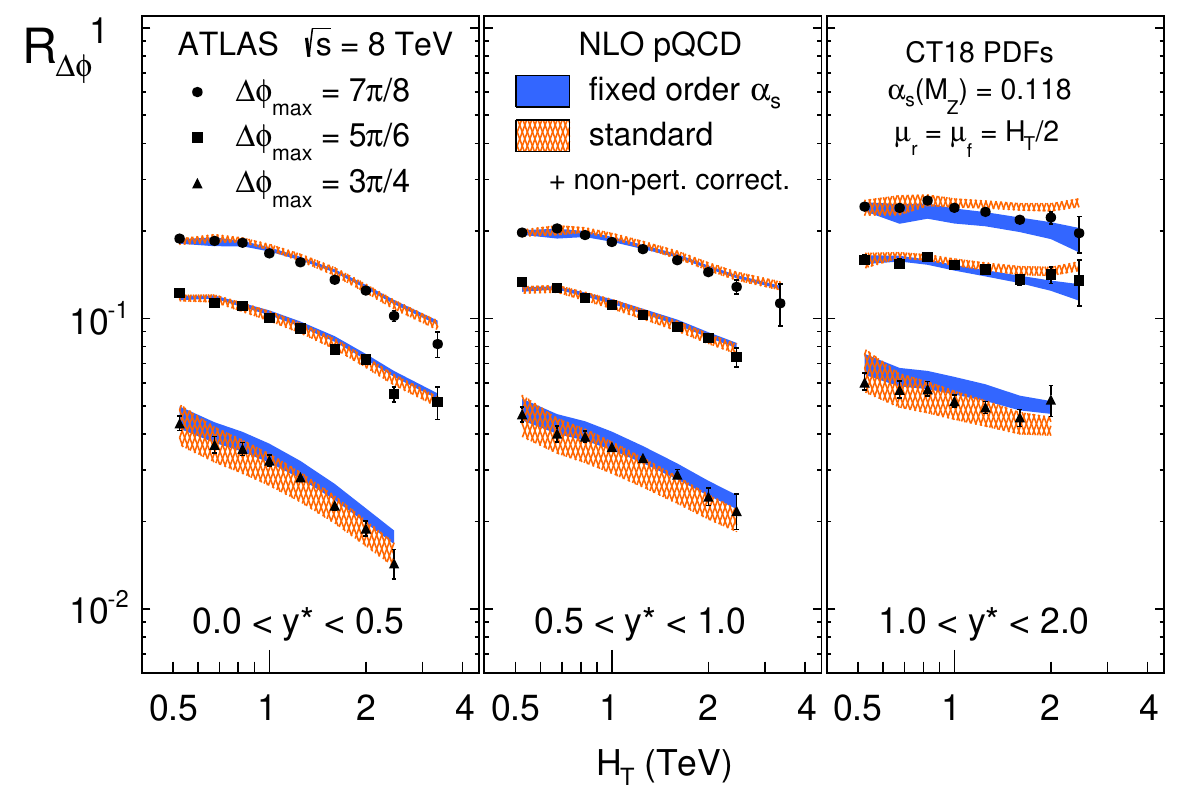}
\includegraphics[width=0.49\textwidth]{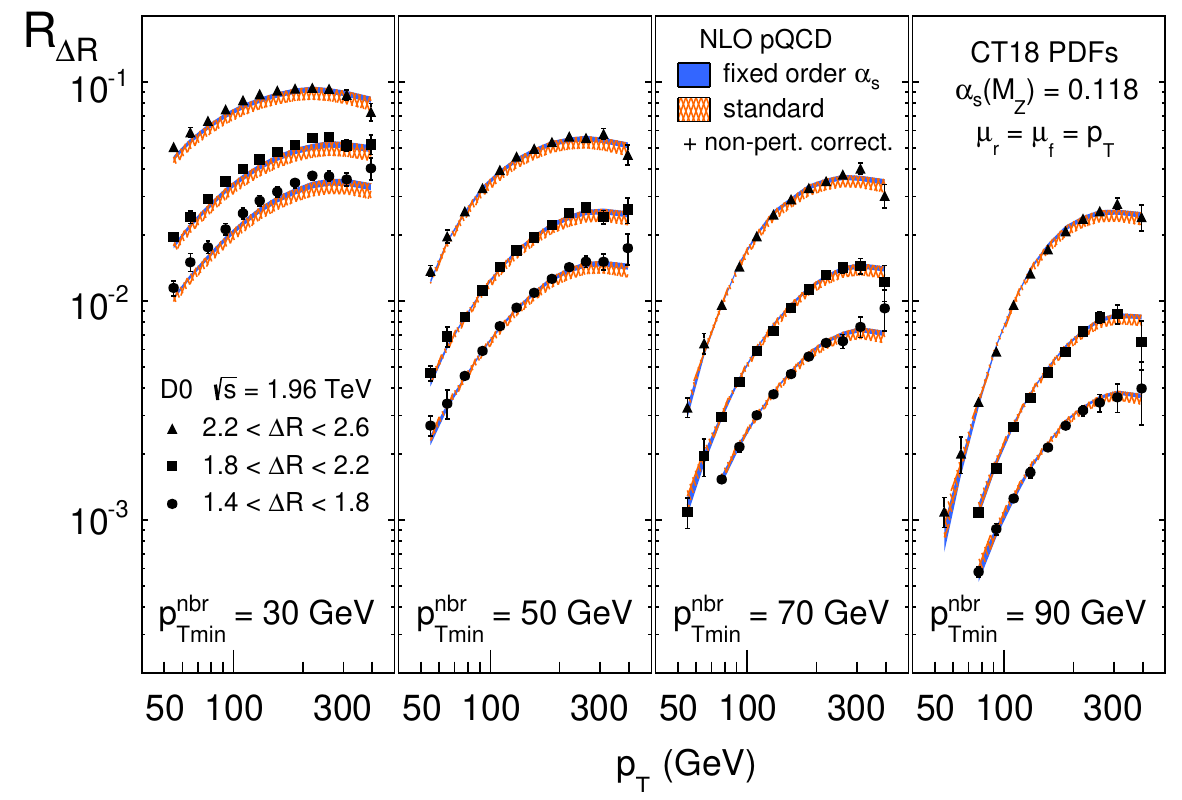}
\caption{The multijet cross section ratio
 $R_{\Delta \phi}$, measured
  in p-p collisions at $\sqrt{s}=8$~TeV
  in the ATLAS experiment~\cite{ATLAS:2018sjf} (left)
  as a function of $\HT$, in three regions of $y^*$ and for
  three values of $\Delta \phi_\mathrm{max}$,
  and the multijet cross section ratio $R_{\Delta R}$, measured
  in p-$\bar{\rm p}$ collisions at $\sqrt{s}=1.96$~TeV
  in the D0 experiment~\cite{D0:2012xif} (right)
  as a function of $\pT$, in four values of $p_{T \rm min}^\mathrm{nbr}$ and in three regions of $\Delta R$.
  Two sets of pQCD predictions,
  corrected for non-perturbative contributions,
  are compared to the data:
  the fixed-order results for $R_{\Delta \phi}$ and $R_{\Delta R}$, 
  and the results from the ``standard'' approach, computed from the ratio of the fixed-order
  results for the two cross sections.
  The shaded areas represent the ranges of the scale dependencies
  of the calculations.
}
\label{fig:ratio-atl-d0}
\end{figure}

%
The D0 Collaboration also published measurements
of a new quantity, $R_{\Delta R}$, which
also probes the ratio of three-jet and two-jet production~\cite{D0:2012xif}.
The starting point is an inclusive jet sample (which probes the
two-jet production process).
The presence of a neighboring jet with $\Delta R < \pi$
is a sign of an event topology with three or more jets.
The fraction of all inclusive jets with a neighboring jet, $R_{\Delta R}$,
is therefore also a three- over two-jet cross section ratio.
The quantity $R_{\Delta R}$ was measured
for different $p_T$ requirements, $p_{T \rm min}^\mathrm{nbr}$,
and different angular separations, $\Delta R$, for the neighboring jets,
as a function of inclusive jet $p_T$ from
50 to 450~GeV.
The results of the fixed-order and the ``standard'' calculations
for $R_{\Delta R}$ are compared to the data in Fig.~\ref{fig:ratio-atl-d0} (right).
In almost all of the phase space the conclusions mirror
those for the theoretical description of the CMS $R_{3/2}$ data in
Fig.~\ref{fig:ratio-cms-d0} (left):
The fixed-order pQCD predictions agree with those from
the ``standard'' method,
and both give a good description of all data with $p_{T \rm min}^\mathrm{nbr} \ge 50$~GeV.
Only in the softer regime, for $p_{T \rm min}^\mathrm{nbr} = 30$~GeV at smaller $p_T$,
they slightly underestimate the experimental measurement results.

%
The measurements of the multijet cross section ratio $R_{\Delta \phi}$ by the D0 and ATLAS
Collaborations~\cite{D0:2012dqa,ATLAS:2018sjf}
probe the azimuthal decorrelations of the two leading $p_T$ jets
in an event, and both analyses follow the recommendations from the original proposal~\cite{Wobisch:2012au}.
The ATLAS result is shown in Fig.~\ref{fig:ratio-atl-d0} (left).
The D0 result is shown in Ref.~\cite{Sawyer:2021bzn}.
Both measurements are performed in the same three
rapidity regions, $y^*$,
and for the same azimuthal decorrelation requirements, $\Delta \phi_\mathrm{max}$.
The ATLAS (D0) data are presented as a function of
the scalar  $p_T$ sum of all jets in an event, $H_T$,
over the range  0.46--4~TeV  (180--900~GeV).
The degree of agreement between the NLO pQCD predictions
from the fixed-order and the ``standard'' calculations
and how they describe the data is pretty much the same
for the ATLAS and D0 data sets.
In different  $y^*$ and $\Delta \phi_\mathrm{max}$ regions, however,
the two calculations exhibit a rather different behavior.
For $0 < y^* < 1$ and $\Delta \phi_\mathrm{max} = 7\pi/8$ and $5\pi/6$,
both calculations agree very well,
exhibit a relatively small scale dependence,
and both describe the data.
At $\Delta \phi_\mathrm{max} = 3\pi/4$,
the two predictions start to deviate from each other,
and in some cases
their larger scale uncertainty bands have only a small overlap.
The data are described by both predictions.
At $1 < y^* < 2$, for $\Delta \phi_\mathrm{max} = 7\pi/8$ and $5\pi/6$,
the two predictions have a different $H_T$ dependence
and disagree at high $H_T$.
In these regions,
the fixed-order calculation gives a better description
of the overall $H_T$ shape for both data sets.

%
The results from these comparisons can be summarized as follows:
In all cases where the results from the two methods
agree with each other
(as seen for the CMS $R_{3/2}$, the D0 $R_{\Delta R}$,
and some regions of the ATLAS and D0 $R_{\Delta \phi}$ measurements),
they also both describe the data.
In all cases where the two methods disagree
(meaning that their scale uncertainty bands do not overlap,
as seen for the D0 $R_{3/2}$ data and the high $H_T$ tails
in some of the ATLAS and D0 $R_{\Delta \phi}$ data at $y^* > 1$),
one of them (but not always the same) describes the data.
In some intermediate cases, where the scale uncertainty bands from the
two methods have little overlap
(as for the ATLAS and D0 $R_{\Delta \phi}$ data with $\Delta \phi_\mathrm{max} = 3\pi/4$),
both predictions are somehow consistent with the data.

It has to be noted that the ``fixed-order'' and ``standard'' method
both stand on the same footing.
In any given order pQCD, the results from both methods are
equally valid representations of the perturbative expansion,
and they only differ in higher-order terms.
Therefore their discrepancy should be regarded as a genuine uncertainty
of a fixed-order calculation, 
in addition to the scale dependence (since the latter
does not always cover the spread of the two methods).
The central value can be chosen from either method; 
there is no fundamental argument, to pick one over the other,
and the choice may depend on the specific goal.
In $\as$ determinations, where one assumes that the pQCD predictions
are able to describe the data, one should possibly pick the
method that gives a better description of the data.
The proposed treatment of the spread of the two methods as additional uncertainty
will provide more realistic estimates of theoretical uncertainties
in future $\as$ determinations and other phenomenological studies.
A small spread of the two methods can also be a criterion
for identifying robust measurable quantities
for which the theoretical approximations are more reliable.

\clearpage

\subsection{Energy range for the RGE test and PDF sensitivity in $\alpha_S$ evaluations from jet cross section ratios
\protect\footnote{A\lowercase{uthors:} B. M\lowercase{alaescu} (LPNHE, P\lowercase{aris})}}

A series of determinations of $\alphas$ are performed at the LHC \eg\ using jet-based observables, which include the three-jet mass~\cite{CMS:2014mna}, $R_{32}$~\cite{ATLAS:2013lla,CMS:2013vbb,CMS:2017tvp}, transverse energy-energy correlations~(TEECs)~\cite{ATLAS:2017qir} and $R_{\Delta\phi}$~\cite{ATLAS:2018sjf}.
We discuss the relevant energy range on which running of the coupling, determined by the renormalization group equation (RGE), is probed through jet cross-section ratio and event shape observables at hadron colliders, as well as the PDF sensitivity in the corresponding $\alpha_S$ evaluations. This contribution is based mainly on remarks made in Refs.~\cite{Gehrmann:2021qex,Bogdan_ScalesWorksop_2017}.\\

For fixed incoming parton kinematics, constructing the ratios between the cross sections for three-jet and two-jet final states can reduce the PDF dependence.
At the same time, the resulting $R_{32}$ ratio is still sensitive to $\alphas$~\cite{ATLAS:2013lla,CMS:2013vbb}.
The angular decorrelation observable $R_{\Delta\phi}$ in two-jet final states is defined as the ratio between the cross section of events with an azimuthal angular separation below some upper limit and respectively the inclusive dijet cross section.
It also allows to probe three-parton final state kinematics and has been used for an $\alphas$ evaluation~\cite{ATLAS:2018sjf}.

TEECs are event shape variables computed as energy-weighted angular distributions of all individual object pairs in the event.
Similarly to \eg\ the case of the inclusive jet cross section, a TEEC distribution receives multiple contributions from each event. 
Since analytical predictions for their distributions can be computed from first principles~\cite{Dixon:2019uzg}, probing hence fundamental symmetries of QCD, the TEECs are particularly attractive observables.
The TEECs and the associated asymmetries (ATEEC) were measured by ATLAS and used for a determination of $\alphas$~\cite{ATLAS:2017qir}. 
It is interesting to note that in this study the theoretical prediction is also provided by a three-jet to two-jet cross-section ratio evaluated using NLOJET++~\cite{Nagy:2001fj,Nagy:2003tz}, complemented by nonperturbative corrections based on PYTHIA8~\cite{Sjostrand:2007gs} and HERWIG++~\cite{Gieseke:2012ft}.

The choice of the scale used in the theoretical calculations for observables like $R_{32}$, $R_{\Delta\phi}$, and (A)TEECs is often based on event-level quantities.
Typical examples are the average transverse momentum of the two leading jets~\cite{CMS:2013vbb,CMS:2017tvp,ATLAS:2017qir}, the transverse momentum of the leading jet~\cite{ATLAS:2013lla}, or half of the scalar sum of the transverse momenta of all the selected jets in the event~\cite{ATLAS:2018sjf}.
The evaluated $\alphas$ values are typically displayed as a function of this same scale, reaching values up to a few TeV~(see \eg\ Ref.~\cite{Zyla:2020zbs}).
However, we note that the sensitivity to $\alphas$ for such observables is actually directly related to the probability for emission of extra radiation (yielding a third or higher order jet).
This implies that these $\alphas$ determinations probe the prediction of the renormalization group equation in QCD at energy scales related to the transverse momentum of the third jet~($p_\mathrm{ T 3}$), rather than to the event-level quantities above.
It is to be noted that the typical values for $p_\mathrm{ T 3}$ are indeed significantly lower than the scale based on such event-level quantities.
Indeed, in the case of the (A)TEEC studies at 8 TeV~\cite{ATLAS:2017qir}, if the average scales used in the theoretical calculations range between $412$ and $810$ GeV~(depending on the bin), the corresponding average $p_\mathrm{ T 3}$ values are between $169$ and $215$ GeV.

It is desirable to achieve consistency between scale used for theory calculation and the scale at which the RGE test is claimed, while taking into account the remarks above.
The MiNLO procedure~\cite{Hamilton:2012np} may indeed provide a way forward towards this goal.

The $\alphas$ determinations from ratios of three-parton-like over two-parton-like final states~(\ie\ from observables like $R_{32}$, $R_{\Delta\phi}$, (A)TEEC) are impacted by residual PDF uncertainties that have been quantified in the corresponding experimental studies~\cite{CMS:2014mna,CMS:2013vbb,ATLAS:2013lla,CMS:2017tvp,ATLAS:2018sjf,ATLAS:2017qir}.
Contrary to what one may have initially expected, the PDF uncertainties~(originating from the PDF eigenvectors/replicas, as well as from the differences among various PDF sets) are found to be nonnegligible, being typically larger than the combined experimental uncertainties, but smaller than the NLO scale uncertainty~(Table~\ref{tab:alphaSjetXsecAndRatios}).
Furthermore, in cases where direct comparisons are possible, it can be noted that the $\alphas$ determinations from ratio observables can have even larger PDF uncertainties compared to the corresponding absolute cross sections.

\begin{table}[htpb!]
\centering
\caption{List of $\alphasmZ$ values obtained from various LHC observables, together with their corresponding uncertainties from various sources (experimental, PDF eigenvectors/replicas, nonperturbative NP, scale variations) or their quadratic sum. The last column lists the range of $\alphasmZ$ values probed for various PDF sets. See the corresponding references for any details, in particular for the ensemble of PDF sets considered in each study. \label{tab:alphaSjetXsecAndRatios} \vspace{0.2cm}}
\resizebox{\textwidth}{!}{%
\begin{tabular}{lccc}\hline
Observable [Ref.]  &  $\alphasmZ$  &  Range PDF variations  \\ \hline
 $R_{32}$~\cite{ATLAS:2013lla} & $ 0.111 \pm 0.006 \,\mathrm{(exp)} ^{+0.016}_{-0.003} \,\mathrm{(PDF, NP, scale)} $ & $ 0.109 - 0.116 $ \\ \hline
 $R_{32}$~\cite{CMS:2013vbb} & $ 0.1148 \pm 0.0014 \,\mathrm{(exp)} \pm 0.0018 \,\mathrm{(PDF)} \pm 0.0050 \,\mathrm{(theory)} $ & $ 0.1135 - 0.1148 $ \\ \hline
 $3$-jet mass~\cite{CMS:2014mna} & $ 0.1171 \pm 0.0013 \,\mathrm{(exp)} \pm 0.0024 \,\mathrm{(PDF)} \pm 0.0008 \,\mathrm{(NP)} ^{+0.0069}_{-0.0040} \,\mathrm{(scale)} $ & $ 0.1143 - 0.1183 $ \\ \hline
 $2$-jets~\cite{CMS:2017tvp} & $ 0.1159 \pm 0.0025 \,\mathrm{(exp, PDF, NP)} $ & $ 0.1159 - 0.1183 $ \\
 $3$-jets~\cite{CMS:2017tvp} & $ 0.1161 \pm 0.0021 \,\mathrm{(exp, PDF, NP)} $ & $ 0.1159 - 0.1179 $ \\
 $2$- \& 3-jets~\cite{CMS:2017tvp} & $ 0.1161 \pm 0.0021 \,\mathrm{(exp, PDF, NP)} $ & $ 0.1161 - 0.1188 $ \\
 $R_{32}$~\cite{CMS:2017tvp} & $ 0.1150 \pm 0.0010 \,\mathrm{(exp)} \pm 0.0013 \,\mathrm{(PDF)} \pm 0.0015 \,\mathrm{(NP)} ^{+0.0050}_{-0.0000} \,\mathrm{(scale)} $ & $ 0.1139 - 0.1184 $ \\ \hline
 TEEC~\cite{ATLAS:2017qir} & $ 0.1162 \pm 0.0011 \,\mathrm{(exp)} \pm 0.0018 \,\mathrm{(PDF)} \pm 0.0003 \,\mathrm{(NP)} ^{+0.0076}_{-0.0061} \,\mathrm{(scale)} $ & $ 0.1151 - 0.1177 $ \\
 ATEEC~\cite{ATLAS:2017qir} & $ 0.1196 \pm 0.0013 \,\mathrm{(exp)} \pm 0.0017 \,\mathrm{(PDF)} \pm 0.0004 \,\mathrm{(NP)} ^{+0.0061}_{-0.0013} \,\mathrm{(scale)} $ & $ 0.1185 - 0.1206 $ \\ \hline
 $R_{\Delta\phi}$~\cite{ATLAS:2018sjf} & $ 0.1127  ^{+0.0019}_{-0.0018} \,\mathrm{(exp)} \pm 0.0006 \,\mathrm{(PDF)}  ^{+0.0003}_{-0.0001} \,\mathrm{(NP)} ^{+0.0052}_{-0.0019} \,\mathrm{(scale)} $ & $ 0.1127 - 0.1156 $ \\
\hline
\end{tabular}
}
\end{table}

Actually, these features reflect the fact that the three-parton and two-parton processes that are used to define the respective cross section ratios have different composition of partonic initial states~(see also contribution by B.M. in Ref.~\cite{Proceedings:2015eho}).
Indeed, the probability of extra radiation~(which is what makes these observables nontrivial) is correlated, through the relevant matrix elements, to the type of partons in the initial state.
Furthermore, both the $\alphas$ and PDF sensitivities of the observables are reduced when taking ratios.
Both aspects are relevant for the corresponding evaluations of $\alphas$, hence the presence of two competing effects resulting in residual PDF uncertainties.
This residual PDF sensitivity is especially relevant in phase-space regions where PDFs are not strongly constrained.
It can hence be related to the slopes observed in some comparisons of the energy dependence for the extracted $\alphas$ values, with the corresponding RGE predictions~(see \eg\ Refs.~\cite{ATLAS:2017qir,ATLAS:2018sjf}).

Up to now, the precision of the $\alphas$ determinations from three-jet-type observables has been limited by the NLO QCD theory uncertainty.
This limitation will be overcome due to the NNLO corrections to three-jet production computed recently~\cite{Czakon:2021mjy}, which will allow to perform precision QCD studies with these observables.
On the timescale of the FCC projects~\cite{FCC:2018evy}, even more progress on the theoretical predictions for these observables is desirable, allowing to further enhance the precision of these $\alphas$ evaluations.


\subsection{New results on \texorpdfstring{$\alphas$}{alphas} and PDFs: QCD and SMEFT interpretation with inclusive jets at $\sqrt{s}=13$~TeV
\protect\footnote{A\lowercase{uthors:} K. L\lowercase{ipka}, T. M\lowercase{{\"a}kel{\"a}} (DESY) \lowercase{on behalf of the} CMS C\lowercase{ollaboration}}}

Jet production in proton-proton collisions is instrumental for the extraction of the strong coupling constant $\alphas$ and the parton distribution functions (PDFs) of the proton. Furthermore, it is sensitive to the presence of physics beyond the standard model (BSM). In the following, the most recent QCD analysis~\cite{CMS:2021yzl} of the inclusive jet cross sections in p-p collisions at the LHC at a center-of-mass energy of 13~TeV is discussed. The data are collected by the CMS Collaboration. The jets are reconstructed using the anti-$k_T$ algorithm~\cite{Cacciari:2008gp} with distance parameters $R=0.4$ and $R=0.7$, and the cross sections are measured double-differentially as a function of the individual jet $\pT$ and the absolute rapidity $|y|$. The measurements using $R=0.7$ are used in the PDF analysis and correspond to an integrated luminosity of 33.5 $\mathrm{fb}^{-1}$.\\
 
In the PDF analysis, the jet cross sections measured for $R=0.7$ are used, together with the HERA combined~\cite{Abramowicz:2015mha} inclusive charged- and neutral-current deep inelastic scattering (DIS) cross sections. The fit is performed at NLO and NNLO. In the NLO version of the analysis, together with DIS and CMS jet measurements, the normalized triple-differential cross section of top quark-antiquark ($t \bar t$) production measured by the CMS collaboration~\cite{Sirunyan:2019zvx} , are utilized. In the NNLO (NLO) versions of the analysis, the PDFs, the value of $\alphas (m_Z)$ (as well as the value of the top quark pole mass,  $m_t^\text{pole}$) are determined simultaneously.

Furthermore, an alternative SMEFT analysis is performed at NLO, where the cross section for the inclusive jet production is extended to include the effective contributions of 4-quark contact interactions (CI), exploring three different CI models. In this version of the analysis, the relevant Wilson coefficient is fitted simultaneously with the PDFs and SM parameters, avoiding the possibility of absorbing new physics in the PDF fit. 
No primary assumptions on the values of the QCD parameters are applied, such that constraints on the SM and BSM parameters are obtained simultaneously, mitigating their possible bias. 

The fixed-order QCD predictions for inclusive jet production in p-p collisions are available at NLO and NNLO, obtained with NLOJet++~\cite{Nagy:2001fj, Nagy:2003tz} and NNLOJET (rev5918)~\cite{Currie:2016bfm, Currie:2018xkj, Gehrmann:2018szu}, respectively, with the NLO calculations implemented in \textsc{FastNLO}~\cite{Britzger:2012bs}.
The NLO cross-section is improved to NLO+NLL by using corrections computed using the \textsc{NLL-Jet} calculation, provided by the authors of Ref.~\cite{Liu:2018ktv}, and the MEKS~\cite{Gao:2012he} code. Electroweak and nonperturbative corrections are applied, and their details are given in~\cite{CMS:2021yzl}.
The factorization scale $\mu_f$ and renormalization scale $\mu_r$ are set to the individual jet $\pT$ for the inclusive jet cross section, and to the four-momentum transfer $Q$ for the DIS data. In the theory predictions at NLO for the used $t\bar t$ production measurements, $\mu_r$ and $\mu_f$ are set to half the sum of the transverse masses of the partons, following Ref.~\cite{Sirunyan:2019zvx}. The QCD analysis is performed using the \textsc{xFitter} QCD analysis framework~\cite{Alekhin:2014irh, Bertone:2017tig} with an interface to \textsc{CIJET}~\cite{Gao:2012qpa, Gao:2013kp} for SMEFT predictions, which are available at NLO.

\begin{figure}[htbp!]
\centering
\includegraphics[width=0.4\textwidth]{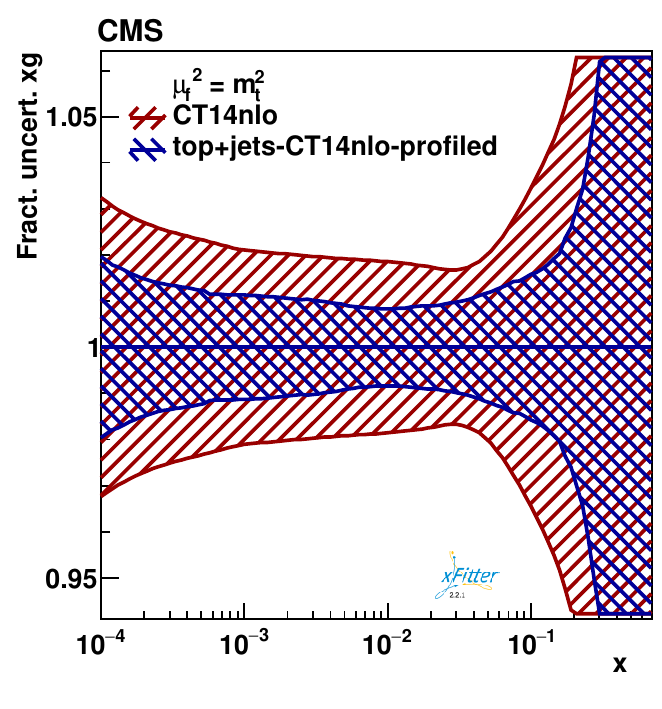}
\includegraphics[width=0.4\textwidth]{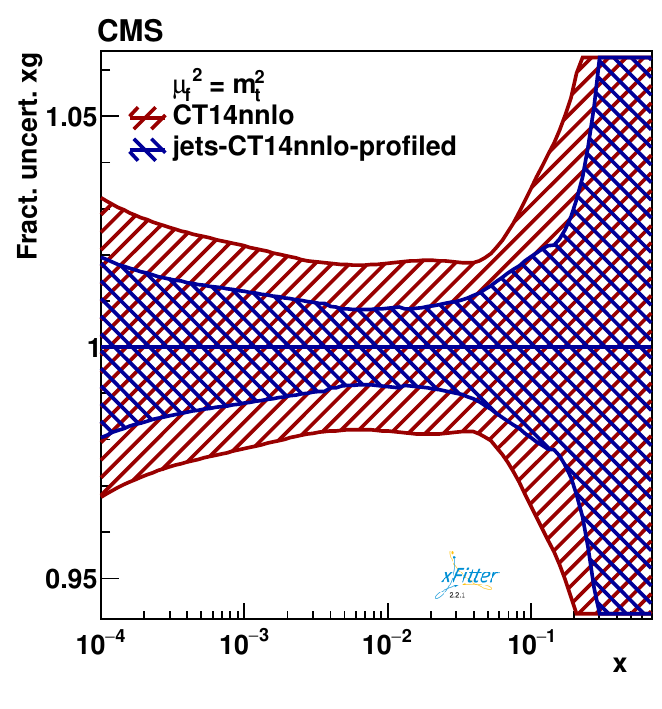}
\caption{Fractional uncertainties in the gluon distribution, shown as functions of $x$ for the scale
$\mu_f$ set to the top quark mass. The profiling is performed at NLO with CT14nlo PDF (left) and at NNLO with CT14nnlo, by using the CMS inclusive jet cross section at $\sqrt{s} = 13$~TeV. The original uncertainty (red) and the profiled result (blue) uncertainty are shown. The figure is taken from~\cite{CMS:2021yzl}.}
\label{profiling_pdfs}
\end{figure}

To illustrate the impact of the 13~TeV data on a global PDF, the profiling procedure~\cite{Paukkunen:2014zia, Schmidt:2018hvu, HERAFitterdevelopersTeam:2015cre} is used. The CT14~\cite{Dulat:2015mca} sets at NLO or NNLO, where appropriate, are chosen. Both the $t \bar t$ and jet cross sections are observed to improve the precision of the gluon PDF significantly, as illustrated in Fig.~\ref{profiling_pdfs}. Profiling of the non-PDF parameters, such as the value of $\alphasmZ$, $m_t^\text{pole}$, and CI Wilson coefficients is also performed and the results are summarized in Ref.~\cite{CMS:2021yzl}. 
The disadvantage of the profiling approach is that the simultaneous extraction of the PDFs and non-PDF parameters is currently not available. 
The full QCD analyses, implying the simultaneous fit of the PDF and non-PDF parameters, are performed using SM predictions at NNLO and NLO, or, alternatively, as a SMEFT fit, assuming three SM\,$+$\,CI models. In the SMEFT analyses, the scale of the BSM interaction $\Lambda$ is assumed and the Wilson coefficient $c_1$ is a free parameter of the fit, together with the PDF parameters and the values of $\alphasmZ$ and $m_t^\text{pole}$. For each of the NLO and NNLO fits, and also for the SMEFT fit, the investigation of the PDF parameterizations is performed independently. The SM fits at NLO and NNLO result in slightly different parameterizations, due to inclusion of the $t \bar {t}$ measurements in the NLO analysis. The SM and SMEFT fits at NLO result in the same solution for the preferred parameterizations.

\begin{figure}[htbp!]
\centering
\includegraphics[width=0.4\textwidth]{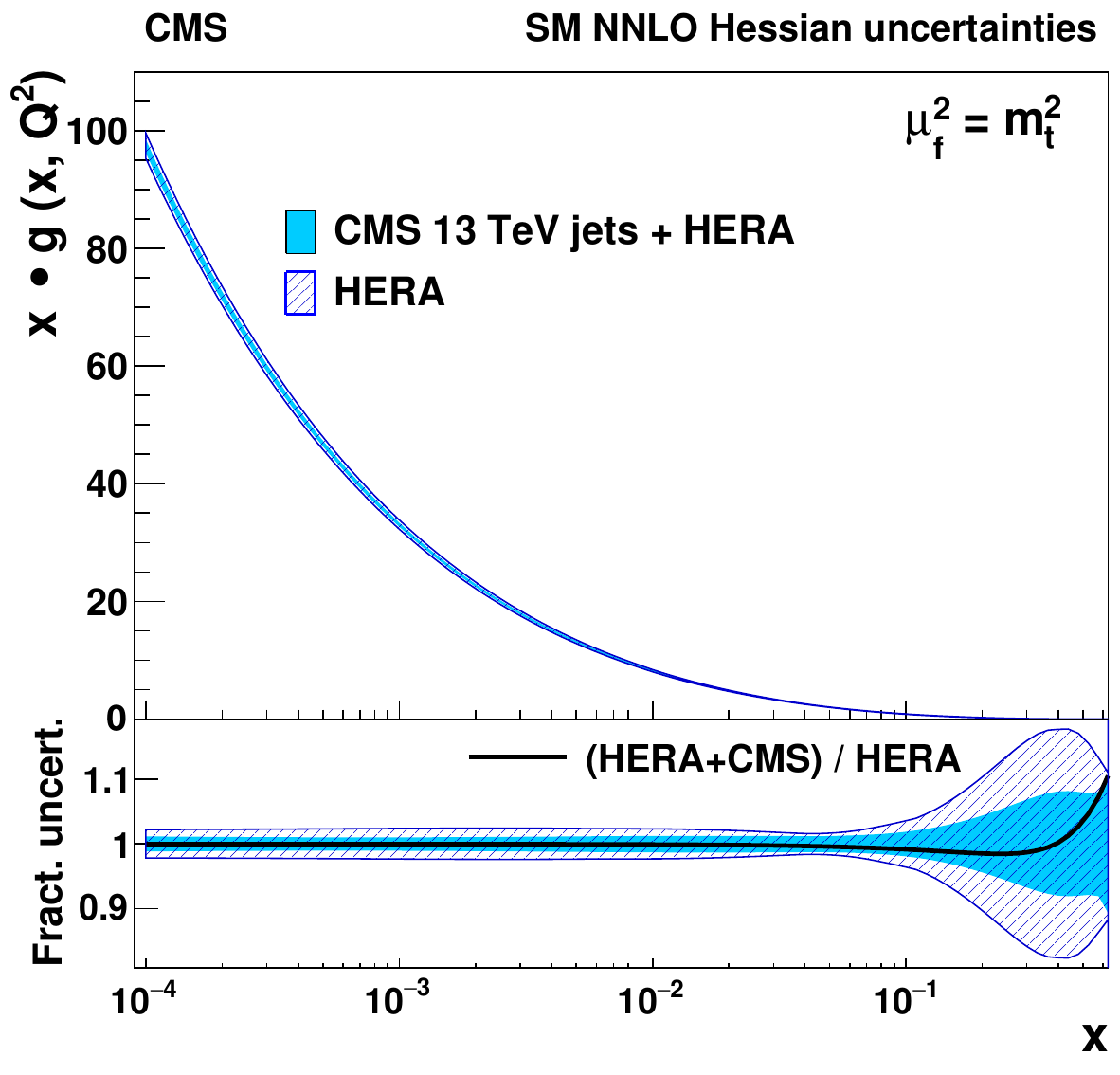}
\includegraphics[width=0.4\textwidth]{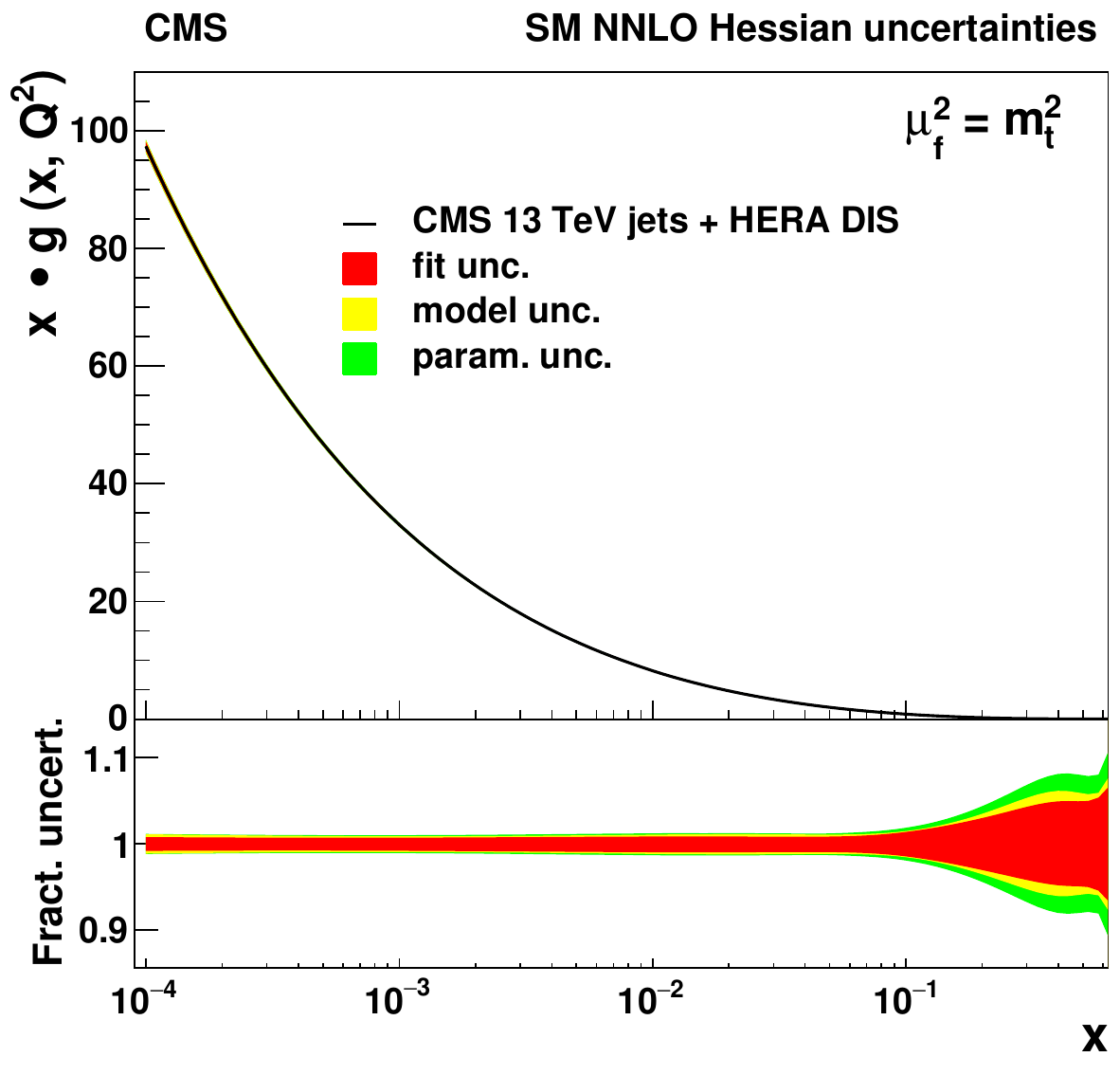}
\caption{Total (left) and individual (right) fractional uncertainties in the gluon distributions resulting form the NNLO fit, shown as functions of $x$ at the scale $\mu_f$ of the top quark mass. 
The filled (hatched) band represents the results of the fit using HERA DIS and the CMS inclusive jet cross
section together (using the HERA DIS data only). The line corresponds to the ratio of the central PDF values of the two variants of the fit. The figure is taken from~\cite{CMS:2021yzl}.}
\label{NNLO_gluon_PDFs}
\end{figure}

The uncertainties are estimated following the HERAPDF approach~\cite{Abramowicz:2015mha}, which accounts for the fit, parameterizations and model uncertainties. The Hessian fit uncertainty emerges from the uncertainties in the experimental measurements and is estimated by Hessian method using the tolerance criterion of $\Delta \chi^2=1$. The uncertainty is also estimated using the Monte Carlo replica method, and the results agree with those obtained by using the Hessian method. The quality of the fit is estimated by $\chi^2$ divided by the number of degrees of freedom of $1321/1118$, with somewhat large value driven by the fit to the inclusive DIS data, investigated in detail in~\cite{Abramowicz:2015mha}.
The model uncertainties are obtained by varying the assumed non-PDF parameter values such as quark masses, the value of the starting evolution scale, strangeness fraction and the minimum value of $Q^2$ of the used DIS data. The theory uncertainties due to missing higher order contributions are obtained by an independent variation of the QCD scales up and down by a factor of two (avoiding cases $\mu_f/\mu_f=4,~1/4$). For each of the independent scale choices, the QCD analysis is performed and the difference of the resulting parameters to those obtained for the central scale choice is taken as an uncertainty, which is treated as a model uncertainty. The parameterizations uncertainty arises from adding and removing additional parameters in the PDF parameterizations, one at a time, and constructing a maximum-difference envelope. The total uncertainty in the PDFs is obtained by adding the fit and the model uncertainties in quadrature, while the parameterizations uncertainty is added linearly. To illustrate the improvement in the PDF uncertainty by adding the CMS jet measurements at 13~TeV, the NNLO fit is also performed using only DIS data. A significant improvement in the uncertainty, in particular for the gluon distribution is observed (Fig.~\ref{NNLO_gluon_PDFs}). In the same figure, also individual contributions of model, parameterizations and fit uncertainties are shown.
In the SM fit at NNLO, the strong coupling constant $\alphasmZ$ is obtained simultaneously with the PDFs and results in
\begin{equation}
\alphasmZ
= 0.1170
\pm 0.0014\,\mathrm{(fit)}
\pm 0.0007\,\mathrm{(model)}
\pm 0.0008\,\mathrm{(scale)}
\pm 0.0001\,\mathrm{(param)},
\end{equation}
which agrees with the previous extractions of the strong coupling constant at NNLO at hadron colliders~\cite{Andreev:2017vxu, Sirunyan:2018goh}, of which it has best precision.

\begin{figure}[htbp!]
\centering
\includegraphics[width=0.5\textwidth]{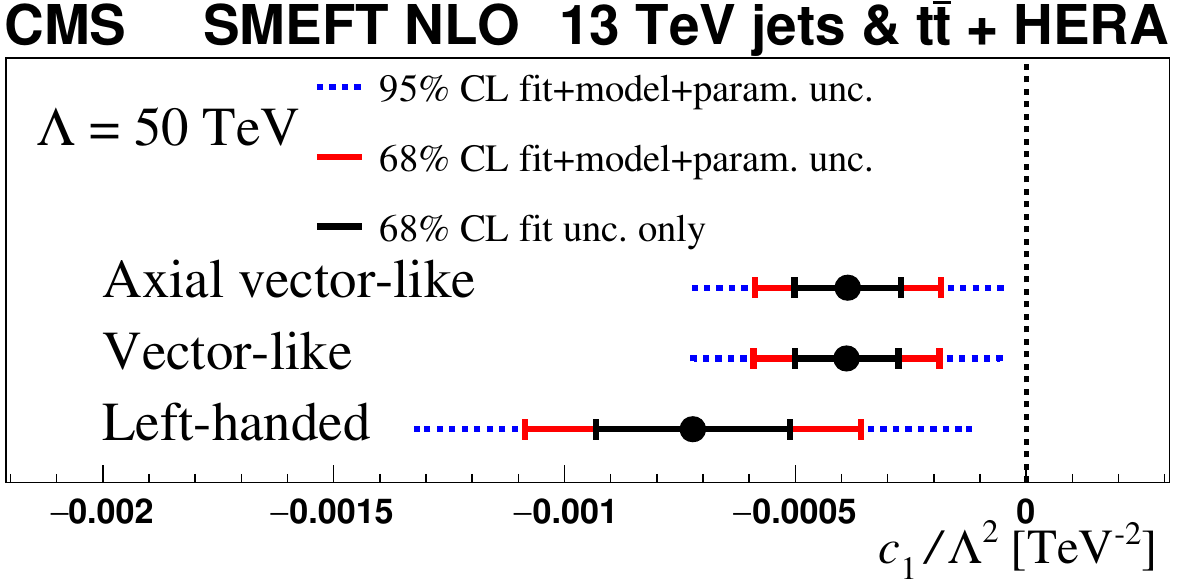}
\caption{The Wilson coefficients $c_1$ obtained in the SMEFT analysis at NLO, divided by $\Lambda^2$, for $\Lambda =50$ TeV. The solid (dashed) lines represent the total uncertainty at 68 (95)\% confidence level (CL). The inner (outer) error bars show the fit (total) uncertainty at 68\% CL.}
\label{ciplot}
\end{figure}

In the SMEFT analysis, the SM Lagrangian is extended with effective operators of dimension 6, introducing vertices with 4 quark legs. The considered operators are colour-singlets and lead to purely left-handed, vector-like and axial vector-like CI models, depending on how the quarks' handedness may change in the interaction. The operators' Wilson coefficients are fitted simultaneously with the PDFs, $\alphasmZ$, and $m_t^\text{pole}$ at NLO. Independent of the value of $\Lambda$, the strong coupling constant and the top quark mass in these SMEFT fits result to $\alphasmZ = 0.1187 \pm 0.0016_\textrm{(fit)} \pm 0.0005_\textrm{(model)}\pm 0.0023_\textrm{(scale)} \pm 0.0018_\textrm{(param.)}$, and $m_t^\text{pole} = 170.4 \pm 0.6_\textrm{(fit)} \pm 0.1_\textrm{(model)} \pm 0.1_\textrm{(scale)} \pm 0.2_\textrm{(param.)}$ GeV. The Wilson coefficients are obtained for different assumed values of $\Lambda$. For illustration, the values of $c_1$ are shown for three investigated CI models in Fig.~\ref{ciplot} for $\Lambda= 50$ TeV. 

\begin{figure}[htbp!]
\centering
\includegraphics[width=0.3\textwidth]{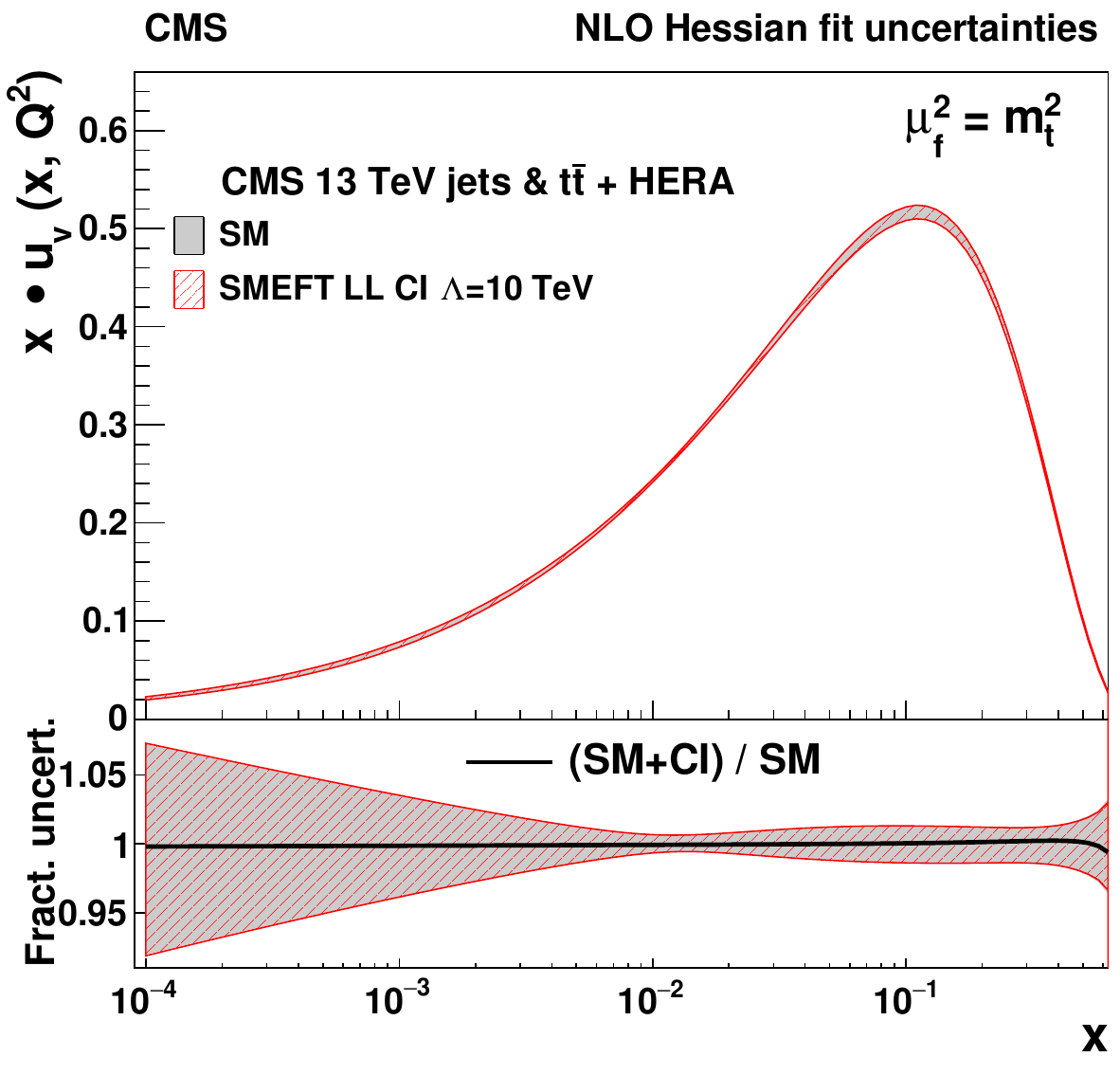}
\includegraphics[width=0.3\textwidth]{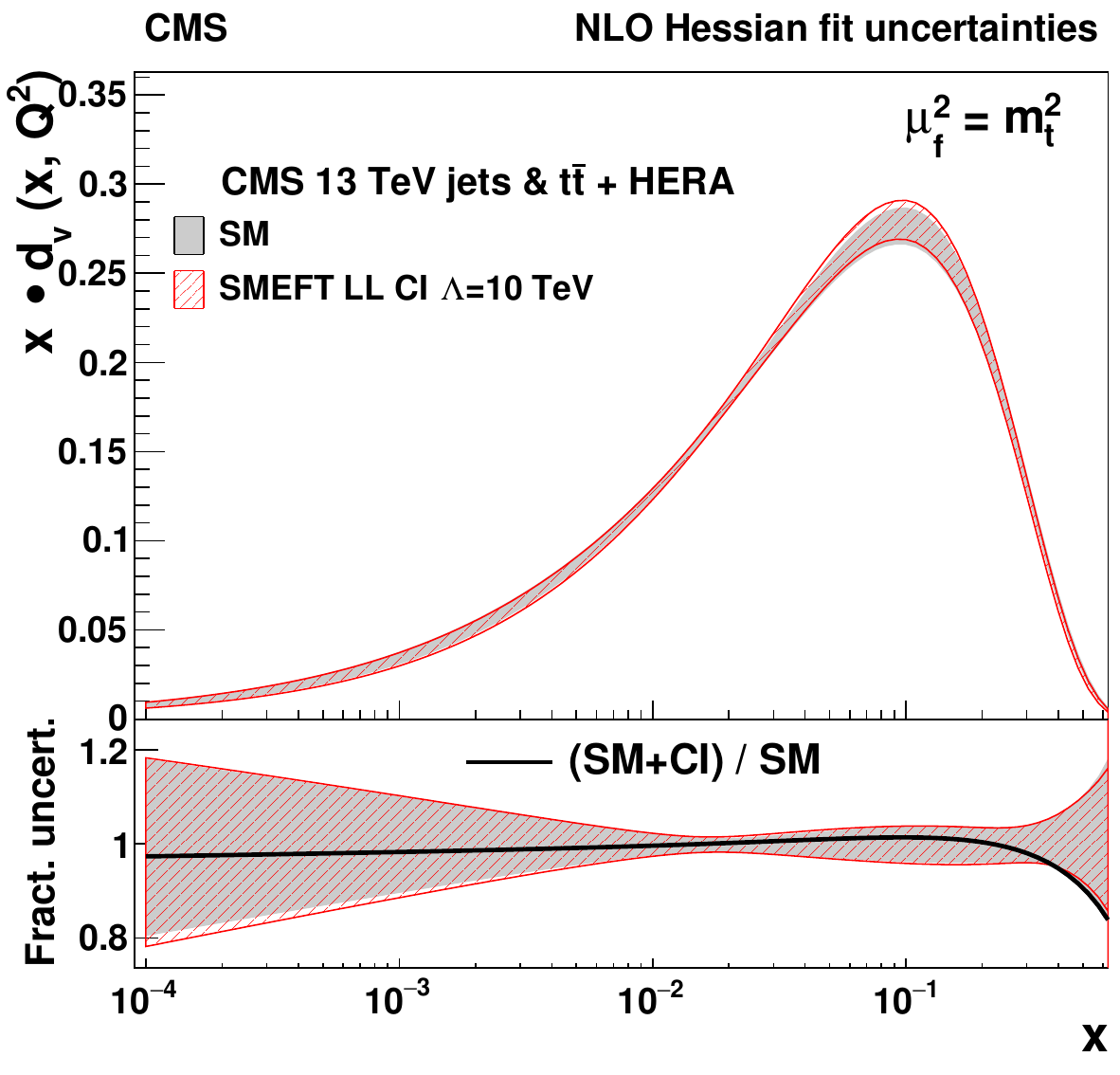}\\
\includegraphics[width=0.3\textwidth]{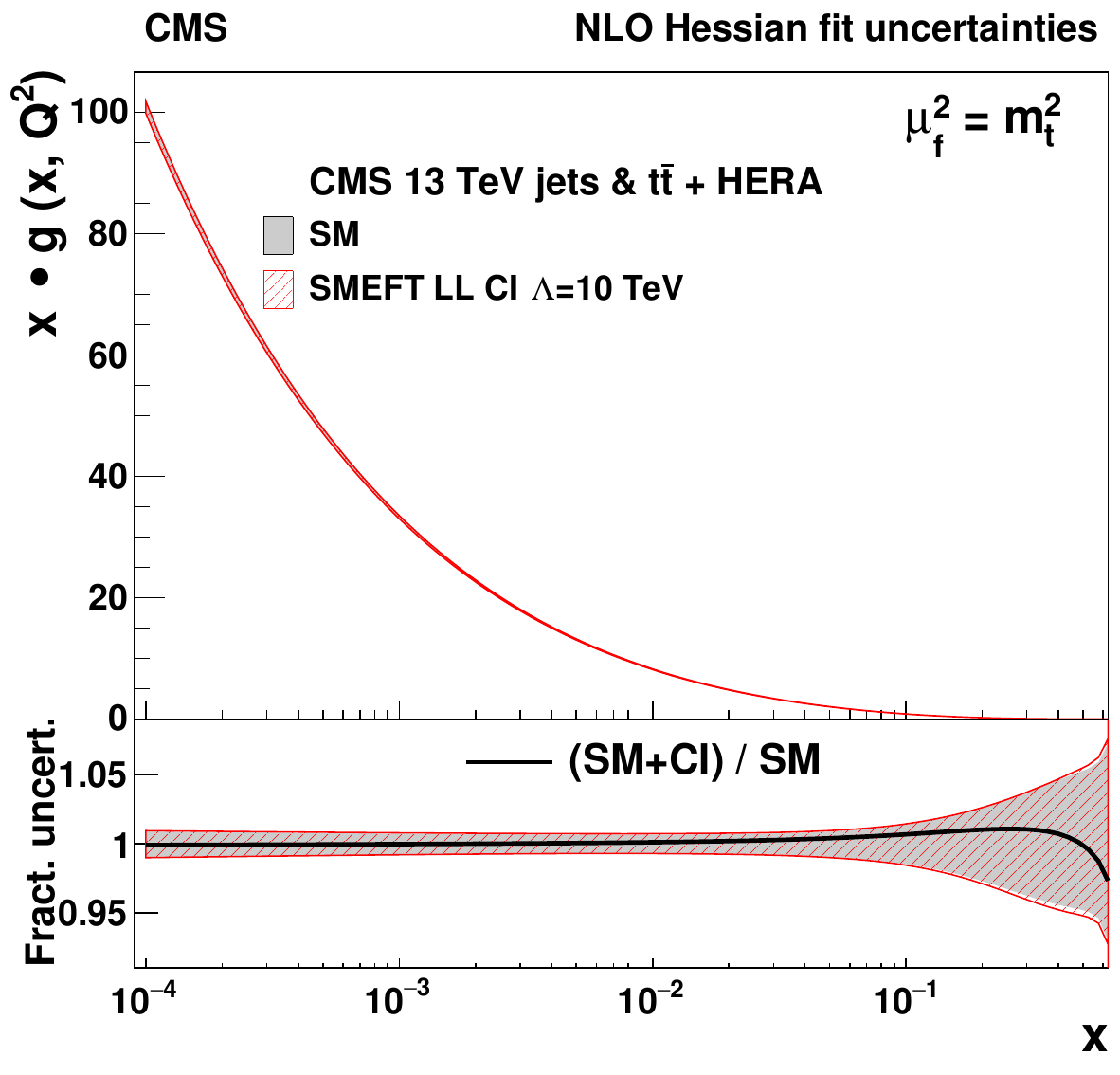}
\includegraphics[width=0.3\textwidth]{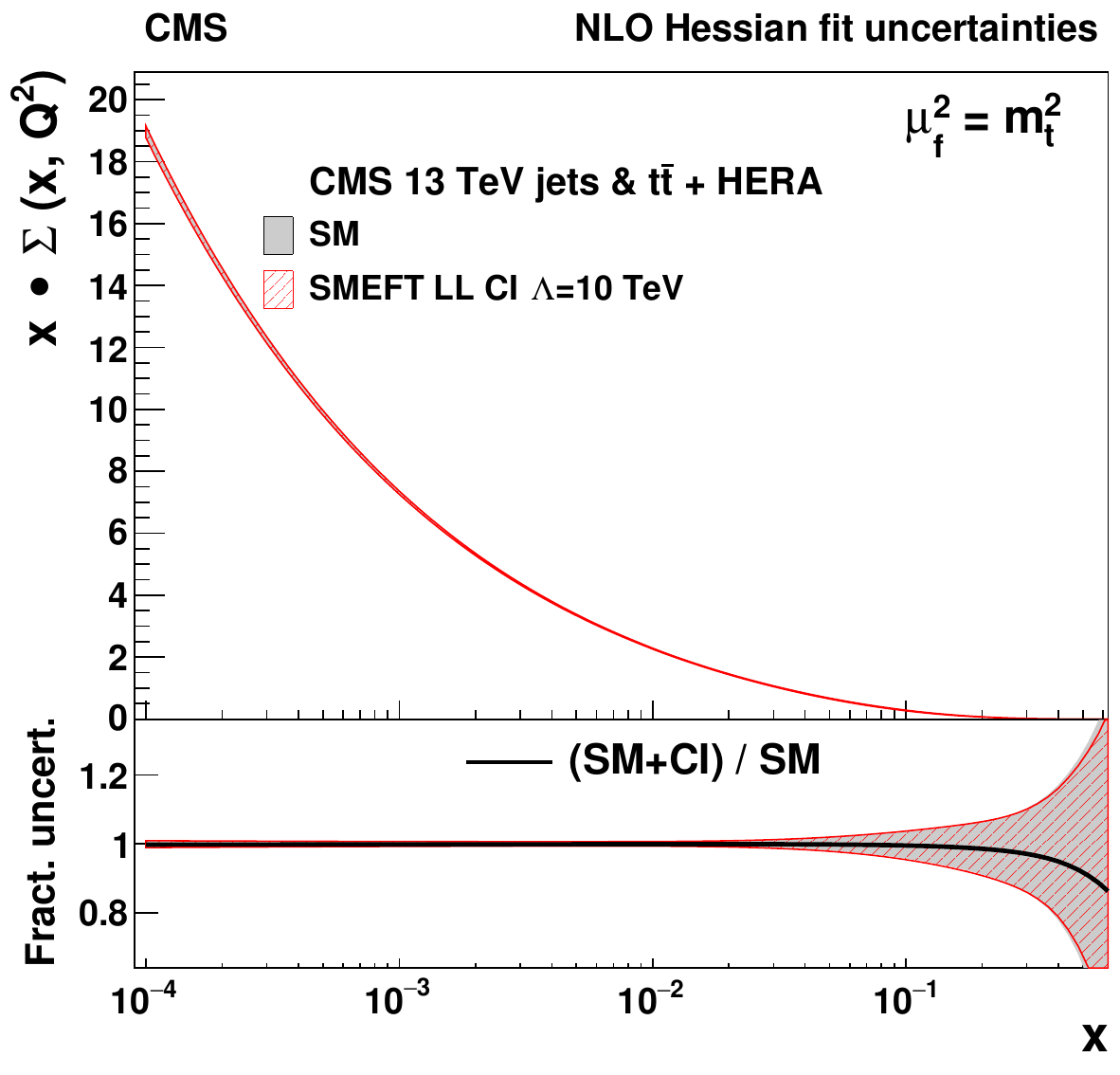}
\caption{The u-valence (upper left), d-valence (upper right), gluon (lower left), and sea quark
(lower right) distributions, shown as functions of $x$ resulting from the
fits with and without the CI terms. The SMEFT fit is performed with the left-handed CI model
with $\Lambda$ = 10 TeV. The figure is taken from~\cite{CMS:2021yzl}.}
\label{smeft_PDFs}
\end{figure}

To compare to a conventional search for CI, the Wilson coefficients obtained in the SMEFT analysis are translated into unbiased 95\% confidence level exclusion limits on $\Lambda$ with $c_1=-1$, resulting in the limits of 24~TeV for left-handed, 32~TeV for vector-like, and 31~TeV for axial vector-like CI. The present analysis provides for the first time such limits using hadron collider data while following an unbiased search strategy. To compare the PDFs and QCD parameters obtained in the SMEFT fit to respective SM results, the NLO analysis is performed considering only the standard model. The resulting QCD parameters obtained in both variants of the fit, SM and SMEFT, agree well, however the SM results have smaller parametrization uncertainty. The fit quality of both fits expressed in $\chi^2$ divided by number of degrees of freedom results in $1411/1141$ for the SM and $1401/1140$ for SMEFT fits, respectively. The PDFs resulting from the SMEFT fit are shown in Fig.~\ref{smeft_PDFs} in comparison to the results of the SM fit. Both results agree within the fit uncertainties. 

For the future developments, the authors of this measurement and interpretation would appreciate the public availability of the fast-grid techniques for the NNLO prediction of the inclusive jet cross section. Furthermore, the used $t\bar t$ data are interpreted at NLO, since the measurements are split into categories according to the presence of an additional jet. In spite of the strong sensitivity of these measurements to the top quark mass, strong coupling constant and to the gluon distribution, these could not be used in the NNLO analysis. A stronger effort in development of the theory calculation for the $t\bar t$ production associated with a jet is highly desirable. 

The performed SMEFT analysis is only a step towards the global SM\,$+$\,BSM interpretation of the LHC measurements and implies only colour-singlet CI contributions. The availability of colour-octet contributions and also of the corresponding EFT corrections for the $t\bar t$ cross section 
predictions at NLO, as well as to the other processes, probing the operators of similar structure, would be necessary. Finally, bringing the EFT corrections to the QCD processes to at least NNLO accuracy would be the ultimate goal for the interpretation of the HL-LHC measurements.

\subsection{The strong coupling constant and quark masses
\protect\footnote{A\lowercase{uthors:} M. V\lowercase{os} (IFIC, V\lowercase{al\`encia})}}

The determination of the strong coupling constant is strongly tied to the determination of other fundamental parameters of the SM and, in particular, with the quark masses in the QCD Lagrangian. Precise determinations of quark masses typically relies on a comparison of QCD predictions beyond leading-order accuracy with the measurement of experimental observables. These measurements are often sensitive to both the strong coupling constant and the mass value, and both parameters should ideally be treated in a simultaneous fit.


Like the strong coupling constant, quark masses are renormalized, scheme-dependent quantities. In the $\MSbar$ scheme the value of the quark mass depends at a given order in perturbation theory on the dimensionful renormalization scale $\mu$. The evolution with this scale, or ``running'' of the mass, forms a testable prediction of the theory. 
QCD yields a precise prescription for the scale evolution: given a value for a quark mass at a reference scale, its value at any other scale can be determined using the renormalization group equation (RGE). RGE calculations for the the running quark masses have by now reached the 5-loop ${\cal O}(\alphas^5)$ level~\cite{Vermaseren:1997fq,Chetyrkin:1997dh,Baikov:2014qja}, and software packages such as RunDec~\cite{Herren:2017osy} and REvolver~\cite{Hoang:2021fhn} provide access to state-of-the-art renormalization evolution and scheme conversions.

A large number of measurements over a broad range of energies characterize the evolution of the strong coupling $\alphas(\mu)$~\cite{Zyla:2020zbs}. Experimental tests of the running of quark masses have been performed for the charm~\cite{Gizhko:2017fiu}, bottom~\cite{Abreu:1997ey,Rodrigo:1997gy,Abe:1998kr,Brandenburg:1999nb,Barate:2000ab,Abbiendi:2001tw,Abdallah:2005cv,Abdallah:2008ac} and top~\cite{Sirunyan:2019jyn} quarks. 

\begin{figure}[htbp!]
\includegraphics[width=0.6\columnwidth]{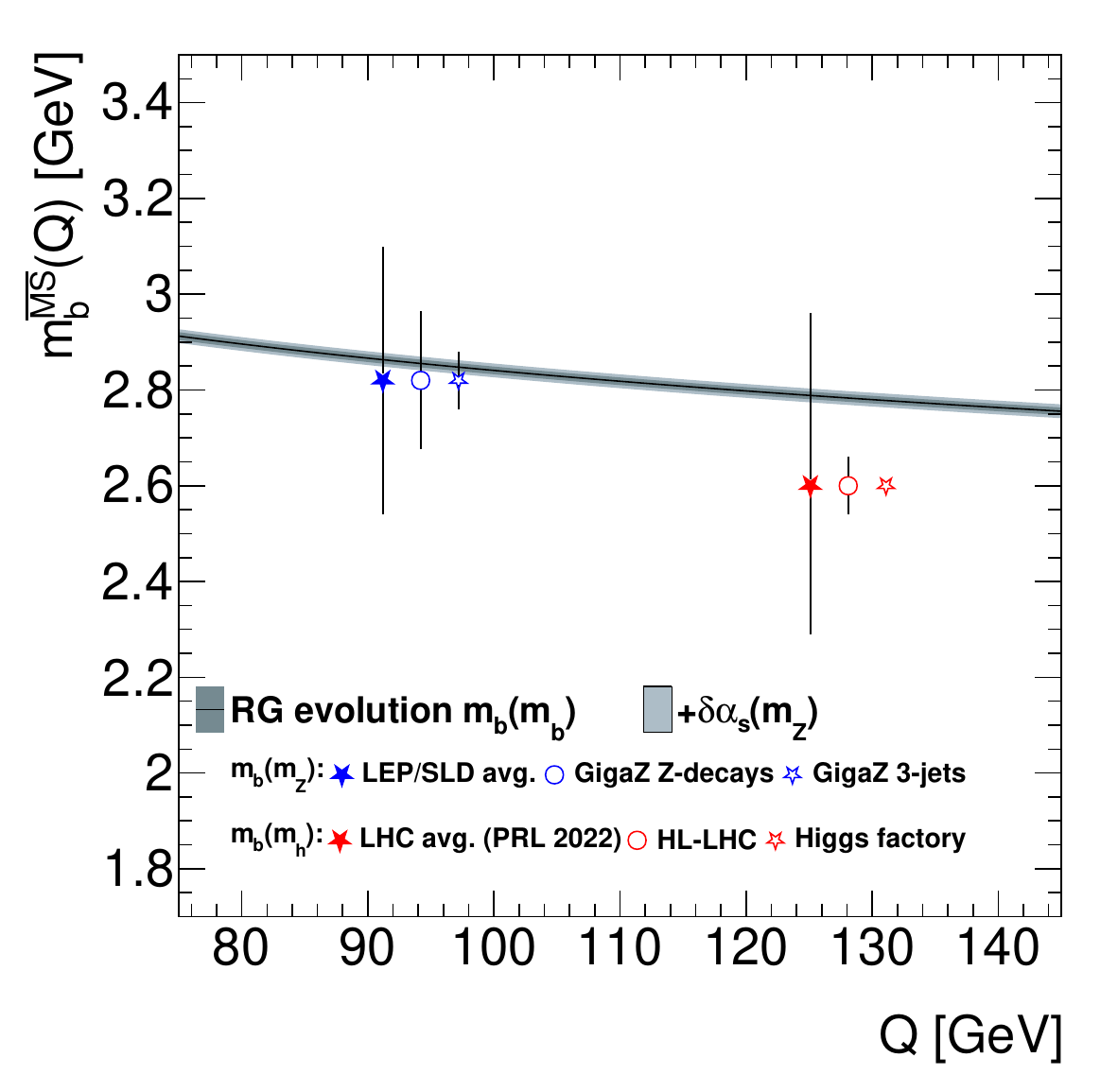}%
\caption{\label{fig:running_mass_projection} The scale evolution of the bottom quark $\MSbar$ mass. The markers are projections for $m_b(m_\mathrm{Z})$ from three-jet rates at the Z-pole and for $m_b(m_\mathrm{H})$ from Higgs boson branching fractions. The prediction of the evolution of the mass is calculated at five-loop precision with REvolver~\cite{Hoang:2021fhn}. The grey error band includes the effect of missing higher orders and the projected parametric uncertainties from $m_b(m_b)$ and $\alphasmZ$. 
[Figure from Ref.~\cite{Snowmass:bprospects}].
}
\end{figure}

Two recent studies revisit the running of the bottom quark mass using Higgs decay measurements at the LHC~\cite{Aparisi:2021tym} and Z-decay rates to bottom quarks at LEP~\cite{Kluth:2022ucw}. The measurement of $m_b(m_\mathrm{H})$ reaches a competitive precision of 14\% using LHC Run-2 measurements. The precision could improve to 2\% after the complete LHC programme, including the high-luminosity phase. The method has the potential for a subpercent-level determination at a future electron-positron ``Higgs factory''. The three-jet rates in a high-luminosity ``GigaZ/TeraZ'' Z-pole run at future ``Higgs/EW/top factory'' electron-positron collider can improve the measurement of $m_b(m_\mathrm{Z})$ by a factor of two~\cite{ILDnote2020}. The determination of $m_b(m_\mathrm{Z})$ from the Z\,$\to b\bar{b}$ decay rate is currently not competitive, but could reach an interesting precision (5\%) with a new Z-pole run. 

Together, these measurements take the test of the scale evolution of quark masses to a new level of precision~\cite{Snowmass:bprospects}. Figure~\ref{fig:running_mass_projection} from Ref.~\cite{Snowmass:bprospects} compares the projections for $m_b(m_\mathrm{Z})$ and $m_b(m_\mathrm{H})$ to the current averages and the predicted evolution of the $m_b(m_b)$ world average. The precision of $m_b(m_\mathrm{H})$ is expected to increase rapidly in Run-3 of the LHC and the HL-LHC programme. A new electron-positron collider operated at the Z-pole and the optimal Higgs-strahlung cross section ($\sim$240--250~GeV) can improve $m_b(m_\mathrm{Z})$ and $m_b(m_\mathrm{H})$ further and take tests of the ``running'' of the bottom quark mass into the precision regime.
 
The scale evolution of the strong coupling and the quark masses is sensitive to the presence of unknown, massive states that carry colour charge~\cite{Llorente:2018wup, Jezabek:1992sq}. A combined analysis of the scale evolution of the bottom quark mass and the strong coupling constant is required to consistently treat the impact of the new state on both quantities. A joint fit could possibly include also the charm and top quark masses, that can also be improved with an electron-positron collider~\cite{Boronat:2019cgt}. 

With data collected in the next decades at the HL-LHC and a Higgs factory operated at $\sqrt{s}=m_\mathrm{Z}$ and $\sqrt{s}\sim$ 250~GeV, the precision of high-scale determinations of the bottom quark mass is expected to increase very significantly, with $m_b(m_\mathrm{H})$ reaching subpercent precision. A joint analysis of the scale evolution of the strong coupling and the quark masses then provides a powerful and model-independent handle on new coloured states in the mass range between $m_b$ and $m_\mathrm{H}$.


\clearpage
\section{\texorpdfstring{\boldmath$\alphasmZ$}{alphasmZ} from quarkonium}
\label{sec:QQbar}

\subsection{\texorpdfstring{$\alphasmZ$}{alphasmZ} from relativistic quarkonium sum rules
\protect\footnote{A\lowercase{uthors:} D. B\lowercase{oito} (U.\,V\lowercase{ienna \&} U.\,S\lowercase{{\~a}o} P\lowercase{aulo}), V. M\lowercase{ateu} (U\lowercase{niv.} S\lowercase{alamanca} \& UAM-CSIC M\lowercase{adrid})}}
\label{sec:RelSumRulCharm}


One of the classical observables in QCD is the inclusive cross section  for $e^+e^-\to {\rm hadrons}$,
which is more conveniently cast in terms of the so-called $R$ ratio defined as
\begin{equation}\label{eq:Rqq}
R_{q\bar{q}}(s) =\! \frac{3s}{4\pi\alpha^2}\sigma_{e^+e^-\to\, q\bar{q}\,+X}(s) \simeq
\dfrac{\sigma_{e^+e^-\to\, q\bar{q}\,+X}(s)}{\sigma_{e^+e^-\to\,\mu^+\mu^-}(s)}\,,
\end{equation}
where $\alpha$ is the fine fine-structure constant,  $\sqrt{s}$ the center-of-mass energy, and  the right-hand side is exact if $\sigma_{e^+e^-\to\,\mu^+\mu^-}(s)$ is calculated at tree level for massless muons. Here we are interested in the case of heavy quarks, $q=c,b$.

Integrated moments of $R_{\bar q q}(s)$ are particularly suitable for phenomenological studies since, as opposed to local  measurements, they
 have significantly smaller errors and
 suffer less from  residual duality
violations, allowing for a direct comparison with computations carried out with
partonic degrees of freedom. Furthermore, they
 can be  computed accurately in perturbation theory and receive
 small nonperturbative corrections, written as an
 expansion of local operators with increasing dimension. 
In particular, the inverse moments of $R_{\bar q q}(s)$, that we denote $M_q^{(n)}$, lead to the following sum rules~\cite{Shifman:1978bx,Shifman:1978by}
\begin{equation}
\label{eq:momentdefvector2}
M_q^{(n)} = \int_{s_0}^{\infty}\!\dfrac{{\rm d}s}{s^{n+1}}R_{q\bar{q}}(s) = \dfrac{12\pi^2 Q_c^2}{n!}\,\dfrac{{\rm d}^n}{{\rm d}s^n}\Pi_q(s)\Big|_{s=0}\,.
\end{equation}
On the left-hand side, obtained from experimental data, $s_0$ must be below the first $\bar qq$ narrow resonance with the same quantum numbers as the photon. On the right-hand side, theoretical moments can be related to derivatives of the
heavy-quark  vector correlator
\begin{equation}
\!\big(g^{\mu\nu}s-p^{\mu}p^{\nu}\big)\Pi_q(s) =\! - i\!\!\int\!\!{\rm d}x\, e^{i\,p\cdot x}
 \langle 0|T\,j_c^{\mu}(x)j_c^{\nu}(0)|0\rangle\,,
\end{equation}
evaluated at $s=0$ [\,$j_c^{\mu}(x) = \bar{q}(x)\gamma^\mu q(x)$\,]. The moments are dominated by a short-distance scale given by $\sim m_q/n>\Lambda_{\rm QCD}$ (where $m_q$ is the quark mass), so restricting $n$ to small values  they can be computed in fixed-order QCD and their expansion is known up to $\mathcal{O}(\alphas^3)$ for $n\leq 4$ \cite{Chetyrkin:1995ii, Chetyrkin:1996cf, Boughezal:2006uu, Czakon:2007qi,Maier:2007yn,Chetyrkin:2006xg, Boughezal:2006px, Sturm:2008eb,Maier:2008he, Maier:2009fz,Maier:2017ypu}. Due to the strong sensitivity
of $M_q^{(n)}$ to the heavy quark mass, which appears as a prefactor $1/{[2\,\overline m_q(\mu_m)]^{2n}}$ in their perturbative expansion,
these sum rules
have been used for many years to extract $m_c$ and $m_b$, the masses of the charm and bottom quarks, with very good precision \cite{Kuhn:2001dm,Kuhn:2007vp,Chetyrkin:2009fv,Chetyrkin:2010ic,Chetyrkin:2017lif,Dehnadi:2011gc,Dehnadi:2015fra} ($\overline{m}_q(\mu_m)$ denotes the $\overline{\rm MS}$ quark mass at scale $\mu_m$).

Here we summarize the main results of Refs.~\cite{Boito:2019pqp,Boito:2020lyp}, where  dimensionless ratios of roots of moments~$M_q^{(n)}$ were  considered. Since the ratios
of charm-quark moments lead to a more precise determination of the strong coupling (mainly because experimental uncertainties are smaller) we restrict the presentation to the charm-quark moment ratios defined as
\begin{equation}\label{eq:ratMM}
R_c^{V,n}\equiv \frac{\big(M_c^{(n)}\big)^\frac{1}{n}}{\big(M_c^{(n+1)}\big)^\frac{1}{n+1}}\, .
\end{equation}
Ratios of this type were first introduced for the analysis of lattice data for the pseudoscalar correlator \cite{Maezawa:2016vgv,Petreczky:2019ozv} (see also the contribution of Petreczky and Weber to this volume, Sec.~\ref{sec:latt:moments}).
In the ratios $R_c^{V,n}$, the quark-mass dependence from the pre-factor of $M_c^{(n)}$ exactly cancels. These dimensionless ratios are suitable for precise $\alphas$ extractions for the following reasons (some of them already stated):
\begin{enumerate}
  \item They have a very small residual dependence on the  quark mass,
  \item are known up to $\mathcal{O}(\alphas^3)$ for $1\leq n \leq3$,
  \item contributions from nonperturbative physics are fairly small,
  \item they can be determined experimentally  from narrow resonance parameters and continuous $R_{\bar q q}(s)$ data.
\end{enumerate}

The QCD fixed-order perturbative expansion of the ratios $R_c^{V,n}$ reads
\begin{align}\label{eq:R2scales}
R^{V,n}_c =
\sum_{i=0} \bigg[\frac{\alphas(\mu_\alpha)}{\pi}\bigg]^i   \times\sum_{k=0}^{[i-1]}\sum_{j=0}^{[i-2]} r^{(n)}_{i,j,k}
\ln^j\biggl[\frac{\mu_m}{\overline{m}_c(\mu_m)}\biggr]
\ln^k\biggl[\frac{\mu_\alpha}{\overline{m}_c(\mu_m)}\biggr]\,,
\end{align}
where $[i-1]\equiv {\rm Max}(i-1,0)$ and we use $\Nf=4$ (the $\Nf$ dependence in $\alphas$ and the perturbative coefficients will be omitted). The running coupling $\alphas(\mu_\alpha)$ and quark  mass $\overline{m}_c(\mu_m)$ are expressed in the $\overline{\rm MS}$ scheme and evaluated at the renormalization scales $\mu_\alpha$ and $\mu_m$, respectively. We do not assume these scales to be the same,  as they account for different physics. This leads to a more conservative theory error estimate as argued in \cite{Dehnadi:2011gc,Dehnadi:2015fra}. We do avoid, however, large logarithms when varying the scales $\mu_\alpha$ and $\mu_m$, as discussed below. Note that the leading dependence of the ratios $R^{V,n}_c$ on the quark mass is  only logarithmic and  $\alphas^2$ suppressed.  Concretely, for  $R_c^{V,2}$ at N$^3$LO one finds
\begin{align}\label{eq:ptexp}
R_c^{V,2} =& 1.0449\Bigl[1+0.57448\,a_s  +\bigl(0.32576 + 2.3937\,L_{\alpha}\bigr)\,a_s^2 \\
-&\bigl(2.1093+ 4.7873 L_m -6.4009 L_\alpha - 9.9736 L_\alpha^2\bigr)\,a_s^3\,\Bigr]\,,\nonumber
\end{align}
with $a_s=\alphas(\mu_\alpha)/\pi$, $L_\alpha = \ln[\,\mu_\alpha/\overline{m}_c(\mu_m)\,]$  and  $L_m = \ln[\,\mu_m/\overline{m}_c(\mu_m)\,]$. The total
$\alphas$ corrections are about $12.5\%$ for $R_c^{V,1}$, $7.2\%$ for $R_c^{V,2}$, and $5.2\%$ for $R_c^{V,3}$, which makes these observables rather sensitive to the strong coupling.
The dominant nonperturbative correction to these observables stems from the gluon-condensate contribution and is known to NLO. We include it in our analysis, but the numerical  impact is small.\\

We summarize now the determination of the moments $R_c^{V,n}$ from experimental data, which consists of three parts: the contribution from the narrow resonances  ($J/\psi$ and $\psi^\prime$), a contribution from threshold data taken from Refs.~\cite{Bai:1999pk,Bai:2001ct,Ablikim:2004ck,Ablikim:2006aj,Ablikim:2006mb,:2009jsa,Osterheld:1986hw,
Edwards:1990pc,Ammar:1997sk,Besson:1984bd,:2007qwa,CroninHennessy:2008yi,Blinov:1993fw,Criegee:1981qx,Abrams:1979cx}, and the contribution for $\sqrt{s}>10.538$~GeV, where there are no measurements anymore, modelled with perturbative QCD (the `contiuum contribution'). Since charm threshold data is inclusive in all flavours, one must also perform the subtraction of the $u$, $d$, $s$ background and of the secondary charm production which is not accounted for in theoretical computations. (Small singlet contributions can be neglected~\cite{Kuhn:2007vp}).
It is important to keep the value of $\alphas$ used in the continuum contribution and the subtraction of the $u$, $d$, $s$ background as a free parameter, otherwise the extraction of the coupling would be biased. For values not too far from the $\alphas$ world average, the dependence of $R_c^{V,n}$ on the strong coupling is linear to an excellent approximation. Parametrizing the results through $\Delta_\alpha\equiv\alphas^{(\Nf=5)}(m_Z)-0.1181$ we find
\begin{eqnarray}\label{eq:Rexp}
R_c^{V,1}&=& (1.770- 0.705\,\Delta_\alpha) \pm 0.017 \,,\\
R_c^{V,2}&=& (1.1173 -0.1330\,\Delta_\alpha)\pm 0.0022\, , \nonumber\\
R_c^{V,3}&=& (1.03535 - 0.04376\,\Delta_\alpha)\pm 0.00084 \,.\nonumber
\end{eqnarray}
The above errors are dominated by threshold data and are quite small. This is in part due to positive correlations between the moments $m_c^{(n)}$ which produces a partial cancellation of the errors in the ratios.\\

We extract $\alphas$ equating the expansions of the type of Eq.~(\ref{eq:ptexp}) with the respective experimental counterparts in Eq.~(\ref{eq:Rexp}) and solving for  $\alphas$ numerically.
It is important to carefully and conservatively estimate the theoretical uncertainties arising from the truncation of perturbation theory. In the case of the determination of quark masses, the work of Ref.~\cite{Dehnadi:2011gc,Dehnadi:2015fra} shows that it is important to vary the two renormalization scales $\mu_\alpha$ and $\mu_m$ independently. We therefore vary them in the interval \mbox{$\overline{m}_c \leq \mu_\alpha,\mu_m \le 4~{\rm GeV}$} applying the constraint \mbox{$1/\xi\leq (\mu_\alpha/\mu_m)\leq \xi$} and using $\xi=2$ for our final results.
A much less conservative choice, often adopted in related works in the literature,  which leads to significantly smaller uncertainties, is to set $\mu_\alpha=\mu_m$, or equivalently $\xi=1$.  We obtain the uncertainties from renormalization scale variation from the spread of values of $\alphas$ in the $\mu_m\times \mu_\alpha$ plane, using a grid with 3025 points.

\begin{figure}[htpb!]
\centering
\includegraphics[width=0.51\textwidth]{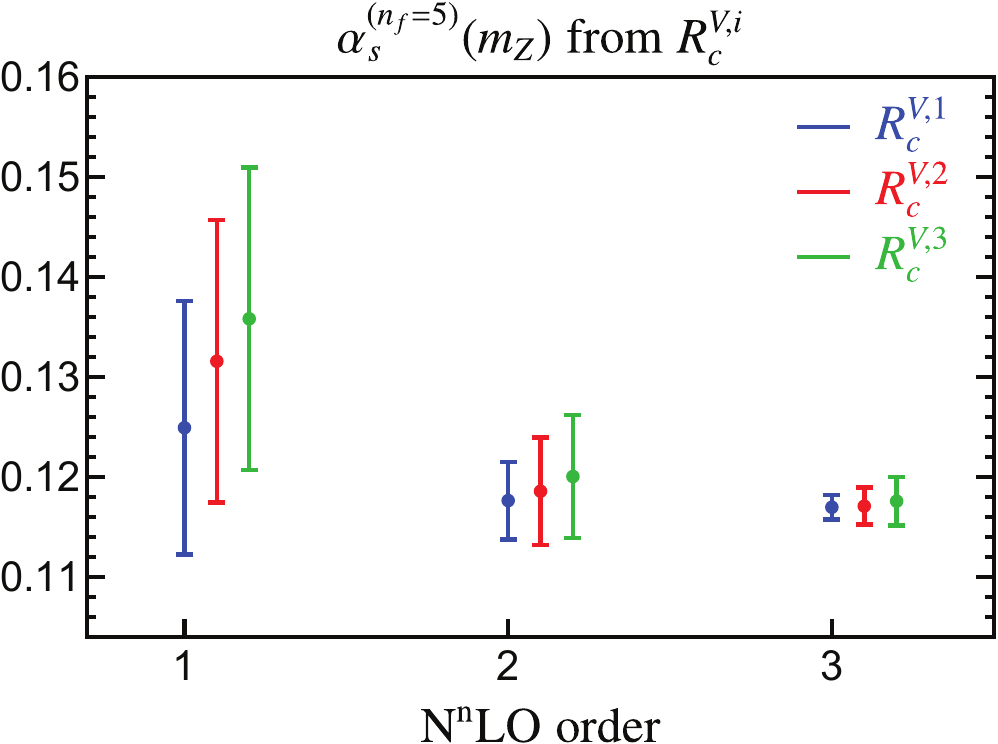}~~~~~~
\includegraphics[width=0.35\textwidth]{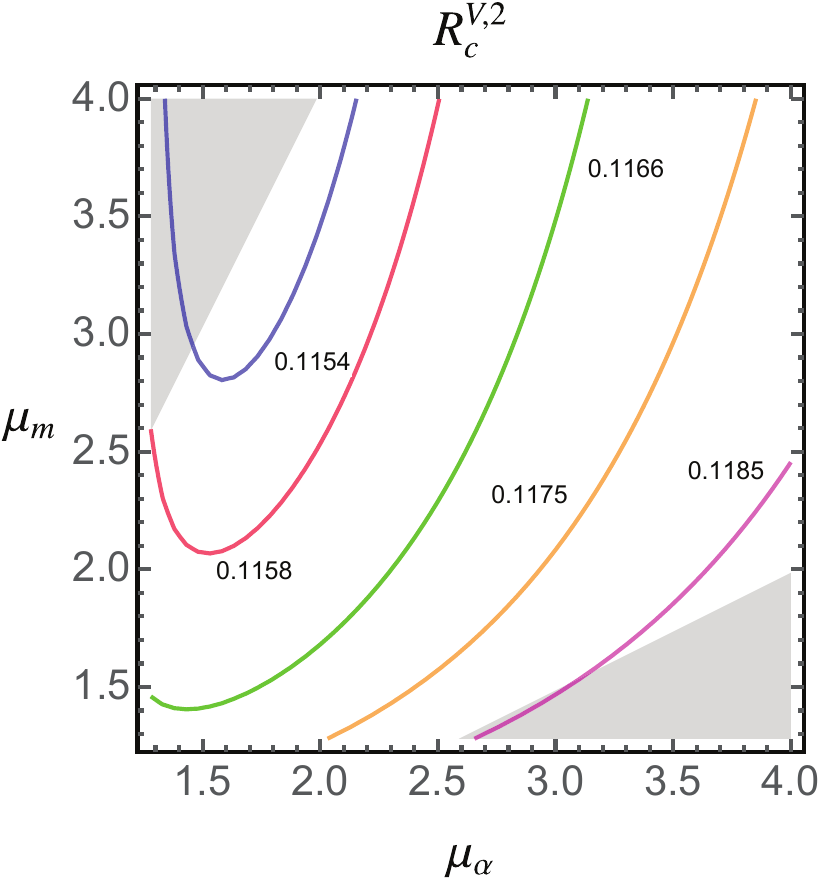}
\caption{Left: Results for $\alphas$ obtained from the analysis of $R_c^{V,n}$ with $n~=1, 2$, and 3, order-by-order in perturbation theory. Only truncation errors are shown. Right: Results for $\alphas$ obtained from $R_c^{V,2}$ in the $\mu_m\times \mu_\alpha$ plane. The shaded region in gray is excluded  from the analysis with $\xi=2$ (see text).}
\label{fig:results}
\end{figure}

First, we verify the convergence of the extraction of $\alphas$ order-by-order in perturbation theory, considering solely the perturbative uncertainty --- estimated through the scale variation described above. In Fig.~\ref{fig:results} (left), we show that the results display a nice convergence, which indicates that the theory error bars from the truncation of perturbation theory are not underestimated in our analysis.
Further light on the perturbative uncertainty can be shed from two-dimensional $\alphas$ contour plots in the $\mu_m\times \mu_\alpha$ plane, as shown in Fig.~\ref{fig:results} (right) for $R_c^{V,2}$. One sees that setting $\mu_\alpha=\mu_m$ (corresponding to $\xi=1$) the error would be underestimated.
Numerically,  this choice leads to $\alphas$ errors obtained from the analysis of $R_c^{V,n}$ that are smaller by factors of $3$ ($n=1$), $2$ ($n=2$), and $1.5$ ($n=1$) than those obtained with $\xi=2$.

We can then extract the final value of $\alphas$ from our analysis of the charmonium ratios $R_c^{V,n}$. Apart from the truncation uncertainty (labelled `pt'), discussed in detail above, we  include in the final value the experimental error (`exp'), the incertitude from the residual charm mass dependence (which happens to be negligible), and the nonperturbative error (`np', estimated from twice the gluon-condensate contribution). Ratios of moments with lower values of $n$ lead to smaller errors in $\alphas$ from the truncation of perturbation theory but, on the other hand, have larger experimental uncertainties. It turns out that the optimal determination is obtained from the analysis of $R_c^{V,2}$, which gives, after evolving to the Z boson mass scale,
\begin{equation}\label{eq:finalAlphaQQrat}
\alphas^{(\Nf=5)}(m_Z) = 0.1168(15)_{\rm pt}(9)_{\rm exp}(7)_{\rm np}=0.1168(19)
\qquad[\,{\rm charm~vector~current}\,]\,.
\end{equation}
 This result is fully compatible with the world average~\cite{Zyla:2020zbs}. We remark that by setting $\mu_\alpha=\mu_m$ the total uncertainty would be reduced to 0.0013, which shows that our error estimate is indeed conservative.

We have applied our method to bottom ratios $R_b^{V,n}$  as well, but due to large experimental uncertainties the extracted value for the strong coupling is not competitive. Furthermore, we have reanalyzed lattice data on the moments of the pseudoscalar correlator. As opposed to the vector current, the pseudoscalar 0-th moment $m_c^{P,0}$, which is scaleless and not sensitive to the charm quark mass, is an observable and therefore a direct comparison to perturbative computations is possible. From higher moments, one can construct ratios $R_c^{P,n}$ in  the same way as displayed in Eq.~\eqref{eq:ratMM}. For either the 0-th moment or the ratios, the exact same program for estimating perturbative errors can be applied, and we find uncertainties that are, in general, larger than what has been quoted
in the  original lattice articles. In particular, we find that the result in Eq.~\eqref{eq:finalAlphaQQrat} is slightly more accurate than  the extractions that we obtain with our method using lattice results quoted in Refs.~\cite{HPQCD:2008kxl,McNeile:2010ji,Maezawa:2016vgv,Petreczky:2019ozv,Nakayama:2016atf}.

A slightly more competitive determination can be obtained from a fit to $R_c^{V,2}$ and $m_c^{P,0}$ (which makes use of the fact that both theoretical and experimental uncertainties for those two observables are uncorrelated), which renders
\begin{equation}\label{eq:fitVecPsudo}
\alphas^{(\Nf=5)}(m_Z) = 0.1170\pm (0.0014)_{\rm total}\qquad [\,{\rm lattice\;+\;continuum}\,]\,.
\end{equation}
This result can be regarded as a hybrid lattice-continuum determination.

Our ongoing efforts to improve these $\alphas$ determinations include a more refined treatment of threshold data using an iterative reclustering procedure based on linear splines, in which the new BES-III data \cite{BESIII:2021wib} on the R-ratio below the charm production threshold has been included.
Recent results for the large-order behavior of the perturbative series in the
large-$\beta_0$ limit of QCD can also guide the design of new moments with reduced perturbative uncertainties, see Ref.~\cite{Boito:2021wbj}. One way of making the bottom sum rules more competitive is taking $n$ large, such that experimental errors virtually vanish. This requires using nonrelativistic QCD (NRQCD) in the theoretical expressions, but the necessary expressions are available since some time already. Finally, as mentioned in the bottomonium section, Sec.~\ref{sec:bottomonium}, the method explained in the preceding paragraphs should apply equally well to the ratios of bottomonium bound-state masses, which again exhibit experimental errors negligibly small.\\

\noindent{\it Acknowledgments---} DB thanks the Universidad de Salamanca, and VM thanks the Sao Carlos Physics Institute at Universidade de S\~ao Paulo, for hospitality. This work was supported by  the FAPESP-USAL SPRINT grant No. 2018/14967-4. DB's work is supported by the S\~ao Paulo Research Foundation (FAPESP) Grants No.~2021/06756-6,  the Coordena\c c\~ao de Aperfei\c coamento de Pessoal de N\'ivel Superior -- Brasil (CAPES) -- Finance Code 001, and by CNPq grant No.~309847/2018-4. VM is supported by the MECD grant PID2019-105439GB-C22, the EU STRONG-2020 project under the program H2020-INFRAIA-2018-1, grant agreement No.\ 824093 and the COST Action CA16201 PARTICLEFACE.

\subsection{\texorpdfstring{$\alphasmZ$}{alphasmZ} determination from bottomonium spectrum
\protect\footnote{A\lowercase{uthors:} V. M\lowercase{ateu} (U\lowercase{niv.} S\lowercase{alamanca} \& UAM-CSIC M\lowercase{adrid})}}

\label{sec:bottomonium}

Early attempts to determine $\alphas$ from quarkonia are based on radiative $\Upsilon(1S)$ decays. The work of~\cite{Brambilla:2007cz} uses the ratio of the radiative branching ratio over the nonradiative one $\Gamma(\Upsilon(1S)\to X\gamma)/\Gamma(\Upsilon(1S)\to X)$, with $X$ the inclusive set of possible final-state hadrons. The relevant computations can be carried out with partonic degrees of freedom in the final state, and taking into account the charge conjugation of $\Upsilon(1S)$ it reduces to $\Gamma(\Upsilon(1S)\to \gamma gg)/\Gamma(\Upsilon(1S)\to gg)$, where interesting cancellations take place. Hadronic matrix elements also play a role, and need to be computed in the lattice or in the continuum. The theoretical prediction was compared to CLEO data~\cite{CLEO:2005mdr} and the value $\alphas^{(\Nf=5)}(m_\mathrm{Z})=0.119^{+0.006}_{-0.005}$ was found.\\

In Ref.~\cite{Mateu:2017hlz}, charmonium and bottomonium spectra were studied within NRQCD, employing the short-distance low-scale scheme for the heavy quark mass known as MSR~\cite{Hoang:2008yj,Hoang:2017suc} to remove the renormalon inherited from the static potential. Since the bottom and charm quarks are no longer dynamical degrees of freedom they have to be integrated out in the MSR mass as well: this prevents the
appearance of large logarithms and the power counting of the theory is not upset. It was shown that, for a conservative estimate of perturbative uncertainties, the renormalization scales of $\alphas$ and the MSR mass should be varied independently in a given range determined by the principal quantum number. Furthermore, for a global analysis the scales for the different excited states should be correlated. Even though the main aim of the article was to determine $m_c$ and $m_b$ from global fits (data for quarkonium masses was taken from the PDG, see Ref.~\cite{Zyla:2020zbs} for updated results), the possibility of obtaining $\alphas$ from a 2-parameter fit to bottomonium masses was also explored. Since experimental errors on $b\bar b$ bound
states are tiny, their impact in $\alphas$ is immaterial. However,
the very strong correlation between $m_b$ and $\alphas$ translates into a sizable perturbative uncertainty: $\alphas^{(\Nf=5)}(m_\mathrm{Z})=0.1178\pm 0.0051$,
very similar to that quoted in Ref.~\cite{Brambilla:2007cz}.\\

A related study~\cite{Peset:2018ria} obtains the strong coupling from a renormalon-free combination of charm and bottom bound-state masses, including $b\bar{c}$, namely $m_{B_c} - m_{\eta_b}/2 - m_{\eta_c}/2$, which is also less sensitive to ultrasoft effects. Using the experimental values for the charm and bottom masses as obtained from fits to the masses of $\eta_c$ and $\eta_b$, respectively, the value $\alphas^{(\Nf=5)}(m_\mathrm{Z})=0.1195\pm0.0053$ was obtained, with experimental values for the quarkonium masses taken from the PDG. The
uncertainty is very similar to that of~\cite{Mateu:2017hlz}, which opens up the possibility of improving the determination through global fits (determining $m_c$, $m_b$ and $\alphas$ simultaneously) to $b\bar b$, $c\bar c$ and $b\bar c$ bound states. Another promising way of obtaining a precise value for the strong coupling is to adapt
the strategy of Ref.~\cite{Boito:2020lyp}, see also Sec.~\ref{sec:RelSumRulCharm}, to bottomonium. Taking the ratio of
bound-states masses with different quantum numbers would cancel most of the heavy-quark mass dependence, along with the leading renormalon: the ratios are dimensionless and almost exclusively sensitive to $\alphas$. Results on this direction
will be reported soon.\\


\noindent \textit{Acknowledgments}--- VM is supported by the MECD grant PID2019-105439GB-C22, the EU STRONG-2020 project under the program H2020-INFRAIA-2018-1, grant agreement No.\ 824093 and the COST Action CA16201 PARTICLEFACE.


\clearpage
\section{\texorpdfstring{\boldmath$\alphasmZ$}{alphasmZ} world average
\protect\footnote{A\lowercase{uthors:} J. H\lowercase{uston} (MSU, E\lowercase{ast} L\lowercase{ansing}), K. R\lowercase{abbertz} (KIT, K\lowercase{arlsruhe}), G. Z\lowercase{anderighi} (MPP, M\lowercase{unich})}}
\label{sec:average}

\subsection{Preliminary considerations}

We summarize here the current procedure used in the Particle Data Group (PDG)~\cite{Zyla:2020zbs} to obtain the world average value of $\alphasmZ$ and its uncertainty, and we discuss future prospects for its improvement.

Any physics observables where the strong interaction is involved (directly, or through loop corrections) depend on the value of the strong coupling constant. This implies that a number of different observables can be used to determine the QCD coupling constant, provided that a suitable pQCD theoretical prediction is available for that observable. 
The following considerations are taken into account to assess if a particular observable is suitable for use in the  determination of the strong coupling constant:
\begin{itemize}
\item[$\bullet$] The observable's sensitivity to $\alphas$ as
  compared to the experimental precision. For example, for the $\epem$
  cross section to hadrons (\eg\ the $R$ ratio), QCD effects are only a small
  correction, since the perturbative series starts at order
  $\alphas^{0}$, but the experimental precision is high.  Three-jet production, or event shapes, in $\epem$
  annihilation are directly sensitive to $\alphas$ since they start
  at order $\alphas$.  Four- and five-jet cross-sections start at $\alphas^2$ 
  and $\alphas^3$ respectively, and hence are very sensitive to $\alphas$.  However, 
  the precision of the measurements deteriorates as the number of jets involved increases. 

\item[$\bullet$] The accuracy of the perturbative prediction, or
  equivalently of the relation between $\alphas$ and the value of the
  observable.
  The minimal requirement is generally considered to be
  an NLO prediction. The PDG imposes now that at least NNLO-accurate 
  predictions be available. 
  %
  %
  %
  In certain cases where phase space restrictions require it, fixed-order predictions are supplemented with
  resummation.
  %
An improved perturbative accuracy is necessary to guarantee that the theoretical uncertainty is assessed in a robust way. 

\item[$\bullet$] The size of nonperturbative effects. Sufficiently
  inclusive quantities, like the $\epem$ cross section to hadrons, have
  small nonperturbative contributions, power corrections of order $\sim {\rm \Lambda}^4/Q^4$. 
  Other quantities,
  such as event-shape distributions, typically have contributions
  $\sim {\rm \Lambda}/Q$. All other aspects being equivalent, observables with smaller nonperturbative 
  corrections are preferable. 

\item[$\bullet$] The scale at which the measurement is performed. An
  uncertainty $\delta$ on a measurement of $\alphas(Q^2)$, at a scale
  $Q$, translates to an uncertainty
  $\delta' = (\alphas^2(m_\mathrm{Z}^2)/\alphas^2(Q^2)) \cdot\delta$ on
  $\alphasmZ$. For example, this enhances the already important
  impact of precise low-$Q$ measurements, such as from $\tau$ decays,
  in combinations performed at the $m_\mathrm{Z}$ scale.
\end{itemize}

The PDG determination of $\alphas$ first separates  measurements into a number of different categories, then calculates an average for each category. Such per-category subaverages are then used as inputs to the world average.
The PDG procedure requires four specifications of:
\begin{description}
    \item 1) the conditions that a determination of $\alphas$ should fulfill in order be included in the average; 
    \item 2) the separations of the different extractions of $\alphasmZ$ into the (approximately) independent categories; 
    \item 3) the procedure within each category to compute the average and its uncertainty;
    \item 4) the manner in which the different subaverages and their uncertainties are combined to determine the final value of $\alphasmZ$ and its uncertainty. 
\end{description}

\subsection{Details of the PDG averaging procedure}

In the following, we summarize the procedure adopted in the last edition of the PDG~\cite{Zyla:2020zbs}. 

\subsubsection{Criteria for determinations to be included in the world average}

In the PDG, the selection of results from which to determine the world average
value of $\alphasmZ$ is restricted to those that are
\begin{description}
\item 1) accompanied by reliable estimates of all experimental and
  theoretical uncertainties;
\item 2) based on the most complete perturbative QCD predictions of at
  least NNLO accuracy;  
\item 3) published in a peer-reviewed journal at the time of writing of the PDG report.
  \end{description}
Note that the second condition to some extent follows from the first. In fact, determinations of the strong coupling from observables in $\epem$ involving \eg\ five or more jets are very sensitive to $\alphas$, and could provide additional constraints. However, these observables are currently described only at leading order (LO) or next-to-leading order (NLO), and the determination of the theoretical uncertainty is thus considered not sufficiently robust. It is also important to note that some determinations are included in the PDG, but the uncertainty quoted in the relevant publications is increased by the PDG authors to fulfil the first condition. Similarly, in some cases the central value used in the PDG differs from the one quoted in some publications, but can be extracted from the analysis performed in that work. 

\subsubsection{Categories of observables}
\label{sec:cat}

All observables used in the determination of $\alphasmZ$ in the PDG exercise are classified in the following 
categories (Table~\ref{tab:averages} and Fig.~\ref{fig:alphas_2022}):
\begin{itemize}
\item ``Hadronic $\tau$ decays and low $Q^2$ continuum'' ($\tau$ decays and low $Q^2$):

Calculations for $\tau$ decays are available at N$^3$LO; there are different approaches to treat the perturbative and nonperturbative contributions, which result in significant differences; the value of $\alphas$ is determined at the $\tau$ mass;
\item ``Heavy quarkonia decays and masses'' ($\QQbar$ bound states):

Calculations are available at NNLO and N$^3$LO;
\item ``PDF fits'' (PDF fits)

Taken both from global PDF fits and analyses of singlet and non-singlet structure functions; for theory uncertainty, half of the difference between results obtained with NNLO and NLO is added in quadrature;

\item ``Hadronic final states of $\epem$ annihilations'' ($\epem$ jets and event shapes):

Taken from measurements at PETRA and LEP; nonperturbative corrections are important and can be estimated either via Monte Carlo simulation or analytic modeling;
\item ``Observables from hadron-induced collisions'' (hadron colliders):

NNLO calculations for $t\bar{t}$, electroweak bosons, and jet production at both the LHC and HERA, and Z\,$+$\,jet production at the LHC have allowed measurements for these processes to be used in $\alphas$ determinations. There is still an ongoing discussion of whether a simultaneous PDF fit has to be carried out to avoid any significant bias;
\item ``Electroweak precision fit'' (electroweak):

$\alphas$ determinations are averaged from electroweak fits to data from the Tevatron, LHC, LEP and the SLC; such fits rely on the strict validity of the Standard Model; and 
\item ``Lattice'':

The average determined by the FLAG group in 2019~\cite{FlavourLatticeAveragingGroup:2019iem} from an input of eight determinations was used; the 2021 result~\cite{Aoki:2021kgd} came out too late to be included, but the 2021 $\alphas$ average is very consistent with that of 2019. 
\end{itemize}
Detailed information about which observables are included in the different categories can be found in Ref.~\cite{Zyla:2020zbs}. 

\subsubsection{Average and uncertainty in each category}

In order to calculate the world average value of $\alphasmZ$, a preliminary step of pre-averaging results within the each category listed in Sec.~\ref{sec:cat} is carried out. 
For each subfield, except for the ``Lattice'' category, 
the {\it unweighted average} of all selected
results is taken as the pre-average value of $\alphasmZ$, and
the unweighted average of the quoted uncertainties is assigned to be
the respective overall error of this pre-average.
An unweighted average is used to avoid the situation in which individual measurements,  
which may be in tension with other measurements and may have underestimated uncertainties, 
can considerably affect the determination of the strong coupling in a given category. 
As an example, the determination of $\alphasmZ$ from $\epem$ jets \& shapes currently averages ten determinations and arrives at $\alphasmZ= 0.1171\pm0.0031$. Since two determinations~\cite{Abbate:2010xh,Hoang:2015hka}, both based on a similar theoretical framework, arrive at a small value of $\alphasmZ$ and have a very small uncertainty, if one were to perform a weighted average one would obtain an $\alphasmZ$ from $\epem$ jets \& shapes of $\alphasmZ= 0.1155\pm0.0006$, which is not compatible with the current world average. This would, in fact,  considerably change the world average because of the very small uncertainties. The current procedure is instead robust against $\alphasmZ$ determinations that are outliers with small uncertainties as compared to the other determinations in the same category. 
For the ``Lattice QCD'' (lattice)
subfield, we do not perform a pre-averaging step. Instead, we adopt 
 the FLAG2019 average value  and uncertainty for this subfield~\cite{FlavourLatticeAveragingGroup:2019iem}. FLAG2019 also requires strict 
conditions on its own for a determination to be included in their average, which are in line with 
those used in the PDG. 
The results of the averages of the categories are given in Table~\ref{tab:averages}. Following the $\alphas(2022)$ workshop, the subfields of ``$\tau$ decays and low $Q^2$'' and ``$Q\overline{Q}$ bound states'' have been updated to account for the latest studies of Ref.~\cite{Boito:2020xli} and Refs.~\cite{Boito:2019pqp,Boito:2020lyp}, respectively. The world average value does not change.

\begin{table}[htpb!]
\centering
\caption{\label{tab:averages} PDG average of the categories of observables [March'22 update of the PDG'21 results]. 
These are the final input to the world average of $\alphasmZ$.\vspace{0.2cm}}
\tabcolsep=4.5mm
\begin{tabular}{lcc}\hline
\rule{0pt}{3ex} category & $\alphasmZ$ & relative $\alphasmZ$ uncertainty \\
\hline
\rule{0pt}{3ex} $\tau$ decays and low $Q^2$ & $0.1178\pm0.0019$ & 1.6\% \\
\rule{0pt}{3ex} $\QQbar$ bound states & $0.1181 \pm 0.0037$  & 3.1\% \\ 
\rule{0pt}{3ex} PDF fits & $0.1162 \pm  0.0020$  & 1.7\% \\ 
\rule{0pt}{3ex} $\epem$ jets \& shapes & $0.1171 \pm  0.0031$  & 2.6\% \\ 
\rule{0pt}{3ex} electroweak & $0.1208 \pm 0.0028$  & 2.3\% \\
\rule{0pt}{3ex} hadron colliders & $0.1165 \pm 0.0028$  & 2.4\% \\
\hline
\rule{0pt}{3ex} lattice & $0.1182 \pm 0.0008$  & 0.7\% \\
\hline
\rule{0pt}{3ex} world average (without lattice) & $0.1176 \pm 0.0010$  & 0.9\% \\
\rule{0pt}{3ex} world average (with lattice) & $0.1179 \pm 0.0009$  & 0.8\% \\
\hline
\end{tabular}
\end{table}

From the table, it is clear that determinations from different categories are compatible with each other and accordingly can be combined to give rise to a final average.

\subsubsection{Final average}

\begin{figure}[htpb!]
\centering
\includegraphics[width=0.5\textwidth]{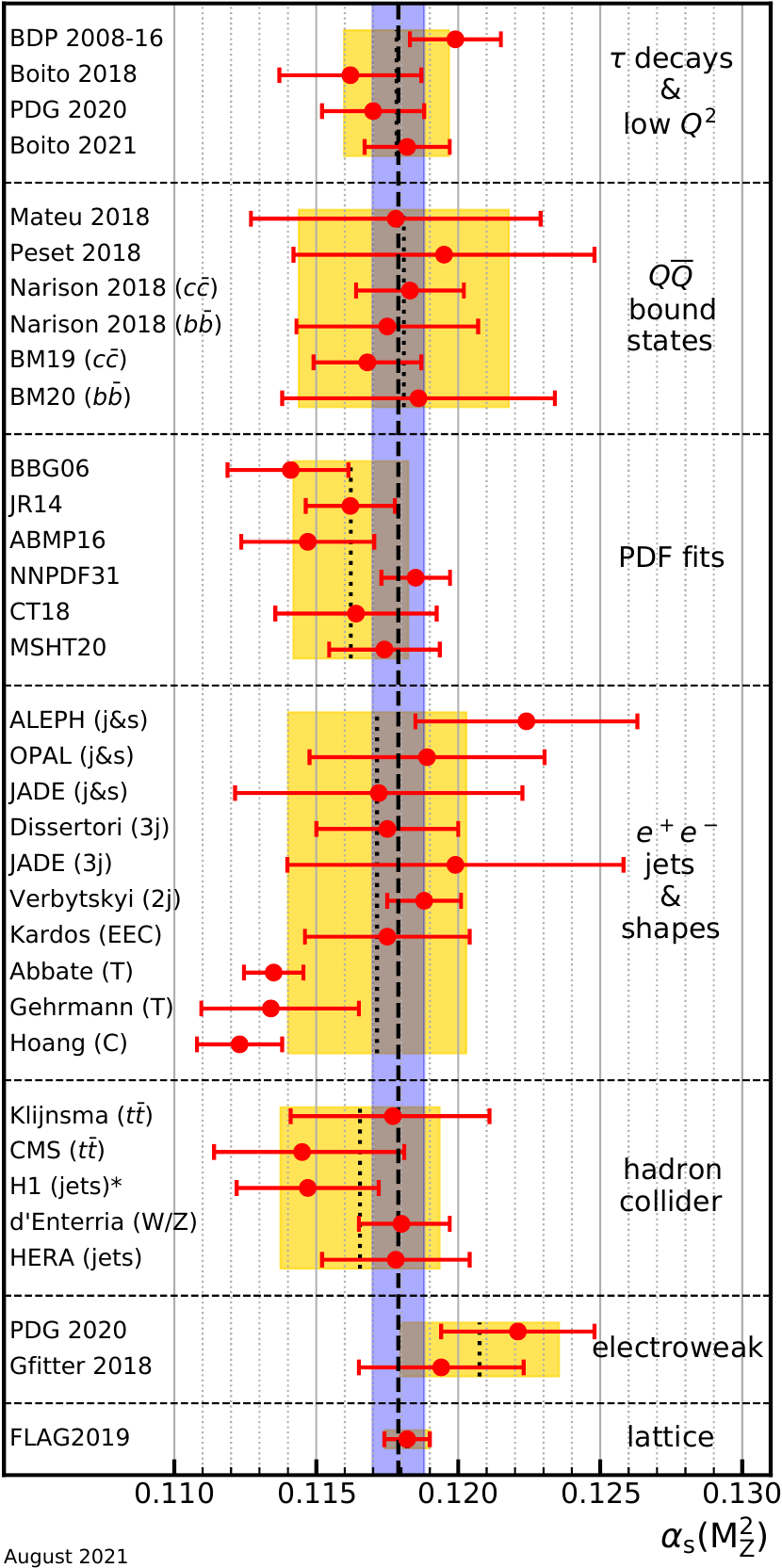}
\caption{Summary of determinations of $\alphasmZ$ from seven subfields. The yellow (light shaded) bands and dotted lines indicate the pre-average values of each subfield. The dashed line and blue (dark shaded) band represent the final $\alphasmZ$ world average [March'22 update of the PDG'21 results~\cite{Zyla:2020zbs}].}
\label{fig:alphas_2022}
\end{figure}

Since the six subfields (excluding lattice) are largely independent of each other, the PDG  determines a nonlattice world average value using a standard `{\it $\chi^2$ averaging}' method. This result in the final average of the six categories of 
\begin{equation}
\alphasmZ = 0.1175 \pm 0.0010\,, \qquad {\rm (without\,\, lattice)}, 
\end{equation}
which is fully compatible with the lattice determination (Fig.~\ref{fig:alphas_2022}).
In a last step the PDG performs an unweighted average of the values and uncertainties of
$\alphasmZ$ from the nonlattice result and the lattice result
presented in the FLAG2019 report~\cite{FlavourLatticeAveragingGroup:2019iem}, which results in the final average of 
\begin{equation}
\alphasmZ = 0.1179 \pm 0.0009\,,  \qquad {\rm  (final \,\, average)}.      
\end{equation}

Performing a weighted average of all seven categories would instead give rise to $\alphasmZ = 0.1180 \pm 0.0006$. The PDG uncertainty is instead more conservative and about 50\% larger.

\subsection{Outlook}

While the strong coupling remains the least well-known gauge coupling, with an uncertainty of about 1\%, it is remarkable that the determinations from all categories agree remarkably well with each other, all within one standard deviation. Future improvements will be driven by categories which have smallest uncertainties, currently lattice and  $\tau$ decays and low $Q^2$. 
The uncertainty quoted in the latter category includes the difference in the extractions that are obtained using contour improved perturbation theory (CIPT) and fixed order perturbation theory (FOPT). Recent arguments suggest that FOPT are to be preferred, see also dedicated discussions on this point in this white paper. If confirmed, this would slightly shift the value of $\alphasmZ$ in this category to lower values, and would allow one to quote a reduced theoretical uncertainty since this additional source of uncertainty would be completely removed. Further improvements could come from a better understanding of nonperturbative effects.
Some progress is also likely to come in the category $\epem$ jets \& shapes where the calculation of power corrections in the 3-jet region~\cite{Luisoni:2020efy,Caola:2021kzt} could have a sizeable impact, and improve fits of the coupling from event shapes. In fact, the region used in the fits are dominated by events with an additional hard emission, therefore the applicability of nonperturbative power corrections computed in the two-jet limit has been questioned and a treatment of these corrections in the three-jet region is certainly more appropriate. The impact of this on $\alphasmZ$ in this category has still to be assessed.
For the hadron collider category, it is an open discussion how to deal with correlations
between PDF parameters and $\alphasmZ$ in the cases where a full PDF\,$+\,\alphas$ fit is not
performed simultaneously. In view of many more NNLO results to come we can expect some
advances here. Particularly, NNLO for 3-jet production will enable to perform fits
of $\alphasmZ$ from ratios with at least partial cancellation of some uncertainties.
Some doubts were raised whether this reduction in uncertainty also holds for the PDF
dependence of such ratio predictions. Moreover, for predictions of ratios of cross sections, 
the central scale definition in numerator
and denominator will require more elucidation. 
Finally, it is important to mention that the last years have seen remarkable advances in the determination of $\alphasmZ$ from lattice gauge theory, also thanks to the FLAG effort which imposes strict quality criteria for lattice determinations to be included in the FLAG average. This is now the single most precise result of
all categories included in the PDG\@. The challenge is on to still beat it.

\clearpage
\section{Summary}
\label{sec:summ}

In this summary section we provide first an outline of the main points that were the subject of discussions during the $\alphas(2022)$ workshop, and secondly we recapitulate the current $\alphasmZ$ uncertainties of each extraction method as well as the expected experimental and theoretical developments that will help reduce those in the coming years (and, in some cases, in the longer term of the next generation of collider facilities).

\subsection{Summary of the discussions}

We provide here a brief summary of the most relevant contributions of the live discussions 
of the $\alphas(2022)$ workshop meeting preparatory for this document.

\paragraph*{Lattice:}\vspace{0.3cm}

The FLAG collaboration uses a set of criteria accepted in the lattice community to classify the analyses. Only analyses ranking high in all criteria are used to derive the final FLAG determination of the strong coupling. This procedure is very similar to that adopted by the PDG. It was suggested that the FLAG and PDG procedures should be harmonized as much as possible. It was also pointed out that there were updates to individual analyses since the current FLAG 2021 report~\cite{Aoki:2021kgd} was published. 

\paragraph*{\texorpdfstring{$\tau$}{tau} decays:}\vspace{0.3cm}

The discussions covered broadly two areas. The first area concerned the treatment of duality violations (DV) (incomplete asymptotic OPE) terms. One defines moments of the tau spectral function to control and suppress DV terms. However, stronger DV term suppression is correlated with more contributions of gluon condensates leading to a truncated OPE. This was referred to as a kind of ``seesaw mechanism'' by S.~Peris.

The presentation by A.~Hoang (Section~\ref{sec:FOCI}) providing an explanation in terms of renormalon effects of the FOPT vs.\ CIPT differences in the pQCD predictions, generated considerable interest. The analysis is not yet complete, and further important information would be the value of the gluon condensate normalization. New data would not help directly. The renormalon effects could possibly be studied on the lattice, but this would need a dedicated effort. One would need input from hadronic $\tau$ decay theory to derive observables to be measured on the lattice.

The second topic of discussion involved uncertainties connected with nonperturbative contributions to predictions for the $\tau$ hadronic width that are not directly related to the FOPT vs.\ CIPT discussion. More and better spectral function data from $\tau$ decays at B-factories or at new $\epem$ colliders, and from $\epem$ annihilation to hadrons at low energies could help to understand and constrain the DV and/or OPE truncation effects. 

\paragraph*{Global PDF analyses:}\vspace{0.3cm}

One focus of the discussion in this session was the treatment of theoretical (missing higher-order) uncertainties in the comparison of the predictions with data. The global fit analyses use $\mathcal{O}(1000)$ data points and thus classical $\chi^2$-probabilities are hard to interpret, and estimation of parameter uncertainties is difficult. Some authors recommend including the prediction uncertainties in the fits. A critique to this procedure is that this influences and possibly biases the fitted parameter values, including $\alphas$.

The ATLAS analysis of PDFs (talk by F.~Giuli, Section~\ref{sec:pp_ATLASpdf21}) emphasized the importance of a detailed analysis of systematic uncertainties and their correlations in the p-p collision data used in the fit. This was welcomed by global PDF fit groups who would like more detailed information on experimental uncertainties and their correlations.

K.~Rabbertz collected a list of open problems\footnote{Link: \href{https://docs.google.com/spreadsheets/d/13KpbCaTpgIP7zrOJL3eveCvLoynBoiT50shzIz1HCzo}{https://docs.google.com/spreadsheets/d/13KpbCaTpgIP7zrOJL3eveCvLoynBoiT50shzIz1HCzo}.} in the area of $\alphasmZ$ determinations from global PDF analyses. This was welcomed in the discussion and could serve as a reference to study and resolve pending and new issues. 

\paragraph*{\texorpdfstring{$\epem$}{e+e-} annihilation to hadronic final states:}\vspace{0.3cm}

The discussions in this area concentrated on the level of understanding of nonperturbative (NP) corrections. The latest results providing a better understanding of power corrections for the $C$-parameter observable (presentation by P.~F.~Monni and P.~Nason, Section~\ref{sec:ee_Cparam}) were seen as an important step. It was pointed out that a deeper understanding of NP corrections is also needed for other event-shape and for jet-based observables. Other items for improvements were raised such as: NNLO calculations for $\epem\to \bbbar g$, NNLL precision parton-shower algorithms and their matching to fixed order predictions and the corresponding retuning of the MC event generators NP parameters (hadronization, colour/flavour/spin correlations, etc.). As a potentially useful approach to reduce the NP corrections, groomed observables were discussed. It was stated that by pursuing such an agenda, a better consistency and precision of $\alphasmZ$ results from $\epem$ annihilation hadronic final states should be possible.

The improved prediction for the $R$-ratio at low energies (A.~Nesterenko, Section~\ref{sec:ee_Rratio}) was discussed as promising, but needs more precise measurements than currently available. 

\paragraph*{Lepton-hadron colliders:}\vspace{0.3cm}

Analyses at the only so-far operating lepton-hadron collider HERA showed high sensitivity to $\alphas$ through a large variety of processes with precision measurements available. These include inclusive cross sections, single- or multi-jet production cross sections, event-shape observables, jet-substructure observables, and heavy-flavour cross sections. Unfortunately, most of these measurements were not performed by the experiments using their larger HERA-II data samples, and the only useful published data from HERA-II that achieves 1--2\%  experimental uncertainties are jet cross sections in neutral-current DIS by H1, and inclusive jet cross section in photoproduction by ZEUS. Since considerable progress on the theory side was made since the HERA times, the H1 and ZEUS collaborations should feel encouraged to continue analyzing their data.

Future e-p experiments provide excellent opportunities for many precision measurements of $\alphasmZ$. The use of polarized PDFs was highlighted as a novel interesting opportunity at the future EIC. Beyond that, the PDFs from LHeC/FCC-ep will also allow improving many $\alphas$ analyses done at the HL-LHC.  

\paragraph*{Hadron colliders:}\vspace{0.3cm}

The analysis presented by D.~d'Enterria (Section~\ref{sec:pp_sigma_WZ}) was discussed for two reasons. First, it was claimed that the use of the MCFM~v.8.0 generator to compute inclusive fiducial NNLO cross sections for W and Z bosons could be sensitive to power corrections connected with its subtraction scheme~\cite{Alekhin:2021xcu}. However, in the publication a cross check with a different generator (FEWZ) gave consistent (within few \%) results. The second topic was the correlation of $\alphasmZ$ results from this analysis with results from global PDF analyses, since a few of the more recent PDF global fits also use differential cross section distributions of W- and Z-bosons measured at LHC.


The discussion on this second point moved on to considering merging the results from hadron collider data and global PDF fits into the same $\alphasmZ$ category, at least before the inclusive hadron collider cross sections ($\ttbar$ and EW bosons at the LHC, and jets at HERA) are not directly into the global PDF\,$+\,\alphas$ fits. The hadron collider extractions use the $\alphasmZ$ dependence of the public PDF sets to derive the QCD coupling itself, but it was pointed out that this may not converge to the same result as a global PDF\,$+\,\alphas$ fit~\cite{Forte:2020pyp}. A complete merging of the ``DIS and PDF'' and ``hadron collider'' categories was seen as problematic, since in a global fit with many data sets the impact of a given data set is difficult to quantify. The ``Lagrange multiplier'' approach of the CT18 group is an advanced example of such studies. One improvement could be to combine a new data set with a PDF fit in order to constrain the PDFs, but report as main result only the obtained $\alphas$. An example of data sets for individual analysis, possibly together with a PDF fit, are ratios of jet production cross sections. A further problem of global analysis including specific data sets for $\alphas$ studies is the missing information on correlations of experimental uncertainties between data sets. In addition, the global PDF fits have an intrinsic theory theoretical uncertainty from missing higher-order terms, and any independent $\alphasmZ$ determination from inclusive cross sections (not used in the fit) provides also a useful cross-check. In any case, since updated PDF (or PDF\,$+\,\alphas$) sets are only released every $\sim$5 years, it was clear that keeping the independent measurements of the ``hadron collider'' $\alphas$ category of the PDG world-average is appropriate as long as those are not directly incorporated into the global PDF\,$+\alphas$ fits.

Other discussions concerned with finding the appropriate choice of renormalization scale in the pQCD predictions, especially in multiscale observables such as event shapes in p-p collisions (e.g. TEEC, ATEEC, transverse thrust) or (multi)jet production. In addition, the problem of properly determining the correlations between measurements of the different LHC experiments was discussed. Jet production measurements are systematics limited and sensitive to many not easy to control experimental effects. M.~Wobisch's presentation (Section~\ref{sec:sigma_ratios}) of a different formulation of cross section ratios was seen as another way to estimate theory uncertainties.

The determination of $\alphas$ from the Z-boson transverse momentum distribution (S.~Camarda, Section~\ref{sec:pp_zpt}) was seen as a good example of an $\alphasmZ$ extraction of the hadron collider category, since its precision is high (1.3\%) and its dependence on PDFs is smaller than other extractions of this subfield. The measurements as well as the theory can further improve in the future. 


\subsection{Prospects and wish-lists for high-precision extractions}

In Table~\ref{tab:alphas_prospects}, we summarize the current precision of the seven extraction methods that contribute today to the PDG world average (Table~\ref{tab:averages}), as well as the expected improvements in the next $\sim$10 years (or in the longer future, in parenthesis) for each one of them. For each category, we list the dominant sources of theoretical and experimental uncertainties that propagate into $\alphasmZ$ today, and the anticipated progress in the next years (or in the longer term of planned future collider facilities) that will lead to a corresponding reduction of the uncertainties. The last row list the relative uncertainty of the current world-average (0.8\%) and of the one expected within the next decade ($\approx$\,0.4\%) obtained from taking a weighted average of the individual per-category uncertainties (in parenthesis, we provide the permil precision expected in the longer term from lattice-QCD and/or electroweak fits at a future high-luminosity $\epem$ facility). Of course, the latter result assumes that no new physics impacts any of the extraction methods and, as a matter of fact, significant inconsistency among independent determinations would indicate either a problem in our theoretical/experimental understanding of any given observable, or provide a potential indirect evidence of BSM physics.

\begin{table}[htbp!]
\centering
\caption{Summary of current and expected future (within the decade ahead or, in parenthesis, longer time scales) uncertainties in the $\alphasmZ$ extractions used today to derive the world average. Acronyms and symbols: 
CIPT=`contour-improved perturbation theory', FOPT=`fixed-order perturbation theory', NP=`nonperturbative QCD', SF=`structure functions', PS=`Monte Carlo parton shower'.
\label{tab:alphas_prospects}\vspace{0.2cm}}
\renewcommand\arraystretch{1.4}
\resizebox{\textwidth}{!}{%
\begin{tabular}{lcc} \hline
        & \multicolumn{2}{c}{Relative $\alphasmZ$ uncertainty}\\ 
Method  & Current &  Near (long-term) future \\
 & theory \& exp.\ uncertainties sources & theory \& experimental progress \\\hline
\multirow{2}{*}{(1) Lattice} & 
\textcolor{red}{$0.7\%$} &  \textcolor{red}{$\approx\,0.3\%~(0.1\%)$} \\
& Finite lattice spacing \& stats. & Reduced latt.\ spacing. Add more observables\\
& N$^{2,3}$LO pQCD truncation      & Add N$^{3,4}$LO, (active charm, QED effects)\\
&   & Higher renorm.\ scale via step-scaling to more observ.\\
\hline
\multirow{2}{*}{(2) $\tau$ decays} 
 & \textcolor{red}{$1.6\%$} &  \textcolor{red}{$<1.\%$} \\
 & N$^3$LO CIPT vs.\ FOPT diffs. & Add N$^4$LO terms. Solve CIPT--FOPT diffs. \\
 & Limited $\tau$ spectral data  & Improved $\tau$ spectral functions at Belle~II \\
\hline
\multirow{2}{*}{(3) $\QQbar$ bound states} & 
 \textcolor{red}{$3.3\%$} &  \textcolor{red}{$\approx\,1.5\%$}  \\
 & N$^{2,3}$LO pQCD truncation & Add N$^{3,4}$LO \& more $(\ccbar)$, $(\bbbar)$ bound states\\
 & $m_{c,b}$ uncertainties & Combined $m_{c,b}+\alphas$ fits\\
\hline
\multirow{2}{*}{(4) DIS \& PDF fits} 
 & \textcolor{red}{$1.7\%$} & \textcolor{red}{$\approx\,1\%$~(0.2\%)} \\
 & N$^{2,(3)}$LO PDF (SF) fits &  N$^3$LO fits. Add new SF fits: $F^{p,d}_2,\,g_i$ (EIC) \\
 & Span of PDF-based results & Better corr.\ matrices. More PDF data (LHeC/FCC-eh) \\
\hline
\multirow{2}{*}{(5) $\epem$ jets \& evt shapes} 
  & \textcolor{red}{$2.6\%$} & \textcolor{red}{$\approx\,1.5\%$~($<1$\%)} \\
 & NNLO+N$^{(1,2,3)}$LL truncation & Add N$^{2,3}$LO+N$^3$LL, power corrections\\
 & Different NP analytical \& PS corrs.  & Improved NP corrs.\ via: NNLL PS, grooming\\
 & Limited datasets w/ old detectors & New improved data at B factories (FCC-ee)\\
\hline
\multirow{2}{*}{(6) Electroweak fits} 
 & \textcolor{red}{$2.3\%$}  & \textcolor{red}{($\approx\,0.1\%$)} \\
 & N$^3$LO truncation & N$^4$LO, reduced param.\ uncerts.\ ($m_\mathrm{W,Z},\,\alpha,$ CKM) \\
 & Small LEP+SLD datasets & Add W boson. Tera-Z, Oku-W datasets (FCC-ee) \\
 \hline
\multirow{2}{*}{(7) Hadron colliders}
 & \textcolor{red}{2.4\%} & \textcolor{red}{$\approx\,1.5\%$} \\
 & NNLO(+NNLL) truncation, PDF uncerts. & N$^3$LO+NNLL (for colour-singlets), improved PDFs \\
 & Limited data sets ($\ttbar$, W, Z, e-p jets) & Add more datasets: Z $\pT$, p-p jets, $\sigma_i/\sigma_j$ ratios,...\\
\hline
World average & \textcolor{red}{$0.8\%$}  & \textcolor{red}{$\approx\,0.4\%$~(0.1\%)} \\
 \hline
 \end{tabular}
}
\end{table}

\paragraph*{Lattice:}\vspace{0.3cm}

The current $\pm0.7\%$ precision of the lattice-QCD extraction of $\alphasmZ$ can be reduced by about a factor of two within the next $\sim$10 years. In order to improve the lattice-QCD based determinations of $\alphas$, it would be important to reach higher renormalization scales for more observables, which can be achieved by implementing elements of the step-scaling technique even in infinite space-time volumes. This requires resources for dedicated lattice simulations. In addition, it would be very helpful to push the corresponding pQCD calculations to higher orders. Depending on the process under study, this will require to calculate N$^3$LO, N$^4$LO, and/or N$^3$LL contributions. In some cases it may also be beneficial to compute lattice artifacts perturbatively, in order to improve the control of the continuum limit. For extractions based on moments of quarkonium correlators, one should perform more accurate lattice calculations with hadron masses $m_h \ge 2m_c$, where the truncation errors are subleading in the current results.
In view of the fully recursive step-scaling strategies, also nonstandard perturbative techniques for selected finite volume renormalization schemes should be developed. 
Aiming for 0.1\% uncertainties will likely require the explicit inclusion of charm quark effects (2+1+1 calculations) and of QED and isospin-breaking effects in the determination of both
the physical scale and the running of $\alphas$.

\paragraph*{\texorpdfstring{$\tau$}{tau} decays:}\vspace{0.3cm}

The present $\alphasmZ$ value of the $\tau$ lepton category has a $\pm 1.6\%$ uncertainty as derived from the pre-averaging of four different determinations.
Solving the CIPT vs.\ FOPT discrepancies, \eg\ through the method proposed in~\cite{Benitez-Rathgeb:2022yqb}, is a basic prerequisite to reduce in about a half the current theoretical uncertainty assigned to this extraction method. Although hard to compute, a calculation of the $\mathcal{O}(\alphas^5)$ pQCD term (N$^4$LO) would be also beneficial, given the sometimes slow convergence of terms with typical weights. On the nonperturbative side, increased data precision would allow more stringent tests of the duality-violation contributions. In the near future, more precise 2-pion and 4-pion exclusive-mode $\tau$ data from Belle~II would also help to disentangle perturbative and nonperturbative contributions.

\paragraph*{\texorpdfstring{$\QQbar$}{QQbar} bound states:}\vspace{0.3cm}

The $\alphasmZ$ value derived from quarkonium decays and masses has today a $\pm 3.1\%$ uncertainty from the pre-averaging of six different determinations at NNLO or N$^3$LO accuracy. Improved determinations can be obtained by performing combined fits of $\alphas$ and charm and/or bottom quark masses, adding one extra degree of pQCD accuracy to the NNLO predictions, and/or adding more $\ccbar$, $\bbbar$ bound-states data.

\paragraph*{Structure functions and global PDF analyses:}\vspace{0.3cm}

The current $\alphasmZ$ value of the parton densities category has a $\pm 1.7\%$ uncertainty as derived from the pre-averaging of six different determinations (two analyses of structure functions, and four from global PDF fits). With complete N$^3$LO predictions for DIS, the residual theoretical uncertainty due to the scale variation and the truncation of the perturbative series will be limited to $\approx$1\% in the range of parton kinematics relevant for the current world DIS data and the EIC. 
In general, progress is expected from analyses at future e-p colliders such as the EIC, LHeC, or FCC-eh. Adding new observables, such as \eg\ deuteron and spin-dependent structure functions, which can be both measured at the EIC, will provide new useful constraints. Ultimately, an experimental precision of $\delta\alphasmZ \sim 0.2\%$ is projected at the LHeC/FCC-eh.

For PDF\,+\,$\alphas$ extractions, future N$^3$LO fits (or, in the shorter term, an improved quantification of the impact of missing higher-order corrections) are required to reach $\approx$1\% precision on $\alphasmZ$. Achieving such a higher degree of pQCD accuracy will likely reduce the present broad span of $\alphasmZ$ central values among global PDF fits. In addition, the availability of LHC experimental data with more complete information on correlations would improve the global analyses.




\paragraph*{\texorpdfstring{$\epem$}{e+e-} annihilation to hadronic final states:}\vspace{0.3cm}

The present precision of the $\alphasmZ$ determination from the $\epem$ category of the PDG world-average is of $\pm2.6\%$, as obtained by pre-averaging 10 different extractions based on event shapes or jet rates. Such a comparatively large uncertainty is driven by the large span among central values derived from measurements where NP corrections have been obtained with Monte Carlo parton showers or analytically. The former (latter) corrections tend to give larger (smaller) QCD couplings than the world average. The main reduction of this uncertainty will therefore come through an improved convergence between the theoretical (analytical and MC) descriptions of NP effects. First, detailed analytical studies of nonperturbative power corrections exists now for the $C$-parameter that need to be applied to other similar observables, such as thrust, to clarify the current discrepancies. Second, improved parton-shower algorithms (\eg\ of the PanXX family) reaching NNLL accuracy are needed, followed by matching to fixed-order (NNLO) predictions plus a full retuning of the hadronization and other final-state parameters (colour reconnection, spin correlations,...) of the event generators. In parallel, improved soft-drop grooming techniques should be applied to the $\epem$ data to evaluate their impact in the reduction of NP effects.

\paragraph*{Electroweak fits:}\vspace{0.3cm}

The $\alphasmZ$ uncertainty of the EW category is of $\pm2.3\%$ and is dominantly driven by the statistical uncertainty in the measurements of the Z-boson pseudoobservables at LEP (the W-boson determination is even much less precise, and features a $\pm30\%$ uncertainty). Such an $\alphasmZ$ determination is extremely clean, and can reach the 0.1\% precision provided one collects the $10^{12}$ and $10^{8}$ Z- and W-bosons data samples expected at a future high-luminosity $\epem$ collider such as the FCC-ee. To reach such a level of precision will require also to compute missing higher-order pQCD (N$^4$LO), EW ${\cal O}(\alpha^{2,3})$, and mixed pQCD+EW $\mathcal{O}(\alpha\alphas^2,\,\alpha\alphas^3,\,\alpha^2\alphas)$ corrections that, today, are negligible compared to the experimental uncertainties.

\paragraph*{Hadron colliders:}\vspace{0.3cm}

The QCD coupling derived from the hadron-collider category, obtained from the pre-averaging of 5 different extractions at NNLO accuracy, has a $\pm2.4\%$ precision today. Theoretical (experimental) uncertainties are driven mostly by the scale and PDF (luminosity) uncertainties. A natural way to reduce the $\alphasmZ$ uncertainties would be to incorporate N$^3$LO corrections, but this seems realistically feasible in the next years only for colour-singlet cross sections, and requires PDFs at the same level of accuracy. A faster way to reduce the current uncertainties by about a half may come from the expected addition of many more upcoming observables (Z $\pT$ peak at N$^3$LO, p-p jets and $\sigma_i/\sigma_j$ ratios at NNLO,...) and/or extended high-precision datasets (integrated luminosity uncertainties at the LHC have been lately reduced to the $\approx$1\% level).

\paragraph*{Final wish-list:}\vspace{0.3cm}

The determination of the strong coupling has the potential to witness considerable improvements in the decade ahead, with an anticipated reduction in the $\alphasmZ$ world average uncertainty by a factor of about two (from 0.8\% down to $\approx$\,0.4\%) over this time scale through well-identified experimental and theoretical developments. Such a progress will have an important impact in the theoretical calculations and associated interpretation of upcoming LHC data and in searches for new physics through high-precision SM studies. Such advances will be facilitated by:
\begin{itemize}
\item Sufficient dedicated computing resources to generate state-of-the-art samples for lattice QCD analyses. Enough person-power to develop perturbation theory for selected observables in a finite space-time volume, and compute identified higher-order pQCD corrections to match improved lattice QCD samples as necessary basis for $\alphas$ extractions. 
\item Beyond lattice-related calculations, person-power is needed for all other important theory efforts. These include the completion of the hadronic $\tau$ decay renormalon analysis, the three-jet power corrections for $\epem$ event shapes and jet algorithms, NNLL accuracy parton shower algorithms and their matching to fixed order, NNLO MC simulations for complex final states in $\epem$, e-p or p-p scattering, and differential NNLO predictions for HERA and LHC multi-jet observables, among others.
\item Incorporation of multiple new precision observables and/or datasets measured at the LHC into NNLO hadron-collider- or PDF- based extractions, with an improved treatment of the experimental correlation uncertainties among measurements, will lead to a visible improvement of the accuracy and precision of the world $\alphasmZ$ average.
\item Measurements of DIS with new high-energy facilities (EIC first and, in the longer term LHeC/FCC-eh) will allow determining PDFs\,$+\,\alphas$ over a large phase space covering LHC kinematics. This would resolve PDF-$\alphas$ ambiguities in hadron-collider analyses, and provide new precision $\alphasmZ$ determinations. A direct extraction of $\alphas$ and studies of its energy evolution will also benefit from high-precision PDFs over a large kinematic range. 
\item Arguably, the only way known to actually reach permil $\alphasmZ$ uncertainty from purely experimental measurements, without lattice-QCD data simulations, is through the study of hadronic Z (and W) decays. A high-energy lepton collider (Higgs factory, ILC, FCC-ee) with a Giga- or Tera-Z program and low-energy ($\sqrt{s}<m_Z$) runs will massively improve $\epem$ determinations of $\alphasmZ$ via Z- and W-boson hadronic decays, $\tau$ decays, jets rates and event shapes, significantly also improving our understanding of parton showers and nonperturbative effects, and providing accurate tests of the $\alphas$ evolution. 
\end{itemize}
\vspace{1cm}

\subsection*{Acknowledgments}

Support from the EU STRONG-2020 project under the program H2020-INFRAIA-2018-1, grant agreement No.\ 824093 is acknowledged.

\clearpage
\bibliographystyle{myutphys}
\bibliography{alphaS}

\end{document}